\documentclass[structabstract]{aa1}  
\usepackage{graphicx}
\usepackage{graphics}
\usepackage{lscape}
\usepackage{longtable}
\usepackage{txfonts}
\usepackage{natbib}
\usepackage{textcomp}
\usepackage{appendix}
\usepackage{gensymb}
\usepackage{color}
\usepackage{txfonts}
\bibpunct{(}{)}{;}{a}{}{,}

\begin{document}

\title{Herschel GASPS spectral observations of T Tauri stars in Taurus\thanks{{\it Herschel} is an ESA space observatory with science instruments provided by European-led Principal Investigator consortia and with important participation from NASA.}}
\subtitle{Unraveling far-infrared line emission from  jets and discs}

\author{M. Alonso-Mart\'inez\inst{\ref{inst1},\ref{inst2}}
, P. Riviere-Marichalar\inst{\ref{inst1},\ref{inst3}}
, G. Meeus\inst{\ref{inst1},\ref{inst2}}
, I. Kamp\inst{\ref{inst4}}
, M. Fang\inst{\ref{inst1},\ref{inst5}}
, L. Podio\inst{\ref{inst6}}
, W. R. F. Dent\inst{\ref{inst7}}
, C. Eiroa\inst{\ref{inst1},\ref{inst2}}}

\institute{Dpto. de F\'isica Te\'orica, Fac. de Ciencias, UAM Campus Cantoblanco, 28049 Madrid, Spain\label{inst1} 
\and Astro-UAM, UAM, Unidad Asociada CSIC\label{inst2} 
\and Dpto. de Astrof\'isica, Centro de Astrobiolog\'ia, ESAC Campus, P.O. Box 78, E-28691 Villanueva de la Ca\~nada, Madrid, Spain \label{inst3} 
\and Kapteyn Astronomical Institute, University of Groningen, Postbus 800, 9700 AV Groningen, The Netherlands\label{inst4}
\and Department of Astronomy, University of Arizona, 933 North Cherry Avenue, Tucson, AZ 85721, USA \label{inst5}
\and INAF - Osservatorio Astrofisico di Arcetri, Largo E. Fermi 5, 50125, Firenze, Italy\label{inst6}
\and ALMA SCO, Alonso de Cordova 3107, Vitacura 763-0355, Santiago, Chile\label{inst7}}

\offprints{M. Alonso-Mart\'inez\\ \email{miguel.alonso@uam.es}}
   \authorrunning{Alonso--Mart\'inez}     

\date{Received:/Accepted:}

\abstract
%context heading (optional)
%{}
{At early stages of stellar evolution young stars show powerful jets and/or outflows that interact with protoplanetary discs and their surroundings. Despite the scarce knowledge about the interaction of jets and/or outflows with discs, spectroscopic studies based on \textit{Herschel} and ISO data suggests that gas shocked by jets and/or outflows can be traced by far-IR (FIR) emission in certain sources.}
{We want to  provide a consistent catalogue of selected atomic ([OI] and [CII]) and molecular (CO, H$_{2}$O, and OH) line fluxes observed in the FIR,  separate and characterize the contribution from the jet and the disc to the observed line emission, and  place the observations in an evolutionary picture.}
{The atomic and molecular FIR (60--190 $\rm \mu m$) line emission of protoplanetary discs around 76 T Tauri stars located in Taurus are analysed. The observations were carried out within the \emph{Herschel} key programme Gas in Protoplanetary Systems (GASPS). The spectra were obtained with the Photodetector Array Camera and Spectrometer (PACS). The sample is first divided in outflow and  non-outflow sources according to literature tabulations. With the aid of archival stellar/disc and jet/outflow tracers and model predictions (PDRs and shocks), correlations are explored to constrain the physical mechanisms behind the observed line emission.} 
{Outflow sources exhibit brighter atomic and molecular emission lines and higher detection rates than non-outflow sources. The line detection fractions decrease with SED evolutionary status (from Class I to Class III). 
We find correlations between [OI] 63.18 $\rm \mu m$ and [OI] 6300 $\mbox{\rm \AA}$, o--H$_{2}$O 78.74 $\rm \mu m$, CO 144.78 $\rm \mu m$, OH 79.12+79.18 $\rm \mu m$, and the continuum flux at 24 $\rm \mu m$. 
The atomic line ratios can be explain either by fast ($V_{\rm shock}>$50 km s$^{-1}$) dissociative J-shocks at low densities ($n\sim10^{\rm 3}$ cm$^{-3}$) occurring along the jet and/or PDR emission ($G_{\rm 0}>10^{\rm 2}$, $n\sim10^{\rm 3}-10^{\rm 6}$ cm$^{-3}$). To account for the [CII] absolute fluxes, PDR emission or UV irradiation of shocks is needed.
In comparison, the molecular emission is more compact and the line ratios are better explained with slow ($V_{\rm shock}<$40 km s$^{-1}$) C-type shocks with high pre-shock densities (10$^{4}$--10$^{6}$ cm$^{-3}$), with the exception of OH lines, that are better described by J-type shocks. Disc models alone fail to reproduce the observed molecular line fluxes, but a contribution to the line fluxes from UV-illuminated discs and/or outflow cavities is expected. 
Far-IR lines dominate disc cooling at early stages and weaken as the star+disc system evolves from Class I to Class III, with an increasing relative disc contribution to the line fluxes.} 
{Models which take into account jets, discs, and their mutual interaction are needed to disentangle the different components and study their evolution. 
The much higher detection rate of emission lines in outflow sources and the compatibility of line ratios with shock model predictions supports the idea of a dominant contribution from the jet/outflow to the line emission, in particular at earlier stages of the stellar evolution as the brightness of FIR lines depends in large part on the specific evolutionary stage.}
%{}

\maketitle

\section{Introduction} 
Protoplanetary discs are ubiquitously found around young stars and are the birth sites of planets. 
They are initially composed of well-mixed gas and dust \citep[e.g.][and references therein]{Williams2011} and are in continuous evolution \citep[e.g.][]{Semenov2011}. Although gas constitutes the bulk of the disc mass, before the advent of ALMA our knowledge  of protoplanetary discs was mainly based on dust studies \citep[e.g.][]{Beckwith1990,Andrews2005,Hartmann2008}.

Different molecular transitions probe a diversity of gas kinetic temperatures and densities; for example,  CO ro-vibrational transitions can be excited in hot (T$\sim$4000 K) and dense ($n>$10$^{10}$ cm$^{-3}$) gas located at $\sim$1 au \citep[e.g.][]{Hamann1988}, whereas purely rotational transitions are excited at much lower temperatures (typically a few hundred) beyond 1 au \citep{Najita2003}.
The mid-IR observations of molecular lines (H$_{2}$ and CO) and forbidden atomic and  ionized lines (S and Fe)  trace warm gas ($T\sim$50--100 K)  at a few radii from the central star up to several tens of au \citep{Pascucci2006}.
In the submillimetre, CO observations \citep{Pietu2007}, as well as HCO$^{+}$, H$_{2}$CO, HCN, and CN \citep[e.g. ][and references therein]{Oberg2011}, probe cold gas (20 K$<T<$50 K) at radii $>$20 au. 
The incident radiation field, depth in the disc, and distance from the central star, etc., govern the chemical reactions and temperature structure of the gas in protoplanetary discs \citep{Dutrey2014a}.

\begin{table*}[htpb]
\setcounter{table}{0}
 \centering
\caption{Observed wavelength ranges, resolution, instrument configuration, amount of sources observed, and properties of the main transitions covered.}
\label{LineRange}
\tiny
  \begin{tabular}{lcclclccc}
\hline
\hline
    \multicolumn{1}{l}{Channel} &
    \multicolumn{1}{c}{$\lambda\lambda$ [$\rm \mu m$]} &
    \multicolumn{1}{c}{R} &
    \multicolumn{1}{l}{Mode} &
    \multicolumn{1}{c}{No. Sources} &
    \multicolumn{1}{l}{Sp} &
    \multicolumn{1}{c}{Transition} &
    \multicolumn{1}{c}{$T_{\rm ex}$ [K]} &
    \multicolumn{1}{c}{$\lambda_{\rm 0}$ [$\rm \mu m$]} \\
\hline
Blue &  62.93--63.43     & 3150 & LineSpec  & 76 & [OI] &$^3$P$_{1}$--$^3$P$_{2}$   &228&63.18\\[0.5mm]
         &                             &         &                  &     & o--H$_{2}$O &8$_{18}$--7$_{07}$ &1293&63.32\\[0.5mm]
Red  & 188.76--190.29    & 1500 & LineSpec  & 76 & DCO$^{+}$ &\emph{J}=22--21 &2068& 189.57\\[0.5mm]
\hline
Blue &  71.82--73.33     & 1800 & RangeSpec & 39 & o--H$_{2}$O& 7$_{07}$--6$_{16}$ &685& 71.94\\[0.5mm]
         &                            &           &                    &     &  CH$^{\rm +}$   & \emph{J}=5--4     & 600 & 72.14\\[0.5mm]
         &                           &          &                      &     &  CO& \emph{J}=36--35                             &3700 & 72.84\\[0.5mm]                                                 
Red  & 143.61--146.66    & 1200 & RangeSpec & 39 & p--H$_{2}$O& 4$_{13}$--3$_{22}$ &396& 144.52 \\[0.5mm]
         &                             &           &                     &     &  CO&\emph{J}=18--17                               & 945& 144.78 \\[0.5mm]
         &                             &           &                     &     &   [OI] &$^3$P$_{0}$--$^3$P$_{1}$ &326& 145.52\\[0.5mm]
\hline
Blue &  78.37--79.73     & 1900 & RangeSpec & 38 & o--H$_{2}$O& 4$_{23}$--3$_{12}$                     &432 &78.74 \\[0.5mm]
         &                             &           &                     &   &p--H$_{2}$O& 6$_{15}$--5$_{24}$                      & 396& 78.92\\[0.5mm]
         &                             &           &                     &   &  OH &$^{2} \Pi _{1/2,1/2}$--$^{2} \Pi _{3/2,3/2}$  &182& 79.12+79.18\\[0.5mm]
         &                             &           &                     &    & CO& \emph{J}=33--32                                                   &3092 &  79.36\\[0.5mm]
Red  & 156.73--159.43    & 1250 & RangeSpec & 38 & [CII]& $^2$P$_{3/2}$--$^2$P$_{1/2}$  &91& 157.74\\[0.5mm]
         &                             &           &                     &    & p--H$_{2}$O& 3$_{31}$--4$_{04}$      &410& 158.31\\[0.5mm]
\hline
Blue &  89.29--90.72     & 2500 & RangeSpec & 30 & p--H$_{2}$O &3$_{22}$--2$_{11}$    &297& 89.99\\[0.5mm]
         &                             &           &                     &    & CO &\emph{J}=29--28                                   &2400& 90.16\\[0.5mm]
Red  & 178.58--181.44    & 1450 & RangeSpec & 30 & o--H$_{2}$O& 2$_{12}$--1$_{01}$ &115& 179.53\\[0.5mm]
         &                             &           &                     &    &  o--H$_{2}$O &2$_{21}$--2$_{12}$ &194& 180.49\\[0.5mm]
\hline
  \end{tabular}
\end{table*}

Young stars produce X-ray and far-ultraviolet (FUV) radiation \citep{Calvet2004,Ingleby2013} either by chromospheric activity  \citep{Robrade2007} or by accretion \citep{Gudel2007b}. This radiation shapes the structure of the disc and its temperature distribution.
In the inner 50 au of the disc surface, the temperature can be up to $\sim$10$^{4}$ K \citep{Jonkheid2004,Kamp2004}, favouring a rich ion--atomic chemistry, while in deeper and colder (\mbox{$\sim$100 K}) regions where the UV/X-ray photons can still penetrate, the chemistry is more ion--molecule rich. 
Determination of where the different lines arise gives insight on accretion, photoevaporation, and planet formation mechanisms \citep[][and references therein]{Frank2014}.  

Jets, outflows, and winds associated with young stellar objects have been observed from  X-ray to radio wavelengths \citep{Hartigan1995,Reipurth2001,Bally2007,Schneider2013,Lynch2013} in scales that range from tens of au \citep{Agra2011} up to several parsecs \citep{McGroarty2004}, and can persist for millions of years \citep{Cabrit2011}. The Herschel Space Observatory \citep[HSO;][]{Pilbratt2010} has revealed that, on average, far-IR (FIR) emission lines are more frequently seen and are stronger in systems with jets and/or outflows \citep[e.g.][]{Podio2012,Howard2013,LeeJE2014} with temperatures of $\sim$100--1000 K \citep{Karska2014b}. 
Traditionally, the role played by jet/outflows and protoplanetary discs in stellar evolution is treated separately, although shocks produced by a jet are important contributors to emission;
hence, they affect the chemical properties of the disc.

The  Gas in Protoplanetary Systems  \citep[GASPS;][]{Mathews2010,Dent2013}  programme observed 
240 stars in different star forming regions in order to probe the evolution of gas and dust 
in protoplanetary discs. The Herschel/PACS  \citep[Photodetector Array Camera and Spectrometer,][]{Poglitsch2010} was used to observe 76 T Tauri stars in the Taurus region.
\cite{Howard2013} concentrated on [OI] 63.18 $\rm \mu m$, but also listed line intensities for [OI] 145.53 $\rm \mu m$ and [CII] 157.74 $\rm \mu m$ and identified other lines in the spectra. \cite{Riviere2012} analysed the o--H$_{2}$O 63.32 $\rm \mu m$ line; \cite{Keane2014} focused on [OI] 63.18 $\rm \mu m$ and o--H$_{2}$O 63.32 $\rm \mu m$ lines in transitional discs; and \cite{Podio2012} focused on the analysis of six well-known jet sources showing extended [OI] 63.18 $\rm \mu m$ emission.

In this work, we make an inventory of atomic and molecular species covered with PACS in the Taurus sample, and present a consistent line flux catalogue. We assess whether the observed emission is dominated by the jet or the disc, and how this depends on the evolutionary status of the source. Observations include atomic [OI] and [CII], and molecular H$_{2}$O, CO, and OH. 
These lines have been attributed to arise in discs in TW Hya \citep{Kamp2013}, HD 163296 \citep{Tilling2012} and HD 100546 \citep{Thi2011}. Indeed disc models \citep[e.g. the DENT grid][]{Woitke2010,Pinte2010,Kamp2011} can reproduce the line ratios but fail to explain high line fluxes.
In addition, FIR line emission with a jet/outflow origin has been spatially resolved for several Class 0/I protostars clearly showing that the line emission is more extended than continuum emission \citep{Herczeg2012}.

The structure of the paper is as follows. Section \ref{sample} describes the sample and observations. The data reduction is explained in Sect. \ref{dataRed}, and the main results are described in Sect. \ref{results}. Relations between FIR lines are explored in Sect. \ref{relations}, and its possible origins and excitation mechanisms are discussed in Sect. \ref{discussion}. The main conclusions are summarized in Sect. \ref{conclusions}.

\section{Sample and observations}
\label{sample}

\subsection{The sample}
\label{sample_parameters}
The sample consists of 76 T Tauri stars of the Taurus region observed by GASPS. 
Spectral types, as given by \cite{Luhman2010} and \cite{Herczeg2014}, range from K0 to M6, except for three earlier type stars: RY Tau (G0), \mbox{SU Aur} (G4),  and HD283573 (G4). More than one-third of the stars in the  sample ($\sim$38\%) are multiple systems \citep{Ghez1993,Daemgen2015}, with separations from 0.1 up to 6 arcsec (14--840 au). In these cases, the companions can contaminate the  Herschel/PACS results since the pixel size is 9.4 arcsec, corresponding to a separation of $\sim$1300 au at the distance of Taurus (140 pc). However, we kept those binary sources in our sample in order not to bias the results \citep[see Sect. 5.5 in][for a discussion]{Howard2013}.

Following the classification by \cite{Lada1987}, the sample includes 5 Class I, 55 Class II  \citep[including 10 transition discs;][]{Strom1989,Najita2007a}, and 16 Class III objects. The SED classification is taken from \cite{Luhman2010} and/or \cite{Rebull2010}. Objects not observed by these authors and with no sign of infrared excess are classified as Class III. 
The sample is divided in outflow and non-outflow sources motivated by the correlation between the 63/70 $\rm \mu m$ continuum emission and the [OI] 63.18 $\rm \mu m$ line flux in Taurus and Chamaelon II stars found by \cite{Howard2013} and \cite{Riviere2014}. The outflow sources are those showing blue-shifted [OI] 6300 $\mbox{\AA}$ emission in \cite{Hartigan1995}.
A more detailed description of the sample is given in Table \ref{stellarParam}, Appendix \ref{stellarPar}, including stellar temperatures, mass accretion rates, stellar X-ray and accretion luminosities, ages, and disc masses.

\subsection{Herschel/PACS observations}
\label{obs_parameters}

\begin{figure}[htpb]
\setcounter{figure}{0}
\centering
\includegraphics[width=0.4\textwidth,trim=19mm 0mm 30mm 3mm,clip]{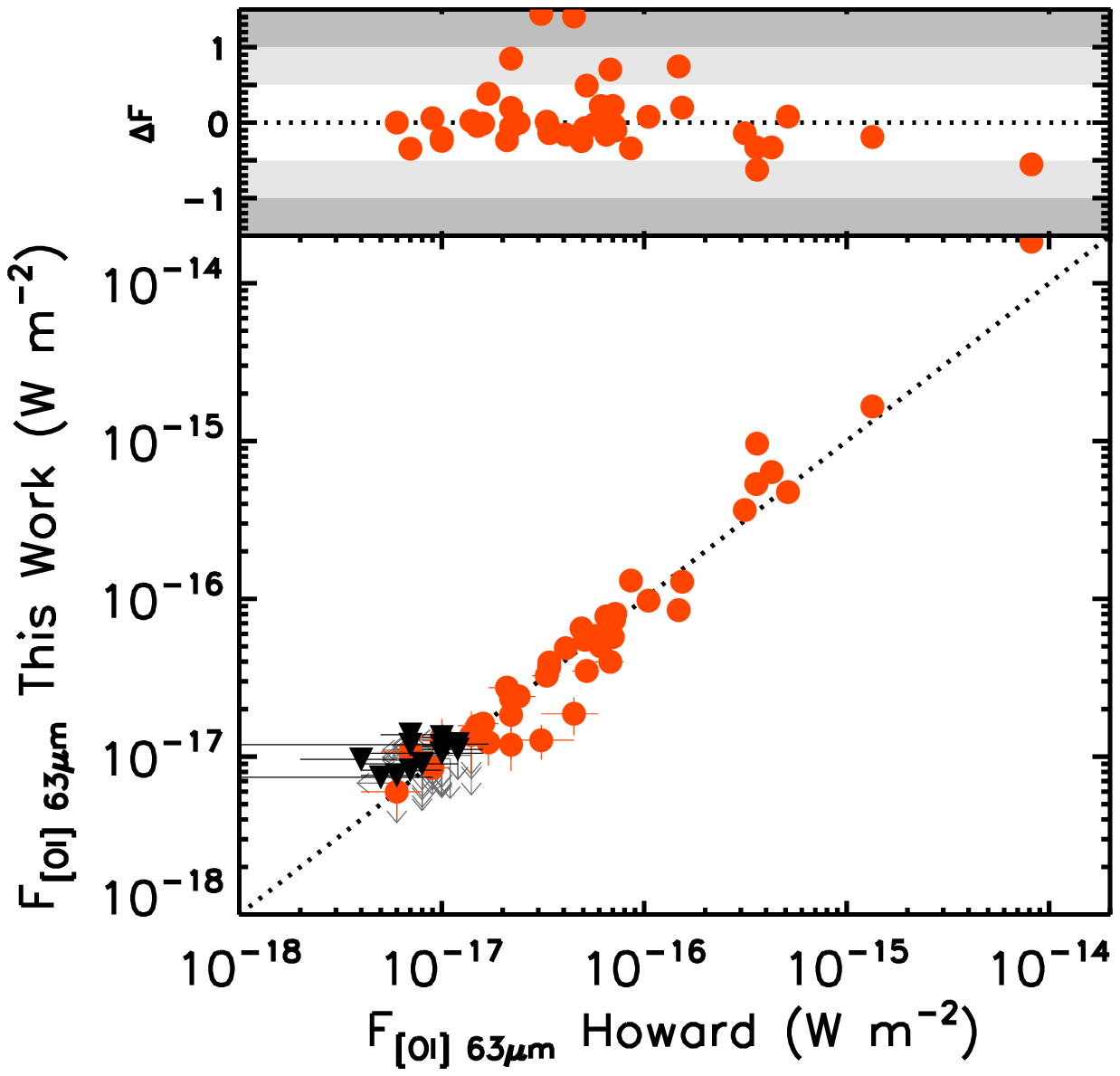}
\includegraphics[width=0.4\textwidth,trim=19mm 0mm 30mm 5mm,clip]{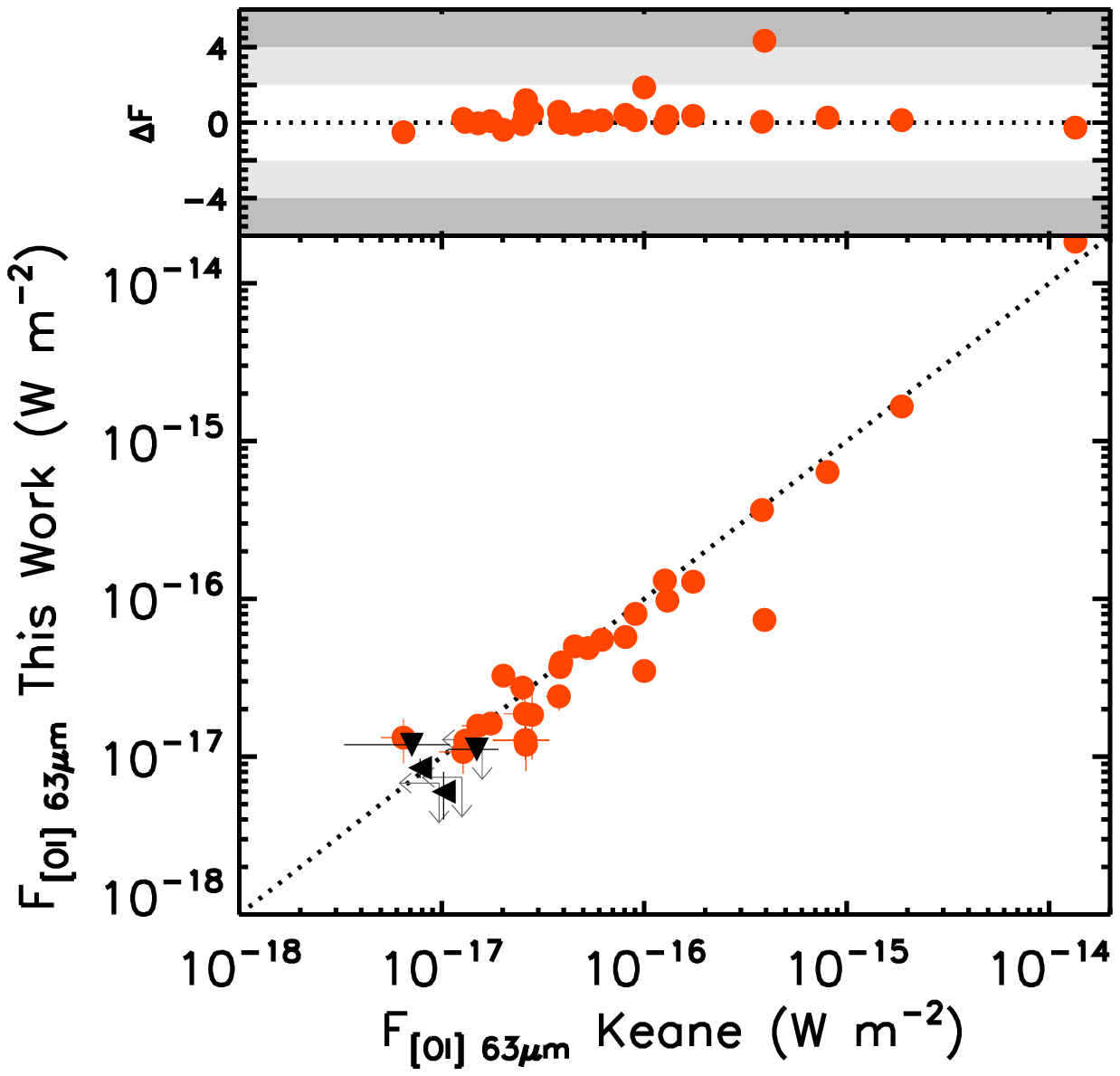}
\includegraphics[width=0.4\textwidth,trim=19mm 0mm 30mm 5mm,clip]{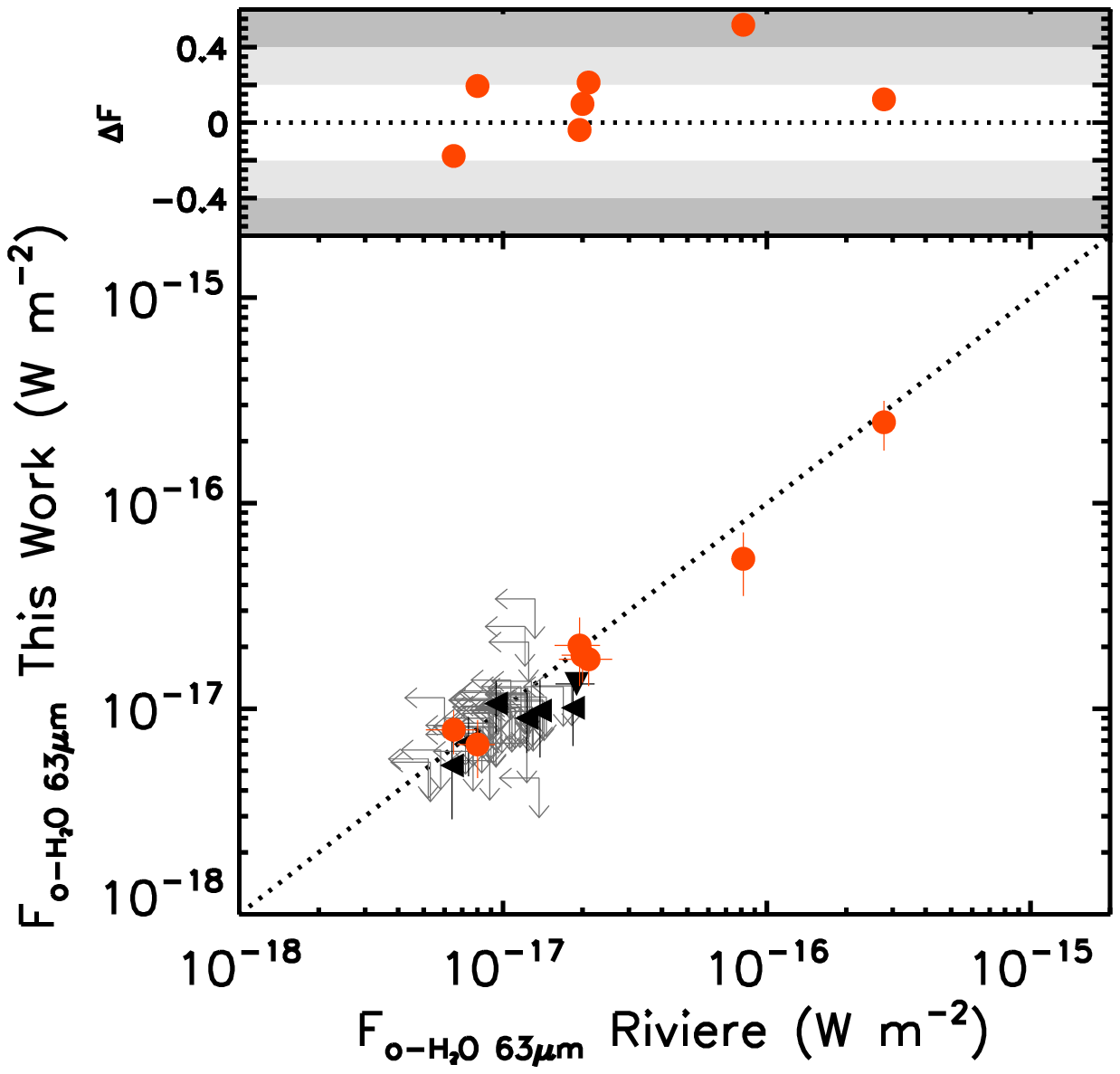}
 \caption{Comparison of our [OI] 63.18 $\rm \mu m$ fluxes with those in \cite{Howard2013} ({\it top}) and \cite{Keane2014} ({\it middle}), and comparison of  o-H$_{2}$O 63.32 $\rm \mu m$ fluxes with those in \cite{Riviere2012} ({\it bottom}). In all the panels the red circles represent the detections, while black down-facing and left-facing triangles are upper limits  in the y-axis and x-axis, respectively. Arrows represent upper limits in both axes. $\Delta F$ ($F_{\rm old}-F_{\rm new} \over F_{\rm new}$) is the fractional difference between the two sets of measurements.} 
  \label{figure:flux_comparison}
\end{figure}

Spectroscopic observations were performed between February 2010 and March 2012. PACS covers the wavelength range 51--220 $\rm \mu m$ in two channels (blue: 51--105 $\rm \mu m$ and red: \mbox{102--220 $\rm \mu m$}). The spatial resolution of the PACS spectrometer is 9.4" at 62-100 $\rm \mu m$, 11.4" at 150 $\rm \mu m$ and 13.1" at \mbox{180 $\rm \mu m$}.
The integral field unit (IFU) images a 47" $\times$ 47" field of view (FOV) in 5$\times$5 spatial pixels (hereafter spaxels) of \mbox{9.4" $\times$ 9.4"} each. For each spaxel, two spectra are obtained simultaneously, one for each channel.

The observations were conducted in chop-nod line (LineSpec) and range (RangeSpec) modes (see Chapter 6.2.6 of the \textit{PACS Observers Manual)} with a small throw (1.5') to remove telescope and background emission.
The observations were performed in one (1152 s on source) or two (3184 s on source) nod cycles with total integration times in the range $\sim$1250--6630 s and $\sim$5140--20555 s for LineSpec and RangeSpec modes, respectively.
The LineSpec mode has a small wavelength coverage (62.93--63.43 $\rm \mu m$) targeting the [OI] 63.18 $\rm \mu m$ and o-H$_{\rm 2}$O 63.32 $\rm \mu m$  lines and the adjacent continuum. 
The RangeSpec observations cover a larger wavelength range, defined by the observer. The lines observed in this mode include several transitions of o-H$_{\rm 2}$O (at 71.94, 78.74, 179.53, and 180.49 $\rm \mu m$), p-H$_{2}$O (at 78.92, 89.99, and 144.52 $\rm \mu m$), CO (at 72.74, 79.36, 90.16, and 144.78 $\rm \mu m$), and a OH doublet (at 79.12 and 79.18 $\rm \mu m$).
The entire sample (76 sources) was observed in LineSpec mode, while for the RangeSpec mode the number of targets observed varies. None of the RangeSpec observations includes Class III objects.
Details of the instrument channel, coverage, spectral resolution, observing mode, number of observed sources, and emission lines covered are summarized in Table \ref{LineRange}.
Identifiers (OBSIDs) and exposure times of the spectroscopic observations are summarized in Table \ref{OBSIDs}.

\section{Data reduction}
\label{dataRed}
The data were reduced using HIPEv10 \citep{Ott2013}. 
The PACS pipeline removes saturated and bad pixels, subtracts the chop on and the chop off nod positions, applies a correction for the spectral response function and flat field, and re-bins at half the instrumental resolution (oversample=2, upsample=1). 
The final spectrum is obtained by the average of the two nod cycles.
The spaxel showing the highest continuum level is extracted and an aperture correction
applied. To estimate the continuum flux, the noisy edges of the spectra are removed, as are   $\pm$3$\sigma$ regions around each line present in the spectral range of interest. Then, a first-order polynomial fit is applied.
The line fluxes are obtained from the continuum subtracted spectra by Gaussian fits to the lines, and considered as real when the signal-to-noise ratio of the emission peak is $>$3$\sigma$. The errors in line fluxes are computed as the integral of a Gaussian with width equal to the fitted value, and peak equal to the RMS noise of the continuum. In case of non-detections, we report 3$\sigma$ upper limits computed as the integral of a Gaussian with a FWHM equal to the instrumental FWHM$_{\rm 0}$ at the wavelength of interest, and amplitude three times the standard deviation of the continuum. 
The line fluxes and upper limits in the \mbox{60--80 $\rm \mu m$} and 90--190 $\rm \mu m$ ranges are given in Tables \ref{lineFluxBlue} and \ref{lineFluxRed}, respectively.

There are a few problems with our approach of only extracting the spaxel with the highest continuum level, as described below. In most cases, the fluxes were extracted from the central spaxel at the location of the star. However, some Taurus observations suffer from large pointing errors, which means that the star lies between two or more spaxels; in these cases the reported fluxes are lower limits to the real flux. 
Previous papers have tried to solve this problem either by reconstructing the PSF to recover the on-source emission \citep{Howard2013} or by integrating all the spaxels (5$\times$5) to recover the extended emission \citep{Podio2012}.
An intermediate solution that we apply is to derive the flux by summing the 3$\times$3 spaxels around the position of the source. When the difference in flux is larger than three times the quadratic sum of the errors, the 3$\times$3 fluxes are considered more accurate. In these cases, the 3$\times$3 fluxes are used instead of the fluxes extracted from a single spaxel. These are listed in Tables \ref{table:blue3x3_1} and \ref{table:red3x3}. For the jet sources showing [OI] 63.18 $\rm \mu m$ extended emission, we obtained lower line fluxes than those given in \cite{Podio2012}.

Figure \ref{figure:flux_comparison} compares the [OI] 63.18 $\rm \mu m$ line flux with those published in \cite{Howard2013} and \cite{Keane2014} and the o--H$_{2}$O 63.32 $\rm \mu m$ in \cite{Riviere2012}.
These three studies used different HIPE versions to reduce the data, but similar line fitting algorithms to estimate line fluxes. The main discrepancies are towards extended, misaligned objects or towards objects displaying very high [OI] fluxes (above $\sim$10$^{-16}$ W m$^{-2}$). For these observations the median differences are between 11\% and 27\%, compatible with the PACS absolute flux accuracy (see pages 40--44 of \emph{PACS Observers' Manual}). 
More recent pipeline versions (HIPEv14) aim to recover the emission from mispointed sources.  A comparative test yields that fluxes from the different HIPE versions are compatible
within errors.

\section{Results}
\label{results}
Atomic ([OI] and [CII]) and molecular (CO, H$_{2}$O, and OH) emission lines are seen in a large number of Taurus sources. Table \ref{detRates} gives the detection fractions for the entire sample, as well as the split in outflow and non-outflow sources. Uncertainties were estimated assuming binomial distributions  \citep[see][]{Burgasser2003}. A clear result (Fig. \ref{figure:statsJets}) is that outflow sources are richer in emission lines, and show systematically higher fluxes (on average \mbox{$\sim$10$^{-16}$ W m$^{-2}$} compared to \mbox{$\sim$10$^{-17}$ W m$^{-2}$}) and detection fractions (on average 42\% compared to 16\%) than non-outflow sources.

This suggests that jets and/or outflows are important contributors to the line emission and that they dominate in sources showing extended emission  \citep{Podio2012}. However, a (partial) disc origin cannot be ruled out. In the following we discuss the atomic and molecular line detections in more detail according to outflow activity, evolutionary status, and spectral types.

\begin{figure}[htpb]
\centering
\setcounter{figure}{1}
\includegraphics[width=0.45\textwidth, trim=0mm 0mm 0mm 0mm, clip]{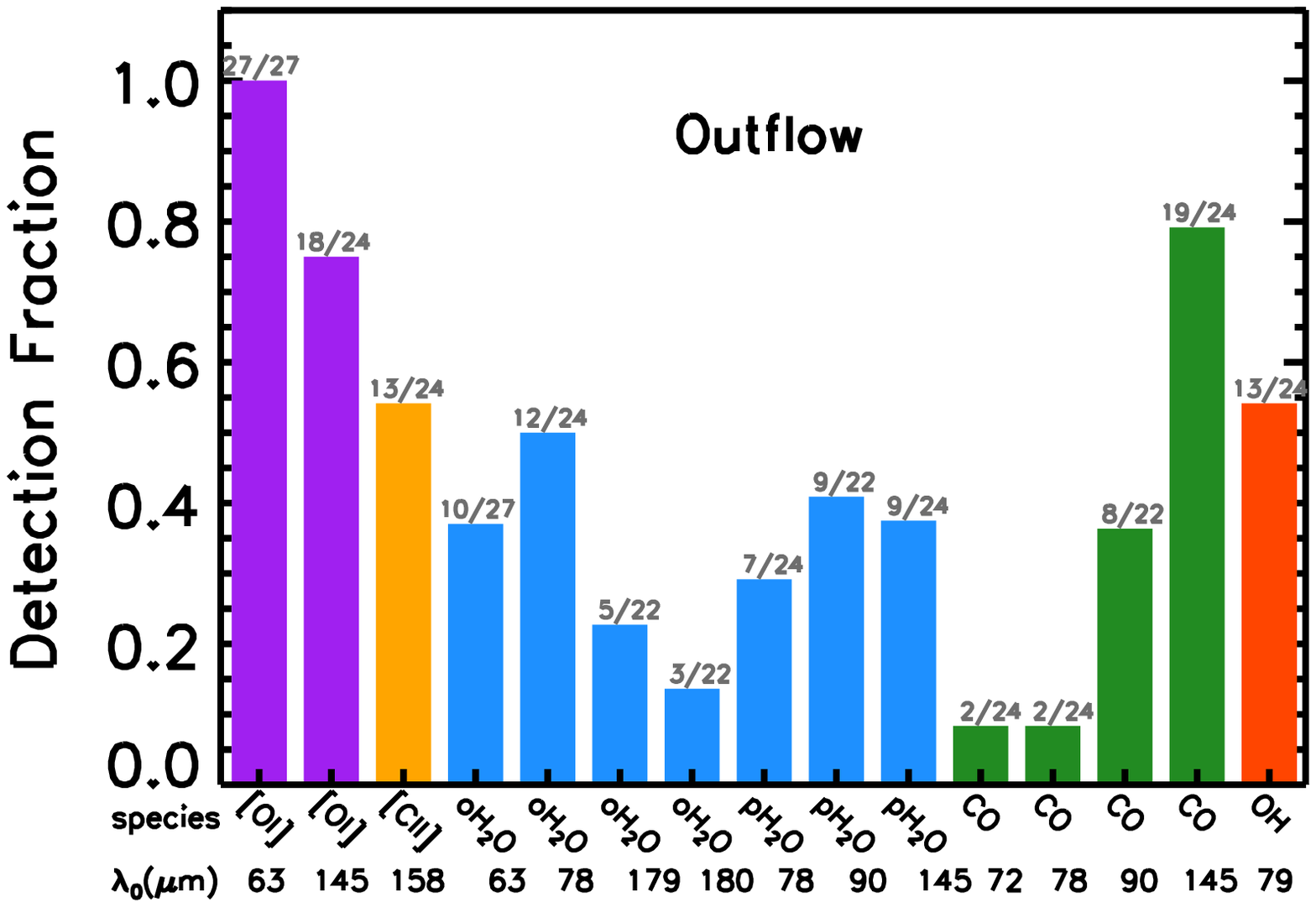}
\includegraphics[width=0.45\textwidth, trim=0mm 0mm 0mm 0mm, clip]{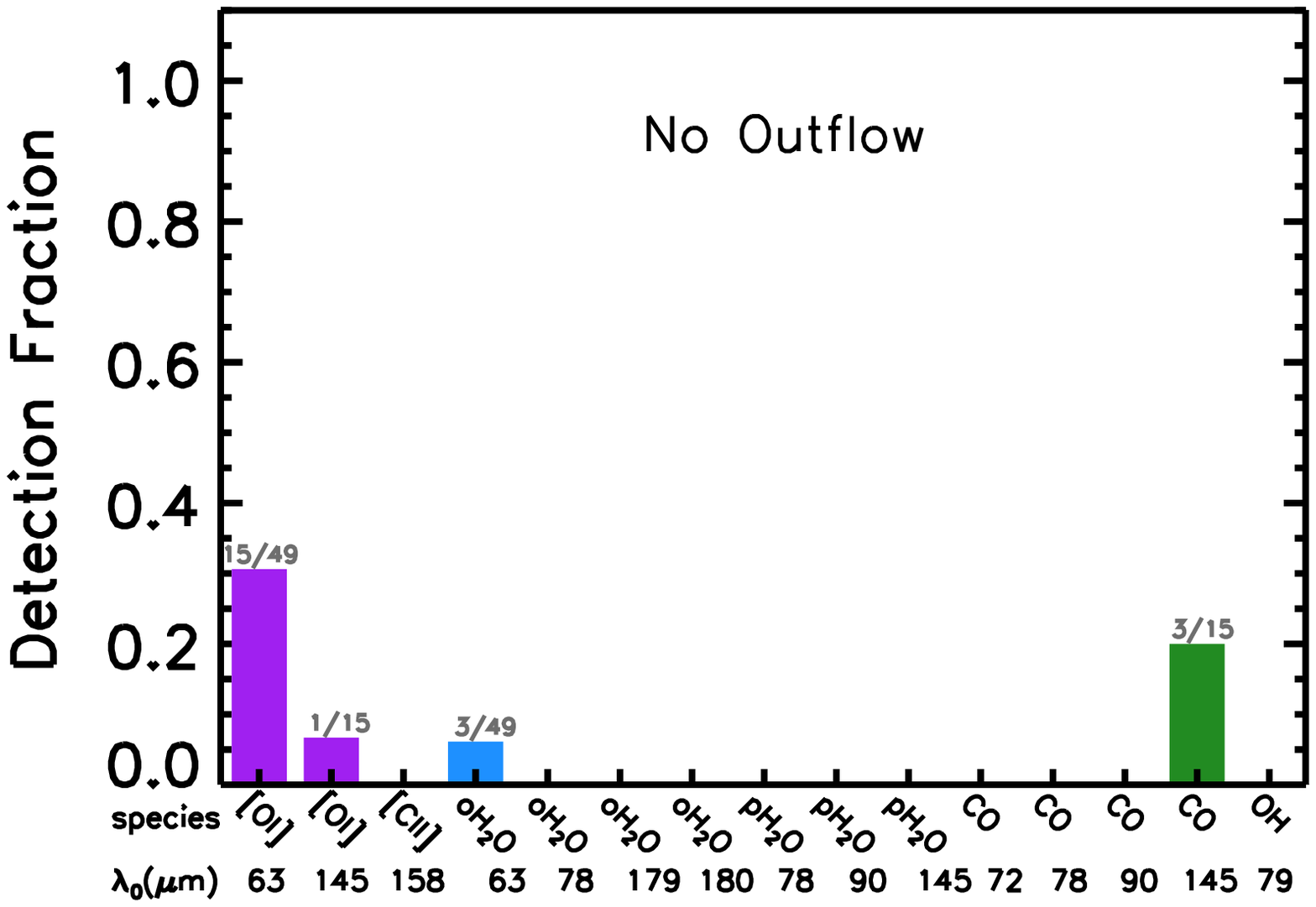}
\caption{Line emission detection fractions for the different species observed within PACS range. Objects with (\textit{top}) and without (\textit{bottom}) an outflow. Each species has a different colour: [OI] (\textit{purple}), [CII]  (\textit{yellow}), H$_{\rm 2}$O  (\textit{blue}), CO  (\textit{green}), and OH  (\textit{red}). The numbers on top of the bars refer to the total detections over the total targets observed. The atomic/molecular species and central wavelengths for the transitions are also indicated.}
\label{figure:statsJets}
\end{figure}

\begin{figure}[htpb]
\centering
\setcounter{figure}{2}
\includegraphics[width=0.55\textwidth, trim=8mm 8mm 20mm 25mm, clip]{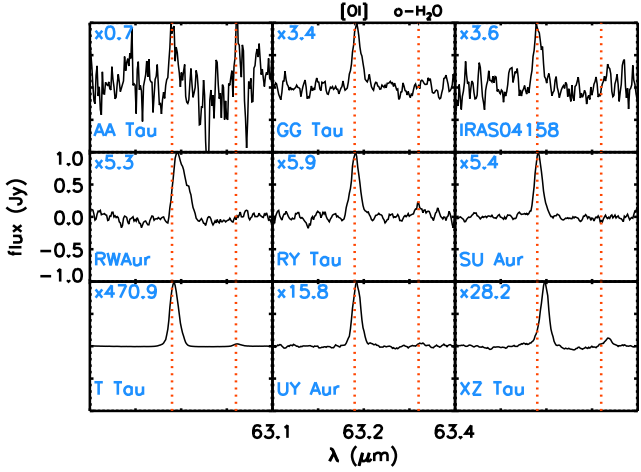}
\caption{Continuum subtracted spectra at 63 $\rm \mu m$ from the spaxel showing the highest continuum level. The scale is the same for all panels. The spectra were divided by the factor indicated in the upper left corner of each panel. The red lines indicate the position of the [OI] 63.18 $\rm \mu m$ and o--H$_{2}$O 63.32 $\rm \mu m$ lines. The offsets seen in the [OI] 63.18 $\rm \mu m$ line can be due to incorrect pointing or to the presence of an additional physical component.}
\label{spectra63}
\end{figure}

\begin{figure}[htpb]
\centering
\setcounter{figure}{3}
\includegraphics[width=0.55\textwidth, trim=8mm 8mm 20mm 25mm, clip]{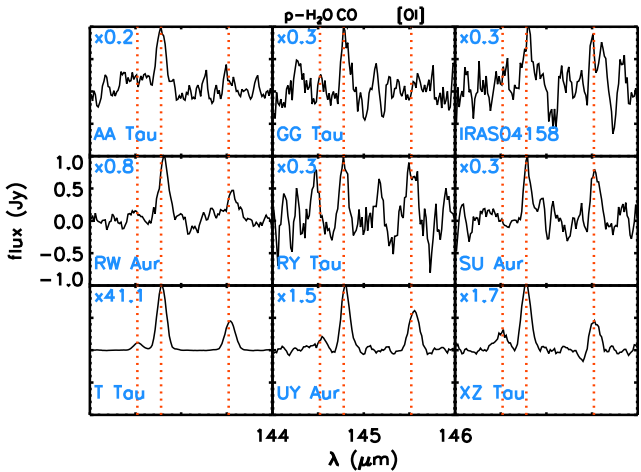}
\caption{Continuum subtracted spectra centred at 145 $\rm \mu m$ from the spaxel showing the highest continuum level. 
The scale is the same for all the panels. The spectra were divided by the factor indicated in the upper left corner of each panel. The red lines indicate the position of the p--H$_{2}$O 144.52 $\rm \mu m$, CO 144.78 $\rm \mu m$, and [OI] 145.53  $\rm \mu m$ lines.}
\label{spectra145}
\end{figure}

\subsection{Atomic emission lines}
\label{subsect:atomicLines}
%\noindent {\it [OI] emission--} 
\subsubsection{[OI] emission}
Figures \ref{spectra63} and \ref{spectra145} show spectra  centred on the [OI] fine-structure lines at 63.18 and 145.53 $\rm \mu m$. 
The [OI] 63.18 $\rm \mu m$ is detected in 42 out of 76 stars (55\%); the line is detected towards all outflow sources, but for non-outflow sources the detection rate drops to 31\%.
Line fluxes vary between 6$\times$10$^{-18}$ and \mbox{2 $\times$10$^{-14}$ W m$^{-2}$};  outflows show  stronger lines ($\sim$1$\times$10$^{-17}$ -- 2$\times$10$^{-14}$ W m$^{-2}$) than non-outflows \mbox{($\sim$6$\times$10$^{-18}$ -- 4$\times$10$^{-17}$ W m$^{-2}$}). Unlike \cite{Howard2013}, we did not detect [OI] 63.18 $\rm \mu m$ in CY Tau and Haro 6--37, probably due to different reduction pipelines and calibration files used.
The profiles (see e.g. Fig. \ref{spectra63}) are mainly Gaussians with some skewness in a few cases (e.g. XZ Tau). A recent and detailed study of [OI] 63.18 $\rm \mu m$ line profiles of young stellar objects (YSOs) by \cite{Riviere2016} suggests that such line profiles can be explained as a combination of disc, jet, and envelope emission.

Table \ref{detRatesClass} shows the fraction of sources where the observed spectral lines are detected as a function of SED class. 
In the following, the statistics of Class II sources do not include transitional discs (TD).
The atomic detection fractions decrease as the sources evolve from Class I down to Class III. The [OI] 63.18 $\rm \mu m$ line is detected in 100\% of Class I objects, 70\% of Class II objects, 60 \% of transitional discs, and 0\% of Class III objects, with average fluxes decreasing from $\sim$3$\times$10$^{\rm -15}$ W m$^{-2}$, through $\sim$2$\times$10$^{\rm -16}$, to 3$\times$10$^{-17}$  W m$^{-2}$ for Class I, Class II, and transitional discs respectively.

Table \ref{detRatesSpT} gives the line detection fractions as a function of spectral types.
The bins were selected so that they have a similar number of targets observed at 63 $\rm \mu m$.
For the [OI] 63.18 $\rm \mu m$ line, a decrease in  the detection fraction with later spectral type is observed.

The [OI] 145.52 $\rm \mu m$ line is harder to excite than [OI] 63.18 $\rm \mu m$ because of its higher energy level (see $T_{\rm ex}$ in Table \ref{LineRange}). Consequently, the line flux is on average ten times weaker than [OI] 63.18 $\rm \mu m$. It is detected in 19 out of 39 objects (49\%) with line fluxes between $\sim$10$^{-18}$ W m$^{-2}$ (the detection limit) and $\sim$8$\times$10$^{-16}$ W m$^{-2}$. The detection rate is 75\% (18 out of 24) for outflow sources, while it is only 7\% (1 out of 15, DE Tau) for non-outflow objects. 
The fluxes decrease according to the evolutionary status of the source from $\sim$2$\times$10$^{\rm -16}$ W m$^{-2}$ for Class I down to $\sim$3$\times$10$^{\rm -18}$ W m$^{-2}$ for transitional discs.

\smallskip
%\noindent {\it [CII] emission--} 
\subsubsection{[CII] emission}
Some spectra centred at the [CII] 157.74 $\rm \mu m$ fine-structure line are shown in Fig. \ref{spectra158}. We note that the [CII] line critical density is two orders of magnitude lower than for [OI] lines, making it easily excited  in the surrounding cloud as well. In most of our targets the line is also detected in the \emph{off-}source positions.
The [CII] fluxes reported in Table \ref{lineFluxRed} are for the \textit{on-} minus the \textit{off-} positions to be consistent with the procedure followed for the rest of the lines.

The [CII] 157.74 $\rm \mu m$ emission line has a detection fraction of 34\% (13 out of 38); only detected in 54\% (13 out of 24) of the outflow sources with average line flux $\sim$ 7$\times$ 10$^{-17}$ W m$^{-2}$. 
In DG Tau, DG Tau B, FS Tau, T Tau, UY Aur, and XZ Tau [CII] is observed as extended with an average line flux of $\sim$6$\times$10$^{-17}$ W m$^{-2}$.
RY Tau (TD)  is associated with a jet, mapped in [OI] 6300 $\mbox{\rm \AA}$ \citep{AgraAmboage2009}, and embedded in diffuse and extended nebulosity \citep[see Fig. 1 in][]{StOnge2008}, suggesting that its [CII] line arises in its immediate surroundings.

We see a clear decline in the detection fractions from Class I (50\%), through Class II (36\%), to transitional discs (17\%), and average line fluxes of  $\sim$2$\times $10$^{-16}$, $\sim$4$\times $10$^{-17}$, and  $\sim$4$\times $10$^{-18}$ W m$^{-2}$, respectively. This trend is similar to that observed for [OI] 63.18 $\rm \mu m$.

\begin{figure}[htpb]
\centering
\setcounter{figure}{4}
\includegraphics[width=0.55\textwidth, trim=8mm 8mm 20mm 25mm, clip]{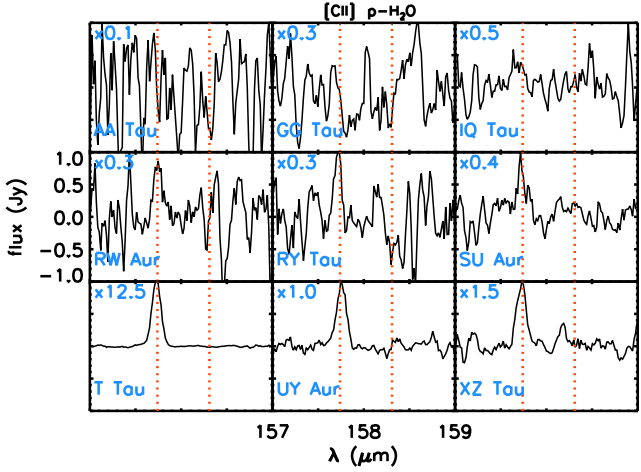}
\caption{Continuum subtracted spectra centred at 158 $\rm \mu m$ from the spaxel showing the highest continuum level. The scale is the same for all the panels. The spectra were divided by the factor in the upper left corner of each panel. The red lines indicate the position of the [CII] 157.74 $\rm \mu m$ and p--H$_{2}$O 158.31 $\rm \mu m$ lines.}
\label{spectra158}
\end{figure}

\subsection{Molecular emission lines}
%\noindent {\it CO emission.--} 
\subsubsection{CO emission}
Spectra of mid- to high-\emph{J} CO transitions (see Table \ref{LineRange}) are shown in Figs. \ref{spectra145}, \ref{spectra72}, \ref{spectra90}, and  \ref{spectra78}.
The CO {\it J=18--17} transition is most often detected in the stars observed (22 out of 39). Outflow sources have a detection rate of 79\% (19 out of 24) with an average flux of $\sim$9$\times$10$^{\rm -17}$ W m$^{-2}$, while the line is only detected in 3 non-outflows sources (CI Tau, DE Tau, and HK Tau), i.e. a rate of 20\%  (3 out of 15) and an average flux of $\sim$3$\times$10$^{-18}$ W m$^{-2}$. With respect to the SED classes, the detection fractions are 100\% (4 out of 4) for Class I objects, decreasing to 59\% (17 out of 29) for Class II objects, and 17\% (1 out of 6) for transitional disc sources with average line fluxes of $\sim$3$\times$10$^{-16}$, $\sim$2$\times$10$^{-17}$, and $\sim$4$\times$10$^{-18}$ W m$^{-2}$, respectively.

The CO \emph{J}=29--28, \emph{J}=33--32, and \emph{J}=36--35 lines are only detected in outflow sources, with detection fractions of 36\% (8 out of 22), 8\% (2 out of 24), and 8\% (2 out of 24), respectively; their average line fluxes are $\sim$6$\times$10$^{-17}$, $\sim$8$\times$10$^{-17}$, and $\sim$6$\times$10$^{-17}$ W m$^{-2}$, respectively.
The \emph{J}=33--32  and \emph{J}=36--35 CO lines detections are in DG Tau (Class II) and T Tau (Class I/II), and are known to drive powerful bipolar jets \citep[e.g.][]{Eisloffel1998}.
None of the CO lines shows a trend with spectral type.

\smallskip
%\noindent {\it H$_{2}$O emission.--}  
\subsubsection{H$_{2}$O emission}
Several transitions of water  were observed (see Figs. \ref{spectra63}, \ref{spectra145}, \ref{spectra90}, \ref{spectra78}, and \ref{spectra180}). 
The o--H$_{2}$O lines at 78.74, 179.53, and 180.49 $\rm \mu m$, and the p--H$_{2}$O lines at 78.92, 89.99, and 144.52 $\rm \mu m$ are only detected in outflow sources.
The highest detection fraction is for o--H$_{2}$O 78.74 $\rm \mu m$ (50\%) with an average $\sim$7$\times$10$^{-17}$ W m$^{\rm -2}$, followed by p--H$_{2}$O 89.99 $\rm \mu m$ (41\%) and average flux of $\sim$5$\times$10$^{-17}$ W m$^{\rm -2}$. The p-H$_{2}$O 144.52 $\rm \mu m$, 78.92 $\rm \mu m$, o--H$_{2}$O 179.53, and 180.49 $\rm \mu m$ lines have respectively a detection fraction of 38\%, 29\%, 23\%, and 14\%; and average fluxes are of $\sim$2$\times$10$^{-17}$, $\sim$2$\times$10$^{-17}$, $\sim$1.4$\times$10$^{-16}$, and $\sim$7$\times$10$^{-17}$ W m$^{\rm -2}$, respectively.
The o--H$_{2}$O 63.32 $\rm \mu m$ line is detected in 10 out of 27 (37\%) of the outflow sources  with an average flux of $\sim$4$\times$10$^{-17}$ W m$^{\rm -2}$. It is the only water line detected in non-outflow sources, seen in 3 of them (6\%), namely \mbox{BP Tau}, GI/GK Tau, and IQ Tau. This (\emph{warm}) water line was first reported in \cite{Riviere2012,Fedele2013b}. 
The p--H$_{2}$O 158.31 $\rm \mu m$ line is undetected in all targets (even in T Tau). 
Table \ref{table:waterClasses} lists the average water line fluxes of Class I, Class II, and TD sources. Water lines are brighter and more often detected (higher detection fractions) towards Class I objects than towards Class II and TD sources.

\begin{table}[htpb]
\centering
\caption{Average water line fluxes at different wavelengths (60--180 $\rm \mu m$) in units of 10$^{-16}$ W m$^{\rm -2}$ for Class I, Class II, and TD objects.}
\label{table:waterClasses}
\tiny
\begin{tabular}{l c c c c}
\hline
\hline
\multicolumn{1}{l}{Source}&
\multicolumn{4}{c}{o-H$_{\rm 2}$O}\\
\multicolumn{1}{l}{}&
\multicolumn{1}{c}{63  $\rm \mu m$}&
\multicolumn{1}{c}{78  $\rm \mu m$}&
\multicolumn{1}{c}{179 $\rm \mu m$}&
\multicolumn{1}{c}{180 $\rm \mu m$}\\[0.5mm]
\hline
Class I   & 1.05  & 2.50  & 6.35  & 2.05  \\
Class II  & 0.10  & 0.14  & 0.20  & 0.07  \\
TD        & 0.20  & 0.13  & \dots & \dots \\[0.5mm]
\hline
\multicolumn{1}{l}{Source}&
\multicolumn{4}{c}{p-H$_{\rm 2}$O}\\
\multicolumn{1}{l}{}&
\multicolumn{1}{c}{78  $\rm \mu m$}&
\multicolumn{1}{c}{90  $\rm \mu m$}&
\multicolumn{1}{c}{145 $\rm \mu m$}&
\multicolumn{1}{c}{158 $\rm \mu m$}\\[0.5mm]
\hline
Class I   & 0.38  & 1.83  & 0.45  & \dots \\
Class II  & 0.06  & 0.08  & 0.08  & \dots \\
TD        & \dots & 0.10  & 0.04  & \dots \\[0.5mm]
\hline
\end{tabular}
\end{table}

\smallskip
%\noindent {\it OH emission.--} 
\subsubsection{OH emission}
Selected spectra of the hydroxyl doublet at 79.11 and 79.18 $\rm \mu m$ are shown in Fig. \ref{spectra78}. The line is only detected in outflow sources (12 out of 38) with line flux values $\sim$10$^{-16}$ W m$^{-2}$, on average. 
It is detected in 50\% of the \mbox{Class I} sources (average flux $\sim$4$\times$10$^{\rm -16}$ W m$^{-2}$), decreasing to 38\% in Class II (average flux $\sim$1.5$\times$10$^{\rm -17}$ W m$^{-2}$), and is undetected in transitional discs. 
We point out that DO Tau and DL Tau show peculiar OH detections: DO Tau shows only the 79.11 $\rm \mu m$ component, while DL Tau only shows the 79.18 $\rm \mu m$ component. We tested whether it could be due to significant pointing errors that translate into wavelength shifts which  yield negative results \citep[see Sect. 3.2.2 in][]{Howard2013}. Thus, the detection of only one component of the OH doublet appears to be real. Such asymmetries of OH lines in a doublet have  already been noticed in ISO data \citep{Goicoechea2011}, in Class 0/I sources \citep{Wampfler2013}, and are discussed by \cite{Fedele2015} for HD100546.

\smallskip
%\noindent{\it CH$^+$ emission.--} 
\subsubsection{CH$^+$ emission}
The CH$^+$ feature remains undetected in almost all targets. Only T Tau shows CH$^+$ emission at 72.14 $\rm \mu m$ (see Fig. \ref{spectra72}) with a line flux value of 1.77$\pm$0.65$\times$10$^{\rm -17}$ W m$^{-2}$. There could also be blends of CH$^+$ with H$_{2}$O lines at 89.99 and 179.53 $\rm \mu m$.

\begin{figure}[htpb]
\centering
\setcounter{figure}{5}
\includegraphics[width=0.55\textwidth, trim=8mm 8mm 20mm 25mm, clip]{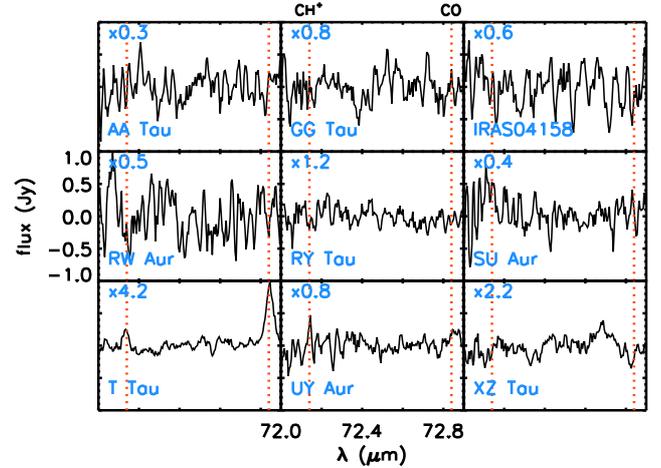}
\caption{Continuum subtracted spectra at 72 $\rm \mu m$. The scale is the same for all the panels. The spectra were divided by the factor in the upper left corner of each panel. The red lines indicate the position of the CH$^{\rm +}$ 72.14 $\rm \mu m$ and CO 72.84 $\rm \mu m$ lines.}
\label{spectra72}
\end{figure}

\begin{figure}[htpb]
\centering
\setcounter{figure}{6}
\includegraphics[width=0.55\textwidth, trim=8mm 8mm 20mm 25mm, clip]{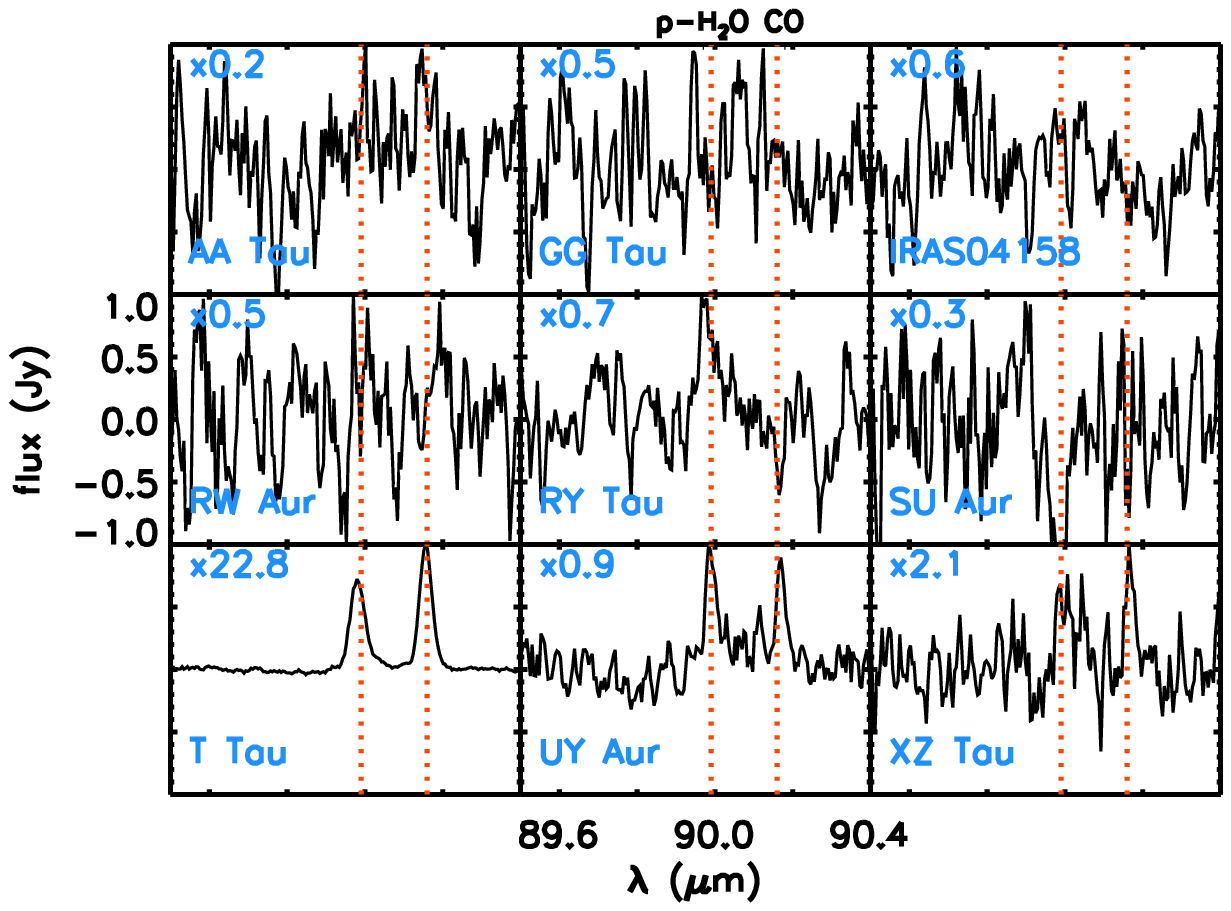}
\caption{Continuum subtracted spectra at 90 $\rm \mu m$ from the spaxel showing the highest continuum level. The scale is the same for all the panels. The spectra were divided by the factor in the upper left corner of each panel. The red lines indicate the position of the p--H$_{2}$O 89.99 $\rm \mu m$ and CO  90.16 $\rm \mu m$ lines.}
\label{spectra90}
\end{figure}

\begin{figure}[htpb]
\centering
\setcounter{figure}{7}
\includegraphics[width=0.55\textwidth, trim=8mm 8mm 20mm 25mm, clip]{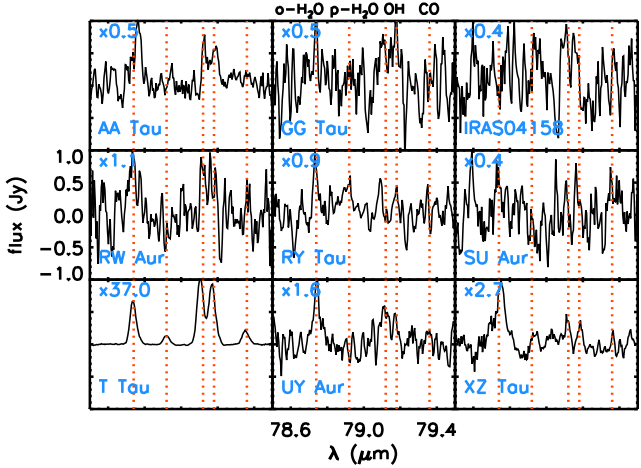}
\caption{Continuum subtracted spectra at 79 $\rm \mu m$ from the spaxel showing the highest continuum level. The scale is the same for all the panels. The spectra were multiplied by the factor in the upper left corner of each panel. The red lines indicate the position of the o--H$_{2}$O 78.74 $\rm \mu m$ line, p--H$_{2}$O 78.92 $\rm \mu m$ line, OH 79.11/79.18 $\rm \mu m$ doublet, and CO 79.36 $\rm \mu m$ lines.}
\label{spectra78}
\end{figure}

\begin{figure}[ht]
\centering
\setcounter{figure}{8}
\includegraphics[width=0.55\textwidth, trim=8mm 8mm 20mm 25mm, clip]{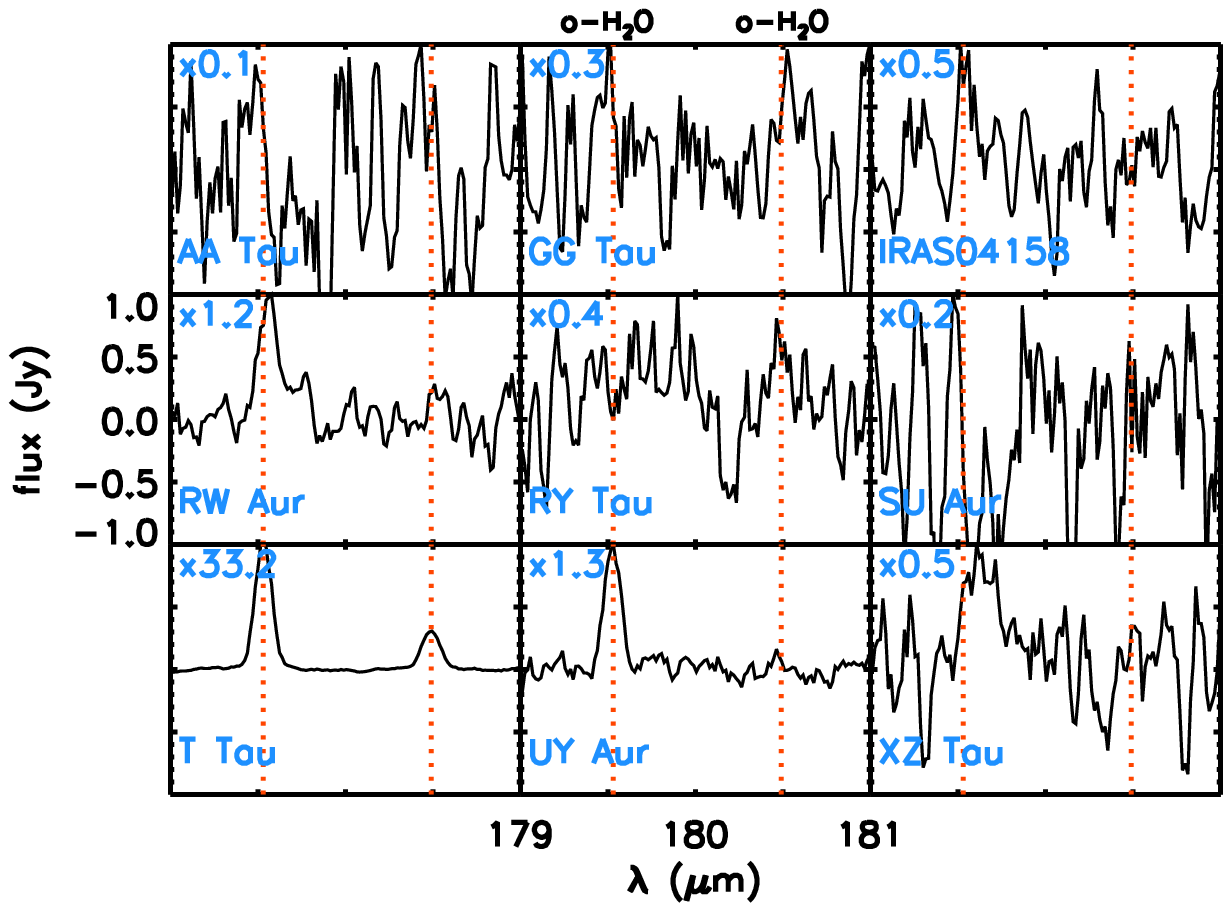}
\caption{Continuum subtracted spectra centred at 180 $\rm \mu m$ from the spaxel showing the highest continuum level. The scale is the same for all the panels. The spectra were divided by the factor in the upper left corner of each panel. The red lines indicate the position of the o--H$_{2}$O 179.53 and 180.49 $\rm \mu m$ lines.}
\label{spectra180}
\end{figure}

%===================================================================================
%===================================================================================
\section{Observational trends of far-IR lines}
\label{relations}
\subsection{Evolution from Class 0/I to Class III}

For the full sample of YSOs  discussed here, the fraction of sources with [OI] 63.18 $\rm \mu m$ detections (55$^{\rm +5}_{\rm -6}$\%) is higher than for older ($>$3 Myr) star forming regions like TW Hya \citep[22\%, age$\sim$8--20 Myr;][]{Riviere2013}, Upper Sco \citep[4\%, age$\sim$5--11 Myr;][]{Mathews2013}, Cha II \citep[37\%, age$\sim$4 Myr;][]{Riviere2014}, and $\eta$ Cha \citep[8\%, age$\sim$5--9 Myr;][]{Riviere2015}. Such a decrease with age is a clear indication of evolution.

Figure \ref{figure:avgfluxes} shows the average fluxes of the [OI] 63.18 $\rm \mu m$, OH 79.12+79.18 $\rm \mu m$, CO 144.78 $\rm \mu m$, o-H$_{2}$O 78.74 $\rm \mu m$, and o-H$_{\rm 2}$O 63.32 $\rm \mu m$ lines of Class I and II sources with jets, and for Class II objects with no jets, for which the OH 79.12+79.18 $\rm \mu m$ and o-H$_{\rm 2}$O at 78.74 $\rm \mu m$ lines are below the detection limit. The atomic and molecular line fluxes decrease rapidly, approximately by an order of magnitude at each stage, suggesting that a physical mechanism related to evolution is the most likely scenario. 
 
One possible explanation could be that FIR line emission is due to a combination of jets shocking the surrounding material and UV radiation.
In general, molecular emission from shocks is expected and is very important at the earliest stages, whereas photodissociation is more effective when the envelope dissipates \citep{Nisini2002}. In particular, H$_{\rm 2}$O in shocks is abundant because neutral-neutral reactions switch on at high temperature \citep[see e.g.][]{vanDishoeck2013}.
Processes involving dust grains are also important. Owing to photodesorption, sputtering, and grain-grain collisions, likely triggered by shocks, H$_{\rm 2}$O is also removed from icy dust grains. The progressive dissipation of gaseous and dusty envelopes from Class 0/I to Class III allows stellar or interstellar FUV fields to penetrate deeper and to dissociate more H$_{\rm 2}$O and OH to produce O. This scenario proposed by \cite{Nisini2002} was followed by \cite{Karska2013} to explain the FIR line weakening of CO and H$_{\rm 2}$O observed from Class 0 to Class I objects.
When the mass accretion and outflow rates drop as the source evolves, the FIR emission originating in shock gas decrease because the strength of the FIR lines is related to the amount of shocked gas \citep{Manoj2016}.
This does not hold for the more evolved Class II sources (`only disc' in Fig. \ref{figure:avgfluxes}), in which FIR line emission is coming from illuminated discs by UV \citep{France2014} and X-ray \citep{Gudel2007}. It is expected that as the disc is accreted and/or dispersed, the strength of the FIR lines will decrease too. This is also suggested by the non detections in Class III objects.

\begin{figure}[htpb]
\centering
\setcounter{figure}{9}  
\includegraphics[width=0.34\textwidth,trim=8mm 0mm 0mm 0mm,clip,angle=90]{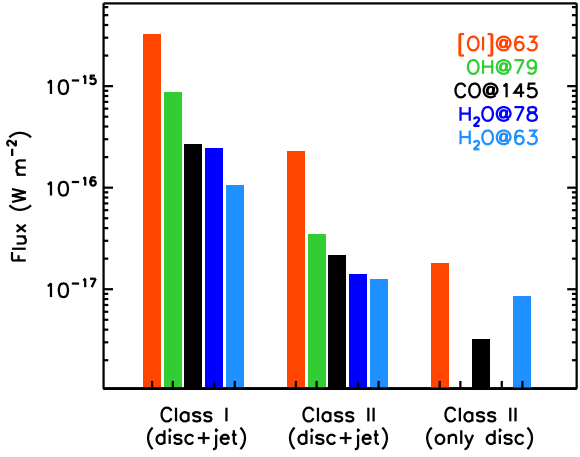}
\caption{Average line flux of [OI] 63.18 $\rm \mu m$ (\textit{red}), OH 79.12+79.18 (\textit{green}), CO 144.78 $\rm \mu m$ (\textit{black}), o-H$_{2}$O 78.74 $\rm \mu m$ (\textit{dark blue}), and o-H$_{\rm 2}$O 63.32 $\rm \mu m$ (\textit{blue}) lines of Class I and II objects with jets and Class II sources with no jets. Most of the TD sources are included within Class II (only disc) objects. The exception is RY Tau (see Sect. \ref{subsect:atomicLines}), included in the Class II (disc+jet) group.}
\label{figure:avgfluxes}
\end{figure}

\subsection{Relations between far-IR lines}
\label{section:correlations}

\begin{figure*}[htpb]
\centering
\setcounter{figure}{10}
(a)\includegraphics[width=0.23\textwidth,trim=0mm 0mm 0mm 0mm,angle=90,clip]{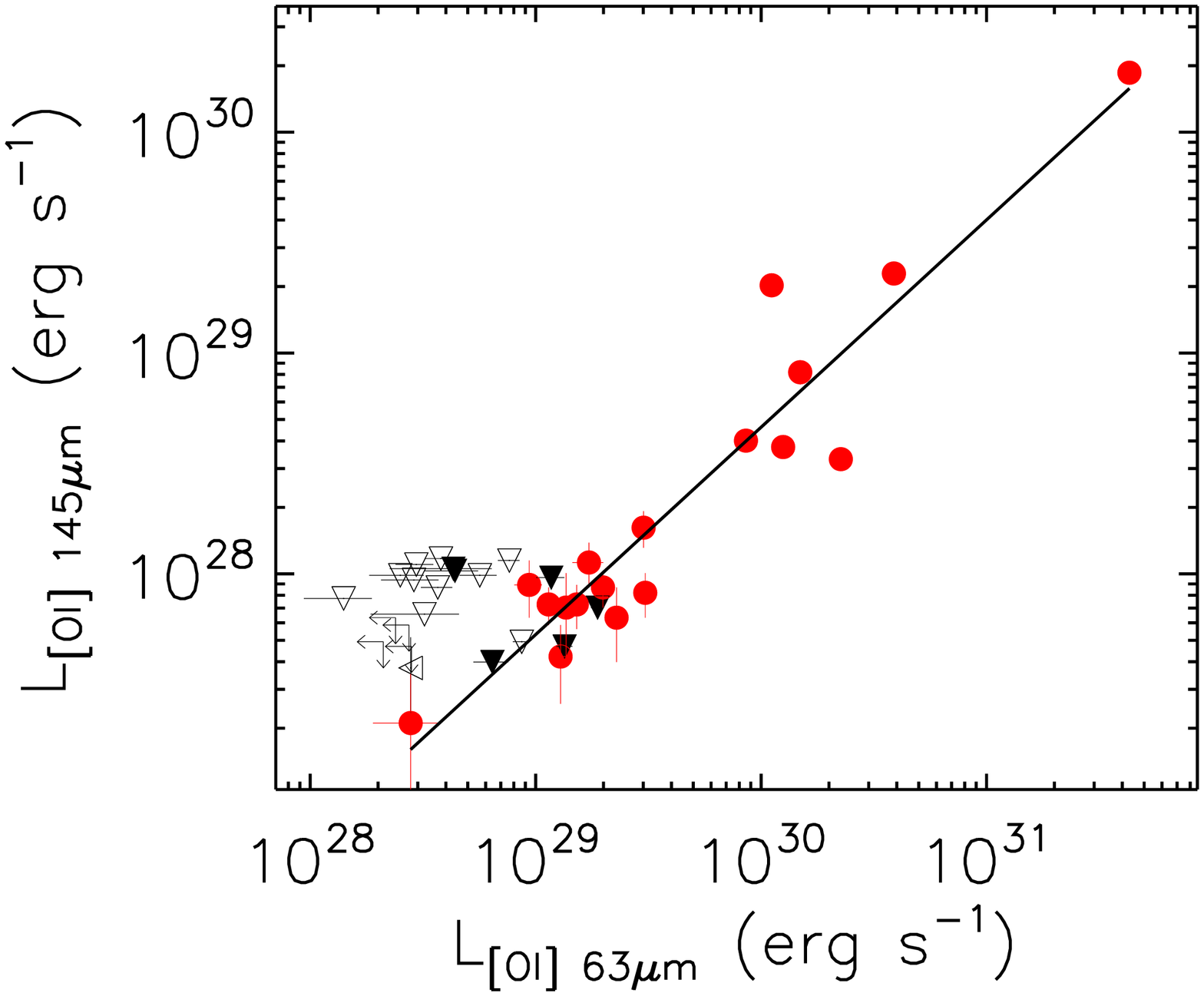}(b)\includegraphics[width=0.23\textwidth,trim=0mm 0mm 0mm 0mm,angle=90,clip]{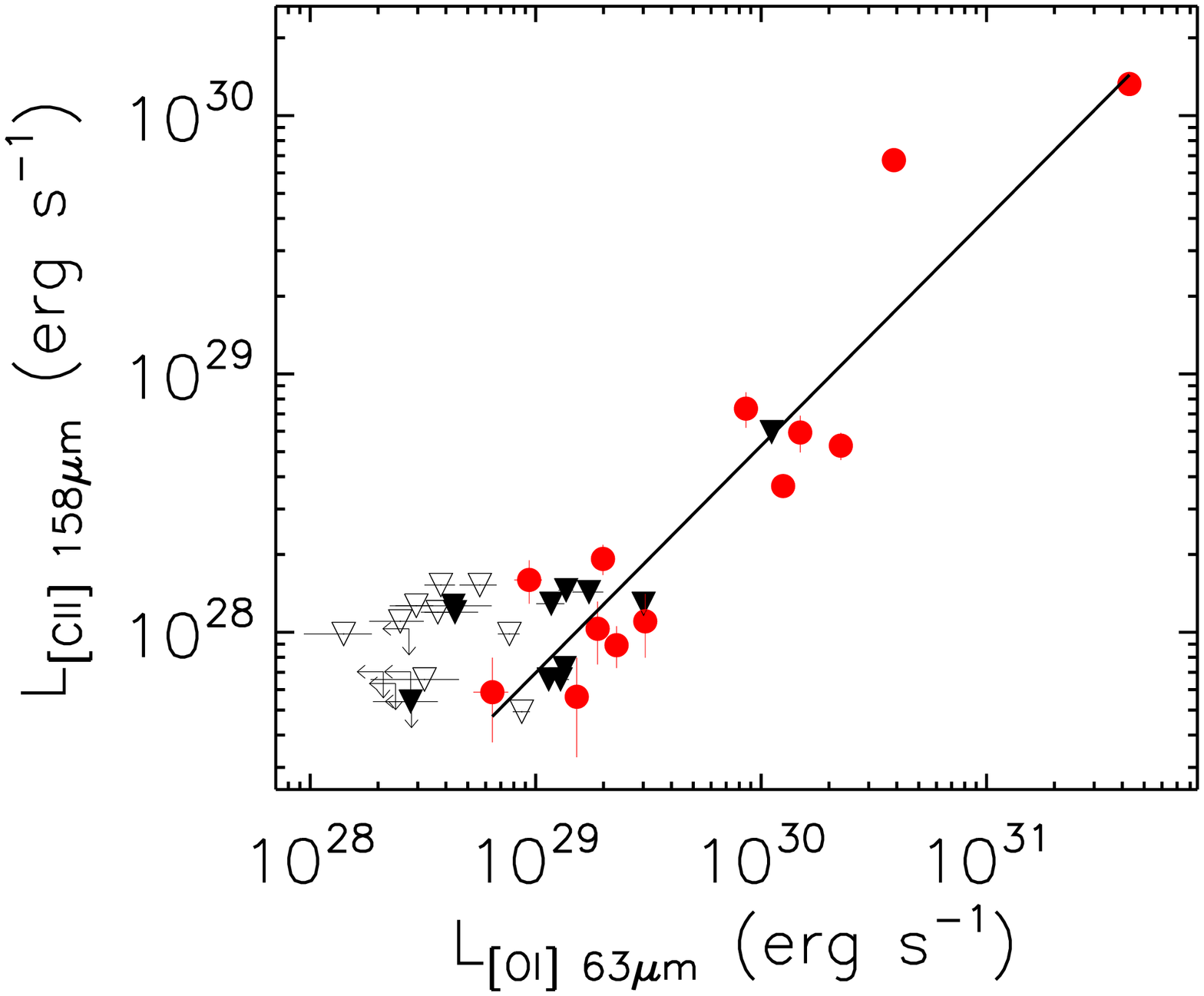}(c)\includegraphics[width=0.23\textwidth,trim=0mm 0mm 0mm 0mm,angle=90,clip]{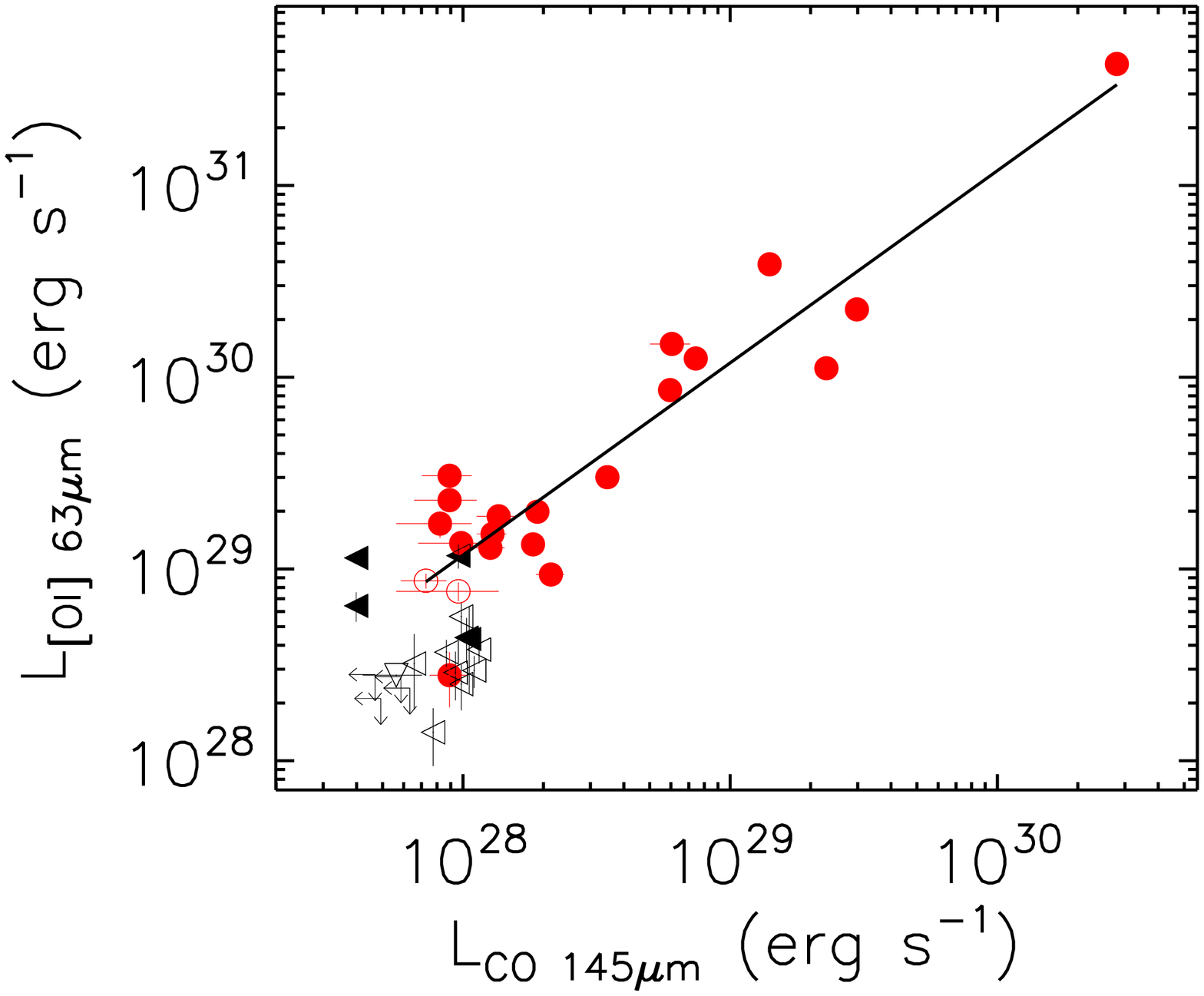}

(d)\includegraphics[width=0.23\textwidth,trim=0mm 0mm 0mm 0mm,angle=90,clip]{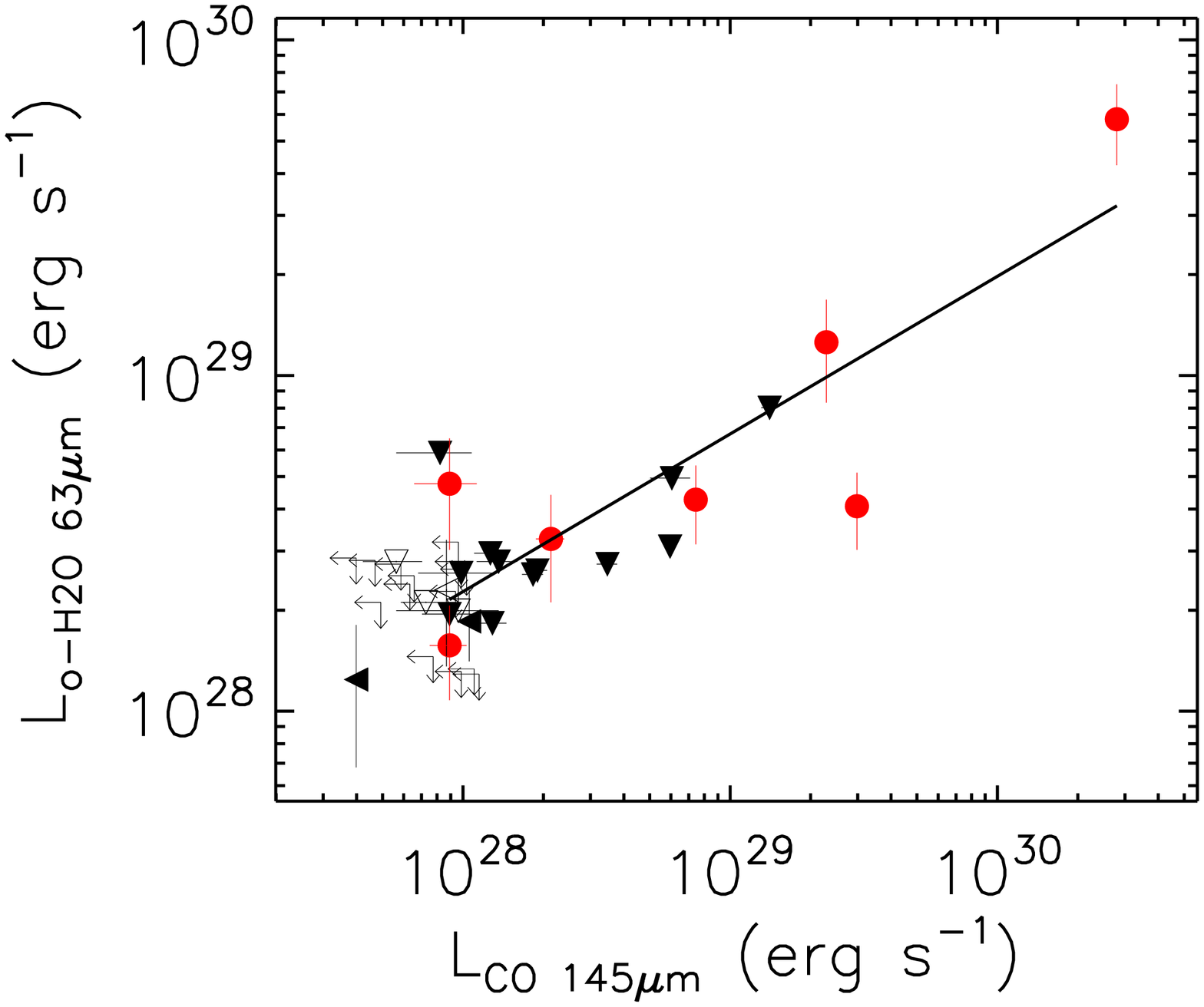}(e)\includegraphics[width=0.23\textwidth,trim=0mm 0mm 0mm 0mm,angle=90,clip]{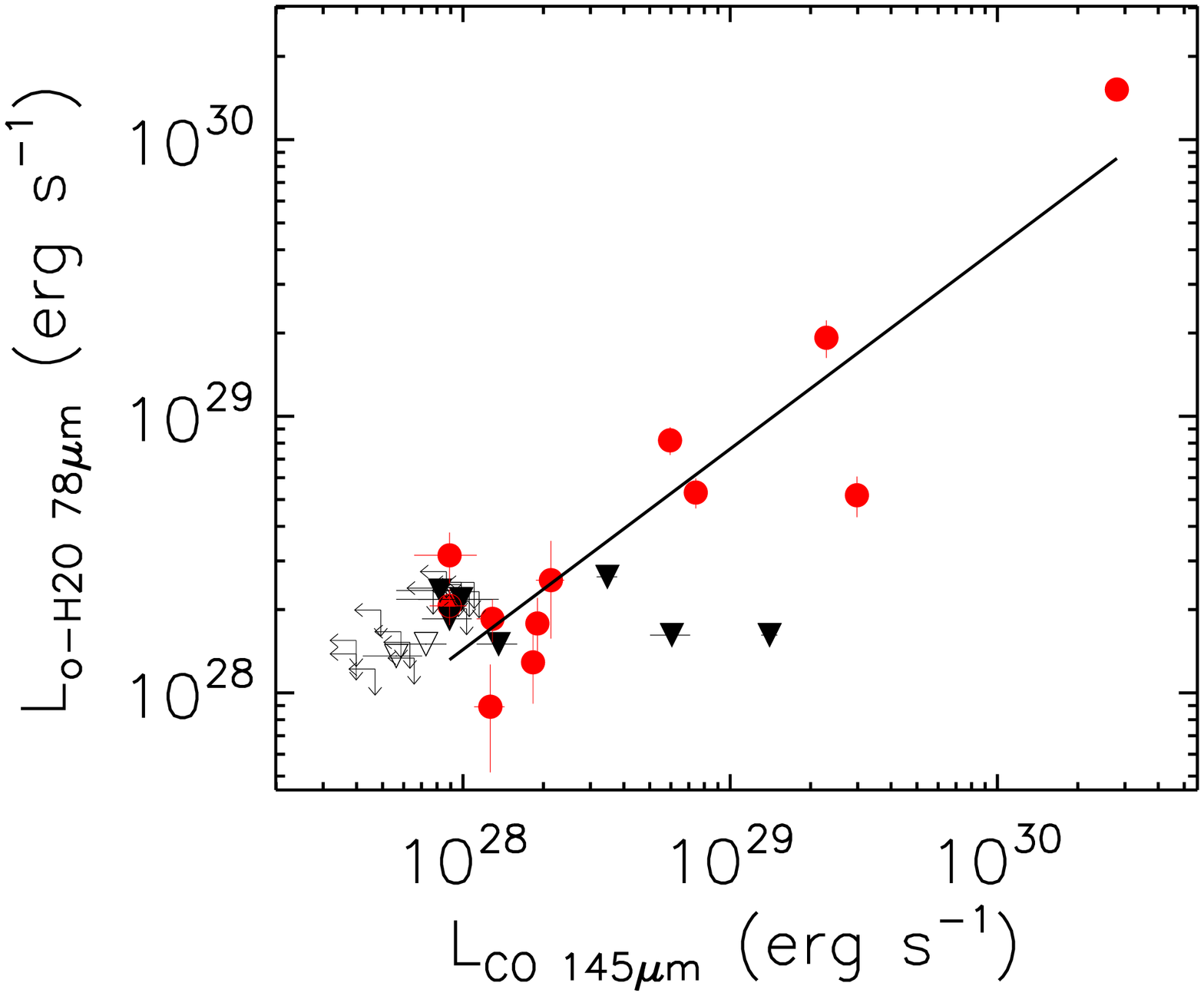}(f)\includegraphics[width=0.23\textwidth,trim=0mm 0mm 0mm 0mm,angle=90,clip]{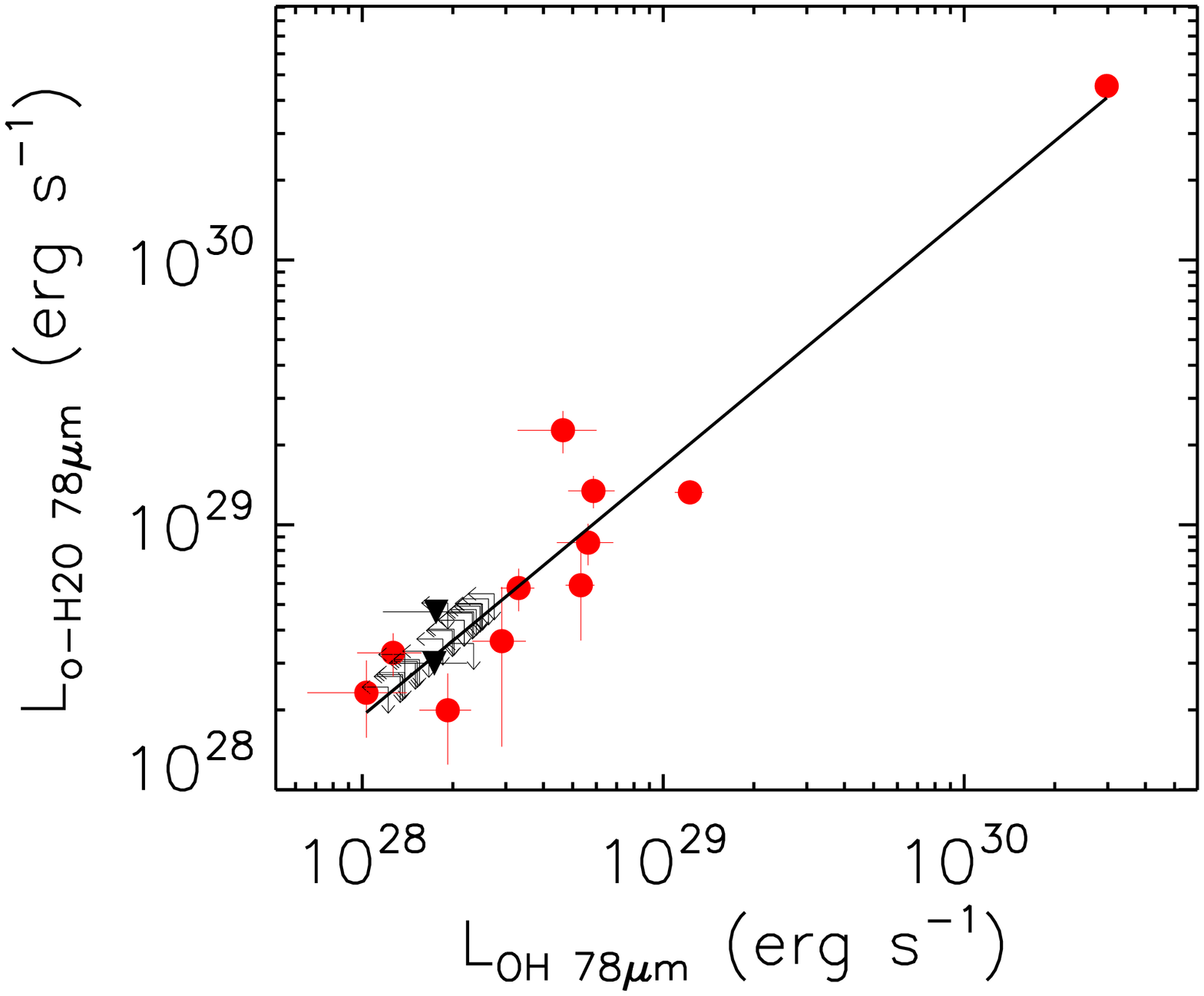}

(g)\includegraphics[width=0.23\textwidth,trim=0mm 0mm 0mm 0mm,angle=90,clip]{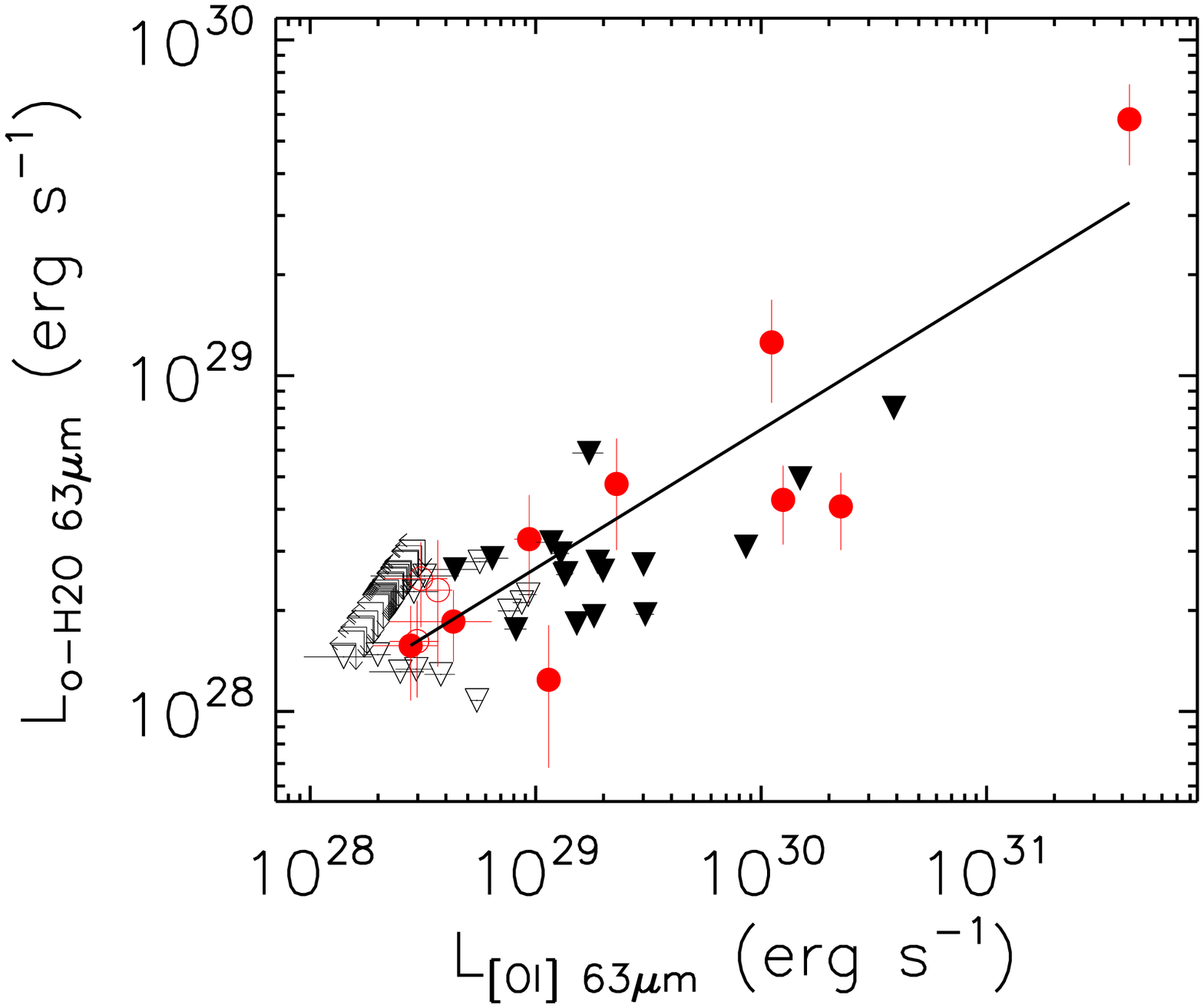}(h)\includegraphics[width=0.23\textwidth,trim=0mm 0mm 0mm 0mm,angle=90,clip]{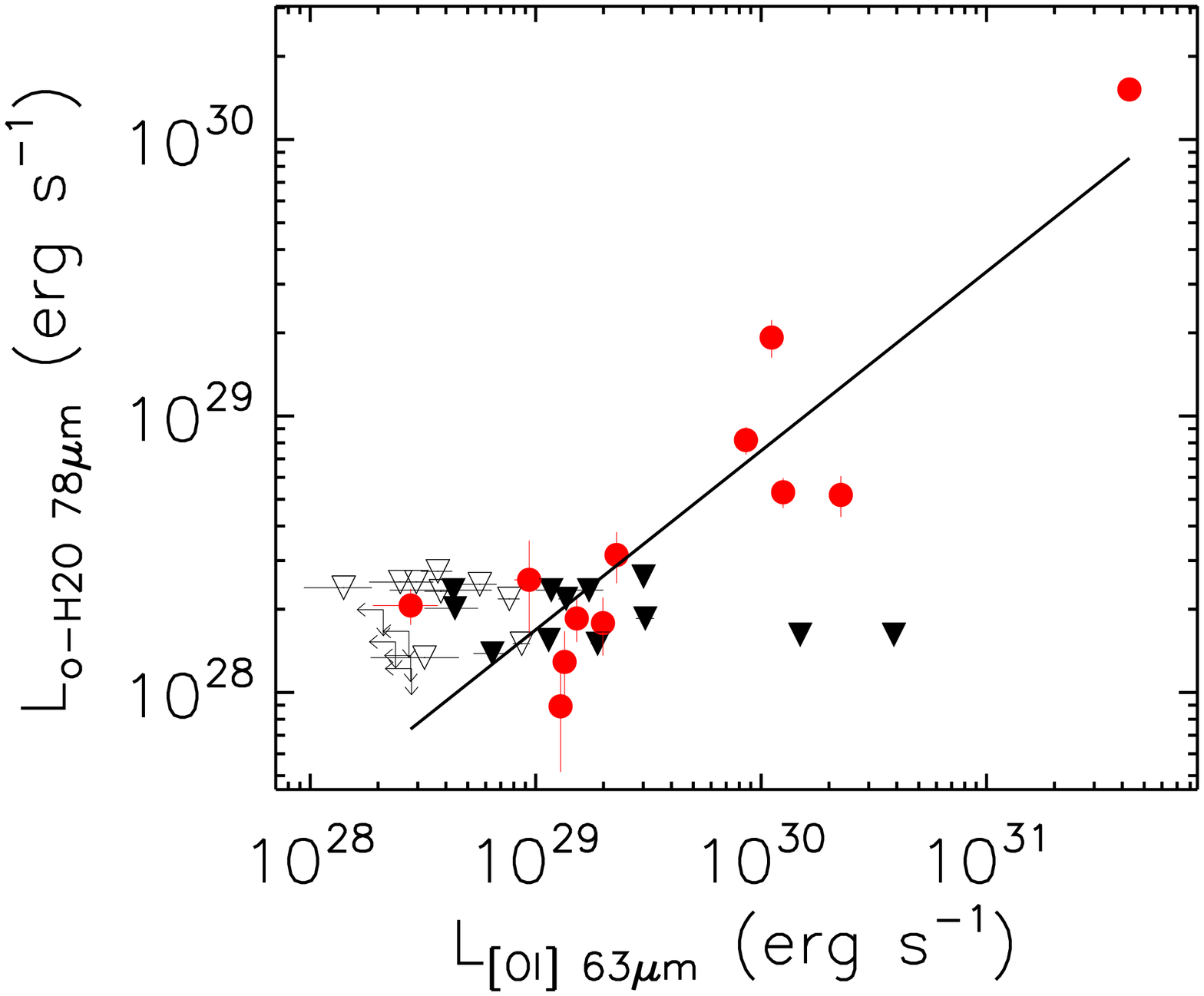}(i)\includegraphics[width=0.23\textwidth,trim=0mm 0mm 0mm 0mm,angle=90,clip]{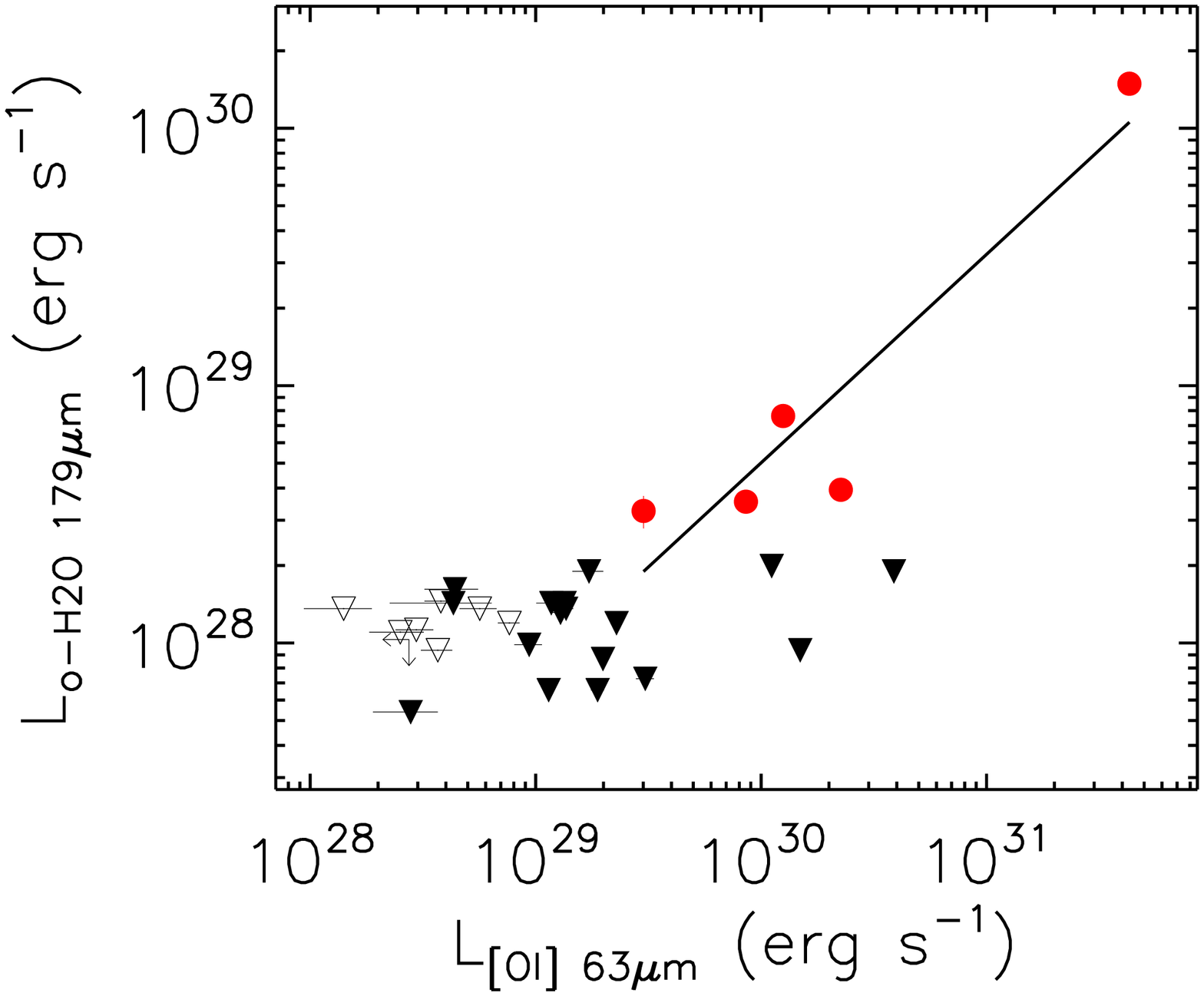}

\caption{Correlation plots between: \textbf{a)} [OI] 63.18 $\rm \mu m$ and [OI] 145.52 $\rm \mu m$; \textbf{b)} [OI] 63.18 $\rm \mu m$ and [CII] 157.74 $\rm \mu m$ \citep[see][]{Howard2013}; \textbf{c)} [OI] 63.18 $\rm \mu m$ and CO 144.78 $\rm \mu m$; \textbf{d)} CO 144.78 $\rm \mu m$ and o-H$_{2}$O 63.32 $\rm \mu m$; \textbf{e)} CO 144.78 $\rm \mu m$ and o-H$_{2}$O 78.74 $\rm \mu m$; \textbf{f)} OH 79.12+79.18 $\rm \mu m$ and o-H$_{2}$O 78.74 $\rm \mu m$; \textbf{g)} [OI] 63.18 $\rm \mu m$ and o-H$_{2}$O 63.32 $\rm \mu m$ \citep[see][]{Riviere2012}; \textbf{h)} [OI] 63.18 $\rm \mu m$ and o-H$_{2}$O 78.74 $\rm \mu m$; and \textbf{i)}  [OI] 63.18 $\rm \mu m$ and o--H$_{2}$O 179.53 $\rm \mu m$ (tentatively). The red circles represent detections, grey triangles are upper limits in their pointing direction, and black arrows are upper limits in both axes. Filled symbols represent outflow sources. The solid line corresponds to a linear fit for detections.}
\label{figure:correlations}
\end{figure*}

We performed an extensive search for correlations to address the possible origins of the FIR lines  discussed here and to see how they are related. Only those atomic and molecular lines with high detection fractions were selected.
Correlation factors $\rho_{\rm x,y}$ \citep[see Appendix A in][]{Marseille2010}, where $\rm x,y$ is the pair of lines considered, are used to validate any possible trends. The 3$\rm \sigma$ correlation corresponds to the threshold coefficient $\rho_{\rm thres} = 3 / \sqrt{N-1}$, where $N$ is the number of detections used in the calculation.
Only those trends with correlation factors above the confidence threshold ($\rho_{\rm thres}$) are taken as statistically real: $|\rho|<0.8$ denotes a lack of correlation, $0.8<|\rho|<0.9$ a weak (3$\rm \sigma$) correlation, and $|\rho|>0.9$ a strong correlation.
To rule out that T Tau is somehow driving the correlations due to its high line fluxes (up to 200 times the median), the analysis is repeated without this star.

Figure \ref{figure:correlations} shows the most promising correlations. Both atomic and molecular lines are observed to correlate. Two new tight ($|\rho|>0.9$) correlations between FIR lines are identified: one between [OI] 63.18 $\rm \mu m$ and CO 144.78 $\rm \mu m$, and the other between [OI] 63.18 $\rm \mu m$ and o-H$_{2}$O 78.74 $\rm \mu m$. Perhaps statistically less meaningful because of T Tau, we also find correlations between CO 144.78 $\rm \mu m$ and o-H$_{2}$O 63.32 $\rm \mu m$, and between OH 79.12+79.18 $\rm \mu m$ and o-H$_{2}$O 78.74 $\rm \mu m$.
The correlation between [OI] 63.18 $\rm \mu m$ and o-H$_{2}$O 63.18 $\rm \mu m$  has already been observed in T Tauri stars \citep{Riviere2012}.

\section{Discussion}
\label{discussion}
The [OI] 63.18 $\rm \mu m$ line can arise in the surface of discs  depending on disc size and spectral type \citep{Gorti2008}. It can be produced in photodissociation regions \citep[PDRs;][]{Tielens1985}, in shocks \citep{Neufeld1994}, or in the envelopes of Class I sources 
\citep{Ceccarelli1996}. Nothing precludes all mechanisms from  contributing simultaneously.
Given that the majority of our objects are Class II (see Sect. \ref{sample}), whose envelopes are likely already dissipated, we dedicate the following sections to a discussion of the most probable ones, i.e. shocks and discs. Each scenario is considered separately;  line fluxes and their line ratios (see Appendix \ref{lineFluxes}) are compared with shock and disc model predictions. We stress that in both scenarios a contribution from PDRs is also expected.

\subsection{Emission in a shock scenario}
\label{subsect:jetScenario}

Jets, outflows, and winds associated with PMS (pre-main sequence) stars can be traced with
forbidden  lines, e.g. [OI] 6300 $\mbox{\rm \AA}$ \citep[e.g.][]{Appenzeller1989,Edwards1993,Hirth1997}.
A correlation between these jet/outflow tracers and FIR lines would suggest a similar origin.
Figure \ref{figure:correlOI6300} shows the [OI] 63.18 $\rm \mu m$ luminosity as a function of the [OI] 6300 $\mbox{\rm \AA}$ line luminosity from \cite{Hartigan1995} integrated over the entire profile.
The lines correlate (\emph{$\rho$}$\sim$0.89; see Sect. \ref{section:correlations}), pointing to a common origin for the two lines.
It is not clear whether the non-outflow sources with relatively bright [OI] emission are associated with fainter unidentified compact outflows located within the central spaxel (9.4").
The observed scatter could be due to the presence of several velocity components.
Indeed, the [OI] 6300 $\mbox{\rm \AA}$ line profile often shows two velocity components \citep{Hartigan1995}: a high velocity component (HVC) shifted by 50--200 km s$^{-1}$ with respect to the stellar velocity, tracing collimated jets, and a low velocity component (LVC) shifted by 5--20 km s$^{-1}$, whose origin is possibly due to a photoevaporative wind \citep[][]{Rigliaco2013,Simon2016}.
Unfortunately, the PACS spectral resolution at 63 $\rm \mu m$ is $\sim$ 88 km s$^{\rm -1}$, not high enough to resolve velocity components, but we note that in several cases the wings are broad. Such broadening is observed in the red wing of the [OI] 63.18 $\rm \mu m$ line profile (see Fig. \ref{figure:allSpec63}) of CW Tau, DO Tau, DQ Tau, FS Tau, Haro6-5 B, HV Tau, and RW Aur. Interestingly, the [OI] 6300 $\mbox{\AA}$ and [OI] 63.18 $\rm \mu m$ line profiles of RW Aur do not show the same line shape, but have roughly the same 300 km s$^{-1}$ broadening. This further suggests the presence of several components like in HH 46 \citep[][]{vanKempen2010} and DK Cha \citep[][]{Riviere2014}.

\begin{figure}[htpb]
\setcounter{figure}{11}
\centering
\includegraphics[width=0.35\textwidth,trim=0mm 0mm 9mm 15mm,angle=90,clip]{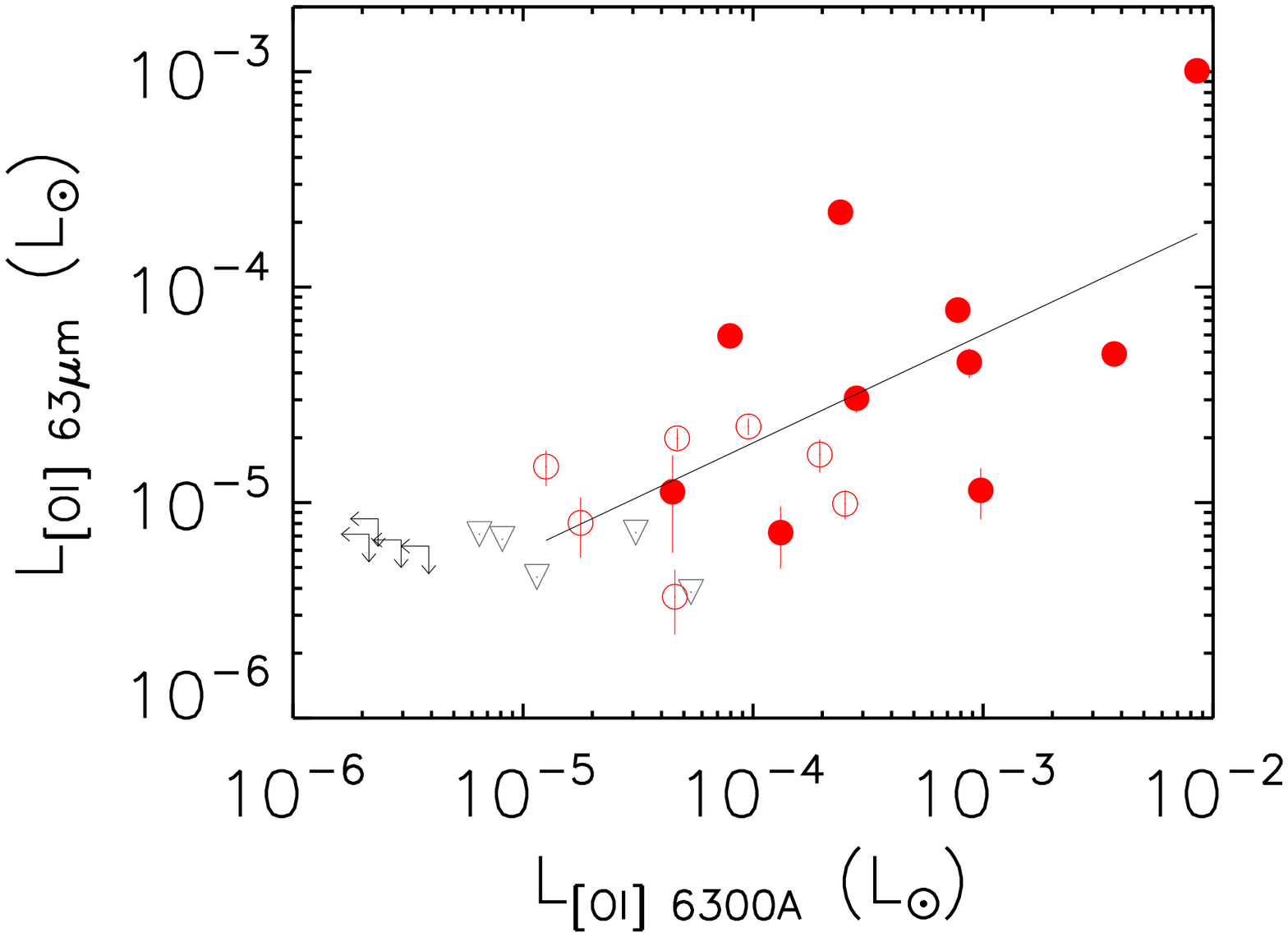}
\caption{ [OI] 63.18 $\rm \mu m$ as a function of [OI] 6300 $\mbox{\rm \AA}$ line luminosities in solar units.  The fit is only for detections and is indicated by the solid line. 
 The red circles represent detections, grey triangles are upper limits in their pointing direction, and black arrows are upper limits in both axes. Filled symbols represent outflow sources.}
\label{figure:correlOI6300}
\end{figure}

\subsubsection{Atomic line ratios as tracers of the excitation conditions}
\label{section:atomicRatios}
The line ratios of [OI]63/[OI]145 combined with [CII]158/[OI]63 can be used as diagnostics of the excitation mechanisms \citep[e.g.][]{Nisini1996,Kaufman1999}.  The [OI]63/145 ratios of our sample are between $\sim$10 and 70 with a median of $\sim$23, therefore compatible with ISO observations \citep{Liseau2006} and the ratios observed in Herbig Ae/Be stars \citep{Meeus2012,Fedele2013b}.
There is no statistical difference in terms of [CII]/[OI] line ratios between extended objects (detected in more than one spaxel but not incorrectly pointed) and compact objects (only detected in one spaxel)  in our sample. 
We note that in those sources with extended outflow emission, the ISO and \textit{Herschel} absolute fluxes are not expected to be similar, due to the much larger beam and a lack of background emission subtraction in ISO.

\begin{figure}
\setcounter{figure}{12}
    \centering
    \includegraphics[width=0.35\textwidth,trim=0mm 0mm 0mm 0mm,angle=90,clip]{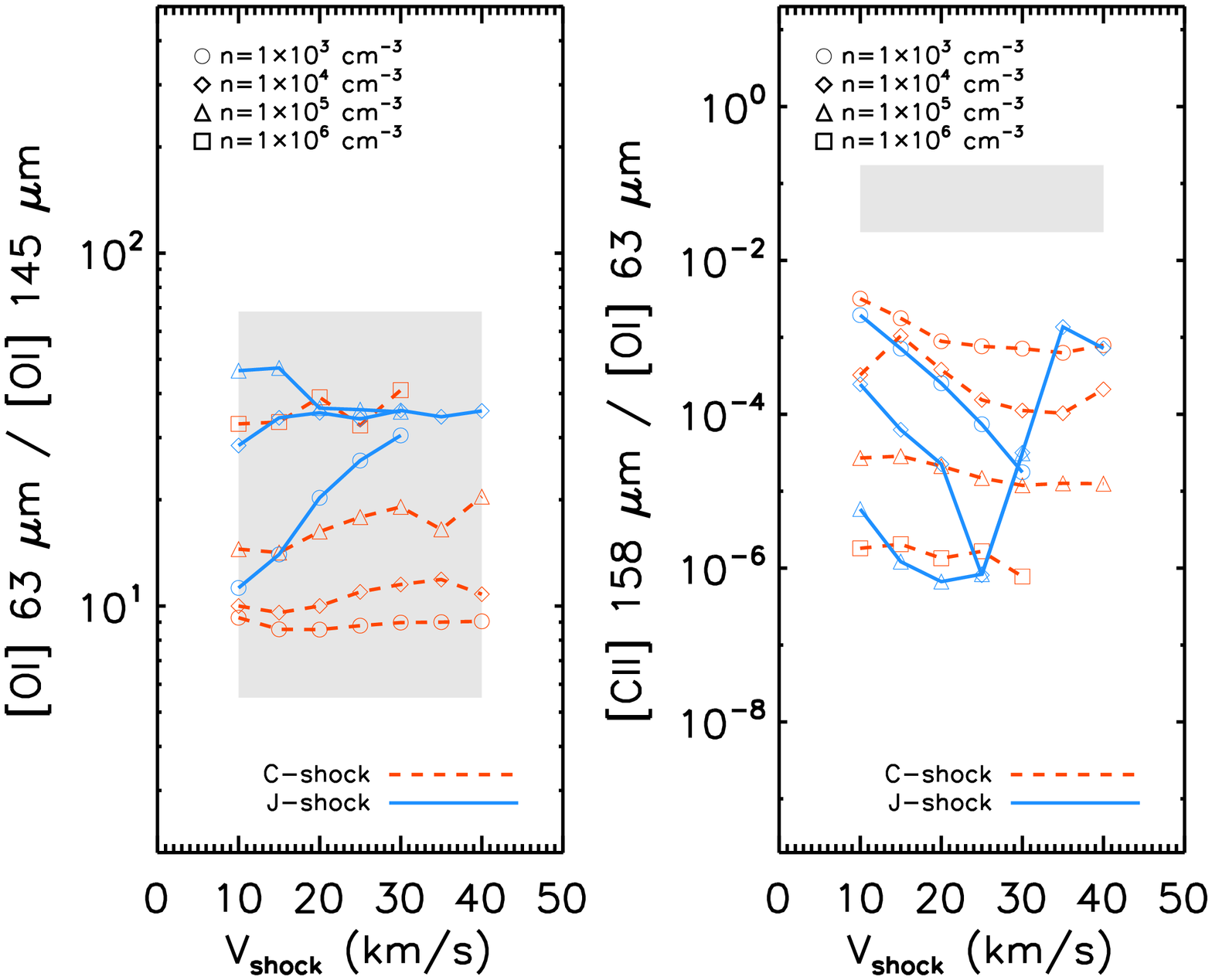}
    \caption{Observed atomic line ratios compared to C-type (red dashed lines) and J-type (blue lines) shock model predictions from \cite{Flower2015} for shock velocities ($V_{\rm shock}$) between 10 and 40 km/s. The circles, diamonds, triangles, and squares correspond to pre-shock densities ($n$) of 10$^{\rm 3}$ cm$^{\rm -3}$, 10$^{\rm 4}$ cm$^{\rm -3}$, 10$^{\rm 5}$ cm$^{\rm -3}$, and 10$^{\rm 6}$ cm$^{\rm -3}$, respectively. The range of the observed atomic ratios is represented by the shaded region.}
    \label{figure:atomicShockModels}
\end{figure}

Figure \ref{figure:atomicShockModels} shows the observed atomic line ratios compared to shock model predictions by \cite{Flower2015}. Both C- and J-type shocks  can reproduce a [OI]63/[OI]145 line ratio for a wide range of shock parameters ($V_{\rm shock}$ and $n$). However, such models fail to explain the observed [CII]158/[OI]63 ratios. This is not surprising as [CII] is thought to arise in PDRs. 
To further disentangle the origin of [OI] and [CII] lines, in Fig. \ref{figure:pdrModelsOI} we combined [OI]63/[OI]145 and [CII]158/[OI]63 atomic line ratios and compared them to the  PDR models of \cite{Kaufman1999} and the higher velocity ($V_{\rm shock}>50$ km/s) J-shock models by \cite{Hollenbach1989}.
Figure \ref{figure:pdrModelsOI} indicates that the observed ratios are all compatible with PDR 
models with densities between  $\sim10^{\rm 3}$ and $10^{\rm 6}$ cm$^{-3}$ and FUV fields $G_{\rm 0}$ $>$ 10$^{2}$;  only a few cases are compatible with fast J-shocks with low pre-shock densities ($V_{\rm shock}$ = 50--130 km s$^{-1}$, $n\sim10^{3}$ cm$^{-3}$), or both.
The sources whose  line ratios are compatible with shocks lie in a region in which the models overlap. Thus, it is impossible to discern which phenomenon is responsible for the emission. Similar PDR and shock parameters were obtained by \cite{Podio2012}.
Although  weak [CII] in shocks is  predicted by models \citep[][]{Flower2010}, the [CII] 157.74 $\rm \mu m$ line is more likely to originate in PDRs, as studies based on SOFIA/GREAT \citep{Heyminck2012} observations suggest \citep[e.g.][]{Sandell2015,Okada2015}.

In the few cases where [CII] 157.74 $\rm \mu m$ has been spectroscopically resolved with \textit{Herschel}/HIFI, it is clear that the line is not due to a disc,  but rather to a remnant envelope
or a diffuse cloud \citep[HD 100546][]{Fedele2013b}, or even to  to PDR emission in the outflow \citep[DG Tau,][]{Podio2013}. 
PACS observations of Upper Scorpius, have revealed low [CII] fluxes in two T Tauri stars \citep{Mathews2013}. These are early K-type protoplanetary systems without any signature of jet/outflow emission, further suggesting that [CII] 157.74 $\rm \mu m$ emission is PDR dominated.

\begin{figure*}[!ht]
\setcounter{figure}{13}
\centering
\includegraphics[width=0.7\textwidth,trim=0mm 0mm 0mm 0mm,angle=90,clip]{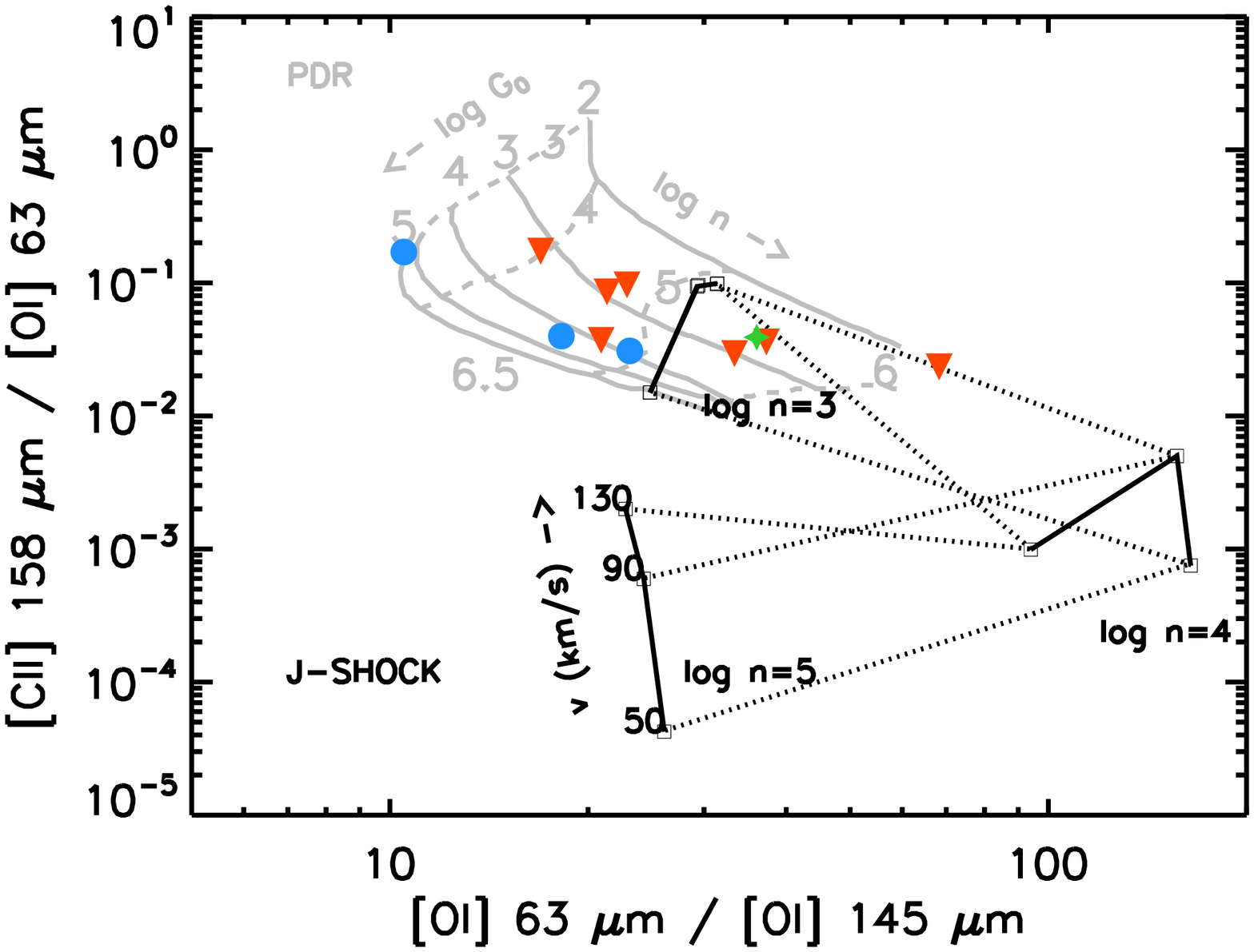}
\caption{Observed [OI]63/[OI]145 and [OI]63/[CII]158 line flux ratios plotted as a function of different emitting conditions. The PDR models (grey) are from \cite{Kaufman1999}; and J-shock models (black) from \cite{Hollenbach1989}. For the PDR models, the labels indicate the gas density ($n$) and the intensity of the FUV field ($G_{0}$), respectively. For the shock models, the labels denote pre-shock densities ($n$) and shock speeds ($V_{\rm shock}$). The data is plotted according to their SED class: blue circles are Class I, red triangles are Class II, and green stars are TD.}
\label{figure:pdrModelsOI}
\end{figure*}

\begin{figure*}[htpb]
\setcounter{figure}{14}
\centering
\includegraphics[width=0.35\textwidth,trim=0mm 0mm 0mm 0mm,angle=90,clip]{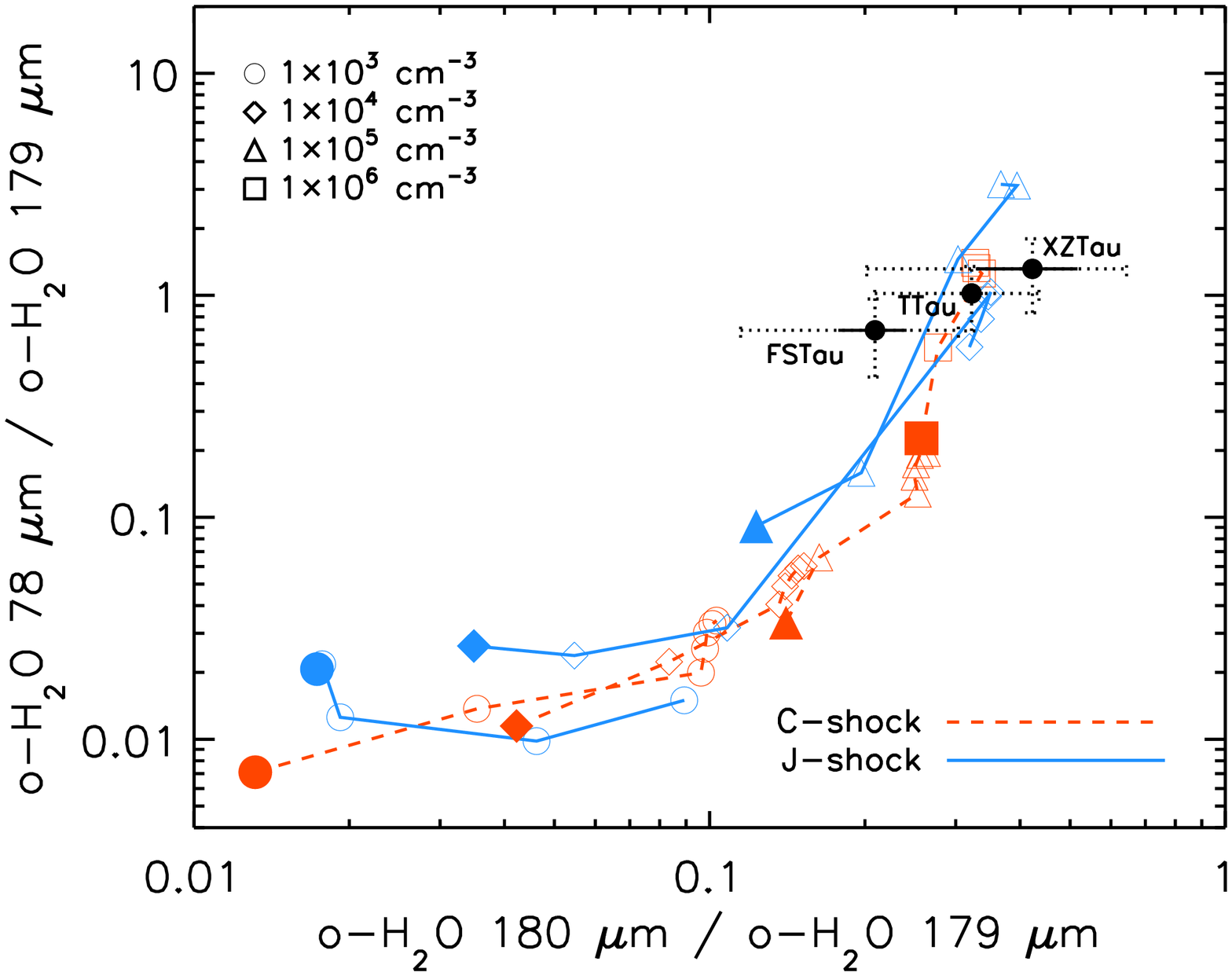}\includegraphics[width=0.35\textwidth,trim=0mm 0mm 0mm 0mm,angle=90,clip]{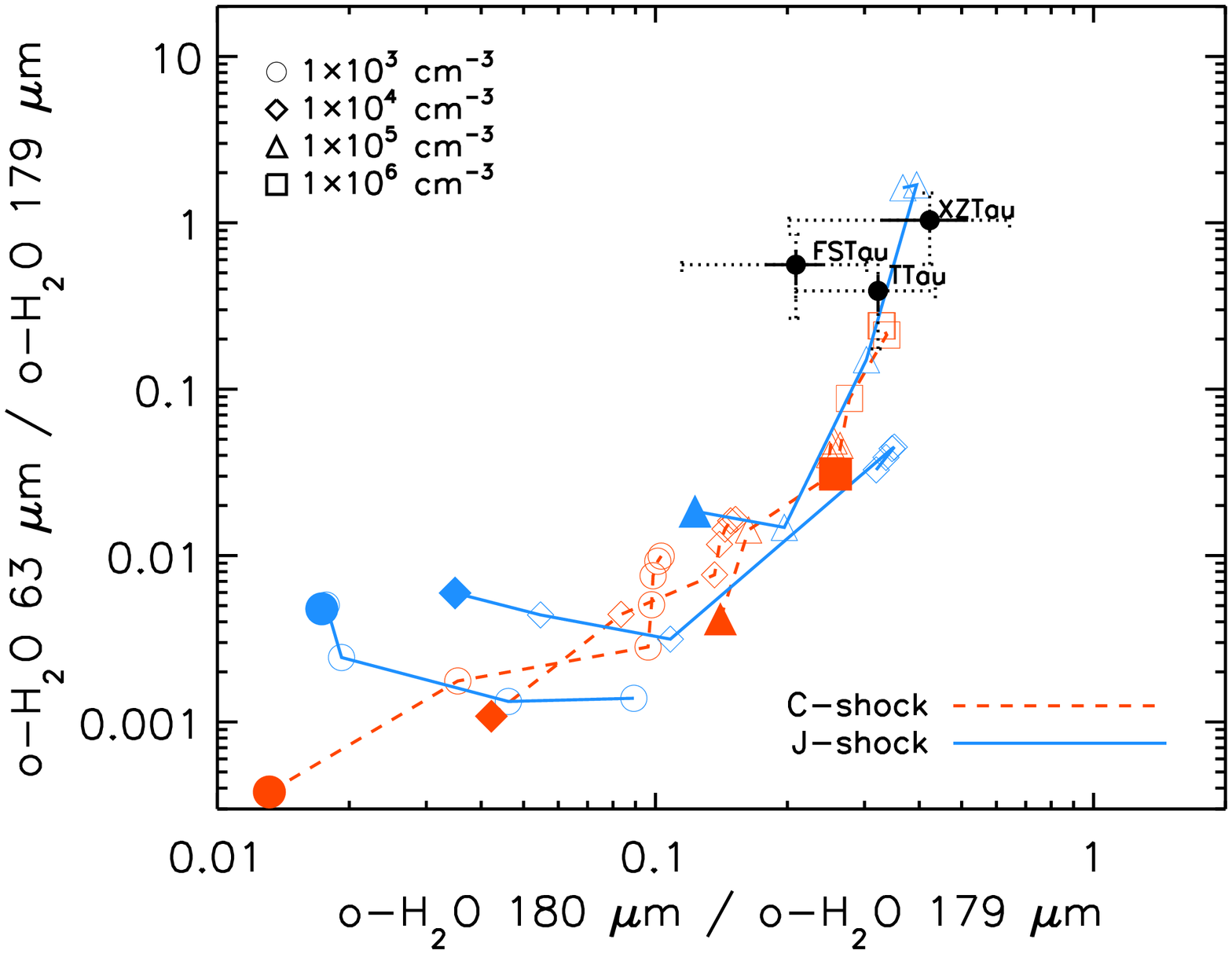}
\includegraphics[width=0.35\textwidth,trim=0mm 0mm 0mm 0mm,angle=90,clip]{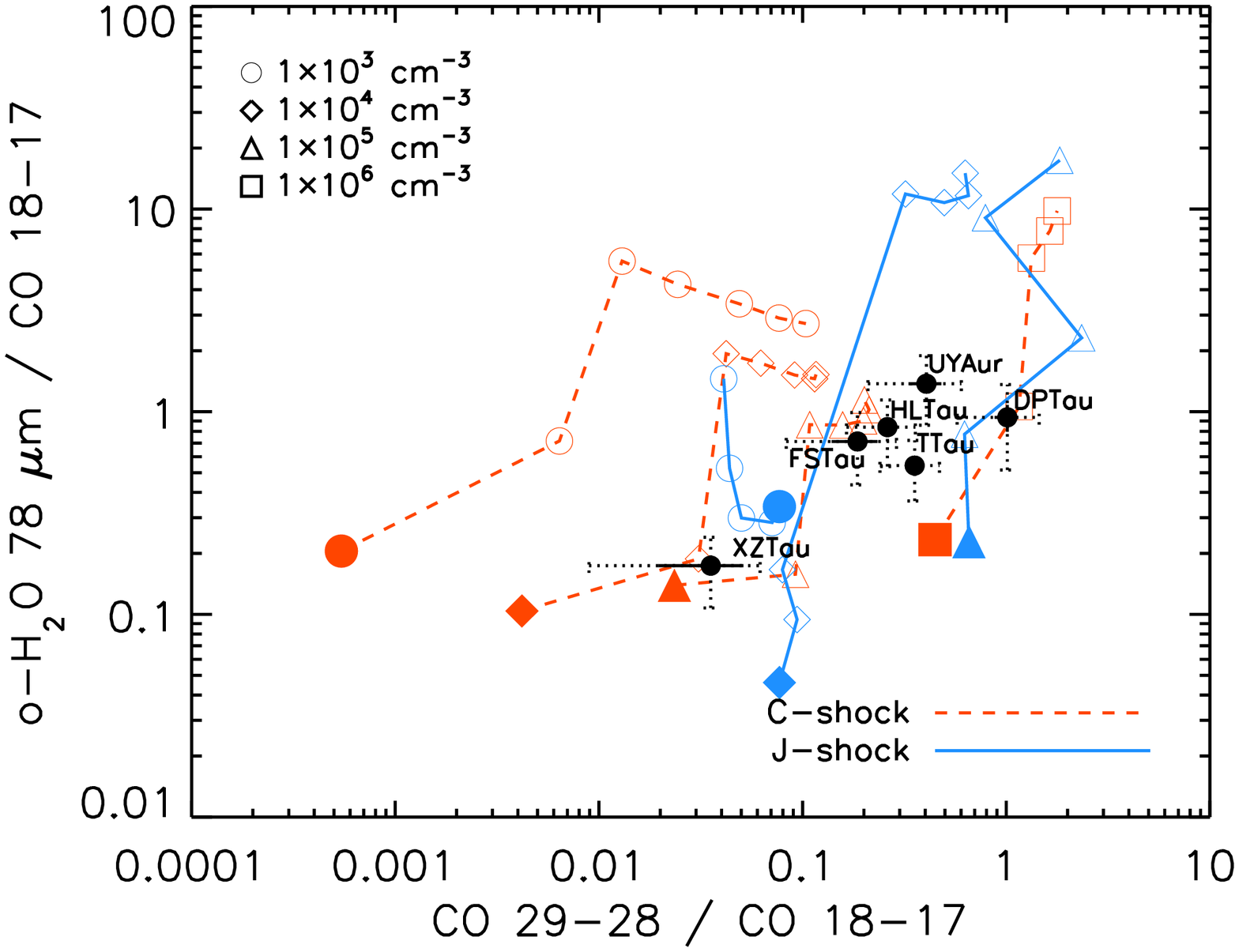}\includegraphics[width=0.35\textwidth,trim=0mm 0mm 0mm 0mm,angle=90,clip]{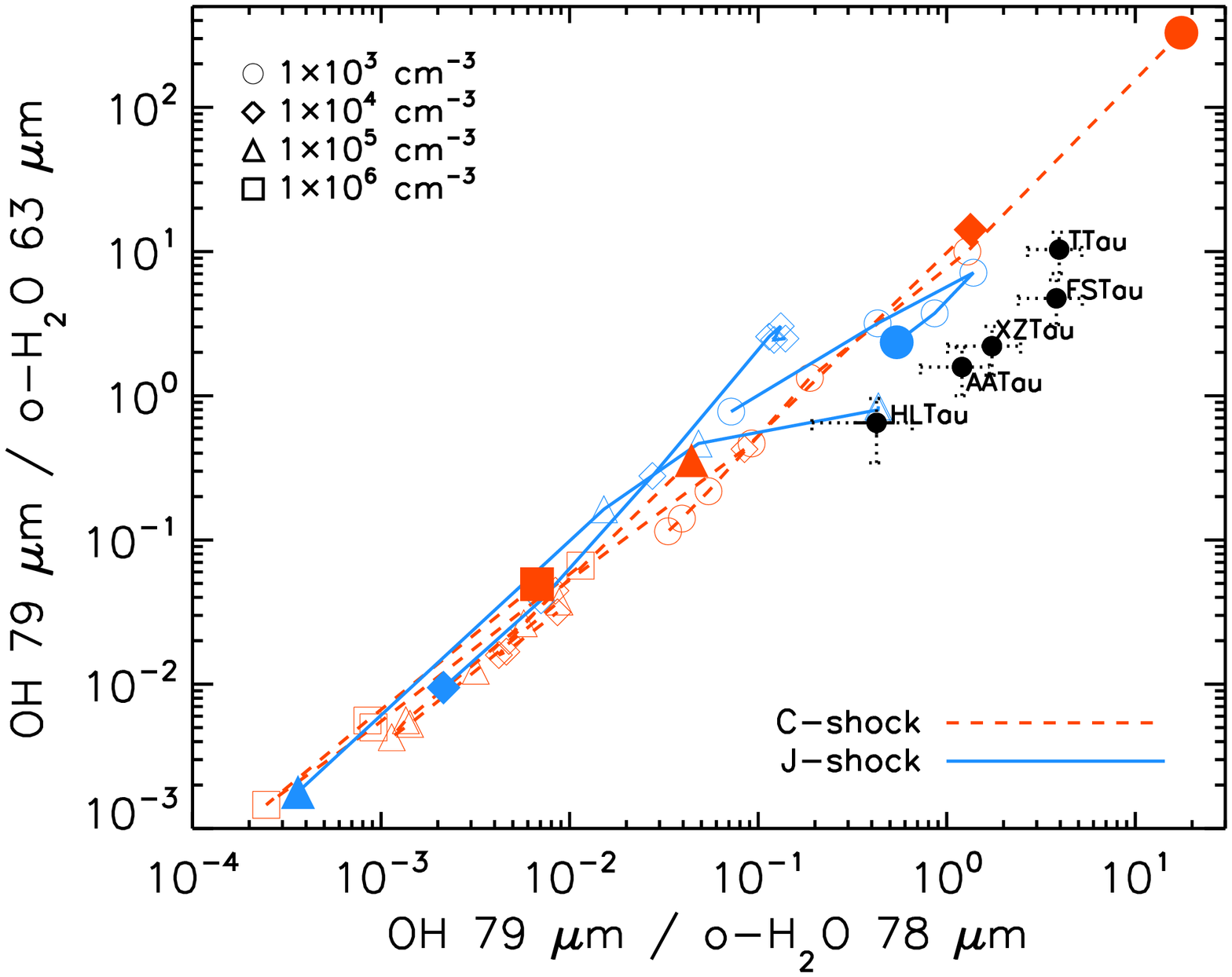}
\caption{Observed molecular line ratios compared with C-type (red dashed lines) and J-type (blue lines) model predictions from \cite{Flower2015}. The circles, diamonds, triangles, and squares correspond to pre-shock densities of 10$^{3}$ cm$^{-3}$, 10$^{4}$ cm$^{-3}$, 10$^{5}$ cm$^{-3}$, and 10$^{6}$ cm$^{-3}$, respectively, with shock velocities ($V_{\rm shock}$) from 10 km s$^{-1}$ to 40 km s$^{-1}$  (C-type) or from 10 km s$^{-1}$ to 30 km s$^{-1}$  (J-type). Filled symbols indicate the position of the lowest velocity ($V_{\rm shock}$=10 km s$^{\rm -1}$). Black filled circles represent detections and arrows are upper limits. Solid error bars are the intrinsic errors of the line fluxes and dotted error bars are obtained by adding  a 30\% error due to the PACS flux calibration.}
\label{figure:molecShockModels}
\end{figure*}

\subsubsection{Molecular line ratios as tracers of  excitation conditions}
\label{section:molecularRatios}
Figure \ref{figure:molecShockModels} shows  a combination of molecular line ratios compared to  C-type and J-type shock models from \cite{Flower2015}. The excitation conditions are, within errors, compatible with both C- and J-type shocks, with pre-shock densities between 10$^{4}$ cm$^{-3}$ and 10$^{6}$ cm$^{-3}$ and $V_{\rm shock}=$ 15--30 km s$^{-1}$ for most of the sources. The agreement between observations and shock model predictions depends on the specific line ratio \citep[see][]{Karska2014b} and the evolutionary stage of each source.
Similar conditions are found in the few cases where individual sources have been compared to shock models \citep{Lee2013,Dionatos2013}. However, we must stress that C-type shocks are probably the main driver of the molecular emission (see below).

Non-outflow sources show low \emph{J} CO and  `hot' ($T_{\rm ex} >$ 1000 K) o-H$_{\rm 2}$O line detections. In addition, outflow sources also show high  \emph{J} CO and `cold' ($T_{\rm ex} <$ 700 K) o-H$_{\rm 2}$O lines.
This indicates that high \emph{J} CO transitions \citep[\emph{J}$_{\rm up} \geq 20$ in][]{Karska2014a} are harder to excite in discs \citep{Woitke2009a}, whereas shocks can account for such emission \citep{vanKempen2010,Visser2012}; in addition, the fact that combinations of hot and cold o-H$_{\rm 2}$O lines are compatible with shock models (upper panels of Fig. \ref{figure:molecShockModels}) with similar parameters suggests that H$_{\rm 2}$O (and CO) can arise in similar regions.

The OH/H$_{\rm 2}$O line ratios (see lower right panel of Fig. \ref{figure:molecShockModels}) can only reproduce the excitation conditions of HL Tau, which shows emission from very hot water lines \citep[][]{Kristensen2016}.
\cite{Karska2014b} could not reproduce those line ratios for 
their sample of Class 0/I source in Perseus.
The discrepancy between observations and models depends on the H$_{\rm 2}$O transition that are used because dissociative radiation (Lyman $\alpha$ photons) may have an impact on the composition of the pre-shocked gas \citep{Flower2015}, hence affecting the abundance of H$_{2}$O \citep{Melnick2015}.
Spitzer mid-IR observations of OH in DG Tau by \cite{Carr2014} support the idea of hot OH emission induced by dissociation of H$_{2}$O by FUV radiation. 
\textit{Herschel} observations of OH in the range 70--160 $\rm \mu m$ indicate that the OH emission originates from dissociative shocks in young stellar objects \citep[Class 0/I in][]{Wampfler2010}.
Further modelling of the OH radical by \cite{Wampfler2013} with radiative transfer codes of spherical symmetric envelopes could not reproduce the OH line fluxes nor the line widths, strongly suggesting that the OH is coming from shocked gas.

There is evidence pointing to C-type rather than to J-type shocks as the main mechanisms driving the excitation of molecular FIR lines \citep[see][for a discussion]{Karska2014b}. 
The [OI]/H$_{\rm 2}$O line ratios can be used to discern between shock types.
We followed the criteria established by \cite{LeeJE2014}. The division between  J- and irradiated C-type shocks occurs when [OI]/H$_{\rm 2}$O$\sim$100, indicative of water photodissociation. 
The division between irradiated C-type and normal C-type occurs when [OI]/H$_{\rm 2}$O$\sim$1. 
Table  \ref{table:oxwaRatios} lists the [OI]/H$_{\rm 2}$O of our sample. The ratios of the outflow sources are compatible with irradiated C- shocks; and only in XZ Tau ([OI]/H$_{2}$O$>$100) a fast J-type shock may also contribute to the emission. The line ratios of the non-outflow sources BP Tau, GI/GK Tau, and IQ Tau are low (1$<$[OI]/H$_{\rm 2}$O$<$2). It is unclear whether such low ratios are compatible with weak outflow activity within the size of the PACS central spaxel (9.4") or a disc.

\begin{table}[htpb]
\setcounter{table}{2}
\begin{center}
\caption{ [OI]/o--H$_{2}$O line ratios.}
\label{table:oxwaRatios}
\begin{tabular}{lcccc}
\hline
\hline
\multicolumn{1}{l}{Target}&
\multicolumn{1}{c}{63/63}&
\multicolumn{1}{c}{63/78}&
\multicolumn{1}{c}{63/179}&
\multicolumn{1}{c}{63/180}\\
(1) & (2) & (3) & (4) & (5)\\
\hline
AA Tau&  1.78&  1.35&   --&   --\\
BP Tau&  1.25&   --&   --&   --\\
DL Tau&  2.33&   --&   --&   --\\
DP Tau&   --& 11.16&   --&   --\\
FS Tau& 29.34& 23.63& 16.43& 78.53\\
GG Tau&   --& 14.47&   --&   --\\
GI/GK Tau&  1.84&   --&   --&   --\\
Haro6-5B&  2.87&  3.66&   --&   --\\
Haro6-13&   --& 10.42&   --&   --\\
HL Tau&  8.86&  5.79&   --&   --\\
HN Tau&  9.19&   --&   --&   --\\
IQ Tau&  1.60&   --&   --&   --\\
IRAS04385+2550&   --&  8.22&   --&   --\\
RW Aur&   --&   --&  9.24&   --\\
RY Tau&  4.80&  7.27&   --&   --\\
T Tau& 74.12& 28.34& 28.88& 89.66\\
UY Aur&   --& 10.47& 24.19&   --\\
XZ Tau& 55.33& 43.56& 57.30&135.59\\
\hline
\end{tabular}
\end{center}
\textbf{Notes.} All the targets listed in Col. (1) have [OI] 63.18 $\rm \mu m$ detected. Columns (2) to (5) show the line ratios between the [OI] line at 63.18 $\rm \mu m$ and o--H$_{2}$O lines at 63.32, 78.74, 179.53, and 180.49 $\rm \mu m$, respectively; otherwise is indicated by `--'. 
\end{table}

\begin{table}[htpb]
\setcounter{table}{3}
\centering
\caption{Diameter (in au) of the emitting area for H$_{2}$O, CO, and OH estimated from \emph{C-} and \emph{J-}type shock models.}
\label{table:emitRegions}
\begin{tabular}{lccc}
\hline
\hline
\multicolumn{1}{l}{Source}&
\multicolumn{1}{c}{$D$(H$_{2}$O)}&
\multicolumn{1}{c}{$D$(CO)}&
\multicolumn{1}{c}{$D$(OH)}\\
\hline
AA Tau  & --        & --       & 102--109 \\
DP Tau  & --        & 56--122  & -- \\
FS Tau A& 33--99    & 184--395 & 261--277 \\
HL Tau  & --        & 339--821 & 174--183 \\
T  Tau  & 160--364  & 307--610 & 1613--1701 \\
UY Aur  & --        & 53--115  & 212--225 \\
XZ Tau  & 33--97    & 225--730 & 170--181 \\
\hline
\end{tabular}
\end{table}

To further test the shock scenario we follow the \cite{Flower2015} models to estimate the size ($D$) of the emitting areas (see Table \ref{table:emitRegions}) necessary to reproduce the observed molecular line fluxes. If the emission is indeed associated with shocked gas, the emitting areas have to be compatible with the observed scales (10 arcsec level) of molecular gas in T Tauri stars.
The OH areas are computed assuming that the same physical conditions ($V_{\rm shock}$ and $n$) as for HL Tau holds for all detected objects. Details of the derivation are in Appendix \ref{app:emitReg}.
We find that the emitting areas range from tens to hundreds of au, consistent with molecular emission being compact and unresolved with PACS at the distance of Taurus. 
The CO emitting areas are larger than those of H$_{\rm 2}$O, found to be between tens of au  and a few hundred. In particular, the size of the H$_{\rm 2}$O emitting area for T Tau is comparable with previous estimates \citep{Spinoglio2000,Podio2012}, and compatible with those obtained by \cite{Mottram2015} for Class 0/I sources.

Concerning the spatial extent of  H$_{2}$O compared to [OI] along the outflow, in the maps presented in \cite{Nisini2015} the two species (o-H$_{\rm 2}$O 179.53 $\rm \mu m$ and [OI] 63.18 $\rm \mu m$) show a similar extent. However, this  depends on the H$_{2}$O line considered because various transitions require different physical conditions for excitation. Even in the case of jet/outflow emission, the water lines can originate in much denser -- and probably more confined -- regions compared to [OI].

\subsection{Emission in a disc scenario}
\label{subsect:discEmission}

\subsubsection{Disc contribution to [OI]}
We now compare the observed fluxes with dust tracers, i.e. infrared continuum, to try to determine the contribution of the disc to the line emission.

Figure \ref{dustCorr} represents the flux of the [OI] 63.18 $\rm \mu m$ line as a function of \emph{Spitzer}/IRAC 3.6, 8.0 $\rm \mu m$, \emph{Spitzer}/MIPS \mbox{24 $\rm \mu m$}, and PACS 70 $\rm \mu m$ \citep[][]{Rebull2010,Luhman2010,Howard2013}.
The outflow sources are clearly brighter than non-outflow ones.  
While there is no obvious trend with 3.6 $\rm \mu m$ and the scatter is large at 8.0 $\rm \mu m$, there is a clear trend at 24 and 70 $\rm \mu m$. 
Longer continuum wavelengths are associated with colder dust, probing deeper in and further out disc regions. At longer wavelengths it is more likely that  continuum and FIR line emission both come from the same radial zone.

The distribution of Taurus sources in the [OI]--infrared diagrams is connected with the evolutionary status of the sources and the presence of outflows. 
The different behaviour of outflow/non-outflow sources in such diagrams has already been pointed out by \cite{Howard2013}. 
The correlation between the line flux and the dust emission for non-outflow sources suggests that both arise in the disc. The contribution of the disc to the gas emission in the outflow sources can be estimated assuming that the [OI]--70 micron correlation for non-outflow sources holds for all discs. In the case that outflow and non-outflow sources show similar trends with continuum emission, this test cannot distinguish clearly between the two origins or whether outflow emission does not have a relevant effect. A linear fit to our data for such correlation is given by

\begin{equation}
\label{eq:pacs}
        {\rm log_{10}} (F_{\rm [OI]63\mu m}) = -16.82 + 0.40\times {\rm log_{10}} (F_{ \rm 70 \mu m})
,\end{equation}

\noindent where $F_{\rm [OI]63\mu m}$ is the [OI] 63.18 $\rm \mu m$ line flux in W m$^{-2}$ and $F_{\rm 70 \mu m}$ is the continuum flux at 70 $\rm \mu m$ in Jy. 
Table \ref{table:discCont} shows the disc contribution in terms of the SED Classes. The relative contribution from the disc increases as the system evolves. In Class I sources the disc contributes $\sim$20\% and it keeps increasing until outflow activity dissipates. This is clear when comparing Class II sources with and without outflows (38\% compared to 100\%). This is in agreement with \cite{Podio2012} who obtained a disc contribution between 3\% and 15\% for Class I and Class II sources with outflows and showing [OI] 63.18 $\rm \mu m$ extended emission. 
In the case of T Tau the disc contribution ($<$1\%) is negligible.

\begin{table}[htpb]
\setcounter{table}{4}
\centering
\caption{Ranges and mean (median) values of disc contribution in percentage to the [OI] 63.18 $\rm \mu m$ line flux for different SED classes.}
\label{table:discCont}
\begin{tabular}{lccc}
\hline
\hline

\multicolumn{1}{l}{SED Class}&
\multicolumn{1}{c}{Range}&
\multicolumn{1}{c}{Mean (Median)}&
\multicolumn{1}{c}{N sources}\\
\hline
Class I       & 1\%--51\% &  19\% (18\%)   &4\\ 
ClassII+Jet& 3\%--100\% &  38\% (24\%)   &11\\
TD+Jet      & 43\%  &  43\% (43\%) &1\\
ClassII       & 65\%--100\% &  92\% (99\%)   &7\\
TD             & 100\% & 100\% (100\%)&2\\ 
\hline
\end{tabular}
\end{table}

\begin{figure*}[htpb]
\setcounter{figure}{15}
\centering
\includegraphics[scale=0.39,trim=0mm 0mm 0mm 0mm,angle=90,clip]{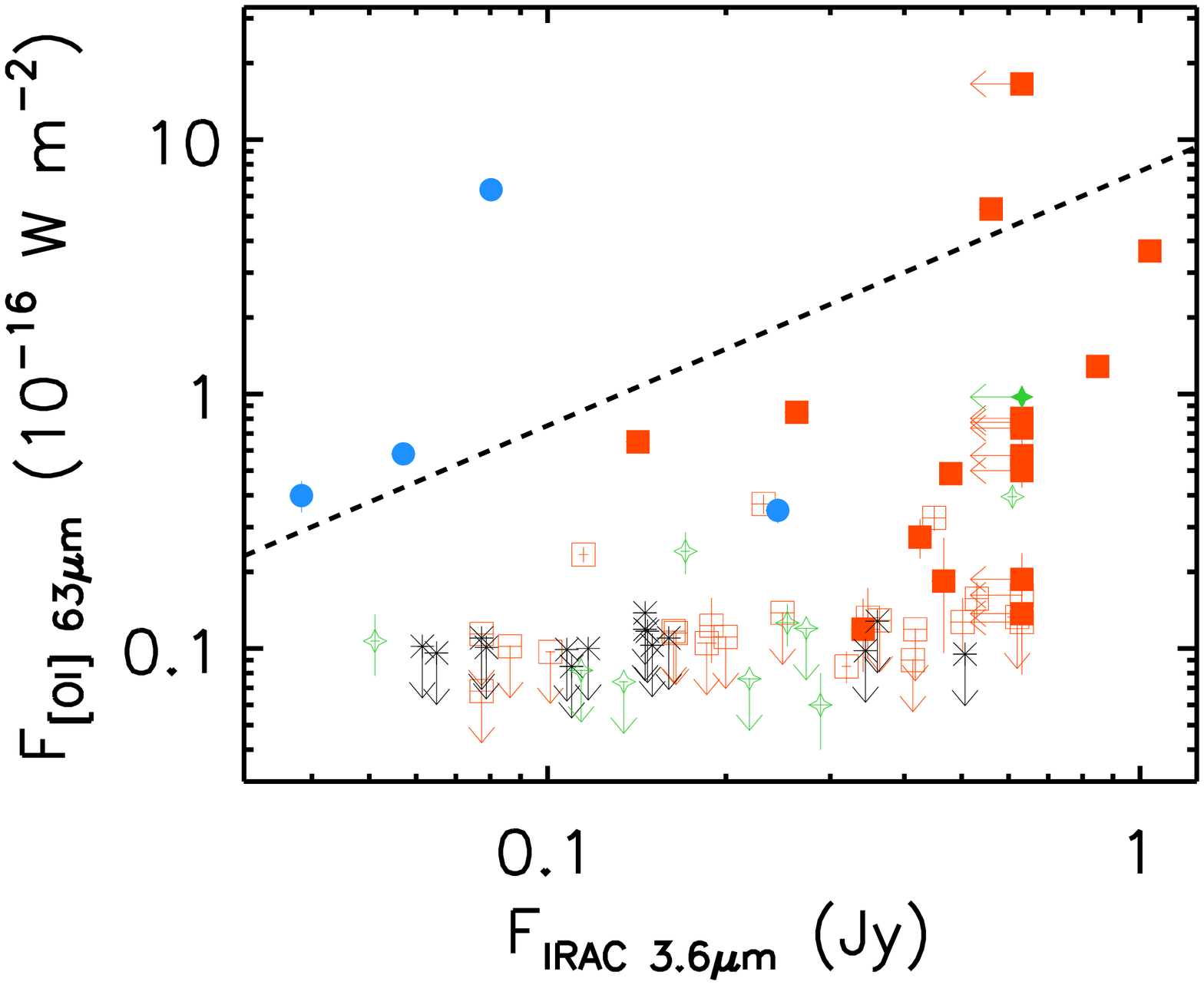}\includegraphics[scale=0.39,trim=0mm 0mm 0mm 0mm,angle=90,clip]{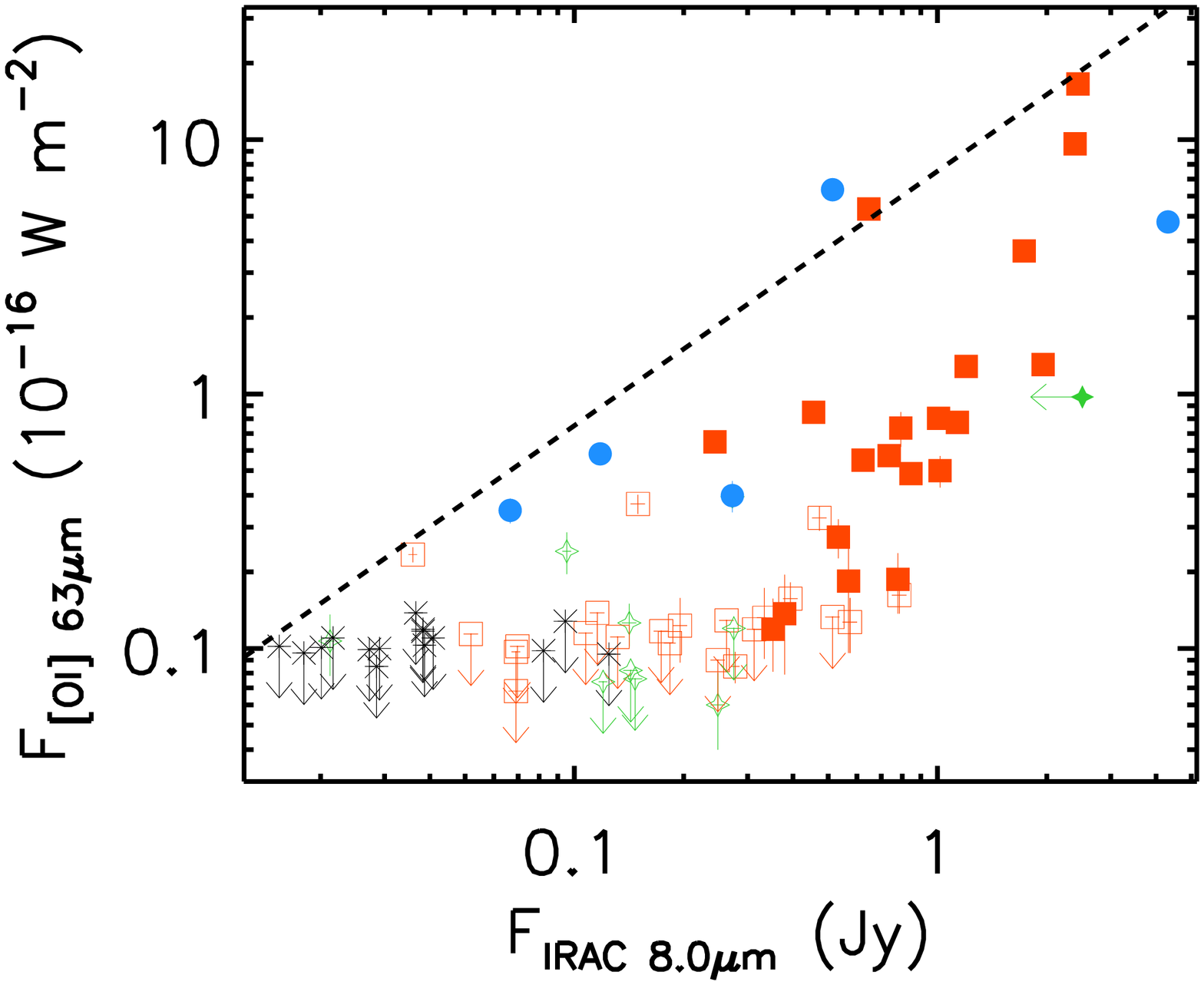}\\
\includegraphics[scale=0.39,trim=0mm 0mm 0mm 0mm,angle=90,clip]{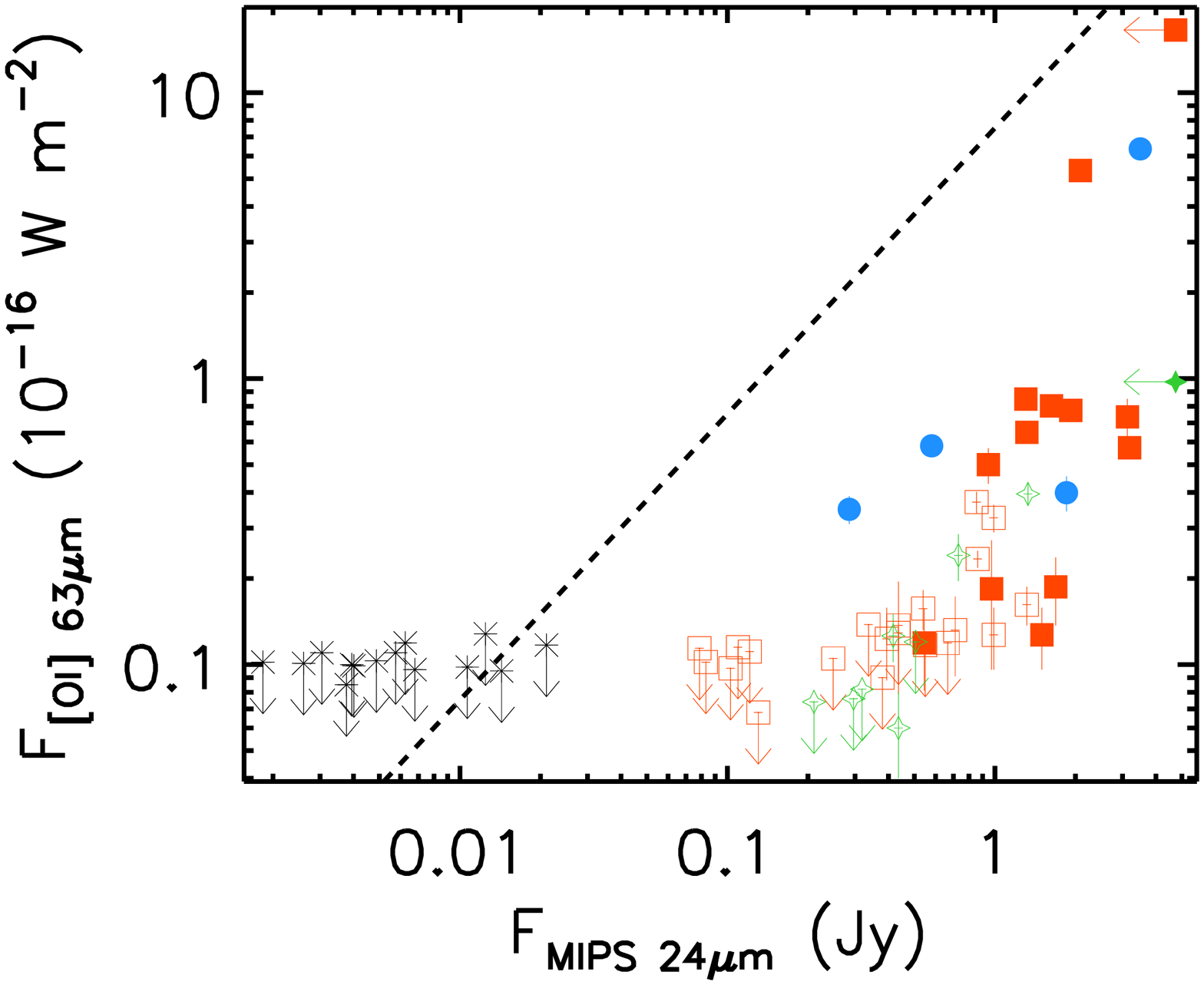}\includegraphics[scale=0.39,trim=0mm 0mm 0mm 0mm,angle=90,clip]{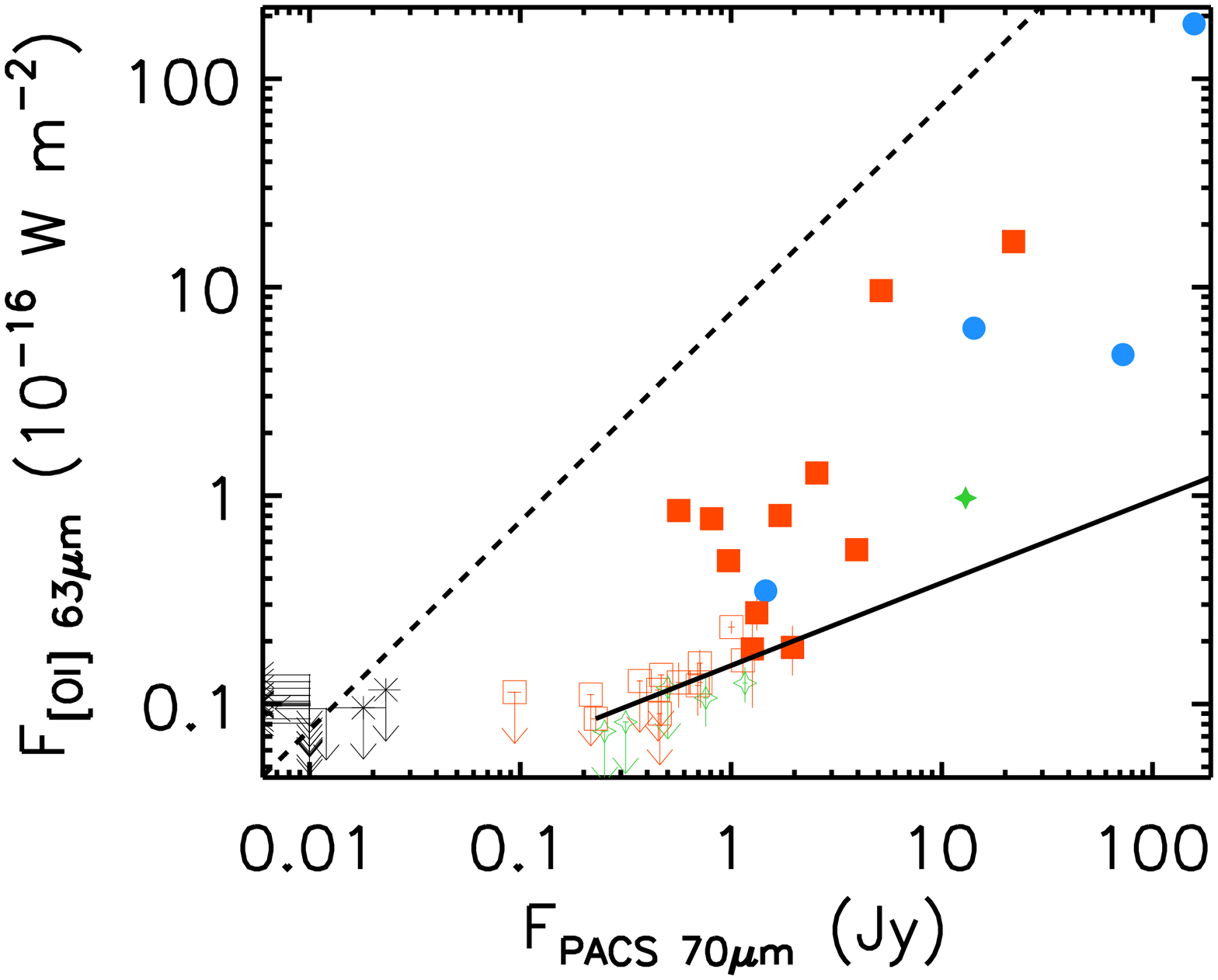}\\
\caption{ [OI] 63.18 $\rm \mu m$ flux versus: IRAC 3.6 $\rm \mu m$ (\emph{top left}); IRAC 8.0 $\rm \mu m$ (\emph{top right}); MIPS 24 $\rm \mu m$ (\emph{bottom left}); and PACS 70 $\rm \mu m$ (\emph{bottom right}) continuum fluxes.
Class I, Class II, Class III, and transition discs (TD) are represented by circles (\textit{blue}), squares (\textit{red}), asterisks (\textit{black}), and stars (\textit{green}), respectively. Filled symbols corresponds to outflow sources. Arrows indicate upper limits. The dashed lines are 1:1 relations and the solid line is a fit to disc sources as explained in the text.}
\label{dustCorr}
\end{figure*}

\subsubsection{ Disc contribution to {H$_{\rm 2}$O}}
We showed (see Sect \ref{section:correlations}) that the o--H$_{2}$O 63.32, 78.74 and 179.53 $\rm \mu m$ line luminosities correlate with  [OI] 63.18 $\rm \mu m$ ($\rho>$ 0.9).
The slopes of the fit become steeper for lower excitation energies of the water transition (Table \ref{table:slopes}).
The hydrostatic equilibrium models by \cite{Aresu2012} showed that the amount UV and X-ray radiation in a disc influences the line luminosities, and predicted the slopes between the different line luminosities. Their models do not match our observed water line luminosities and fail to reproduce the slopes we observe. 
The predictions from the fully parametrized DENT disc grid do reproduce the observed H$_{2}$O/[OI] and CO/[OI] ratios, but fail to explain the high H$_{2}$O and CO fluxes \citep{Podio2012}, suggesting that  an additional component and/or different gas-to-dust ratios are needed to account for these high
fluxes. In order to account for these discrepancies, higher disc masses and/or low dust-to-gas ratios, high FUV fluxes, or discs heated by X-rays \citep{Aresu2011,Aresu2012} are needed.
Indeed, \cite{Podio2013} showed that a model of DG Tau with a massive gaseous disc associated with strong UV and X-ray radiation reproduces the H$_{2}$O lines well. 
Other options include a more complicated inner disc structure, such as a puffed up inner rim \citep[][]{Aresu2011,Aresu2012}, gap, or hole.

\begin{table}[htpb]
\centering
\setcounter{table}{5}
\caption{Comparison of the slopes (with errors) between the strength of the water lines and OI observed, and the predictions by \cite{Aresu2012} for two different models. }
\label{table:slopes}
\begin{tabular}{ccccc}
\hline
\hline
\multicolumn{1}{l}{$\lambda$ ($\rm \mu m$)}&
\multicolumn{1}{l}{Transition}&
\multicolumn{1}{c}{Observed}&
\multicolumn{1}{c}{UV}&
\multicolumn{1}{c}{UV+X-ray}\\

\multicolumn{1}{c}{}&
\multicolumn{1}{c}{}&
\multicolumn{1}{c}{}&
\multicolumn{1}{c}{model}&
\multicolumn{1}{c}{model}\\
\hline
  63.32& 8$_{18}$--7$_{07}$ &  0.41(0.06) & 0.96(0.07)  & 0.73(0.12)  \\
  78.74& 4$_{23}$--3$_{12}$ &  0.65(0.10) &  -- &  -- \\
  179.53&2$_{12}$--1$_{01}$ &  0.81(0.18) & 0.02(0.01)  & 0.32(0.07)  \\
\hline
\end{tabular}
\end{table}

\subsection{Jet or disc?}
From our analysis described above, a jet/outflow origin is favoured for the strong FIR lines because of (1) the observed extended emission in outflow sources \citep[see maps in e.g.][]{Nisini2010,Podio2012,Nisini2015}, (2) the detections predominantly in outflow sources and the correlations between emission lines, and (3) the compatibility of line ratios with shock models especially for Class I sources for which the disc contribution is estimated to be smaller than 20\%. Nonetheless, we note the following: (1) several detections of [OI] and o--H$_{2}$O at 63 $\rm \mu m$ in non-outflow sources; (2)  correlations between the line fluxes and continuum  for  24 and 70 $\rm \mu m, which$ point to a disc origin;  (3)  compact molecular emission   within the PACS beam also observed in Class I sources, which have (small) outflows; and (4)  gas line ratios reproduced satisfactorily by the disc models \citep[e.g.][]{Kamp2011,Aresu2012}. Therefore, the excitation mechanism in discs should not be the main problem. Different dust-to-gas ratios, larger scale heights, or inner gaps would make the disc likely hotter than a normal continuous disc model, but other agents may only increase the emitting surface area.
This particular issue cannot be fully understood without spectroscopically and spatially resolved observations, so that the location and dynamics of the emitting gas can be pin-pointed; however, for FIR observations this is hard to obtain. 
A promising alternative to  this problem could be to use lines which are likely co-spatial with some of the FIR lines, such as ro--vibrational lines of CO \citep[see e.g.][]{Banzatti2015}. This way, one could determine whether the lines are purely associated with a Keplerian disc or coming from shocks via jets and/or winds. Within a disc, gas kinematics derived from resolved line profiles may help disentangle various regions, especially at shorter wavelengths where the spatial resolution is higher.

\section{Summary and conclusions}
\label{conclusions}
We provide a catalogue of line fluxes in the FIR (63--190 $\rm \mu m$) for T Tauri stars located in Taurus, surveyed with Herschel/PACS as part of the GASPS project. The species 
observed include [OI], [CII], CO, $\rm H_{2}O,$ and OH. The origin of the atomic and molecular emission is investigated by comparing line fluxes and line ratios to shock and disc models. The main conclusions are as follows:

\begin{itemize}

\item Outflow sources exhibit brighter atomic and molecular emission lines and higher detection rates than non-outflow sources. In agreement with previous studies, atomic and molecular FIR line emission in T Tauri systems is observed to decrease with evolutionary stage.

\smallskip
\item The [OI] line emission is brighter and more often detected in systems with signs of jet/outflows and in early phases of stellar evolution (Classes I and II), suggesting a dominant contribution from shocks in young outflow systems. In these systems the emission is spatially extended (10 arcsec level). 
Slow ($V_{\rm shock}$=10--50 km s$^{-1}$) J- and C-type shocks with densities between $n$=10$^{3}$ and 10$^{6}$ cm$^{-3}$) can reproduce the observed [OI]63/[OI]145 ratios. These models do not reproduce the [CII]158/[OI]63 line ratios well.
Fast ($V_{\rm shock}$=50--130 km s$^{-1}$) J-shocks at low densities ($n$=10$^{3}$ cm$^{-3}$) and/or PDR models ($n>$10$^{\rm 3}$ cm$^{\rm -3}$, $G_{\rm 0}>$10$^{\rm 2}$)  can reproduce the combined [OI]63/[OI]145 and [CII]158/[OI]63 line ratios.

\smallskip
\item The main contribution to the [CII] 157.74 $\rm \mu m$ line  most probably come from PDR emission from the disc and its surroundings. A (small) contribution from shocks is also expected, as detections only in outflow sources and shock models suggest. The precise origin of the [CII] line has to be better constrained with spatially and spectroscopically resolved observations.

\smallskip
\item The observed correlations support the interpretation of jet/outflows (when present) as the dominant contributor to the FIR line emission, and points to a common excitation mechanism, i.e. shocks. The broad wings of the [OI] 63.18 $\rm \mu m$ line and its correlation with [OI] 6300 $\mbox{\rm \AA}$ suggest the presence of several velocity components, i.e. a jet origin for the HVC and a combination of disc, envelope, and wind for the LVC.

\smallskip
\item Molecular line emission (H$_{2}$O, CO, and OH) is also mainly detected in outflow systems. Slow ($V_{\rm shock}$=15--30 km s$^{-1}$) C-type shocks with densities between $n$=10$^{4}$ and 10$^{6}$ cm$^{-3}$ can account for these fluxes; with emitting areas ranging from tens to hundreds of au. 
The OH/H$_{\rm 2}$O line ratios are typically overestimated in J-type shocks.

\smallskip
\item The correlations with photometric bands (24 and 70 $\rm \mu m$) indicate that the contribution from the disc to the [OI] 63.18 $\rm \mu m$ line flux may be up to $\sim$50\% for the jet/outflow sources, strongly depending on the evolutionary status of the source. When the jet/outflow activity decreases, the disc contribution relative to the line fluxes increases significantly ($>$65\% in Class II sources).

\smallskip
\item Although there are clear indications that the emission is dominated by the outflow, massive discs and/or low dust-to-gas ratios may also explain the observed high molecular line fluxes.
\end{itemize}

\noindent The low spectral and spatial resolution from PACS are not sufficient to unambiguously determine what fraction of the line emission comes from the disc, outflow, or the surrounding envelope. Spatially and/or velocity resolved observations are needed to pin-point the origin of the emission lines. In this regard, the instruments on board SOFIA may have the potential to resolve the brightest lines. Models which include both a disc and a jet/outflow and their interaction are needed to accurately interpret the multiwavelength observations of young T Tauri stars.

\begin{acknowledgements}
M. Alonso-Martinez, C. Eiroa, and G. Meeus are partially supported by AYA2011--26202 and AYA2014--55840--P. G.M. is supported by RyC--2011--07920. P.R.M. acknowledges funding from the ESA Research Fellowship Programme. I.K. acknowledges funding from the European Union Seventh Framework Programme FP7-2011 under grant agreement No. 284405. L.P. has received funding from the European Union Seventh Framework Programme (FP7/2007−2013) under grant agreement No. 267251. We would also like to thank the anonymous referee for the constructive comments which certainly helped to substantially improve the quality of the paper.
This research has made use of the SIMBAD database, operated at CDS, Strasbourg, France.
\end{acknowledgements}

\bibliographystyle{aa1} 
\bibliography{malonso_herschel_V3.bbl}

\begin{appendix}
\section{Stellar parameters}
\label{stellarPar}
In Table \ref{stellarParam} we list the literature properties of the sample including SED classes, spectral types, effective temperatures, mass accretion rates, stellar luminosities, X-ray luminosities, accretion luminosities, ages, disc masses, and separation if the sources are multiple systems. The targets are split into outflows and non-outflows depending on their signs of jet/outflow activity as explained in the text (see Sect. \ref{sample_parameters}).

\begin{landscape}
 \begin{table}[htpb]
 \setcounter{table}{0}
 \tiny
  \caption{Sample Properties}
 \begin{center}
\label{stellarParam}
 \begin{tabular}{lcccccccccc}
 \hline
 \hline
 [1]& [2]& [3]& [4]& [5]& [6]& [7]& [8]& [9]& [10]& [11]\\
 \hline
\multicolumn{1}{l}{Target} &
 \multicolumn{1}{c}{SED Class} &
 \multicolumn{1}{c}{SpT$^{a}$} &
 \multicolumn{1}{c}{T$^{a}$} &
 \multicolumn{1}{c}{$\dot{M}_{\rm acc}^{d}$} & 
 \multicolumn{1}{c}{$L_{*}^{c}$} &
 \multicolumn{1}{c}{$L_X^{e}$} &
 \multicolumn{1}{c}{$L_{\rm acc}^{d}$} &
 \multicolumn{1}{c}{Age$^{c}$} &
 \multicolumn{1}{c}{$M_{\rm disc}^{f}$} &
 \multicolumn{1}{c}{Sep. (Pair)$^{g}$} \\
 
  [--] &[--]&[--]&  [K]& [10$^{-8} $M$_\odot$ yr$^{-1}$]& [L$_\odot$]& [L$_{\odot}$] & [L$_\odot$]& [Myr]& [M$_\odot$]& [arcsec]\\
 \hline
 \multicolumn{11}{c}{Outflows}\\
 \hline
  AA Tau&II&M0.6  &3770&2.51&0.74&1.240&0.21&2.7&0.01  &...\\
  CW Tau&II&K3  &4470&5.27  &1.35&2.844&1.01  &5.8&0.002 &...\\
  DF Tau&II&M2.7  &3450&10.05&1.60&...  &0.22&0.1&0.0004&0.09 (A--B)\\
  DG Tau&II&K7 &4020&25.25  &0.90&...  & 1.26 &0.6&0.02  & ... \\
  DG Tau B&I& M?$^{b}$&$<$4000&... &... &...&...&...&...  & ...  \\
  DL Tau&II&K5.5  &4190&2.48  &0.70&...  &0.24  &2.9&0.09  &...\\
  DO Tau&II&M0.3  &3830&3.12&1.20&...  &0.17&0.6&0.007 &...\\
  DP Tau&II&M0.8&3730&0.04  &0.41&...  &0.01  &2.6&$<$0.0005&...\\
  DQ Tau&II&M0.6&3770&0.59&...&...  &0.03&...& 0.02 &...\\
  FS Tau&II/Flat&M2.4&3510&... &0.32&3.224&...&2.8 &0.002&0.23 (A--B)\\
  GG Tau&II&K7.5(A)+M5.8(B)&3930(A)+2890(B)&7.95&0.84(A)+0.71(B)&...&0.41&2.3(A)+1.5(B)&0.2&0.25 (Aa--Ab)\\
  Haro 6--5 B&I&M4&3340$^{c}$&...&0.02&17.551&...&...&...&0.23 (A--B)\\
  Haro 6--13&II&M1.6(E)+K5.5(W)&3670(E)+4190(W)&...&1.30&0.799&...&...&0.01&...\\
  HL Tau&I&K3&4470&0.35&6.60&3.838&0.05&0.1&...&...\\
  HN Tau&II&K3(A)+M4.8(B)&4470(A)+3050(B)&0.49&0.70&...&0.11&22.0&0.0008&3.11 (A--B)\\
  HV Tau&I?&M4.1&3120&...&1.23&...&...&0.3&0.002&3.98 (A--C)\\
  IRAS 04158+2805&I&M3&3470$^{c}$&...&0.41&0.882&...&0.9&...&...\\
  IRAS 04358+2550&II&M0&3850$^{c}$&...&0.36&0.401&...&...&...&...\\
  RW Aur&II&K0(A)+K6.5(B)&5110(A)+4160(B)&...&3.20&...&...&2.8&0.004&1.42 (A--B)\\
  RY Tau&T?&G0&5930&78.18&7.60&5.520&10.21&1.8&0.02&...\\
  SU Aur&II&G4&5560&18.03&12.90&9.464&3.17&2.6&0.0009&...\\
  T Tau&II/I&K0&5250$^{c}$&23.62& 15.50&8.048&3.89&1.2&0.008&0.70 (A--B)\\
  UY Aur&II&K7&4020&6.74&3.10&...&0.43&0.8&0.002&0.88 (A--B)\\
  UZ Tau&II&M1.9+M3.1&3560(A)+3410(B)&0.45&1.60(A)+0.52(B)&0.890&0.02&0.2(A)+0.6(B)&0.02&3.54 (A--Ba)\\
  V773 Tau&II&K4&4460&11.95 &2.31&9.488&1.03&4.1&0.0005&0.06 (A--B) \\
  XZ Tau&II&M3.2&3370&0.71&3.90&0.962&0.02&0.6&...&0.30 (A--B)\\
 \hline
 \multicolumn{11}{c}{Non-outflows}\\
 \hline  
  BP Tau&II&M0.5&3510&2.80&0.96&1.365&0.22&1.8&0.02&...\\
  CIDA 2&III&M5.5&3200$^{c}$&...&0.29&0.178&...&0.5&$<$0.0007&...\\
  CI Tau&II&K5.5&4190&12.35&0.87&0.195&2.53&2.1&0.03&...\\
  Coku Tau/4&II&M1.1&3690&...&0.39&...&...&3.0&0.0005&...\\  
     CX Tau&T&M2.5&3580&0.04&0.41&...&0.002&1.3&0.001&...\\
  CY Tau&II&M2.3&3515&0.18&0.47&1.94&0.01&1.8&0.006&...\\
  DE Tau&II&M2.3&3515&1.32&0.81&...&0.06&0.4&0.005&...\\
  DH Tau&II&M2.3&3515&0.77&0.68&8.458&0.05&1.0&0.003&...\\
  DI Tau&III&M0.7&3680&0.01&0.62&1.568&0.001&1.9&...&...\\
  DK Tau&II&K8.5(A)+M1.7(B)&3960(A)+3650(B)&15.40&1.32&0.916&0.98&1.0&0.005&2.30 (A--B)\\
  DM Tau&T&M3&3410&0.09&0.25&...&0.01&2.8&0.02&...\\
  DN Tau&T&M0.3&3830&0.43&0.91&1.155&0.03&0.9&0.03&...\\
  DS Tau&II&M0.4&3860&2.21&0.72&...& 0.33&7.8&0.006&...\\
   
 FF Tau&III&K8&3990&0.02&0.81&0.796&0.002&2.3&...&...\\
 FM Tau&II&M4.5&3100&0.99&0.32&0.532&0.11&4.5&0.002&...\\
 FO Tau&T&M3.9&3190&1.05&0.88&0.065&0.03&0.4&0.0006&0.15 (A--B)\\
 FQ Tau&II&M4.3&3110&0.21&0.32&0.049&0.01&1.7&0.001&0.76 (A--B)\\
 FT Tau&II&M2.8&3430&...&0.38&...&...&...&0.01&...\\
 FW Tau&III&M5.8&2870&...&0.23&...&...&1.2&0.0002&0.08 (A--B)\\
 FX Tau&II&M2.2&3510&0.44&1.02&0.502&0.02&0.5&0.0009&0.89 (A--B)\\
 \hline
 \end{tabular}
 \end{center}
 \end{table}
 \end{landscape}

 \begin{landscape}
 \begin{table}[htpb]
 \setcounter{table}{0}
%\label{stellarParam}
 \tiny
  \caption{(Continued)}
 \begin{center}
 \begin{tabular}{lcccccccccc}
 \hline
 \hline
 [1]& [2]& [3]& [4]& [5]& [6]& [7]& [8]& [9]& [10]& [11]\\
 \hline
\multicolumn{1}{l}{Target} &
 \multicolumn{1}{c}{SED Class} &
 \multicolumn{1}{c}{SpT$^{a}$} &
 \multicolumn{1}{c}{T$^{a}$} &
 \multicolumn{1}{c}{$\dot{M}_{\rm acc}^{d}$} & 
 \multicolumn{1}{c}{$L_{*}^{c}$} &
 \multicolumn{1}{c}{$L_X^{e}$} &
 \multicolumn{1}{c}{$L_{\rm acc}^{d}$} &
 \multicolumn{1}{c}{Age$^{c}$} &
 \multicolumn{1}{c}{$M_{\rm disc}^{f}$} &
 \multicolumn{1}{c}{Sep. (Pair)$^{g}$} \\
 
  [--] &[--]&[--] & [K]& [10$^{-7} $M$_\odot$ yr$^{-1}$]& [L$_\odot$]& [L$_{\odot}$] & [L$_\odot$]& [Myr]& [M$_\odot$]& [arcsec]\\
 \hline
 \multicolumn{11}{c}{Non-outflows}\\
 \hline
 GH Tau&II&M2.3&3515&...&0.79&0.109&...&0.5&0.0007&0.31 (A--B)\\
 GI/GK Tau&II&M0.4/K6.5(A)&3770/4160(A)&2.96/1.99&0.85/1.17&0.833/1.471&0.31/0.13&3.5/1.4&...& 13.15\\
 GM Aur&T&K6&4115&0.60&1.00&...&0.12&9.6&0.03&...\\
 GO Tau&II&M2.3&3515&0.22&0.27&0.249&0.03&5.2&0.07&...\\
 Haro 6--37&II&K8(A)+M0.9(B)&4050(A)+3720(B)&1.41&...&...&...&0.06&0.01&2.62 (A--B)\\
 HBC 347&III&K1&5080$^{c}$&...&...&...&...&...&$<$0.0004&...\\
 HBC 356&III&K2&4905$^{c}$&...&0.17&...&0.01&...&$<$0.0004& ...\\
 HBC 358&III&M3.9&3200& 0.001&0.28&...&0.0001&3.0&$<$0.0005& ...\\
 HD 283572&III&G4&5560&6.12&6.46&13.003$^{c}$&1.16&5.8&...&...\\
 HK Tau&II&M1.5&3670&0.32&0.47&0.079&0.03&2.2&0.004&2.34 (A--B)\\
 HO Tau&II&M3.2&3365&0.10&0.20&0.047&0.02&5.4&0.002&...\\
 IP Tau&T&M0.6&3770&0.06&0.43&...&0.01&3.1&0.03&...\\
 IQ Tau&II&M1.1&3690&0.82&0.65&0.416&0.04&1.4&0.02&...\\
 J1--4872&III&M0.6(A)+M3.7(B)&3770(A)+3250(B)&...&0.72&...&...&2.8&...&3.40 (A--B)\\
 LkCa 1&III&M3.6&3350&...&0.37&0.232&...&0.6&...&...\\
 LkCa 3&III&M2.4&3510&0.01&1.66&...&0.0004&0.2&$<$0.0004&0.48 (A--B)\\
 LkCa 4&III&M1.3&3680&0.20&0.85&...&0.02&2.2&$<$0.0002& ...\\
 LkCa 5&III&M2.2&3520&...&0.29&0.432&...&1.9&$<$0.0002&...\\
 LkCa 7&III&M1.2&3680&0.26&0.89&...&0.02&2.1&$<$0.0004&1.02 (A--B)\\
 LkCa 15&T&K5.5&4190&0.44&0.83&...&0.06&5.2&0.05&...\\
 UX Tau&T&K0(E)+M1.9(W)&4870(E)+3560(W)&1.35&1.30(A)+0.35(B)&...&0.13& 8.0(A)+7.5(B)&0.005&5.86 (A--B)\\
 V710 Tau&II&M3.3(A)+M1.7(B)&3350(A)+3650(B)&0.22&0.63(A)+0.36(B)&1.378&0.02&1.8(A)+1.8(B)&0.007&3.17 (A--B)\\
 V807 Tau&II&K7.5&3930&4.26&4.10&1.049&0.27&0.2&...&0.30 (A--B)\\
 V819 Tau&II/III&K8&3990&0.27&0.81&2.205&0.03&2.3&0.0004&...\\
 V836 Tau&T&M0.8&3740&0.1&0.51&...&0.01&7.9&...&...\\
 V927 Tau&III&M4.9&2990&...&0.33&...&...&0.3&$<$0.0005&0.27 (A--B)\\
 V1096 Tau&III&M0&3850$^{c}$&...&1.30&4.139&...&0.5&$<$0.0004&...\\ 
 VY Tau&II&M1.5&3670&0.05&0.42&...&0.01&3.0&$<$0.0005&0.66 (A--B)\\
 ZZ Tau&II&M4.3&3100&0.04&0.69&...&0.001&0.4&...&...\\
 \hline
 \end{tabular}
 \end{center}
 \textbf{Notes.} [1]  Sample targets, [2] SED Class, [3] spectral type, [4] temperature, [5] accretion rate, [6] stellar luminosities, [7] X-ray luminosities, [8] accretion luminosities, [9] stellar age, [10] disc masses, [11] Separation in arcsec and components. The mass accretion rates were derived from the U band excess as follows:  first the U band photometry from \cite{Kenyon1995} was derredened using the \cite{Cardelli1989} extinction curve. Then the U band excess was derived by subtracting the photospheric contribution, and converted to accretion luminosity according to the empirical relation $\log{(L_{\rm acc}/L_{\rm \odot})} = 1.09 \times \log{(L_{\rm U} /L_{\rm \odot})} + 0.98$ in \cite{Gullbring1998}. The mass accretion rates were finally obtained using the relation $M_{\rm acc} = {{L_{\rm acc} R_{\rm \star}} \over {GM_{\star}(1-R_{\rm \star} / R_{\rm in})}}$, following the assumptions in \cite{Gullbring1998}.\\
 
\textbf{References.}$^{a}$\cite{Herczeg2014}, $^{b}$ see \cite{Howard2013} for a discussion, $^{c}$ \cite{Palla2002}, $^{d}$ $\dot{M}_{\rm acc}$ and $L_{\rm acc}$ as explained above
, $^{e}$ \cite{Gudel2007b}, $^{f}$ \cite{Andrews2005}, $^{g}$ \cite{White2001}.
\end{table}
\end{landscape}
\end{appendix}

\clearpage
\begin{appendix}
\section{Observations}
\label{obs}
Table \ref{OBSIDs} shows the complete Taurus GASPS list of observations. The object coordinates, identifiers (OBSIDs) of line (LineSpec) and range (RangeSpec) observing modes along with exposure times are given. 

\begin{table*}[htpb]
\setcounter{table}{0}
\begin{center}
\caption{Overview of the spectroscopic OBSIDs that were observed. (D) stands for deeper observations than the regular settings. (D1): observations at 72 and 145 $\rm \mu m$; (D2): observations at 72, 79, 145, and 158 $\rm \mu m$; (D3): observations at 72, 78, 90, 145, 158, and 180 $\rm \mu m$.}
\label{OBSIDs}
\begin{tabular}{lccccc}
\hline
\hline
Star & RA[h:m:s] & DEC[d:m:s] & LineSpec ID & RangeSpec ID & $\rm t_{exp}[s]$ \\
\hline
AA Tau & 04:34:55.42& +24:28:53.20& 1342190357 & 1342190356 (D3)  & 1252 / 5141\\
& & & 1342225758 & 1342225759 (D3) & 6628 / 20555\\
BP Tau& 04:19:15.83& +29:06:26.90& 1342192796 & & 1252\\
& & & 1342225728 &  & 3316\\
CIDA 2 & 04:15:05.16 & +28:08:46.20& 1342216643 & & 1252\\
CI Tau& 04:33:52.00& +22:50:30.20& 1342192125 & 1342192124 (D3) & 1252 / 5141\\
CW Tau& 04:14:17.00& +28:10:57.80& 1342216221 & 1342216222 (D3) & 1252 / 10279\\
CX Tau& 04:14:47.86& +26:48:11.01& 1342225729 & & 3316\\
CY Tau & 04:17:33.73& +28:20:46.90& 1342192794 &  & 1252\\
Coku Tau/4& 04:41:16.81& +28:40:00.60& 1342191360 &  & 1252\\
& & &1342225837 & & 6628\\
DE Tau& 04:21:55.64& +27:55:06.10& 1342192797 & 1342216648 (D2)& 1252 / 8316\\
DF Tau& 04:27:03.08& +25:42:23.30 & 1342190359 & 1342190358 (D3)& 1252 / 5141\\
DG Tau& 04:27:04.70& +26:06:16.30& 1342190382 & 1342190383 (D3)& 1252 / 5141\\
DG Tau B& 04:27:02.56& +26:05:30.70& 1342192798 & 1342216652 (D3)& 1252 / 10279\\
DH Tau$^{+}$& 04:29:42.02& +26:32:53.20& 1342225734 &  & 1252\\
DK Tau& 04:30:44.24& +26:01:24.80& 1342192132 & 1342192133 (D3) & 1252 / 5141\\ 
& & &1342225732 & & 3316\\
DL Tau& 04:33:39.06& +25:20:38.20& 1342190355 & 1342190354 (D3) & 1252 / 5141\\
& & &1342225800 &   & 6628\\
DM Tau& 04:33:48.72& +18:10:10.00& 1342192123 & 1342192122 (D3) & 1252 / 5141\\
& & &1342225825 &  & 6628\\
DN Tau& 04:35:27.37& +24:14:58.90& 1342192127 & 1342192126 (D3) & 1252 / 5141\\
& & & 1342225757 &  & 3316\\
DO Tau& 04:38:28.58& +26:10:49.40& 1342190385 & 1342190384 (D3) & 1252 / 5141\\
& & & & 1342225802 (D2) & 12516\\
DP Tau& 04:42:37.70& +25:15:37.50& 1342191362 & 1342225827 (D3) & 1252 / 10279\\
DQ Tau& 04:46:53.05& +17:00:00.02& 1342225806 & 1342225807 (D2) & 1252 / 8316\\
DS Tau& 04:47:48.11& +29:25:14.45& 1342225851 & & 3316\\
FF Tau& 04:35:20.90& +22:54:24.20& 1342192802 &  & 1252\\
FM Tau& 04:14:13.58& +28:12:49.20& 1342216218 & 1342216219 (D3) & 1252 / 10279\\
FO Tau& 04:14:49.29& +28:12:30.60& 1342216645 & 1342216644 (D2) & 1252 / 8316\\
FQ Tau& 04:19:12.81& +28:29:33.10& 1342192795 & 1342216650 (D2) & 1252 / 8316\\
FS Tau$^{*}$& 04:22:02.18& +26:57:30.50& 1342192791 & 1342214358 (D3) & 1252 / 10279\\
FT Tau& 04:23:39.19& +24:56:14.10& 1342192790 & 1342243501 (D1) & 1252 / 1637\\
FW Tau& 04:29:29.71& +26:16:53.20& 1342225735 & & 1252\\
FX Tau& 04:30:29.61& +24:26:45.00& 1342192800 &  & 1252\\
GG Tau& 04:32:30.35& +17:31:40.60& 1342192121 & 1342192120 (D3) & 1252 / 5141\\
& & &  & 1342225738 (D2)& 12516\\
GH Tau$^{\star}$& 04:33:06.43& +24:09:44.50& 1342192801 & 1342225762 (D2) & 1252 / 8316\\
GI/K Tau$^{\dagger}$& 04:33:34.31& +24:21:11.50& 1342225760 & & 3316\\
GM Aur& 04:55:10.99& +30:21:59.25& 1342191357 & 1342191356 (D3) & 1252 / 5141\\
 GO Tau& 04:43:03.09& +25:20:18.80& 1342191361 &  & 1252\\
& & & 1342225826 & & 3316\\
Haro 6--13& 04:32:15.41& +24:28:59.70& 1342192128 & 1342192129 (D3) & 1252 / 5141\\
& & &  & 1342225761 (D2)& 12516\\
Haro 6--37& 04:46:58.98& +17:02:38.20& 1342225805 & & 1252\\
HBC 347& 03:29:38.37& +24:30:38.00& 1342192136 &  & 1252\\
HBC 356& 04:03:13.99& +25:52:59.90& 1342214359 &  & 1252\\
& & & 1342204134 &  & 1252\\
HBC 358& 04:03:50.84& +26:10:53.20& 1342214680 & & 1252\\
& & & 1342204347 &  & 1252\\
HD 283572& 04:21:58.85& +28:18:06.64& 1342216646 &  & 1660\\
HK Tau& 04:31:50.57& +24:24:18.07& 1342225736 & 1342225737 (D2) & 3316 / 8316\\
HN Tau& 04:33:39.35& +17:51:52.37& 1342225796 & 1342225797 (D3) & 3316 / 10279\\ 
HO Tau& 04:35:20.20& +22:32:14.60& 1342192803 & & 1252\\
HV Tau& 04:38:35.28& +26:10:38.63& 1342225801 &  & 3316\\
IP Tau& 04:24:57.08& +27:11:56.50& 1342225756 & & 3316\\
IQ Tau& 04:29:51.56& +26:06:44.90& 1342225733 & 1342192134 (D3) & 3316 / 5141\\
& & & 1342192135 & & 1252\\
IRAS 04158+2805& 04:18:58.14& +28:12:23.50& 1342192793 & 1342192792 (D3) & 1252 / 5141\\
IRAS 04385+2550& 04:41:38.82& +25:56:26.75& 1342225828 & 1342225829 (D2) & 3316 / 8316\\ 
J1-4872& 04:25:17.68& +26:17:50.41& 1342216653 & & 1660\\
\hline
\end{tabular}
\end{center}
\normalsize
\end{table*}

\begin{table*}[htpb]
\setcounter{table}{0}
\begin{center}
\caption{\textit{(Continued)}}
\begin{tabular}{lccccc}
\hline
\hline
Star & RA[h:m:s] & DEC[d:m:s] & LineSpec ID & RangeSpec ID & $\rm t_{exp}[s]$ \\
\hline

LkCa 1& 04:13:14.14& +28:19:10.84& 1342214679 &  & 1660\\
LkCa 3& 04:14:47.97& +27:53:34.65& 1342216220 &  & 1660\\
LkCa 4& 04:16:28.11& +28:07:35.81& 1342216642 &  & 1660\\
LkCa 5& 04:17:38.94& +28:33:00.51& 1342216641 &  & 1660\\
LkCa 7& 04:19:41.27& +27:49:48.49& 1342216649 &  & 1660\\
LkCa 15& 04:39:17.80& +22:21:03.48& 1342190387 & 1342190386 (D3) & 1252 / 5141\\
& & &1342225798 & & 6628\\
RW Aur& 05:07:49.54& +30:24:05.07& 1342191359 & 1342191358 (D3) & 1252 / 5141\\
RY Tau& 04:21:57.40& +28:26:35.54& 1342190361 & 1342190360 (D3) & 1252 / 5141\\
& & &  & 1342216647 (D2)&  12516\\
SU Aur& 04:55:59.38& +30:34:01.56& 1342217844 & 1342217845 (D3) & 3316 / 10279\\
T Tau& 04:21:59.43& +19:32:06.37 & 1342190353 & 1342190352 (D3) & 1252 / 5141\\
UX Tau& 04:30:03.99& +18:13:49.40& 1342214357 &  & 1252\\
& & & 1342204350 &  & 1252\\
UY Aur& 04:51:47.38& +30:47:13.50& 1342215699 & 1342226001 (D3) & 1252 / 10279\\
& & &1342193206 &  & 1252\\
UZ Tau& 04:32:42.89& +25:52:32.60& 1342192131 & 1342192130 (D3) & 1252 / 5141\\
V710 Tau& 04:31:57.80& +18:21:35.10& 1342192804 &  & 1252\\
V773 Tau& 04:14:12.92& +28:12:12.45& 1342216217 & & 3316\\
V819 Tau& 04:19:26.26& +28:26:14.30& 1342216651 &  & 1660\\
V836 Tau& 05:03:06.60& +25:23:19.71& 1342227634 &  & 3316\\
V927 Tau& 04:31:23.82& +24:10:52.93& 1342225763 &  & 3316\\
V1096 Tau& 04:13:27.23& +28:16:24.80& 1342214678 & & 1252\\
VY Tau& 04:39:17.41& +22:47:53.40& 1342192989 & &  1252\\
XZ Tau$^{\ddagger}$& 04:31:39.48& +18:13:55.70& 1342190351 & 1342190350 (D3) & 1252 / 5141\\
ZZ Tau& 04:30:51.38& +24:42:22.30& 1342192799 &  & 1252\\
\hline
\end{tabular}
\end{center}
\textbf{Notes.} \\
$^{+}$ DI Tau was observed in the same FOV as DH Tau\\
\hspace{1.72cm}$^{*}$ Haro 6--5B was observed in the same FOV as FS Tau\\
\hspace{1.72cm}$^{\star}$ V807 Tau was observed in the same FOV as GH Tau\\
\hspace{1.72cm}$^{\dagger}$ HL Tau was observed in the same FOV as XZ Tau\\
\hspace{1.72cm}$^{\ddagger}$ GI/K Tau were spectroscopically unresolved\\
\end{table*}
\end{appendix}

\clearpage
\begin{appendix}
\section{Line fluxes}
\label{lineFluxes}
Tables \ref{lineFluxBlue} and \ref{lineFluxRed} give the aperture corrected line fluxes integrated over one spaxel in the 60--80 $\rm \mu m$ and 90--180 $\rm \mu m$ ranges, respectively. The same holds for Tables \ref{table:blue3x3_1} and \ref{table:red3x3}, but for the fluxes integrated over 3$\times$3 spaxels.

\begin{landscape}
\begin{table}[htpb]
\setcounter{table}{0}
\tiny
\caption{Line fluxes for species in the  60-80 $\rm \mu m$ range integrated over one spaxel.}
\begin{center}
\label{lineFluxBlue}
\begin{tabular}{lcccccccccccccccc}
\hline
\hline
\multicolumn{1}{l}{Target}&
\multicolumn{1}{c}{SED} &
\multicolumn{1}{c}{[OI]        63.18$\rm \mu m$}&
\multicolumn{1}{c}{o--H$_2$O   63.32$\rm \mu m$}&
\multicolumn{1}{c}{CO          72.84$\rm \mu m$}&
\multicolumn{1}{c}{o--H$_2$O   78.74$\rm \mu m$}&
\multicolumn{1}{c}{p--H$_2$O   78.92$\rm \mu m$}&
\multicolumn{1}{c}{OH          79.11$\rm \mu m$}&
\multicolumn{1}{c}{OH          79.18$\rm \mu m$}&
\multicolumn{1}{c}{CO          79.36$\rm \mu m$}\\
%\hline
\multicolumn{1}{l}{--}&
\multicolumn{1}{l}{Class}&
\multicolumn{1}{c}{(10$^{-16}$ W/m$^{2}$)}&
\multicolumn{1}{c}{(10$^{-16}$ W/m$^{2}$)}&
\multicolumn{1}{c}{(10$^{-16}$ W/m$^{2}$)}&
\multicolumn{1}{c}{(10$^{-16}$ W/m$^{2}$)}&
\multicolumn{1}{c}{(10$^{-16}$ W/m$^{2}$)}&
\multicolumn{1}{c}{(10$^{-16}$ W/m$^{2}$)}&
\multicolumn{1}{c}{(10$^{-16}$ W/m$^{2}$)}&
\multicolumn{1}{c}{(10$^{-16}$ W/m$^{2}$)}\\
\hline
\multicolumn{10}{c}{Outflows}\\
\hline
AA Tau         &II     &0.119 (0.038) &0.067 (0.021) &$<$0.032     &0.088 (0.013)&$<$0.040     &0.054 (0.013)      &0.052 (0.013)      &$<$0.040     \\
CW Tau         &II     &0.803 (0.054) &$<$0.119      &$<$0.048     &$<$0.064     &0.055 (0.021)&0.074 (0.021)      &$<$0.064           &$<$0.064     \\
DF Tau         &II     &0.500 (0.071) &$<$0.136      &$<$0.074     &$<$0.100     &$<$0.100     &$<$0.100           &$<$0.100           &$<$0.100     \\
DG Tau         &II     &\textbf{5.831 (0.152)} &$<$0.342      &0.134 (0.034)&$<$0.069     &$<$0.069     &0.227 (0.025)      &0.183 (0.027)      &0.058 (0.023)\\ 
DG Tau B       &I      &\textbf{3.265 (0.096)} &$<$0.211      &$<$0.059     &$<$0.069     &0.059 (0.024)&0.124 (0.025)      &0.086 (0.024)      &$<$0.069     \\
DL Tau         &II     &0.184 (0.088) &0.079 (0.019) &$<$0.078     &$<$0.100     &$<$0.100     &$<$0.100           &0.052 (0.016)&$<$0.100     \\
DO Tau         &II     &0.735 (0.115) &$<$0.251      &$<$0.058     &$<$0.100     &$<$0.100     &0.075 (0.025)      &$<$0.100           &$<$0.100     \\
DP Tau        &II     &0.848 (0.042) &$<$0.112      &$<$0.064     &0.076 (0.018)&$<$0.057     &0.141 (0.018)      &0.170 (0.027)      &$<$0.057     \\
DQ Tau         &II     &0.274 (0.048) &$<$0.122      &$<$0.045     &$<$0.059     &0.058 (0.020)&$<$0.059           &$<$0.059           &$<$0.059     \\
FS Tau         &II/Flat&\textbf{3.725 (0.051)} &0.182 (0.048) &$<$0.042     &0.226 (0.028)&$<$0.084     &0.523 (0.057)      &0.339 (0.028)      &$<$0.10      \\
GG Tau         &II     &0.550 (0.049) &$<$0.126      &$<$0.034     &0.038 (0.016)&$<$0.048     &0.082 (0.016)      &0.047 (0.016)      &$<$0.048     \\
Haro 6--5 B    &I      &0.399 (0.056) &0.139 (0.049) &$<$0.061     &0.109 (0.042)&0.075 (0.030)& $<$0.106          &$<$0.106           &$<$0.106     \\
Haro 6--13     &II     &0.573 (0.047) &$<$0.109      &$<$0.033     &0.055 (0.016)&0.041 (0.016)&0.044 (0.016)      &0.044 (0.016)      &$<$0.047     \\
HL Tau         &I      &4.749 (0.198) &0.536 (0.182) &$<$0.141     &0.820 (0.127)&$<$0.140      &0.198 (0.058)      & 0.150 (0.049)     &$<$0.198     \\
HN Tau         &II     &0.487 (0.027) &0.053 (0.024) &$<$0.049     &$<$0.066     &$<$0.066     &$<$0.066           &$<$0.066           &$<$0.066     \\
HV Tau         &I?     &0.349 (0.039) &$<$0.075      &--           &--           &--           &--                 &--                 &--           \\
IRAS 04158+2805&I      &0.582 (0.044) &$<$0.110      &$<$0.074     &$<$0.093     &$<$0.093     &$<$0.093           &$<$0.093           &$<$0.093     \\
IRAS 04385+2550&II     &0.649 (0.029) &$<$0.078      &$<$0.040     &0.079 (0.014)&$<$0.059     &$<$0.059           &$<$0.059           &$<$0.059     \\
RW Aur         &II     &1.284 (0.058) &$<$0.117      &$<$0.076     &$<$0.112     &$<$0.112     &$<$0.112           &$<$0.112           &$<$0.112     \\
RY Tau         &T?     &0.974 (0.057) &0.203 (0.074) &$<$0.0452    &0.134 (0.028)&$<$0.082     &$<$0.082           &$<$0.082           &$<$0.082     \\
SU Aur        &II     &\textbf{0.833 (0.030)} &$<$0.083      &$<$0.042     &$<$0.079     &$<$0.079     &$<$0.079           &$<$0.079           &$<$0.079     \\
T Tau          &II/I   &\textbf{82.171 (0.800)} &2.475 (0.670) &1.088 (0.141)&\textbf{5.119 (0.085)}&0.990 (0.085)&\textbf{12.685 (0.170)}     &\textbf{8.386 (0.099)}      &1.683 (0.240)\\
UY Aur         &II     &\textbf{1.585 (0.052)} &$<$0.132      &$<$0.057     &0.349 (0.040)&0.100 (0.044)&0.250 (0.044)      &0.223 (0.040)      &$<$0.096     \\
UZ Tau         &II     &0.187 (0.050) &$<$0.113      &$<$0.051     &$<$0.086     &$<$0.086     &$<$0.086           &$<$0.086           &$<$0.086     \\
V773 Tau      & II    & 0.774 (0.040)&$<$0.082 &-- &-- &-- &-- &-- &--\\

XZ Tau         &II     &\textbf{2.409 (0.055)} &0.174 (0.045) &$<$0.071     &0.221 (0.037)&$<$0.113     &0.240 (0.051)&0.144 (0.028)&$<$0.113     \\
\hline
\multicolumn{10}{c}{Non-outflows}\\
\hline
BP Tau & II &0.132 (0.041) & 0.106 (0.030) &-- &-- &-- &-- &-- &--\\
CI Tau &II&0.326 (0.036)&$<$0.085     &$<$0.076&$<$0.093&$<$0.093&$<$0.093&$<$0.093&$<$0.093\\
CIDA 2& III &$<$0.110 &$<$0.110 &-- &-- &-- &-- &-- &--\\
Coku Tau/4 & II & 0.234 (0.016) & $<$0.046 &-- &-- &-- &-- &-- \\
CX Tau & T & $<$0.082 & $<$0.082&-- &-- &-- &-- &-- &--\\
CY Tau & II & $<$0.111 &$<$0.111 &-- &-- &-- &-- &-- &--\\
DE Tau &II&$<$0.119     &$<$0.119&$<$0.045&$<$0.058&$<$0.058&$<$0.058&$<$0.058&$<$0.058\\
DH Tau& II &$<$0.138 &$<$0.138 &-- &-- &-- &-- &-- &--\\
DI Tau& III &$<$0.138 &$<$0.138 &-- &-- &-- &-- &-- &--\\
DK Tau &II&0.162 (0.025)&$<$0.055     &$<$0.069&$<$0.099&$<$0.099&$<$0.099&$<$0.099&$<$0.099\\
DM Tau &T &0.107 (0.029)&$<$0.062     &$<$0.078&$<$0.107&$<$0.107&$<$0.107&$<$0.107&$<$0.107\\
DN Tau &T &0.060 (0.020)&$<$0.062     &$<$0.072&$<$0.102&$<$0.102&$<$0.102&$<$0.102&$<$0.102\\
DS Tau & II & 0.085 (0.012) & $<$0.063&-- &-- &-- &-- &-- &--\\
FF Tau & III & $<$0.100&$<$0.100 &-- &-- &-- &-- &-- &--\\
FM Tau &II&$<$0.117     &$<$0.117     &$<$0.051&$<$0.071&$<$0.071&$<$0.071&$<$0.071&$<$0.071\\
FO Tau &T &$<$0.120     &$<$0.120     &$<$0.034&$<$0.052&$<$0.052&$<$0.052&$<$0.052&$<$0.052\\
FQ Tau &II&$<$0.102     &$<$0.102     &$<$0.042&$<$0.065&$<$0.065&$<$0.065&$<$0.065&$<$0.065\\
FT Tau &II&0.123 (0.035)&$<$0.097     &$<$0.074&--      &--      &--      &--      &--      \\
FW Tau& III &$<$0.096 & $<$0.096&-- &-- &-- &-- &-- &--\\
FX Tau& II & $<$0.129& $<$0.129&-- &-- &-- &-- &-- &--\\
\hline
\end{tabular}
\end{center}
\end{table}
\end{landscape}

\newpage
\begin{landscape}
\begin{table}[htpb]
\setcounter{table}{0}
\tiny
\caption{\textit{(Continued)}}
\begin{center}
\begin{tabular}{lcccccccccccccccc}
\hline
\hline
\multicolumn{1}{l}{Target}&
\multicolumn{1}{c}{SED} &
\multicolumn{1}{c}{[OI]        63.18$\rm \mu m$}&
\multicolumn{1}{c}{o--H$_2$O   63.32$\rm \mu m$}&
\multicolumn{1}{c}{CO          72.84$\rm \mu m$}&
\multicolumn{1}{c}{o--H$_2$O   78.74$\rm \mu m$}&
\multicolumn{1}{c}{p--H$_2$O   78.92$\rm \mu m$}&
\multicolumn{1}{c}{OH          79.11$\rm \mu m$}&
\multicolumn{1}{c}{OH          79.18$\rm \mu m$}&
\multicolumn{1}{c}{CO          79.36$\rm \mu m$}\\
%\hline
\multicolumn{1}{l}{--}&
\multicolumn{1}{l}{Class}&
\multicolumn{1}{c}{(10$^{-16}$ W/m$^{2}$)}&
\multicolumn{1}{c}{(10$^{-16}$ W/m$^{2}$)}&
\multicolumn{1}{c}{(10$^{-16}$ W/m$^{2}$)}&
\multicolumn{1}{c}{(10$^{-16}$ W/m$^{2}$)}&
\multicolumn{1}{c}{(10$^{-16}$ W/m$^{2}$)}&
\multicolumn{1}{c}{(10$^{-16}$ W/m$^{2}$)}&
\multicolumn{1}{c}{(10$^{-16}$ W/m$^{2}$)}&
\multicolumn{1}{c}{(10$^{-16}$ W/m$^{2}$)}\\
\hline
\multicolumn{10}{c}{Non-outflows}\\
\hline

GH Tau &II&$<$0.090     &$<$0.090     &$<$0.044&$<$0.085&$<$0.085&$<$0.085&$<$0.085&$<$0.085\\
GI/GK Tau & II &0.127 (0.031) & 0.069 (0.022)&-- &-- &-- &-- &-- &--\\
GM Aur &T &0.241 (0.045)&$<$0.119     &$<$0.056&$<$0.105&$<$0.105&$<$0.105&$<$0.105&$<$0.105\\
GO Tau & II &$<$0.068 &$<$0.068 &-- &-- &-- &-- &-- &--\\
Haro 6--37& II & $<$0.133 & $<$0.133  &-- &-- &-- &-- &-- &--\\

HBC 347& III &$<$0.128 &$<$0.128 &-- &-- &-- &-- &-- &--\\
HBC 356& III &$<$0.081 & $<$0.081&-- &-- &-- &-- &-- &--\\
HBC 358& III & $<$0.102& $<$0.102&-- &-- &-- &-- &-- &--\\
HD 283572 & III &$<$0.095 &$<$0.095 &-- &-- &-- &-- &-- &--\\
HK Tau &II&0.370 (0.032)&$<$0.090     &$<$0.031&$<$0.064&$<$0.064&$<$0.064&$<$0.064&$<$0.064\\
HO Tau & II &$<$0.114 & $<$0.114&-- &-- &-- &-- &-- &--\\
IP Tau & T &$<$0.076 &$<$0.076 &-- &-- &-- &-- &-- &--\\
IQ Tau &II&0.157 (0.025)&0.098 (0.040)&$<$0.072&$<$0.117&$<$0.117&$<$0.117&$<$0.117&$<$0.117\\
J1--4872& III &$<$0.119 &$<$0.119 &-- &-- &-- &-- &-- &--\\
LkCa 1& III &$<$0.099 & $<$0.099&-- &-- &-- &-- &-- &--\\
LkCa 3& III &$<$0.098 & $<$0.098&-- &-- &-- &-- &-- &--\\

LkCa 4& III &$<$0.103 &$<$0.103 &-- &-- &-- &-- &-- &--\\
LkCa 5& III & $<$0.101&$<$0.101 &-- &-- &-- &-- &-- &--\\
LkCa 7& III & $<$0.110&$<$0.110 &-- &-- &-- &-- &-- &--\\
LkCa 15&T &0.126 (0.024)&$<$0.057     &$<$0.078&$<$0.107&$<$0.107&$<$0.107&$<$0.107&$<$0.107\\
UX Tau& T &0.395 (0.035) &$<$0.095 &-- &-- &-- &-- &-- &--\\
V710 Tau& II &$<$0.105 & $<$0.105&-- &-- &-- &-- &--&-- \\
V807 Tau       &II     &0.137 (0.058) &$<$0.108      &$<$0.042     &$<$0.057     &$<$0.057     &$<$0.057           &$<$0.057           &$<$0.071     \\
V819 Tau& II/III & $<$0.117&$<$0.117 &-- &-- &-- &-- &-- &--\\
V836 Tau& T &$<$0.074 &$<$0.074 &-- &-- &-- &-- &-- &--\\
V927 Tau& III &$<$0.085 &$<$0.085 &-- &-- &-- &-- &-- &--\\
V1096 Tau& III & $<$0.128& $<$0.128&-- &-- &-- &-- &-- &--\\
VY Tau& II & $<$0.097&$<$0.097 &-- &-- &-- &-- &-- &--\\
ZZ Tau& II & $<$0.115& $<$0.115&-- &-- &-- &-- &-- &--\\
\hline
\end{tabular}
\end{center}
\textbf{Notes.} Fluxes in bold text represent misaligned and/or extended sources. In these cases the 3x3 fluxes (see Tables \ref{table:blue3x3_1} and \ref{table:red3x3}) are more accurate and are used in the calculations.
\end{table}
\end{landscape}

\begin{landscape}
\begin{table}[htpb]
\setcounter{table}{1}
\tiny
\caption{Line fluxes for species in the  90--180 $\rm \mu m$ range integrated over one spaxel.}
\begin{center}
\label{lineFluxRed}
\begin{tabular}{lccccccccccccccccccc}
\hline
\hline
\multicolumn{1}{l}{Target}&
\multicolumn{1}{c}{SED} &
\multicolumn{1}{c}{p--H$_2$O   89.99$\rm \mu m$}&
\multicolumn{1}{c}{CO          90.16$\rm \mu m$}&
\multicolumn{1}{c}{p--H$_2$O 144.52$\rm \mu m$}&
\multicolumn{1}{c}{CO         144.78$\rm \mu m$}&
\multicolumn{1}{c}{[OI]       145.52$\rm \mu m$}&
\multicolumn{1}{c}{[CII]      157.74$\rm \mu m$}&
\multicolumn{1}{c}{o--H$_2$O  179.53$\rm \mu m$}&
\multicolumn{1}{c}{o--H$_2$O  180.49$\rm \mu m$}\\
%\hline
\multicolumn{1}{l}{--}&
\multicolumn{1}{l}{Class}&
\multicolumn{1}{c}{(10$^{-16}$ W/m$^{2}$)}&
\multicolumn{1}{c}{(10$^{-16}$ W/m$^{2}$)}&
\multicolumn{1}{c}{(10$^{-16}$ W/m$^{2}$)}&
\multicolumn{1}{c}{(10$^{-16}$ W/m$^{2}$)}&
\multicolumn{1}{c}{(10$^{-16}$ W/m$^{2}$)}&
\multicolumn{1}{c}{(10$^{-16}$ W/m$^{2}$)}&
\multicolumn{1}{c}{(10$^{-16}$ W/m$^{2}$)}&
\multicolumn{1}{c}{(10$^{-16}$ W/m$^{2}$)}\\
\hline
\multicolumn{10}{c}{Outflows}\\
\hline
AA Tau         &II     &$<$0.033           &$<$0.047            &$<$0.016           &0.038 (0.006)       &0.009 (0.006) &$<$0.023            &$<$0.023           &$<$0.023\\
CW Tau         &II     &0.045 (0.018)      &0.089 (0.027)       &$<$0.030           &0.058 (0.010) &$<$0.030            &0.044 (0.012)       &$<$0.028           &$<$0.028\\
DF Tau         &II     &$<$0.071           &$<$0.071            &$<$0.041           &$<$0.041            &$<$0.041            &$<$0.055            &$<$0.061           &$<$0.061\\
DG Tau         &II     &0.066 (0.027)&0.096 (0.024) &$<$0.044           &\textbf{0.255 (0.014)}       &\textbf{0.396 (0.014)}       &\textbf{0.948 (0.014)}       &$<$0.081           &$<$0.081\\
DG Tau B       &I      &$<$0.055           &$<$0.061      &0.027 (0.011)      &\textbf{0.105 (0.011)}       &\textbf{0.167 (0.011)}       &\textbf{0.082 (0.016)}$^\ddagger$& $<$0.040           &$<$0.040\\
DL Tau         &II     &0.076 (0.027)      &$<$0.082            &$<$0.045           &$<$0.045            &$<$0.045            &$<$0.054            &$<$0.061           &$<$0.061\\
DO Tau         &II     &$<$0.054           &$<$0.070            &$<$0.034           &0.035 (0.011)&0.048 (0.011)&$<$0.061            &$<$0.081           &$<$0.081\\
DP Tau         &II     &$<$0.057           &0.082 (0.018)&0.024 (0.007)      &0.081 (0.007)       &0.037 (0.007)       &0.082 (0.011)$^\ddagger$       &$<$0.037           &$<$0.037\\
DQ Tau         &II     &--                 &--                  &$<$0.017           &$<$0.017            &$<$0.017            &0.025 (0.009) &--                 &--      \\
FS Tau         &II/Flat&0.056 (0.014)      &0.059 (0.017)       &$<$0.035           &0.317 (0.011)       &0.160 (0.011)       &0.157 (0.013)$^\ddagger$    &0.325 (0.013)     &0.068 (0.013)\\
GG Tau         &II     &$<$0.082           &$<$0.082            &$<$0.023           &0.054 (0.007) &0.018 (0.007) &$<$0.028            &$<$0.056           &$<$0.056\\
Haro 6--5 B    &I      &$<$0.079           &$<$0.079            &0.047 (0.011)      &0.091 (0.011)       & 0.038 (0.011)      &0.068 (0.013)$^\ddagger$       &$<$0.042           &$<$0.042\\
Haro 6--13     &II     &$<$0.079           &$<$0.079            &0.025 (0.007)&0.078 (0.007)       &$<$0.020            &$<$0.031            &$<$0.061           &$<$0.061\\
HL Tau         &I      &0.255 (0.042)      &0.255 (0.042)       &$<$0.250            &0.977 (0.092)       &0.864 (0.079)       &$<$0.254            &$<$0.085           &$<$0.085\\
HN Tau         &II     &$<$0.057           &$<$0.057            &0.199 (0.006)      &$<$0.017            &0.031 (0.006)       &$<$0.028            &$<$0.028           &$<$0.028\\
IRAS 04158+2805&I      &$<$0.085           &$<$0.085            &$<$0.028           &0.042 (0.013)       &0.030 (0.013) &$<$0.062            &$<$0.058           &$<$0.058\\
IRAS 04385+2550&II     &--                 &--                  &$<$0.020           &0.0548 (0.007)      &0.031 (0.007)       &0.024 (0.010)$^\ddagger$ &--                 &--      \\
RW Aur         &II     &$<$0.076           &$<$0.076            &$<$0.040           &0.148 (0.013)       &0.069 (0.013)       &$<$0.055            &0.139 (0.020)      &$<$0.059\\
RY Tau         &T?     &0.096 (0.028)&0.083               &0.037 (0.010)      &0.038 (0.010)       &0.027 (0.010)       &0.038 (0.007)$^\ddagger$       &$<$0.051           &$<$0.051\\
SU Aur         &II     &$<$0.048           &$<$0.049            &$<$0.027           &0.038 (0.008)       &0.035 (0.008)       &0.047 (0.013)$^\ddagger$ &$<$0.031           &$<$0.031\\
T Tau          &II/I   &3.408 (0.057) &\textbf{3.411 (0.064)}  &\textbf{0.846 (0.035)}   &\textbf{7.210 (0.033)} &\textbf{3.534 (0.035)} &\textbf{1.892 (0.023)}   &\textbf{3.780 (0.038)}&\textbf{1.254 (0.042)}\\
UY Aur         &II     &0.098 (0.028)      &0.103 (0.023)       &0.048 (0.010)      &0.254 (0.010)       &0.171 (0.010)       &\textbf{0.151 (0.011)}$^\ddagger$       &0.151 (0.010)      &$<$0.031\\
UZ Tau         &II     &$<$0.065           &$<$0.065            &$<$0.044           &$<$0.044            &$<$0.044            &$<$0.051            &$<$0.069           &$<$0.069\\

XZ Tau         &II     &0.136 (0.027)&0.045 (0.024)  &0.089 (0.013)      &\textbf{0.311 (0.013)}       &0.141 (0.013)       &0.225 (0.027)$^\ddagger$ &0.168 (0.017)&0.071 (0.023)\\
\hline
\multicolumn{10}{c}{Non-outflows}\\
\hline
CI Tau &II&$<$0.076&$<$0.107&$<$0.049&0.041 (0.017)  &$<$0.049             &$<$0.042         &$<$0.051&$<$0.051\\
DE Tau &II&--      &--      &$<$0.017&0.024 (0.006)&0.016 (0.006)&$<$0.030         &--      &--\\
DK Tau &II&$<$0.078&$<$0.078&$<$0.049&$<$0.049             &$<$0.0495            &$<$0.065         &$<$0.062&$<$0.062\\
DM Tau &T &$<$0.072&$<$0.072&$<$0.042&$<$0.042             &$<$0.042             &$<$0.047         &$<$0.047&$<$0.047\\
DN Tau &T &$<$0.075&$<$0.075&$<$0.033&$<$0.033             &$<$0.033             &$<$0.042         &$<$0.058&$<$0.058\\
FO Tau &T &--      &--      &$<$0.020&$<$0.020             &$<$0.020             &$<$0.023         &--      &--\\
FM Tau &II&$<$0.059&$<$0.059&$<$0.025&$<$0.025             &$<$0.025             &$<$0.044         &$<$0.044&$<$0.044\\
FQ Tau &II&--      &--      &$<$0.027&$<$0.027             &$<$0.027             &$<$0.027         &--      &--\\
FT Tau &II&--      &--      &$<$0.040&$<$0.040            &$<$0.040             &--               &--      &--\\
GH Tau &II&--      &--      &$<$0.021&$<$0.021             &$<$0.021             &$<$0.030         &--      &--\\
GM Aur &T &$<$0.085&$<$0.057&$<$0.042&$<$0.042             &$<$0.042             &$<$0.065         &$<$0.058&$<$0.058\\
HK Tau &II&--      &--      &$<$0.018&0.031 (0.006)        &$<$0.021            &$<$0.021         &--      &--\\
IQ Tau &II&$<$0.079&$<$0.079&$<$0.037&$<$0.037             &$<$0.037             &$<$0.051         &$<$0.040&$<$0.040\\
LkCa 15&T &$<$0.078&$<$0.078&$<$0.047&$<$0.047             &$<$0.047             &$<$0.054         &$<$0.048&$<$0.048\\
V807 Tau       &II     &--                 &--                  &$<$0.028           &$<$0.028            &$<$0.028            &$<$0.028            &--                 &--      \\
\hline
\end{tabular}
\end{center}
\textbf{Notes.} Fluxes in bold text represent misaligned and/or extended sources. In these  cases the 3x3 fluxes (see Tables \ref{table:blue3x3_1} and \ref{table:red3x3}) are more accurate and are used in the calculations. ($^{\ddagger}$) Objects shown in Fig. \ref{figure:pdrModelsOI} with [CII] contamination in the \emph{off-}source positions.
\end{table}
\end{landscape}

\newpage
\begin{landscape}
\begin{table}
\setcounter{table}{2}
\tiny
\caption{Line fluxes for species in the  60-80 $\rm \mu m$ range integrated over 3$\times$3 spaxels.}
\label{table:blue3x3_1}
\begin{center}
\begin{tabular}{lccccccccc}
\hline
\hline
\multicolumn{1}{l}{Target}&
\multicolumn{1}{l}{SED}&
\multicolumn{1}{c}{[OI] 63.18$\rm \mu m$}&
\multicolumn{1}{c}{o-H$_{\rm 2}$O 63.32$\rm \mu m$}&
\multicolumn{1}{c}{CO 72.84$\rm \mu m$}&
\multicolumn{1}{c}{o-H$_{\rm 2}$O 78.74$\rm \mu m$}&
\multicolumn{1}{c}{p-H$_{\rm 2}$O 78.92$\rm \mu m$}&
\multicolumn{1}{c}{OH 79.11$\rm \mu m$}&
\multicolumn{1}{c}{OH 79.18$\rm \mu m$}&
\multicolumn{1}{c}{CO 79.36$\rm \mu m$}\\

\multicolumn{1}{l}{--}&
\multicolumn{1}{l}{Class}&
\multicolumn{1}{c}{(10$^{-16}$ W/m$^{\rm 2}$)}&
\multicolumn{1}{c}{(10$^{-16}$ W/m$^{\rm 2}$)}&
\multicolumn{1}{c}{(10$^{-16}$ W/m$^{\rm 2}$)}&
\multicolumn{1}{c}{(10$^{-16}$ W/m$^{\rm 2}$)}&
\multicolumn{1}{c}{(10$^{-16}$ W/m$^{\rm 2}$)}&
\multicolumn{1}{c}{(10$^{-16}$ W/m$^{\rm 2}$)}&
\multicolumn{1}{c}{(10$^{-16}$ W/m$^{\rm 2}$)}&
\multicolumn{1}{c}{(10$^{-16}$ W/m$^{\rm 2}$)}\\
\hline
\multicolumn{10}{c}{Outflows}\\
\hline
AA Tau&II&  0.21 (0.06) &   $<$0.18  &   $<$0.10 &  $<$0.15&   $<$0.15&   $<$0.15&   $<$0.15 &   $<$0.15\\ 
CW Tau&II&  1.09 (0.28) &   $<$0.50&   $<$0.16 &   $<$0.25 &   $<$0.25 &   $<$0.25 &   $<$0.25 &   $<$0.25 \\ 
DF Tau&II&  $<$0.53 &   $<$0.53  &   $<$0.21 &   $<$0.36 &   $<$0.36 &   $<$0.36 &   $<$0.36 &   $<$0.36 \\ 
DG Tau&II& 16.56 (0.59) &   $<$1.02  &   $<$0.20 &   $<$0.40 &   $<$0.40 &   $<$0.40 &   $<$0.40 &   $<$0.40 \\ 
DG Tau B&I&  6.36 (0.37) &   $<$0.71  &   $<$0.17 &   $<$0.34 &   $<$0.34 &   $<$0.34 &   $<$0.34&   $<$0.34 \\ 
DL Tau&II& $<$0.56 &   $<$0.56 &   $<$0.26 &   $<$0.42 &   $<$0.42 &   $<$0.42 &   $<$0.42 &   $<$0.42 \\ 
DO Tau&II&  2.00 (0.47) &   $<$0.56  &   $<$0.29 &   $<$0.43 &   $<$0.43 &   $<$0.43 &   $<$0.43 &   $<$0.43 \\ 
DP Tau&II&  1.37 (0.20) &   $<$0.44 &   $<$0.16 &   $<$0.27 &   $<$0.27 &   $<$0.27 &   $<$0.27 &   $<$0.27 \\ 
DQ Tau&II&  $<$0.35 &   $<$0.35  &   $<$0.14 &   $<$0.25 &   $<$0.25 &   $<$0.25 &   $<$0.25 &   $<$0.25 \\ 
FS Tau&II/Flat&  5.34 (0.19) &   $<$0.43  &   $<$0.16 &   $<$0.31 &   $<$0.31 &   0.64 (0.20) &   0.83 (0.24) &   $<$0.31 \\ 
GG Tau&II&  0.53 (0.12) &   $<$0.34  &   $<$0.29 &   $<$0.40 &   $<$0.40 &   $<$0.40 &   $<$0.40 &   $<$0.40 \\ 
Haro 6--5B&I&--&--&--&--&--&--&--&--\\
Haro 6-13&II&  0.51 ( 0.14) &   $<$0.41 &   $<$0.27 &   $<$0.44 &   $<$0.44 &   $<$0.44 &   $<$0.44 &   $<$0.44 \\ 
HL Tau&I&--&--&--&--&--&--&--&--\\
HN Tau&II&  0.60 (0.09) &   $<$0.23  &   $<$0.18 &   $<$0.27 &   $<$0.27 &   $<$0.27 &   $<$0.27 &   $<$0.27 \\ 
HV Tau&I?&  0.87 (0.18) &   $<$0.31 &    -- &    -- &    -- &    -- &    -- &    -- \\  
IRAS04158+2805&I&  0.71 (0.19) &   $<$0.47 &   $<$0.21 &   $<$0.36 &   $<$0.36 &   $<$0.36 &   $<$0.36 &   $<$0.36 \\ 
IRAS04385+2550&II&  0.77 (0.12) &   $<$0.27 &    $<$0.14 &   $<$0.26 &   $<$0.26 &   $<$0.26 &   $<$0.26 &   $<$0.26 \\ 
RW Aur&II&  2.33 (0.45) &   $<$0.68 &   $<$0.19 &   $<$0.40 &   $<$0.40 &   $<$0.40 &   $<$0.40 &   $<$0.40 \\ 
RY Tau&T?&  0.99 (0.17) &   $<$0.46  &   $<$0.25 &   $<$0.34 &   $<$0.34 &   $<$0.34 &   $<$0.34 &   $<$0.34 \\ 
SU Aur&II&  1.31 (0.12) &   $<$0.27  &   $<$0.19 &   $<$0.37 &   $<$0.37 &   $<$0.37 &   $<$0.37 &   $<$0.37 \\ 
T Tau&II/I&183.45 (3.87) &   $<$7.85  &   1.48 (0.29) &   6.47 (0.27) &   0.65 (0.18) &  11.02 (0.26) &  12.87 (0.33) &   1.95 (0.28) \\ 
UY Aur&II&  3.65 (0.16) &   $<$0.40  &   $<$0.16 &   0.39 (0.13) &   $<$0.29 &   $<$0.29 &   $<$0.29 &   $<$0.29 \\ 
UZ Tau&II&  $<$0.48 &   $<$0.48  &   $<$0.22 &   $<$0.47 &   $<$0.47 &   $<$0.47 &   $<$0.47 &   $<$0.47 \\ 
V773 Tau&II&  0.91 (0.12) &   $<$0.24 &        -- &    -- &    -- &    -- &    -- &    -- \\
XZ Tau&II&  9.63 (0.58) &   $<$1.07 &   $<$0.36 &   $<$0.75 &   $<$0.75 &   $<$0.75 &   $<$0.75 &   $<$0.75 \\ 
\hline
\multicolumn{10}{c}{Non-outflows}\\
\hline
BP Tau&II&  $<$0.20 &   $<$0.20 &     -- &    -- &    -- &    -- &    -- &    -- \\ 
CI Tau&II&  $<$0.37 &   $<$0.37  &  $<$0.26 &   $<$0.37 &   $<$0.37 &   $<$0.37&   $<$0.37&   $<$0.37 \\ 
CIDA2&III&  $<$0.48 &   $<$0.48 &      -- &    -- &    -- &    -- &    -- &    -- \\ 
CoKuTau/4&II&  0.18 (0.04) &   $<$0.17  &    -- &    -- &    -- &    -- &    -- &    -- \\
CX Tau&T&  $<$0.22 &   $<$0.22 &    -- &    -- &    -- &    -- &    -- &    -- \\ 
CY Tau&II&  $<$0.43 &   $<$0.43 &       -- &    -- &    -- &    -- &    -- &    -- \\ 
DE Tau&II&  $<$0.44 &   $<$0.44 &      $<$0.18 &   $<$0.33 &   $<$0.33 &   $<$0.33 &   $<$0.33 &   $<$0.33 \\ 
DH Tau&II&  $<$0.38 &   $<$0.38 &       -- &    -- &    -- &    -- &    -- &    -- \\ 
DI Tau&III&--&--&--&--&--&--&--&--\\
DK Tau&II&  $<$0.33&   $<$0.33  &   $<$0.18 &   $<$0.37 &   $<$0.37 &   $<$0.37 &   $<$0.37&   $<$0.37 \\ 
DM Tau&T&  $<$0.25 &   $<$0.25  &   $<$0.29 &   $<$0.43 &   $<$0.43 &   $<$0.43 &   $<$0.43 &   $<$0.43\\ 
DN Tau&T&  $<$0.24 &   $<$0.24 &       $<$0.29 &   $<$0.41 &   $<$0.41 &   $<$0.41 &   $<$0.41&   $<$0.41 \\ 
DS Tau&II&  $<$0.25 &   $<$0.25 &       -- &    -- &    -- &    -- &    -- &    -- \\ 
FF Tau&III&  $<$0.37 &   $<$0.37 &       -- &    -- &    -- &    -- &    -- &    -- \\ 
FM Tau&II&  $<$0.40 &   $<$0.40 &      $<$0.20 &   $<$0.26 &   $<$0.26 &   $<$0.26 &   $<$0.26 &   $<$0.26 \\ 
FO Tau&T&  $<$0.43 &   $<$0.43 &     $<$0.14 &   $<$0.21 &   $<$0.21 &   $<$0.21 &   $<$0.21 &   $<$0.21 \\ 
FQ Tau&II&  $<$0.43 &   $<$0.43 &     $<$0.15 &   $<$0.26 &   $<$0.26 &   $<$0.26 &   $<$0.26 &  $<$0.26 \\ 
FT Tau&II&  $<$0.37 &   $<$0.37 &    $<$0.26 &    -- &    -- &    -- &    -- &    -- \\ 
FW Tau&III&  $<$0.36 &   $<$0.36 &       -- &    -- &    -- &    -- &    -- &    -- \\ 
FX Tau&II&  $<$0.37 &   $<$0.37 &        -- &    -- &    -- &    -- &    -- &    -- \\ 
 \hline
\end{tabular}
\end{center}
\end{table}

\begin{table}
\setcounter{table}{2}
\tiny
\caption{\textit{(Continued)}}
%\label{table:blue3x3_1}
\begin{center}
\begin{tabular}{lccccccccc}
\hline
\hline
\multicolumn{1}{l}{Target}&
\multicolumn{1}{l}{SED}&
\multicolumn{1}{c}{[OI] 63.18$\rm \mu m$}&
\multicolumn{1}{c}{o-H$_{\rm 2}$O 63.32$\rm \mu m$}&
\multicolumn{1}{c}{CO 72.84$\rm \mu m$}&
\multicolumn{1}{c}{o-H$_{\rm 2}$O 78.74$\rm \mu m$}&
\multicolumn{1}{c}{p-H$_{\rm 2}$O 78.92$\rm \mu m$}&
\multicolumn{1}{c}{OH 79.11$\rm \mu m$}&
\multicolumn{1}{c}{OH 79.18$\rm \mu m$}&
\multicolumn{1}{c}{CO 79.36$\rm \mu m$}\\

\multicolumn{1}{l}{--}&
\multicolumn{1}{l}{Class}&
\multicolumn{1}{c}{(10$^{-16}$ W/m$^{\rm 2}$)}&
\multicolumn{1}{c}{(10$^{-16}$ W/m$^{\rm 2}$)}&
\multicolumn{1}{c}{(10$^{-16}$ W/m$^{\rm 2}$)}&
\multicolumn{1}{c}{(10$^{-16}$ W/m$^{\rm 2}$)}&
\multicolumn{1}{c}{(10$^{-16}$ W/m$^{\rm 2}$)}&
\multicolumn{1}{c}{(10$^{-16}$ W/m$^{\rm 2}$)}&
\multicolumn{1}{c}{(10$^{-16}$ W/m$^{\rm 2}$)}&
\multicolumn{1}{c}{(10$^{-16}$ W/m$^{\rm 2}$)}\\
\hline
\multicolumn{10}{c}{Non-outflows}\\
\hline

GH Tau&II&  $<$0.35 &   $<$0.35 &   $<$0.14 &   $<$0.23 &   $<$0.23 &   $<$0.23 &   $<$0.23 &   $<$0.23 \\ 
GI/GK Tau&II&  $<$0.28 &   $<$0.28  &    -- &    -- &    -- &    -- &    -- &    -- \\ 
GM Aur&T&  $<$0.41 &   $<$0.41 &     $<$0.27 &   $<$0.46 &   $<$0.46 &   $<$0.46 &   $<$0.46 &   $<$0.46 \\ 
GO Tau&II&  $<$0.52 &   $<$0.52 &        -- &    -- &    -- &    -- &    -- &    -- \\ 
Haro 6-37&II&  $<$0.32 &   $<$0.32 &       -- &    -- &    -- &    -- &    -- &    -- \\ 
HBC 347&III&  $<$0.52 &   $<$0.52 &      -- &    -- &    -- &    -- &    -- &    -- \\ 
HBC 356&III&$<$0.47 & $<$0.47 &       -- &    -- &    -- &    -- &    -- &    -- \\ 
HBC 358&III&  $<$0.30 &   $<$0.30 &       -- &    -- &    -- &    -- &    -- &    -- \\ 
HD 283572&III&  $<$0.32 &   $<$0.32 &       -- &    -- &    -- &    -- &    -- &    -- \\ 
HK Tau&II&  0.33 (0.10) &   $<$0.29  &   $<$0.14 &   $<$0.25 &   $<$0.25 &   $<$0.25 &   $<$0.25 &   $<$0.25 \\ 
HO Tau&II&  $<$0.45 &   $<$0.45 &      -- &    -- &    -- &    -- &    -- &    -- \\ 

IP Tau&T&  $<$0.28 &   $<$0.28 &        -- &    -- &    -- &    -- &    -- &    -- \\ 
IQ Tau&II&  $<$0.48 &   $<$0.48 &     $<$0.24 &   $<$0.44 &   $<$0.44 &   $<$0.44 &   $<$0.44 &   $<$0.44 \\ 
J1-4872&III&  $<$0.34 &   $<$0.34 &        -- &    -- &    -- &    -- &    -- &    -- \\ 

LkCa 1&III&  $<$0.30 &   $<$0.30 &       -- &    -- &    -- &    -- &    -- &    -- \\ 
LkCa 3&III&  $<$0.37 &   $<$0.37 &     -- &    -- &    -- &    -- &    -- &    -- \\ 
LkCa 4&III& $<$0.33 &   $<$0.33 &        -- &    -- &    -- &    -- &    -- &    -- \\ 
LkCa 5&III&  $<$0.34 &   $<$0.34 &      -- &    -- &    -- &    -- &    -- &    -- \\ 
LkCa 7&III&  $<$0.39 &   $<$0.39 &       -- &    -- &    -- &    -- &    -- &    -- \\ 
LkCa 15&T&  $<$0.52 &   $<$0.52  &   $<$0.25 &   $<$0.37 &   $<$0.37 &   $<$0.37&   $<$0.37 &   $<$0.37 \\
UX Tau&T&  0.44 (0.14) &   $<$0.32 &      -- &    -- &    -- &    -- &    -- &    -- \\ 
V710 Tau&II&  $<$0.36 &   $<$0.36 &        -- &    -- &    -- &    -- &    -- &    -- \\  
V819 Tau&II/III&  $<$0.34 &   $<$0.34 &        -- &    -- &    -- &    -- &    -- &    -- \\ 
V836 Tau&T&  $<$0.28 &   $<$0.28 &        -- &    -- &    -- &    -- &    -- &    -- \\ 
V927 Tau&III&  $<$0.24 &   $<$0.24 &       -- &    -- &    -- &    -- &    -- &    -- \\ 
V1096 Tau&III&  $<$0.45 &   $<$0.45 &        -- &    -- &    -- &    -- &    -- &    -- \\ 
VY Tau&II&  $<$0.41 &   $<$0.41 &    -- &       -- &    -- &    -- &    -- &    -- \\ 
ZZ Tau&II&  $<$0.35 &   $<$0.35 &    -- &       -- &    -- &    -- &    -- &    -- \\ 
\hline
\end{tabular}
\end{center}
\end{table}
\end{landscape}

\newpage
\begin{landscape}

\begin{table}
\setcounter{table}{3}
\tiny
\caption{Line fluxes for species in the  90--180 $\rm \mu m$ range integrated over 3$\times$3 spaxels.}
\label{table:red3x3}
\begin{center}
\begin{tabular}{lccccccccc}
\hline
\hline
\multicolumn{1}{l}{Target}&
\multicolumn{1}{l}{SED}&
\multicolumn{1}{c}{p-H$_{\rm 2}$O 89.99$\rm \mu m$}&
\multicolumn{1}{c}{CO 90.16$\rm \mu m$}&
\multicolumn{1}{c}{p-H$_{\rm 2}$O 144.52$\rm \mu m$}&
\multicolumn{1}{c}{CO 144.78$\rm \mu m$}&
\multicolumn{1}{c}{[OI] 145.52$\rm \mu m$}&
\multicolumn{1}{c}{[CII] 157.74$\rm \mu m$}&
\multicolumn{1}{c}{o-H$_{\rm 2}$O 179.53$\rm \mu m$}&
\multicolumn{1}{c}{o-H$_{\rm 2}$O 180.49$\rm \mu m$}\\

\multicolumn{1}{l}{--}&
\multicolumn{1}{l}{Class}&
\multicolumn{1}{c}{(10$^{-16}$ W/m$^{\rm 2}$)}&
\multicolumn{1}{c}{(10$^{-16}$ W/m$^{\rm 2}$)}&
\multicolumn{1}{c}{(10$^{-16}$ W/m$^{\rm 2}$)}&
\multicolumn{1}{c}{(10$^{-16}$ W/m$^{\rm 2}$)}&
\multicolumn{1}{c}{(10$^{-16}$ W/m$^{\rm 2}$)}&
\multicolumn{1}{c}{(10$^{-16}$ W/m$^{\rm 2}$)}&
\multicolumn{1}{c}{(10$^{-16}$ W/m$^{\rm 2}$)}&
\multicolumn{1}{c}{(10$^{-16}$ W/m$^{\rm 2}$)}\\
\hline
\multicolumn{10}{c}{Outflows}\\
\hline
AA Tau&II&  $<$0.14 &   $<$0.14 &   $<$0.05 &   $<$0.04 &   $<$0.05  &   $<$0.07 &   $<$0.08 &   $<$0.08 \\ 
CW Tau&II&  $<$0.19 &   $<$0.19 &   $<$0.07 &   $<$0.07 &   $<$0.07  &   $<$0.10 &   $<$0.08 &   $<$0.08 \\ 
DF Tau&II&  $<$0.27 &   $<$0.27 &   $<$0.15 &   $<$0.15 &   $<$0.15  &   $<$0.17 &   $<$0.13 &   $<$0.13 \\ 
DG Tau&II&  $<$0.35 &   $<$0.35 &   $<$0.11 &   0.60 (0.04) &   0.98 (0.04) &   2.87( 0.09) &   $<$0.20 &   $<$0.20 \\ 
DG Tau B&I&  $<$0.22 &   $<$0.22 &   $<$0.11 &   0.26 (0.04) &   0.35 (0.04) &   0.25 (0.04)  &   $<$0.12 &   $<$0.12 \\ 
DL Tau&II&  $<$0.31 &   $<$0.31 &   $<$0.12 &   $<$0.12 &   $<$0.12 &   $<$0.18 &   $<$0.20 &   $<$0.20 \\ 
DO Tau&II&  $<$0.34 &   $<$0.34 &   $<$0.14 &   $<$0.14 &   $<$0.14 &   $<$0.13  &   $<$0.21 &   $<$0.21 \\ 
DP Tau&II&  $<$0.26 &   $<$0.26 &   $<$0.11 &   $<$0.11 &   $<$0.11 &    -- &      -- &    -- \\ 
DQ Tau&II&   -- &    -- &   $<$0.07 &   $<$0.07 &   $<$0.07 &   $<$0.08  &    -- &    -- \\ 
FS Tau&II/Flat&  $<$0.20 &   $<$0.20 &   $<$0.07 &   0.20 (0.03) &   0.14 (0.03) &   0.22 (0.04)  &   $<$0.11 &   $<$0.11 \\ 
GG Tau&II&  $<$0.38 &   $<$0.38 &   $<$0.07 &   0.14(0.05) &   $<$0.07 &   $<$0.07 &   $<$0.17 &   $<$0.17 \\ 
Haro 6--5 B &I& -- &-- &-- &--&--&--&--&--\\
Haro 6-13&II&  $<$0.25 &   $<$0.25 &   $<$0.13 &   $<$0.13 &   $<$0.13 &   $<$0.12  &   $<$0.17 &   $<$0.17 \\ 
HL Tau &I& -- &-- &-- &--&--&--&--&--\\
HV Tau &II& -- &-- &-- &--&--&--&--&--\\
HN Tau&II&  $<$0.17 &   $<$0.17 &   $<$0.10 &   $<$0.10 &   $<$0.10 &   $<$0.11  &   $<$0.13 &   $<$0.13 \\ 
IRAS04158+2805&I&  $<$0.28 &   $<$0.28 &   $<$0.14 &   $<$0.14 &   $<$0.14 &   $<$0.19 &   $<$0.18 &   $<$0.18 \\ 
IRAS04385+2550&II&   -- &    -- &   $<$0.08 &   $<$0.0 &   $<$0.08  &   $<$0.10 &    -- &    -- \\ 
RW Aur&II&  $<$0.34 &   $<$0.34 &   $<$0.10 &   0.17(0.05) &   $<$0.10 &   $<$0.14 &   $<$0.18 &   $<$0.18 \\ 
RY Tau&T?&   $<$0.31 &    $<$0.31&    $<$0.13 &    $<$0.13 &    $<$0.13 &    $<$0.13  &    $<$0.14 &    $<$0.14 \\ 
SU Aur&II&  $<$0.20 &   $<$0.20 &   $<$0.09 &   $<$0.09 &   $<$0.09 &   $<$0.08  &   $<$0.11 &   $<$0.11 \\ 
T Tau&II/I&  3.89 (0.17) &   4.24 (0.15) &   1.28 (0.13) &  11.93 (0.12) &   7.93 (0.13) &   5.65  (0.17)  &   6.35 (0.12) &   2.05 (0.14) \\ 
UY Aur&II&  $<$0.25 &   $<$0.25 &   $<$0.12 &   0.25 (0.04) &   0.17 (0.05) &   0.31 (0.05)  &   0.14 (0.05) &   $<$0.14 \\ 
UZ Tau&II&  $<$0.27 &   $<$0.27 &   $<$0.12 &   $<$0.12 &   $<$0.12 &   $<$0.14  &   $<$0.22 &   $<$0.22 \\ 
V773 Tau &II& -- &-- &-- &--&--&--&--&--\\
XZ Tau&II&  $<$0.46 &   $<$0.46 &  $<$0.31 &   1.27 (0.11) &   0.43 (0.10) &   0.43 (0.11)  &   $<$0.27 &   $<$0.27 \\ 
\hline
\multicolumn{10}{c}{Non-outflows}\\
\hline
CI Tau&II&  $<$0.33 &   $<$0.33 &   $<$0.09 &   $<$0.09 &   $<$0.09 &   $<$0.17  &   $<$0.21 &   $<$0.21 \\ 
DE Tau&II&   -- &    -- &   $<$0.08 &   $<$0.08 &   $<$0.08 &   $<$0.11  &    -- &    -- \\ 
DK Tau&II&  $<$0.34 &   $<$0.34 &    -- &    -- &    -- &    -- &        -- &    -- \\ 
DM Tau&T&  $<$0.25 &   $<$0.25 &   $<$0.12 &   $<$0.12 &   $<$0.12 &   $<$0.18  &   $<$0.16 &   $<$0.16 \\ 
DN Tau&T&  $<$0.35 &   $<$0.35 &   $<$0.13 &   $<$0.13 &   $<$0.13 &   $<$0.19 &   $<$0.16 &   $<$0.16 \\ 

FO Tau &T& -- &-- &-- &--&--&--&--&--\\
FM Tau&II&  $<$0.24 &   $<$0.24 &   $<$0.07 &   $<$0.07&   $<$0.07 &   $<$0.12  &   $<$0.13 &   $<$0.13 \\ 
FQ Tau&II&   -- &    -- &   $<$0.09 &   $<$0.09 &   $<$0.09 &   $<$0.08  &    -- &    -- \\ 
FT Tau&II&   -- &    -- &   $<$0.14 &   $<$0.14 &   $<$0.14 &    -- &       -- &    -- \\ 
GH Tau&II&   -- &    -- &   $<$0.10 &   $<$0.10&   $<$0.10 &   $<$0.12  &    -- &    -- \\ 
GM Aur&T&  $<$0.30 &   $<$0.30 &   $<$0.14 &   $<$0.14&   $<$0.14  &   $<$0.17 &   $<$0.21 &   $<$0.21 \\ 
HK Tau&II&   -- &    -- &   $<$0.07 &   $<$0.07 &   $<$0.07 &   $<$0.10  &    -- &    -- \\ 
IQ Tau&II&  $<$0.42 &   $<$0.42 &   $<$0.12 &   $<$0.12 &   $<$0.12 &   $<$0.21 &   $<$0.17 &   $<$0.17 \\ 
LkCa 15&T&  $<$0.31 &   $<$0.31 &   $<$0.09 &   $<$0.09&   $<$0.09 &   $<$0.17  &   $<$0.19 &   $<$0.19 \\ 
\hline
\end{tabular}
\end{center}
\end{table}
\end{landscape}
\end{appendix}

\clearpage
\begin{appendix}
\section{Detection fractions}
\label{detFrac}
Tables \ref{detRates}, \ref{detRatesClass}, and \ref{detRatesSpT} show respectively the detection fractions statistics in terms of outflow/non-outflow, SED classes and spectral types.

\begin{table*}[htpb]
\setcounter{table}{0}
\begin{center}
\caption{Detection fractions for the entire sample, outflow sources only, and non-outflow sources only.}
\label{detRates}
\begin{tabular}{lrrrr}
\hline
\hline\\[-0.2mm]
\multicolumn{1}{l}{Species}&
\multicolumn{1}{c}{$\lambda(\mu m)$}&
\multicolumn{1}{c}{All}&
\multicolumn{1}{c}{Outflow}&
\multicolumn{1}{c}{Non-outflow}\\
\hline\\
\multicolumn{5}{c}{Atomic}\\
\hline\\
$[$OI]&63.18       &55$^{+5}_{-6}$\%(42/76) &100$^{+1}_{-6}$\%(27/27)&31$^{+7}_{-6}$\%(15/49)\\[0.5mm]
$[$OI]&145.53      &49$^{+8}_{-8}$\%(19/39) &75$^{+7}_{-11}$\%(18/24)&7$^{+12}_{-2}$\%(1/15)$^{\star}$\\[0.5mm]
$[$CII]&157.74     &34$^{+8}_{-7}$\%(13/38) &54$^{+9}_{-10}$\%(13/24)&--(0/14)$^{\dagger}$\\[0.5mm]
\hline\\
\multicolumn{5}{c}{Molecular}\\
\hline\\
o--H$_{2}$O&63.32  &17$^{+5}_{-3}$\%(13/76) &37$^{+10}_{-8}$\%(10/27)&6$^{+5}_{-2}$\%(3/49)$^{\ddagger}$\\[0.5mm]
o--H$_{2}$O&78.74  &32$^{+8}_{-6}$\%(12/38) &50$^{+10}_{-10}$\%(12/24)&--(0/14)\\[0.5mm]
o--H$_{2}$O&179.53 &17$^{+9}_{-5}$\%(5/30) &23$^{+11}_{-6}$\%(5/22)&--(0/8)\\[0.5mm]
o--H$_{2}$O&180.49 &10$^{+8}_{-3}$\%(3/30) &14$^{+10}_{-4}$\%(3/22)&--(0/8)\\[0.5mm]
\hline\\[0.2mm]
p--H$_{2}$O&78.92  &18$^{+8}_{-5}$\%(7/38) &29$^{+11}_{-7}$\%(7/24)&--(0/14)\\[0.5mm]
p--H$_{2}$O&89.99  &30$^{+9}_{-7}$\%(9/30) &41$^{+11}_{-9}$\%(9/22)&--(0/8)\\[0.5mm]
p--H$_{2}$O&144.52 &23$^{+8}_{-5}$\%(9/39) &38$^{+10}_{-9}$\%(9/24)&--(0/15)\\[0.5mm]
\hline\\
CO&72.84           &5$^{+6}_{-2}$\%(2/39) &8$^{+9}_{-3}$\%(2/24)&--(0/15)\\[0.5mm]
CO&79.36           &5$^{+6}_{-2}$\%(2/38) &8$^{+9}_{-3}$\%(2/24)&--(0/14)\\[0.5mm]
CO&90.16           &27$^{+9}_{-6}$\%(8/30) &36$^{+11}_{-9}$\%(8/22)&--(0/8)\\[0.5mm]
CO&144.78          &56$^{+7}_{-8}$\%(22/39) &79$^{+6}_{-10}$\%(19/24)&20$^{+14}_{-7}$\%(3/15)\\[0.5mm]
\hline\\
OH&79.11           &34$^{+8}_{-7}$\%(13/38) &54$^{+9}_{-10}$\%(13/24)&--(0/14)\\[0.5mm]
OH&79.18           &32$^{+8}_{-6}$\%(12/38) &50$^{+10}_{-10}$\%(12/24)&--(0/14)\\[0.5mm]
\hline\\
\end{tabular}
\end{center}
\textbf{Notes.} $^{(\star)}$ Detections in DE Tau. $^{(\ddagger)}$ Detections in: BP Tau (deeper observations), GI/GK Tau (GK Tau is outflow GI Tau is not), IQ Tau. The parentheses indicate the ratio of targets where a certain line is detected over the total number of targets observed. The hyphens mean no detections.
\end{table*}

\begin{table*}[htpb]
\setcounter{table}{1}
\centering
\caption{Detection fractions in terms of SED Class.}
\label{detRatesClass}
\begin{tabular}{lrrrrr}
\hline
\hline\\[-0.2mm]
\multicolumn{1}{l}{Species}&
\multicolumn{1}{c}{$\lambda(\mu m)$}&
\multicolumn{1}{c}{Class I}&
\multicolumn{1}{c}{Class II}&
\multicolumn{1}{c}{Class TD}&
\multicolumn{1}{c}{Class III}\\
\hline\\
\multicolumn{6}{c}{Atomic}\\
\hline\\
$[$OI]&63.18       &100$^{+3}_{-26}$\%(5/5) &70$^{+6}_{-8}$\%(31/44)&60$^{+12}_{-16}$\%(6/10)&--(0/17)\\[0.5mm]
$[$OI]&145.53      &100$^{+3}_{-31}$\%(4/4) &48$^{+9}_{-9}$\%(14/29)&17$^{+23}_{-6}$\%(1/6)&\\[0.5mm]
$[$CII]&157.74     &50$^{+20}_{-20}$\%(2/4) &36$^{+10}_{-8}$\%(10/28)&17$^{+23}_{-6}$\%(1/6)&\\[0.5mm]
\hline\\
\multicolumn{6}{c}{Molecular}\\
\hline\\
o--H$_{2}$O&63.32  &40$^{+21}_{-16}$\%(2/5) &23$^{+7}_{-5}$\%(10/44)&10$^{+17}_{-3}$\%(1/10)&--(0/17)\\[0.5mm]
o--H$_{2}$O&78.74  &50$^{+20}_{-20}$\%(2/4) &32$^{+10}_{-7}$\%(9/28)&17$^{+23}_{-6}$\%(1/6)&\\[0.5mm]
o--H$_{2}$O&179.53 &--(0/4) &24$^{+11}_{-7}$\%(5/21)&--(0/5)&\\[0.5mm]
o--H$_{2}$O&180.49 &--(0/4) &14$^{+11}_{-5}$\%(3/21)&--(0/5)&\\[0.5mm]
\hline\\[0.2mm]
p--H$_{2}$O&78.92  &50$^{+20}_{-20}$\%(2/4) &18$^{+9}_{-5}$\%(5/28)&--(0/6)&\\[0.5mm]
p--H$_{2}$O&89.99  &25$^{+27}_{-10}$\%(1/4) &33$^{+11}_{-8}$\%(7/21)&20$^{+25}_{-8}$\%(1/5)&\\[0.5mm]
p--H$_{2}$O&144.52 &50$^{+20}_{-20}$\%(2/4) &21$^{+9}_{-5}$\%(6/29)&17$^{+23}_{-6}$\%(1/6)&\\[0.5mm]
\hline\\
CO&72.84           &--(0/4) &7$^{+8}_{-2}$\%(2/29)&--(0/6)&\\[0.5mm]
CO&79.36           &--(0/4) &7$^{+8}_{-2}$\%(2/28)&--(0/6)&\\[0.5mm]
CO&90.16           &25$^{+27}_{-10}$\%(1/4) &33$^{+11}_{-8}$\%(7/21)&--(0/5)&\\[0.5mm]
CO&144.78          &100$^{+3}_{-31}$\%(4/4) &59$^{+8}_{-9}$\%(17/29)&17$^{+23}_{-6}$\%(1/6)&\\[0.5mm]
\hline\\
OH&79.11           &50$^{+20}_{-20}$\%(2/4) &39$^{+10}_{-8}$\%(11/28)&--(0/6)&\\[0.5mm]
OH&79.18           &50$^{+20}_{-20}$\%(2/4) &36$^{+10}_{-8}$\%(10/28)&--(0/6)&\\[0.5mm]
\hline\\
\end{tabular}
\end{table*}

\begin{table*}[htpb]
\setcounter{table}{2}
\centering
\caption{Detection fractions in terms of SpT.}
\label{detRatesSpT}
\begin{tabular}{lrrrrr}
\hline
\hline\\[-0.2mm]
\multicolumn{1}{l}{Species}&
\multicolumn{1}{c}{$\lambda(\mu m)$}&
\multicolumn{1}{c}{G0-K7}&
\multicolumn{1}{c}{K7-M1}&
\multicolumn{1}{c}{M1-M3}&
\multicolumn{1}{c}{M3-M6}\\
\hline\\
\multicolumn{6}{c}{Atomic}\\
\hline\\

$[$OI]&63.18       &84$^{+5}_{-12}$\%(16/19) &61$^{+10}_{-12}$\%(11/18)&35$^{+12}_{-9}$\%(7/20)&42$^{+11}_{-10}$\%(8/19)\\[0.5mm]
$[$OI]&145.53      &62$^{+11}_{-14}$\%(8/13) &50$^{+16}_{-16}$\%(4/8)&38$^{+18}_{-13}$\%(3/8)&40$^{+16}_{-12}$\%(4/10)\\[0.5mm]
$[$CII]&157.74     &46$^{+13}_{-12}$\%(6/13) &25$^{+19}_{-9}$\%(2/8)&25$^{+19}_{-9}$\%(2/8)&33$^{+17}_{-11}$\%(3/9)\\[0.5mm]
\hline\\
\multicolumn{6}{c}{Molecular}\\
\hline\\
o--H$_{2}$O&63.32  &32$^{+12}_{-8}$\%(6/19) &17$^{+12}_{-5}$\%(3/18)&10$^{+11}_{-3}$\%(2/20)&11$^{+11}_{-4}$\%(2/19)\\[0.5mm]
o--H$_{2}$O&78.74  &31$^{+15}_{-9}$\%(4/13) &38$^{+18}_{-13}$\%(3/8)&38$^{+18}_{-13}$\%(3/8)&22$^{+18}_{-8}$\%(2/9)\\[0.5mm]
o--H$_{2}$O&179.53 &23$^{+15}_{-8}$\%(3/13) &--(0/5)&20$^{+25}_{-8}$\%(1/5)&14$^{+21}_{-5}$\%(1/7)\\[0.5mm]
o--H$_{2}$O&180.49 &8$^{+14}_{-3}$\%(1/13) &--(0/5)&20$^{+25}_{-8}$\%(1/5)&14$^{+21}_{-5}$\%(1/7)\\[0.5mm]
\hline\\[0.2mm]
p--H$_{2}$O&78.92  &23$^{+15}_{-8}$\%(3/13) &13$^{+20}_{-4}$\%(1/8)&13$^{+20}_{-4}$\%(1/8)&22$^{+18}_{-8}$\%(2/9)\\[0.5mm]
p--H$_{2}$O&89.99  &54$^{+12}_{-13}$\%(7/13) &--(0/5)&20$^{+25}_{-8}$\%(1/5)&14$^{+21}_{-5}$\%(1/7)\\[0.5mm]
p--H$_{2}$O&144.52 &31$^{+15}_{-9}$\%(4/13) &--(0/8)&25$^{+19}_{-9}$\%(2/8)&30$^{+17}_{-10}$\%(3/10)\\[0.5mm]
\hline\\
CO&72.84           &15$^{+15}_{-5}$\%(2/13) &--(0/8)&--(0/8)&--(0/10)\\[0.5mm]
CO&79.36           &15$^{+15}_{-5}$\%(2/13) &--(0/8)&--(0/8)&--(0/9)\\[0.5mm]
CO&90.16           &38$^{+14}_{-11}$\%(5/13) &--(0/5)&40$^{+21}_{-16}$\%(2/5)&14$^{+21}_{-5}$\%(1/7)\\[0.5mm]
CO&144.78          &69$^{+9}_{-15}$\%(9/13) &50$^{+16}_{-16}$\%(4/8)&63$^{+13}_{-18}$\%(5/8)&40$^{+16}_{-12}$\%(4/10)\\[0.5mm]
\hline\\
OH&79.11           &38$^{+14}_{-11}$\%(5/13) &38$^{+18}_{-13}$\%(3/8)&38$^{+18}_{-13}$\%(3/8)&22$^{+18}_{-8}$\%(2/9)\\[0.5mm]
OH&79.18           &38$^{+14}_{-11}$\%(5/13) &25$^{+19}_{-9}$\%(2/8)&38$^{+18}_{-13}$\%(3/8)&22$^{+18}_{-8}$\%(2/9)\\[0.5mm]
\hline\\
\end{tabular}
\end{table*}

\end{appendix}

\clearpage
\begin{appendix}
\section{Diameter of emitting regions}
\label{app:emitReg}

As shown in \cite{Flower2010}, the relation between the emergent flux, $F_e$, in erg cm$^{\rm -2}$ s$^{-1}$ and $T{\rm d}V$ in K km s$^{\rm -1}$ is
\begin{equation}
F_e = {{8\times10^5 \pi k_B}\over{\lambda^3}} T dV
,\end{equation}
\noindent where $\lambda$ is the wavelength of the transition in cm, and $T{\rm d}V$ as provided in \cite{Flower2015}.

The flux observed at the Earth, $F_0$, then is
\begin{equation}
F_0 = {{F_e\Omega}\over{4\pi}}
,\end{equation}
\noindent where $\Omega$ str is the solid angle subtended by the source.
Given that $\Omega \sim D^2/r^2$, where $r$ and $D$ are the distance to the source and the diameter of the emitting area in au, respectively, then
\begin{equation}
D \sim \sqrt{{{4\pi r^2 F_0} \over{F_e}}}
.\end{equation}

The diameter of the emitting regions provided (see Table \ref{table:emitRegions}) are inferred using only the brightest lines among the ratios (i.e. o-H$_{\rm 2}$O 78.74 $\rm \mu m$, CO 144.78 $\rm \mu m$, and OH 79.12 $\rm \mu m$), besides being compatible with the shock velocities ($V_{\rm shock}$) and pre-shock densities ($n$) observed in Figs. \ref{figure:zoomShock1}, \ref{figure:zoomShock2}, \ref{figure:zoomShock3}, and \ref{figure:zoomShock4}.

\begin{figure}[htpb]
\setcounter{figure}{0}
\centering
\includegraphics[width=0.17\textwidth, trim= 0mm 0mm 0mm 0mm, angle=90]{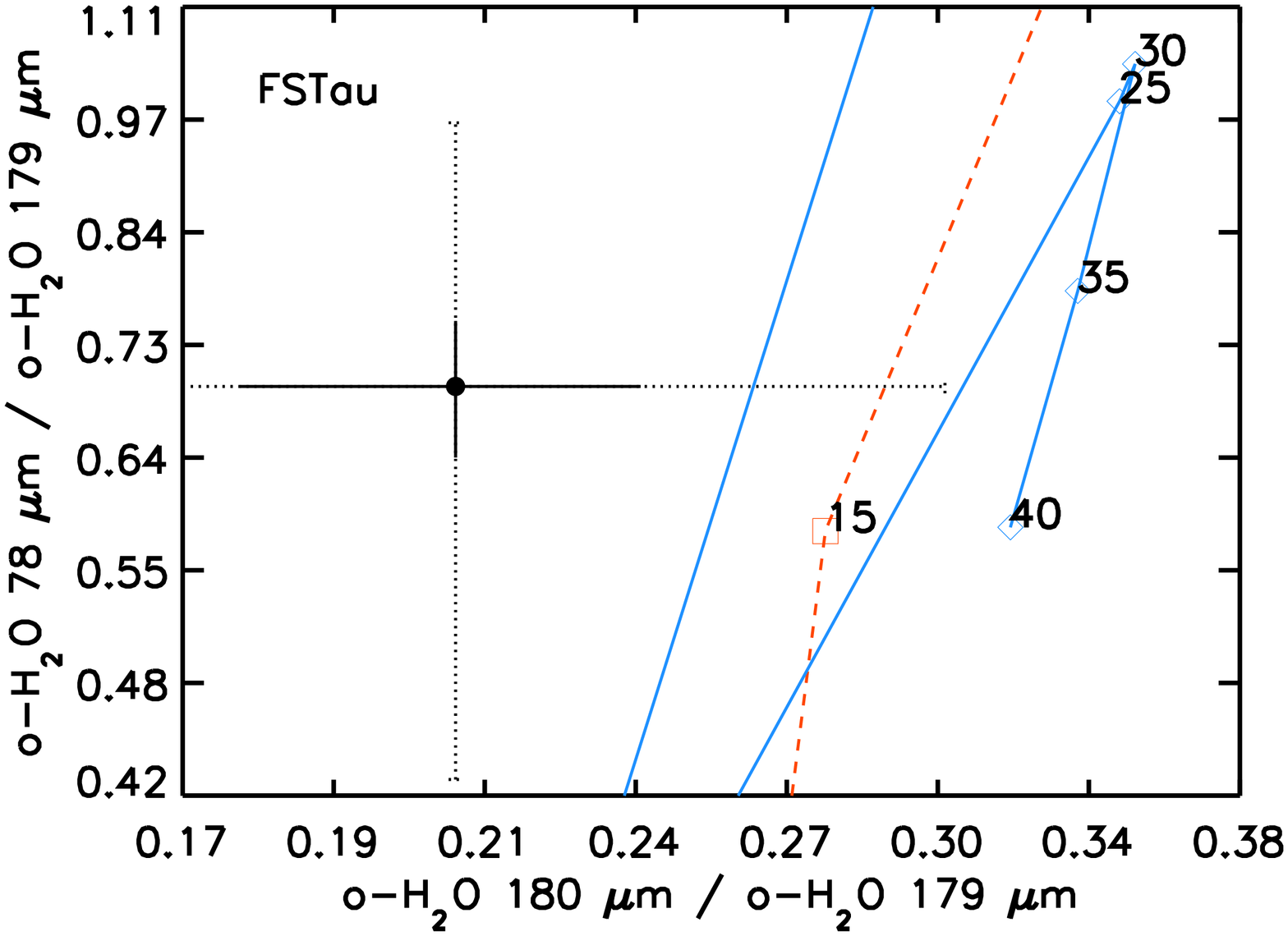}\includegraphics[width=0.17\textwidth, trim= 0mm 0mm 0mm 0mm, angle=90]{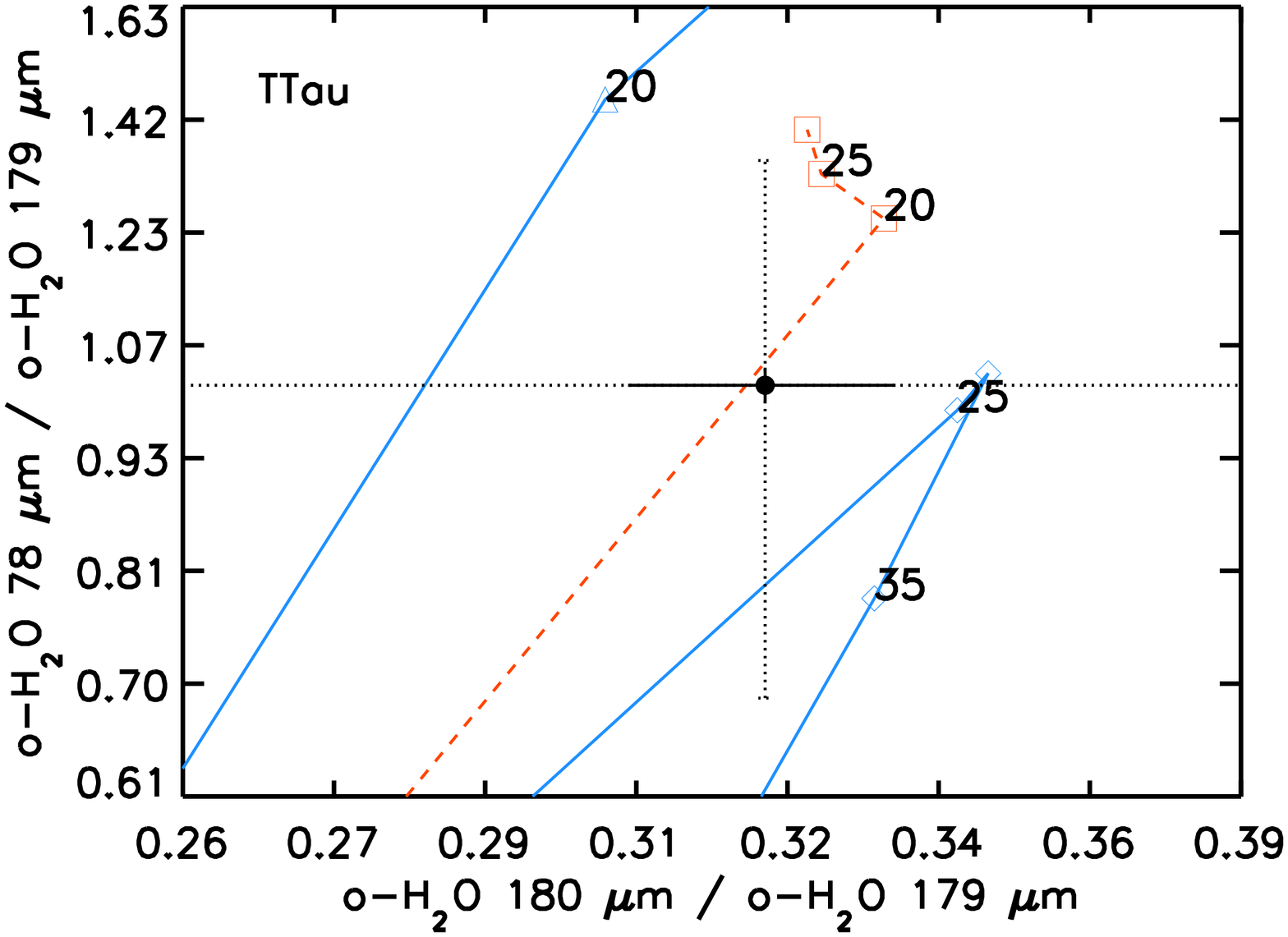}
\includegraphics[width=0.17\textwidth, trim= 0mm 0mm 0mm 0mm, angle=90]{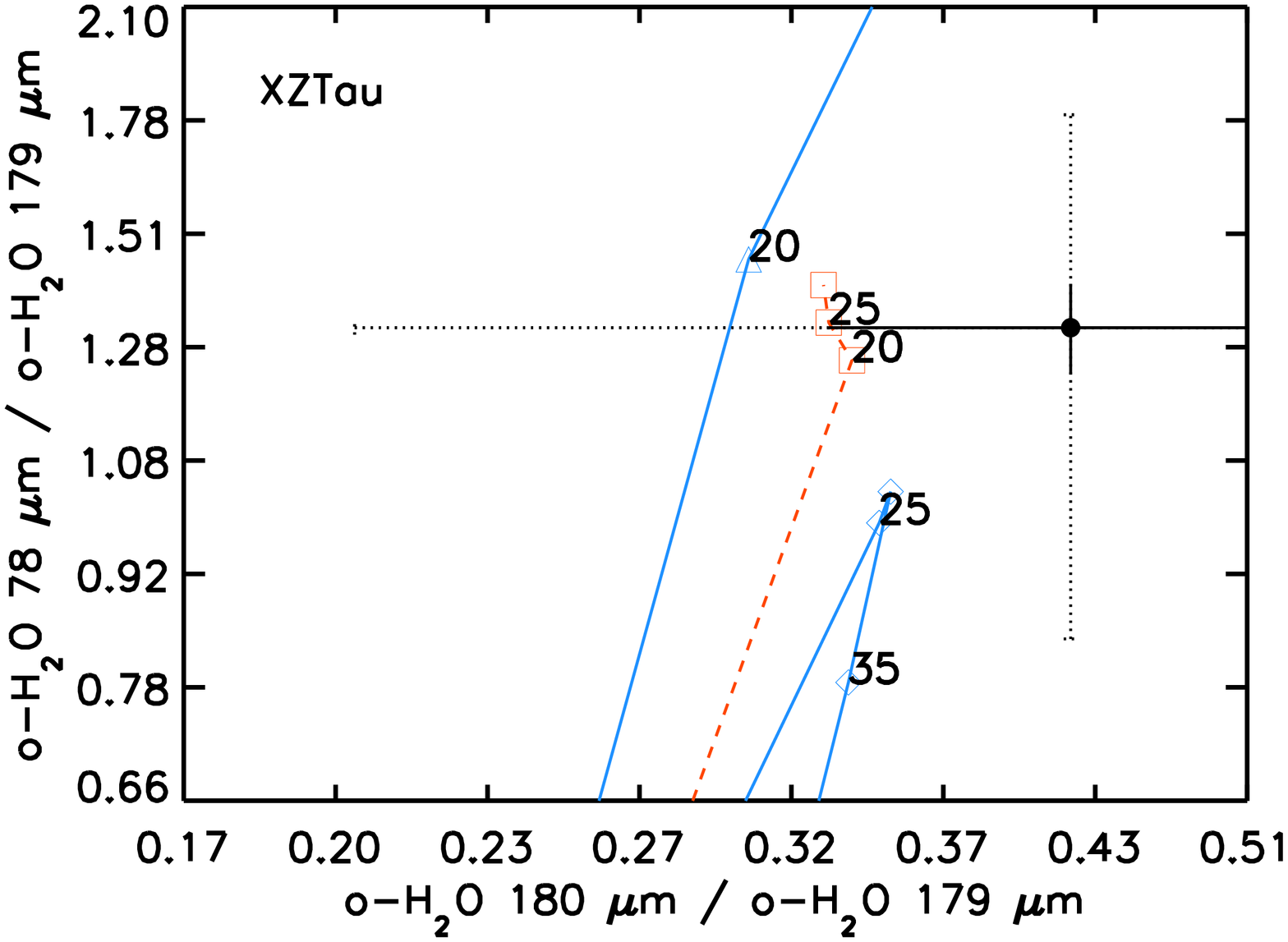}\includegraphics[width=0.17\textwidth, trim= 0mm 0mm 0mm 0mm, angle=90]{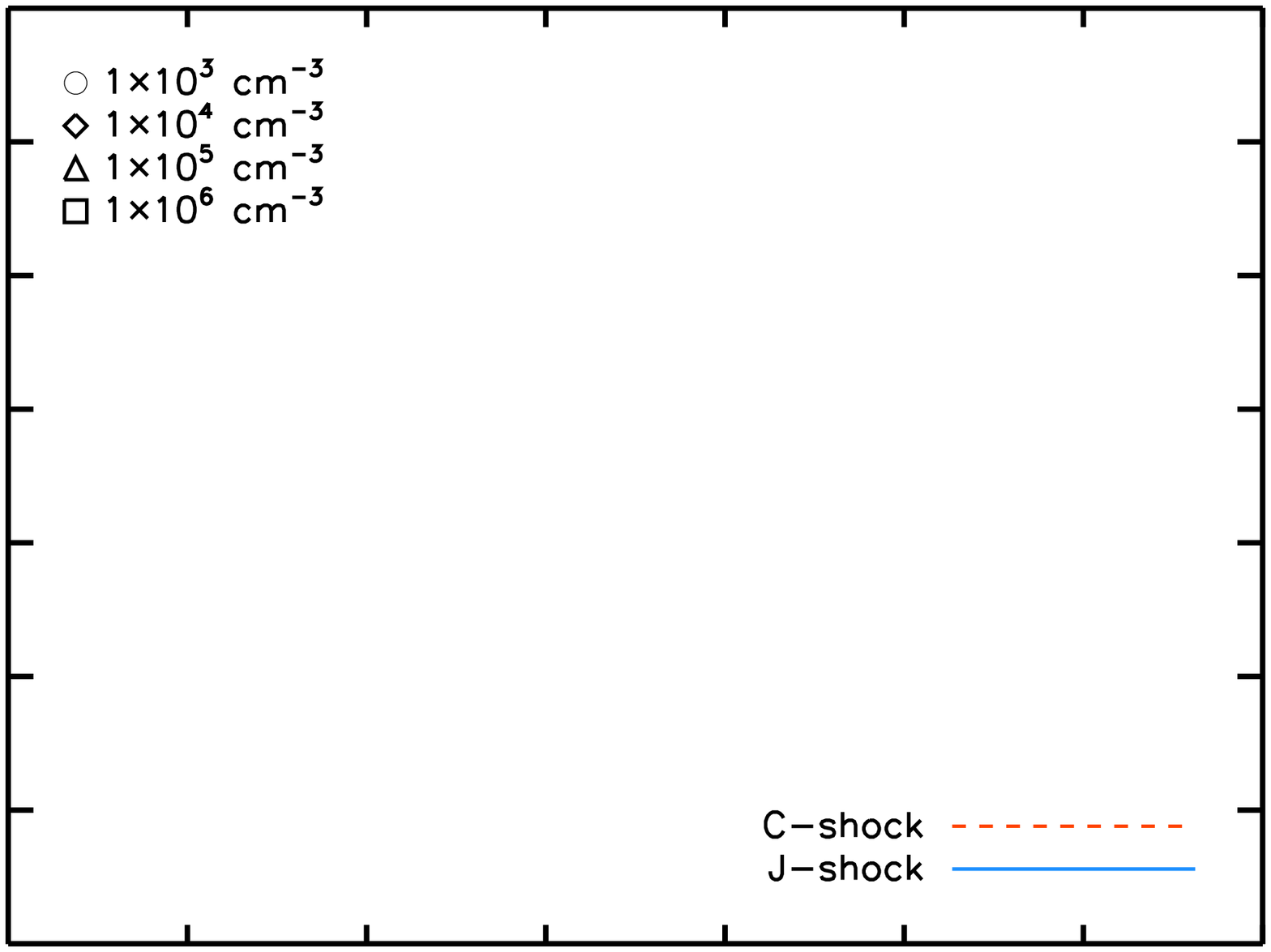}
\caption{Detail of the observed o-H$_{2}$O molecular line ratios (78/179 and 180/179) compared to J-type (\textit{blue}) and C-type (\textit{red}) shock models from \cite{Flower2015} for individual sources. The numbers refer to shock velocities ($V_{\rm shock}$). Different symbols refer to pre-shock densities ($n$). See legend for details.}
\label{figure:zoomShock1}
\end{figure}

\begin{figure}[htpb]
\setcounter{figure}{1}
\centering
\includegraphics[width=0.17\textwidth, trim= 0mm 0mm 0mm 0mm, angle=90]{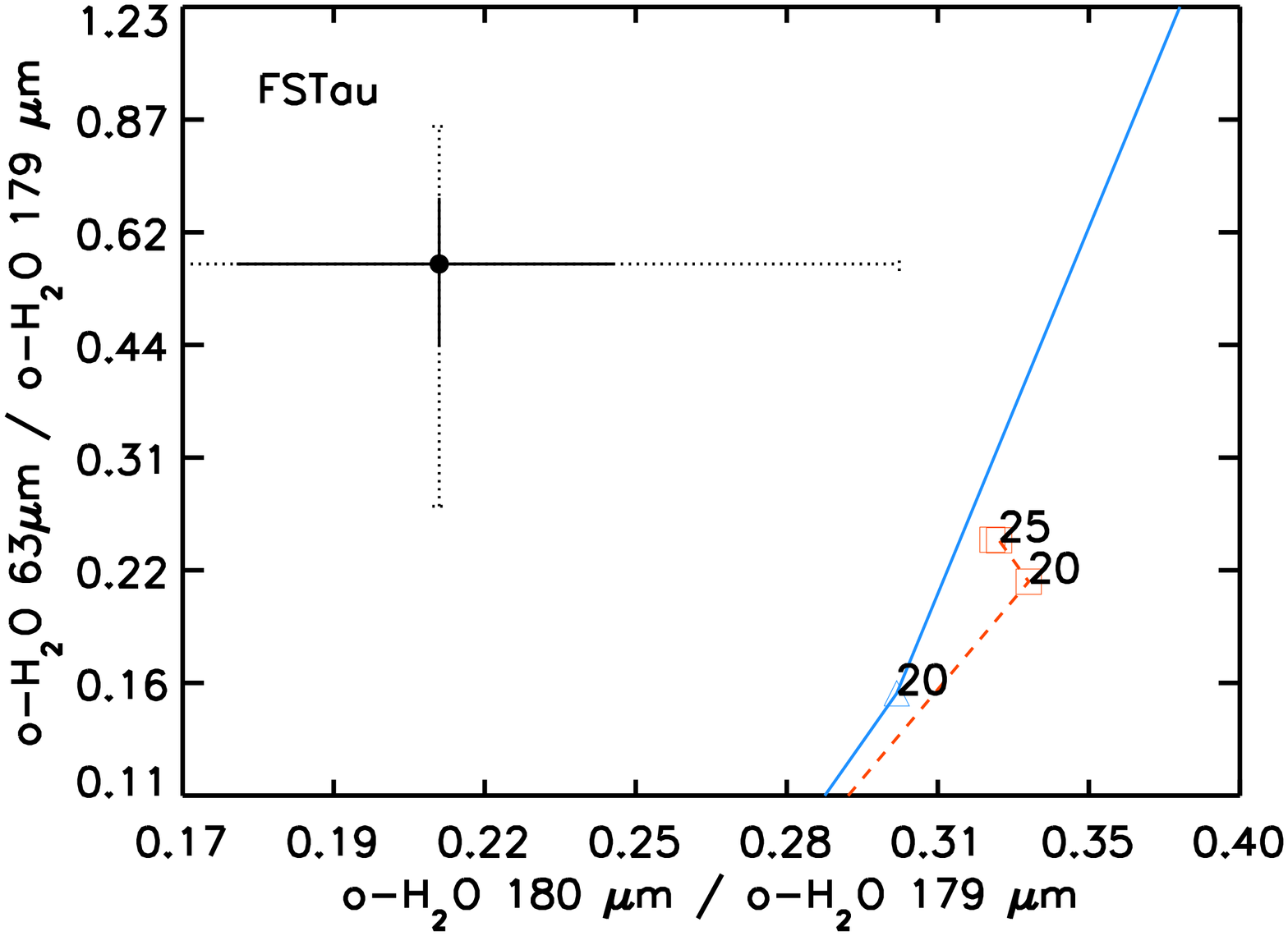}\includegraphics[width=0.17\textwidth, trim= 0mm 0mm 0mm 0mm, angle=90]{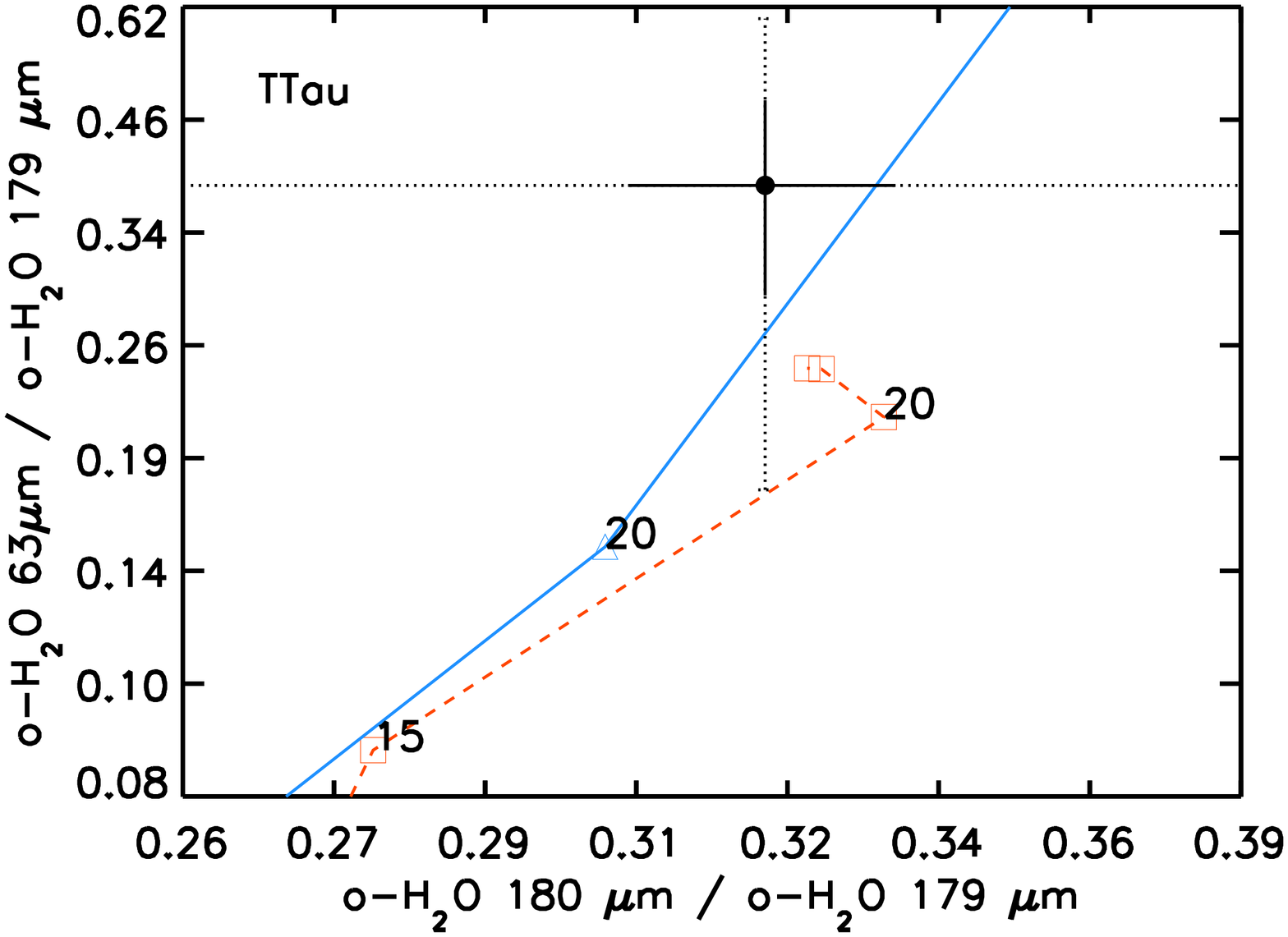}
\includegraphics[width=0.17\textwidth, trim= 0mm 0mm 0mm 0mm, angle=90]{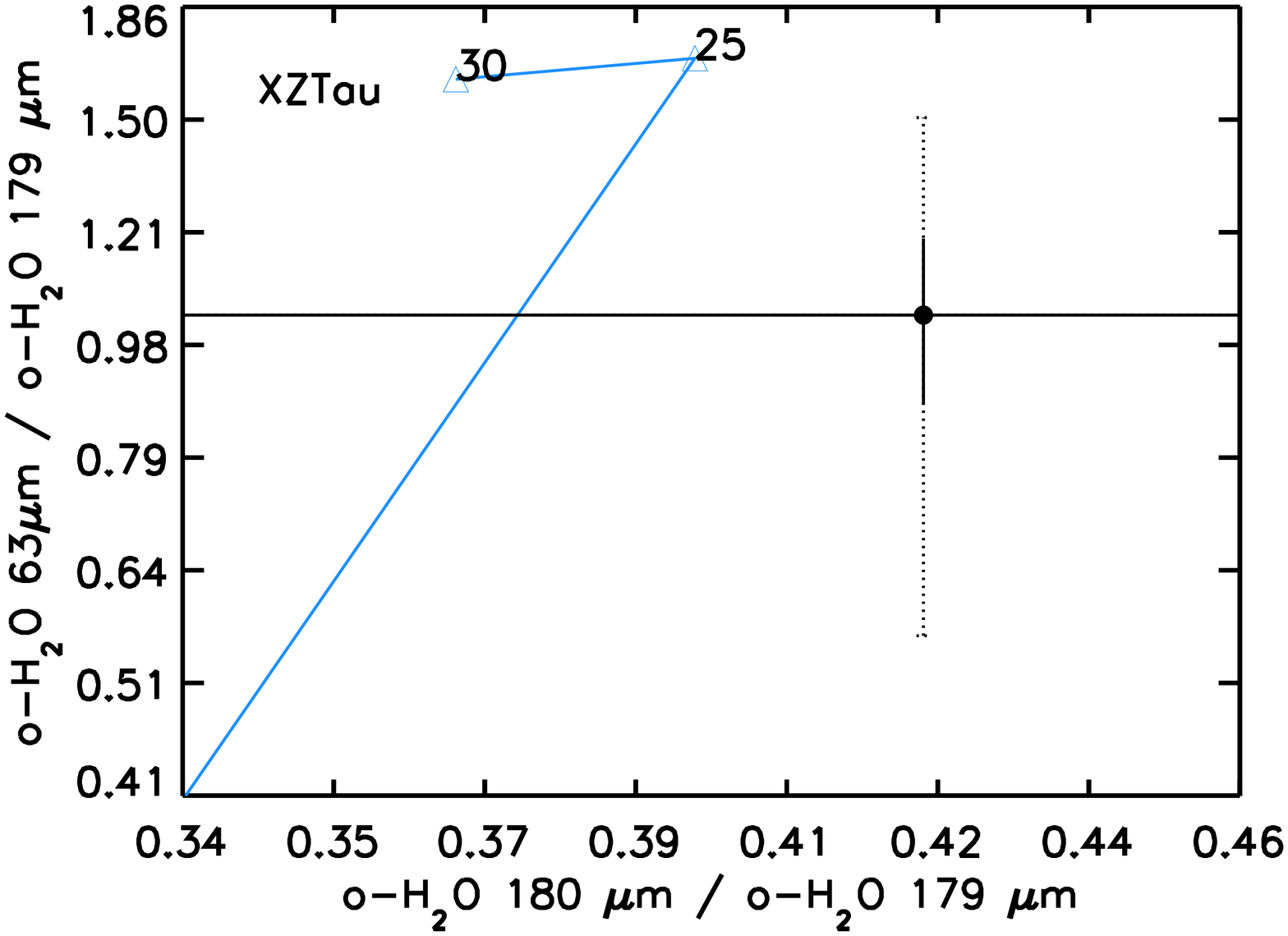}\includegraphics[width=0.17\textwidth, trim= 0mm 0mm 0mm 0mm, angle=90]{em_legend_H2O_1.eps}
\caption{Same as Fig. \ref{figure:zoomShock1}, but for CO and o-H$_{\rm 2}$O/CO ratios.}
\label{figure:zoomShock2}
\end{figure}

\begin{figure}[htpb]
\setcounter{figure}{2}
\centering
\includegraphics[width=0.17\textwidth, trim= 0mm 0mm 0mm 0mm, angle=90]{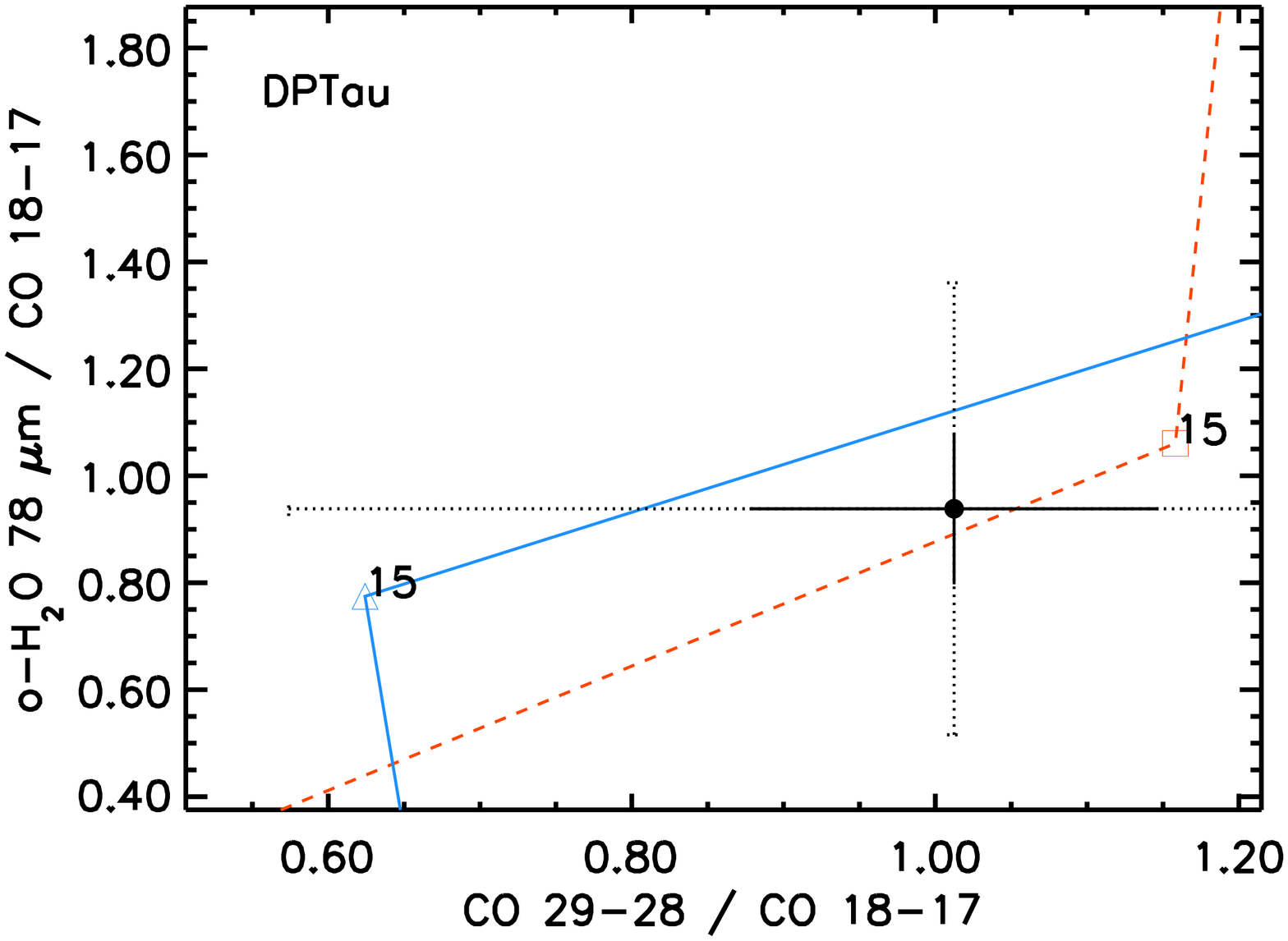}\includegraphics[width=0.17\textwidth, trim= 0mm 0mm 0mm 0mm, angle=90]{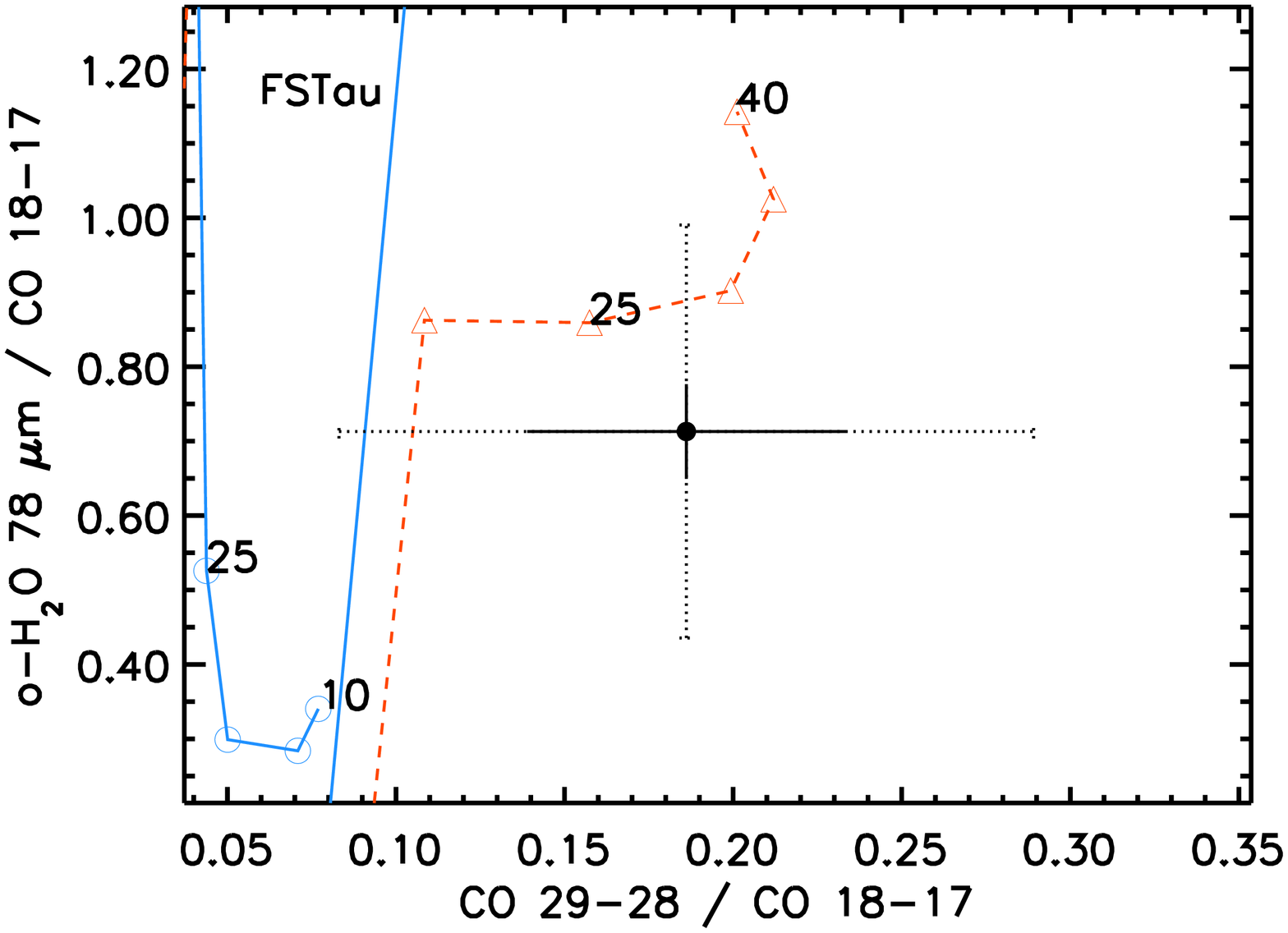}
\includegraphics[width=0.17\textwidth, trim= 0mm 0mm 0mm 0mm, angle=90]{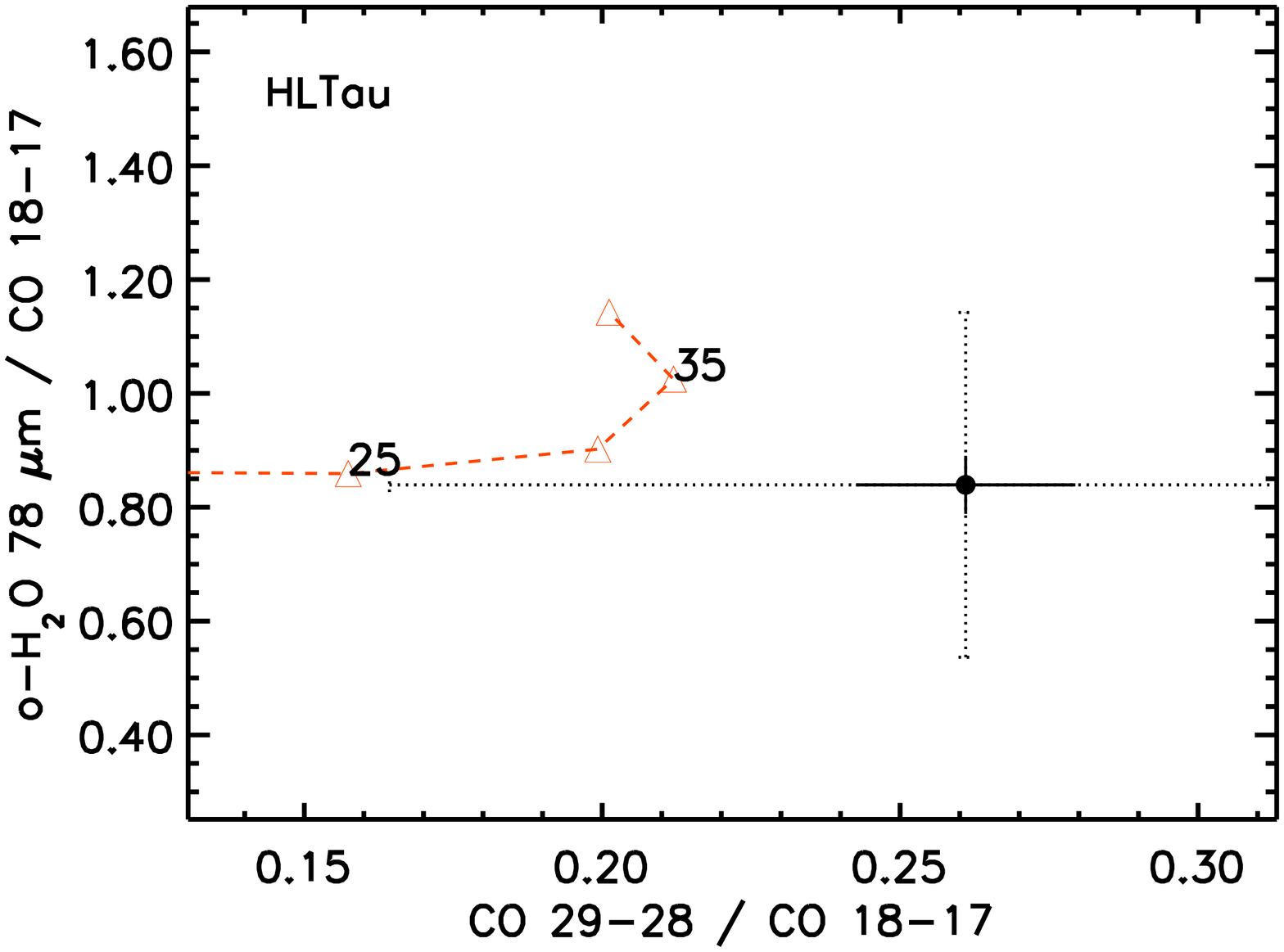}\includegraphics[width=0.17\textwidth, trim= 0mm 0mm 0mm 0mm, angle=90]{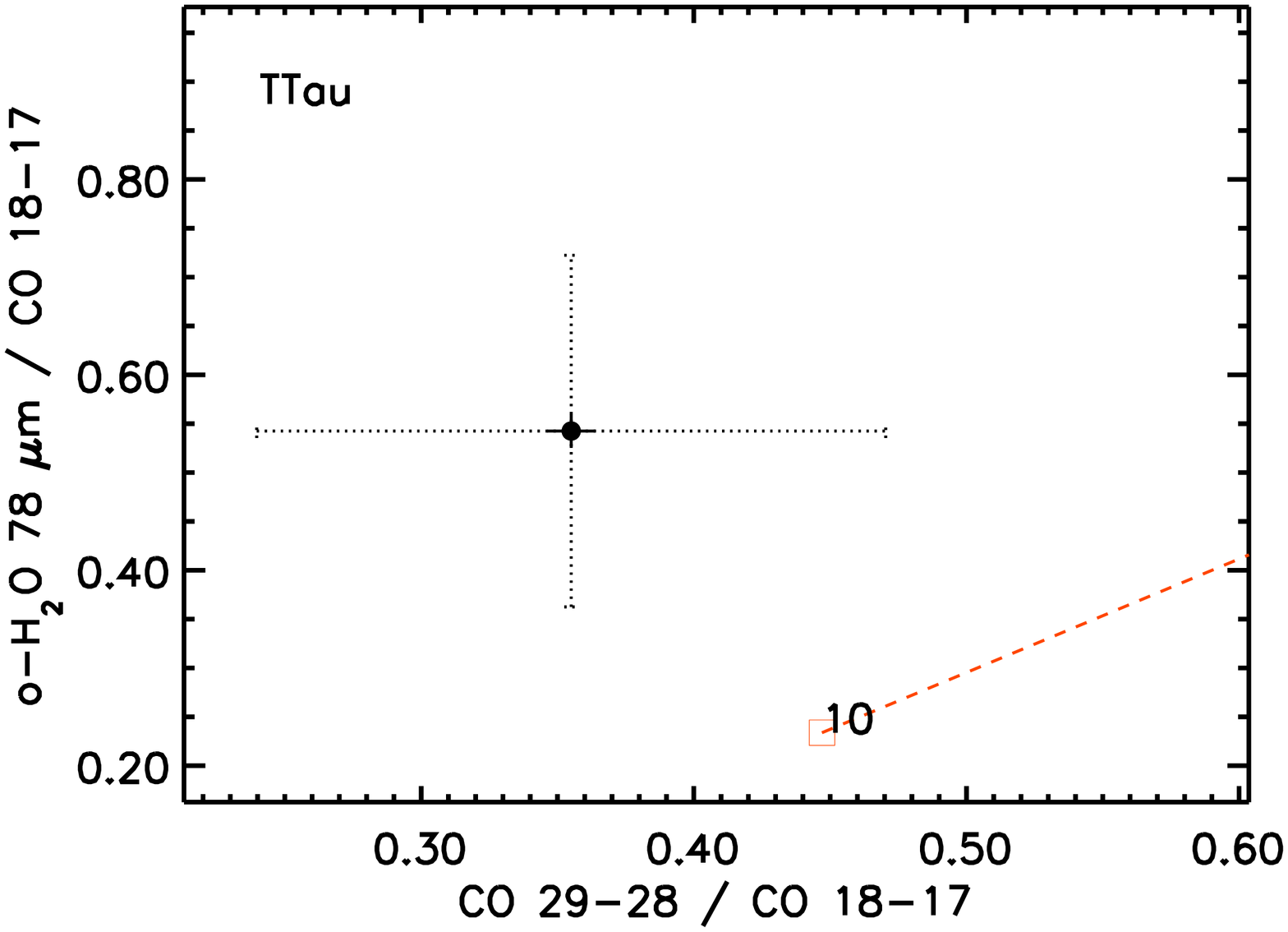}
\includegraphics[width=0.17\textwidth, trim= 0mm 0mm 0mm 0mm, angle=90]{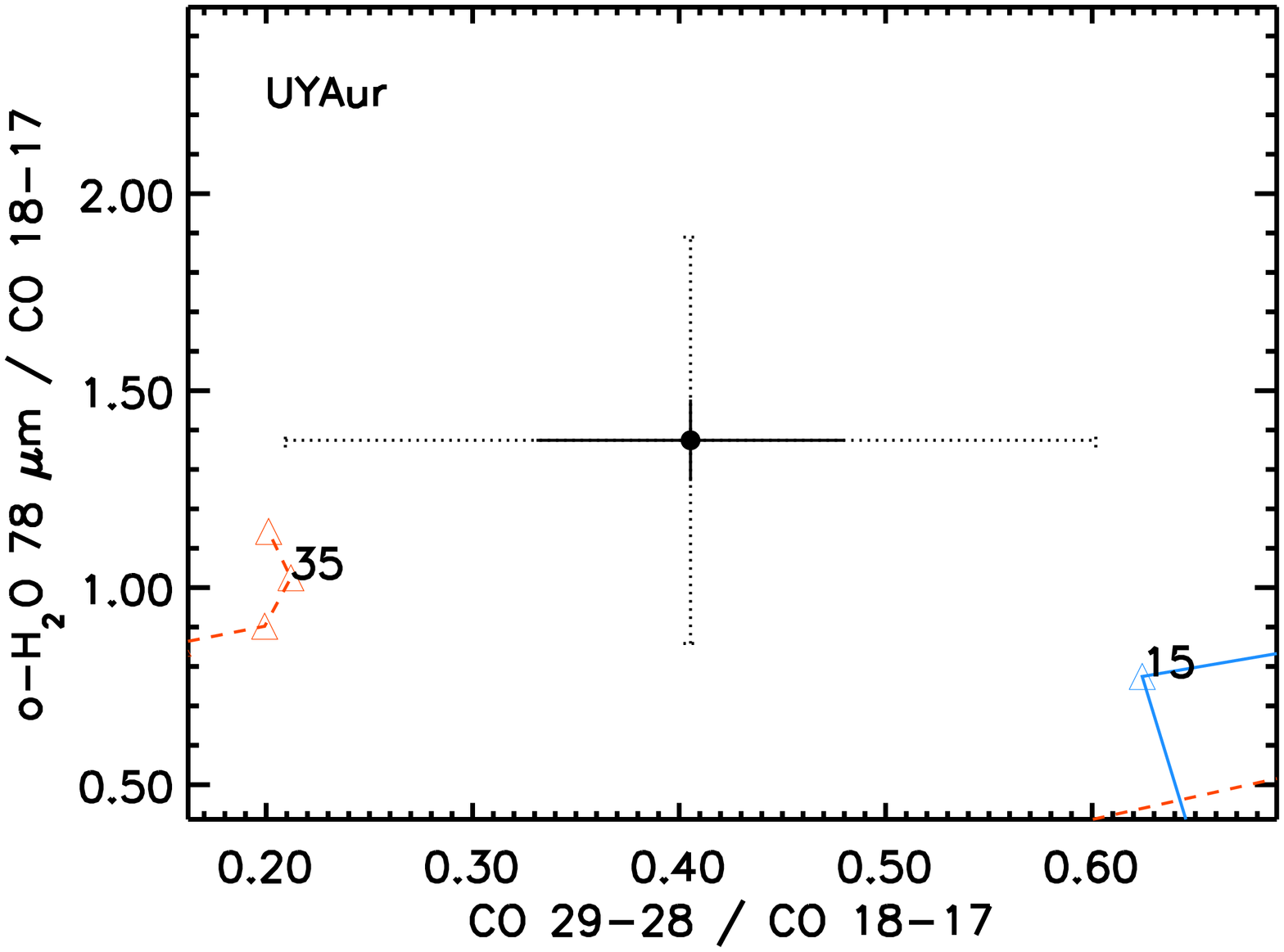}\includegraphics[width=0.17\textwidth, trim= 0mm 0mm 0mm 0mm, angle=90]{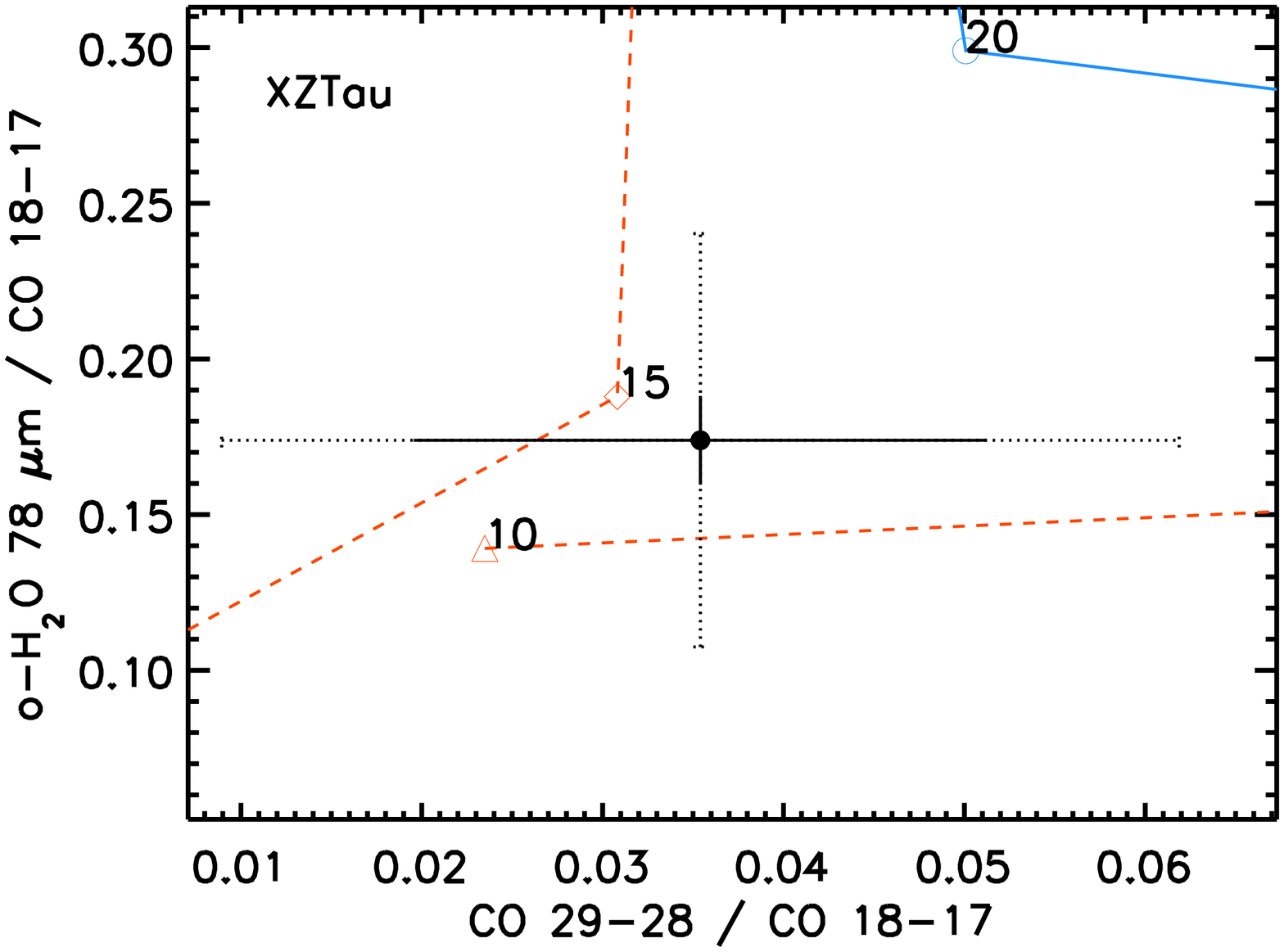}
\caption{Same as Fig. \ref{figure:zoomShock1}, but for CO and o-H$_{\rm 2}$O/CO line ratios.}
\label{figure:zoomShock3}
\end{figure}

\begin{figure}[htpb]
\setcounter{figure}{3}
\centering
\includegraphics[width=0.17\textwidth, trim= 0mm 0mm 0mm 0mm, angle=90]{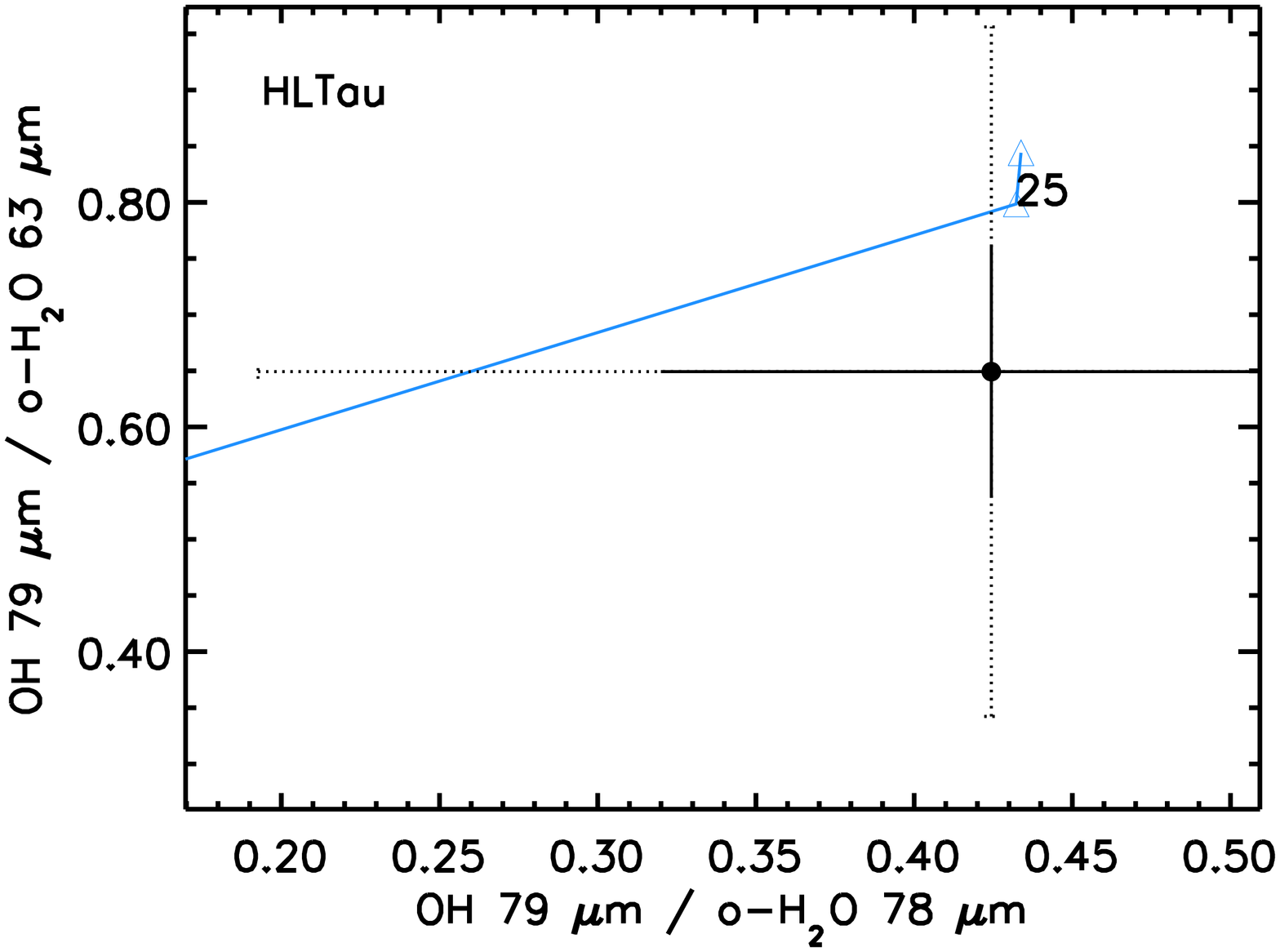}\includegraphics[width=0.17\textwidth, trim= 0mm 0mm 0mm 0mm, angle=90]{em_legend_H2O_1.eps}
\caption{Same as Fig. \ref{figure:zoomShock1}, but for OH/o-H$_{\rm 2}$O line ratios.}
\label{figure:zoomShock4}
\end{figure}
\end{appendix}

%########################################################63um
\clearpage
\begin{appendix}
\section{Spectra 60-190 $\rm \mu m$}

Figures \ref{figure:allSpec63} to \ref{figure:allSpec180} show the objects in the sample with detections in the range 60--190 $\rm \mu m$.

\begin{figure*}[htpb]
\centering
\setcounter{figure}{0}
\includegraphics[width=0.16\textwidth, trim= 0mm 0mm 0mm 0mm, angle=90]{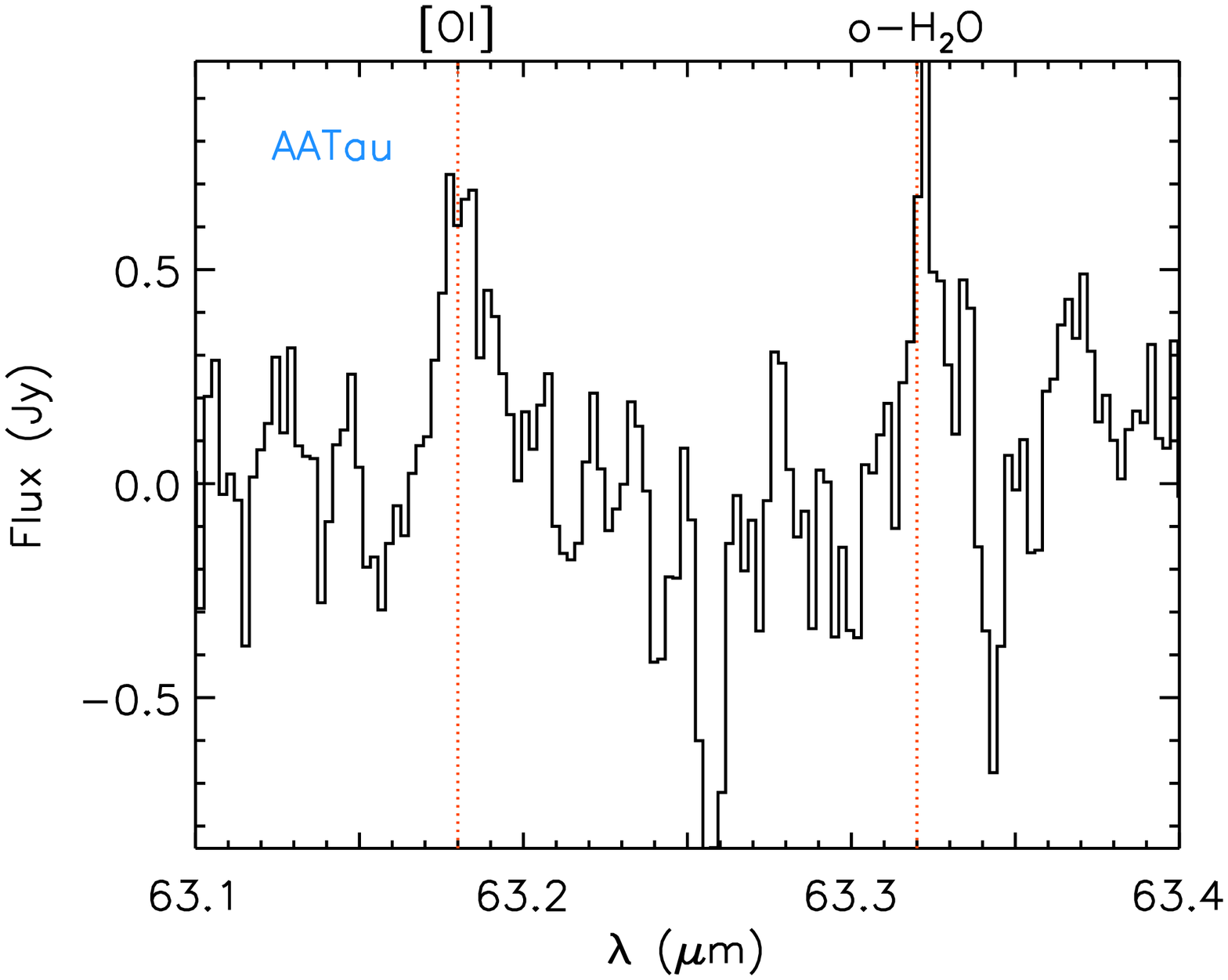}\includegraphics[width=0.16\textwidth, trim= 0mm 0mm 0mm 0mm, angle=90]{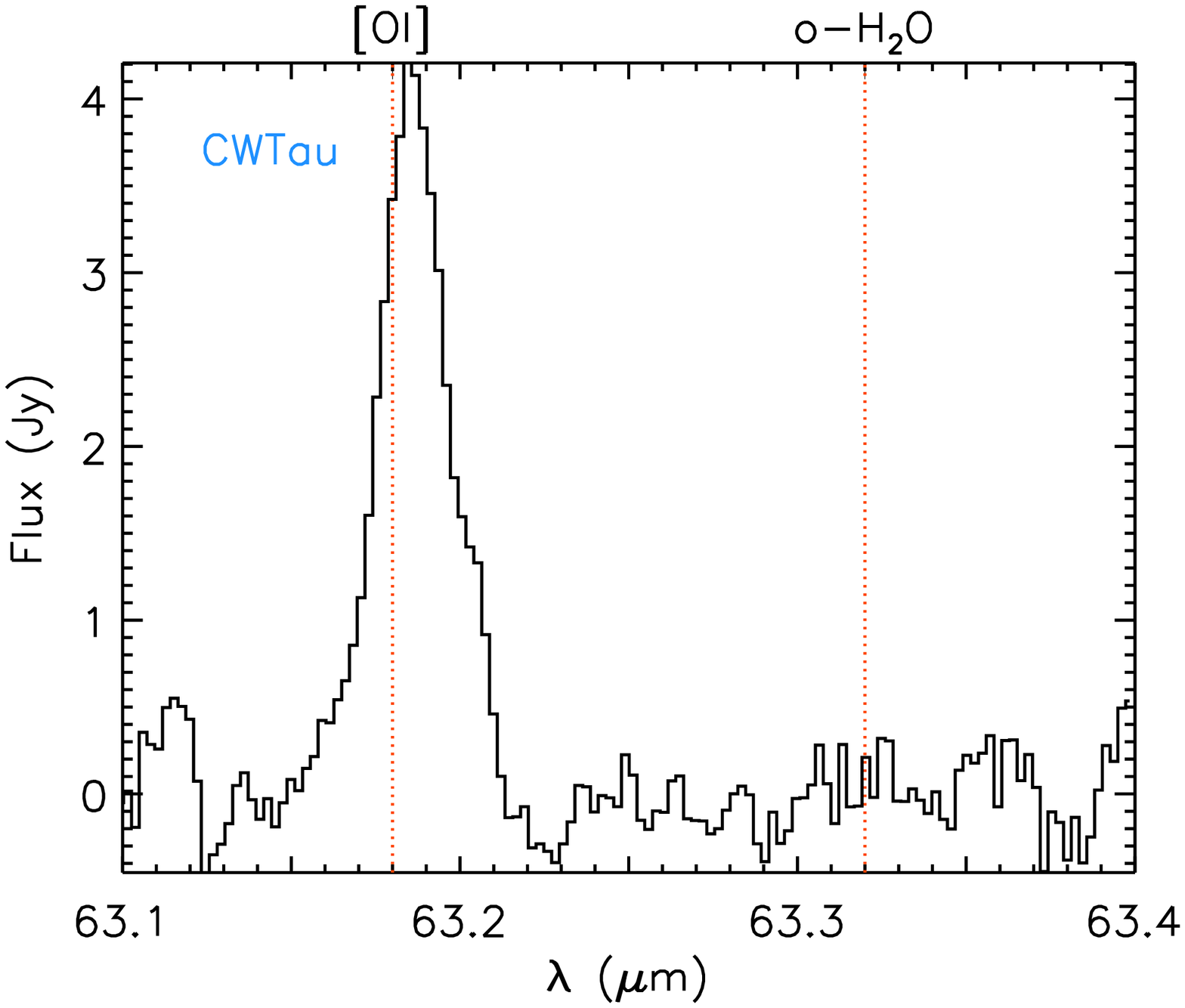}\includegraphics[width=0.16\textwidth, trim= 0mm 0mm 0mm 0mm, angle=90]{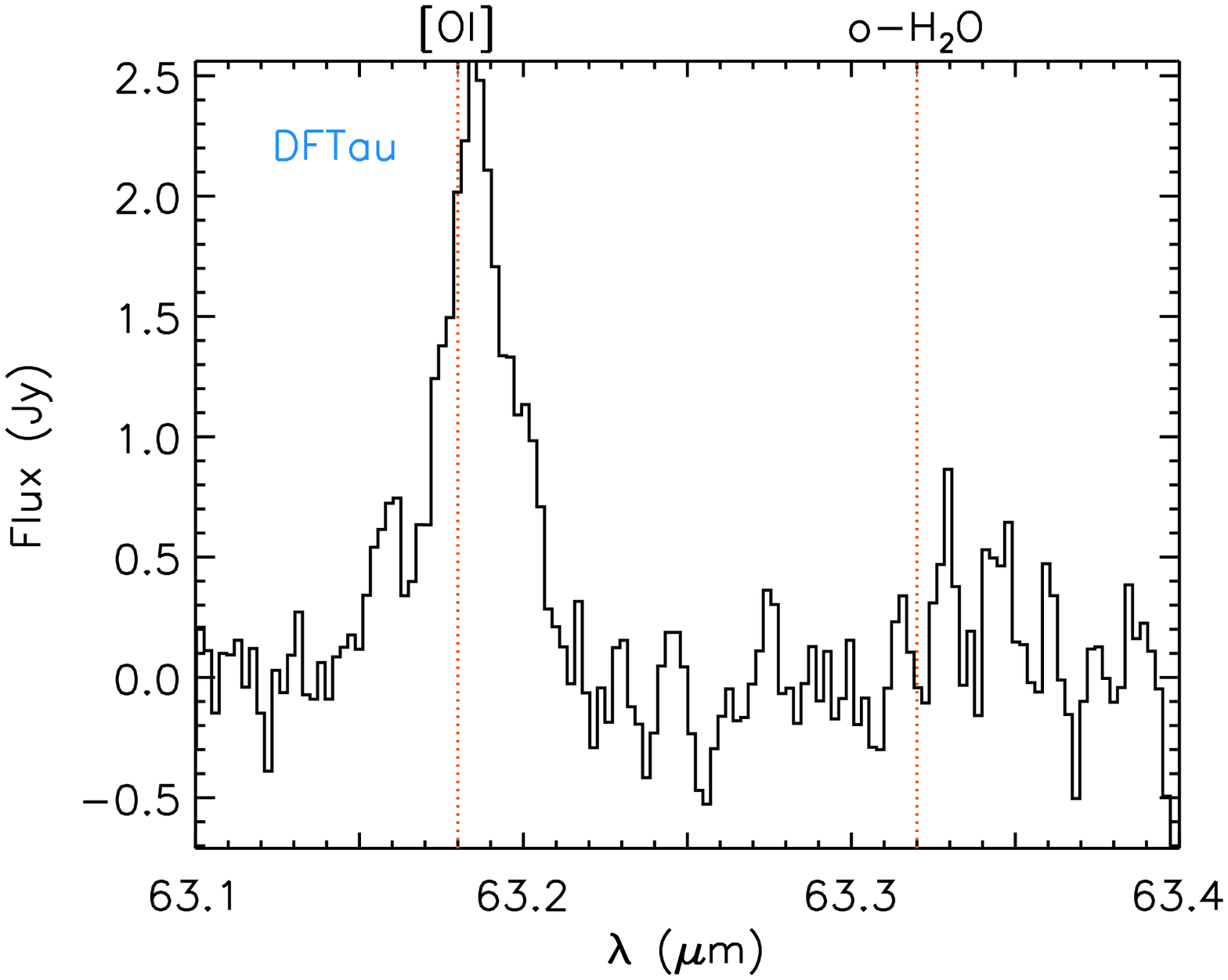}\includegraphics[width=0.16\textwidth, trim= 0mm 0mm 0mm 0mm, angle=90]{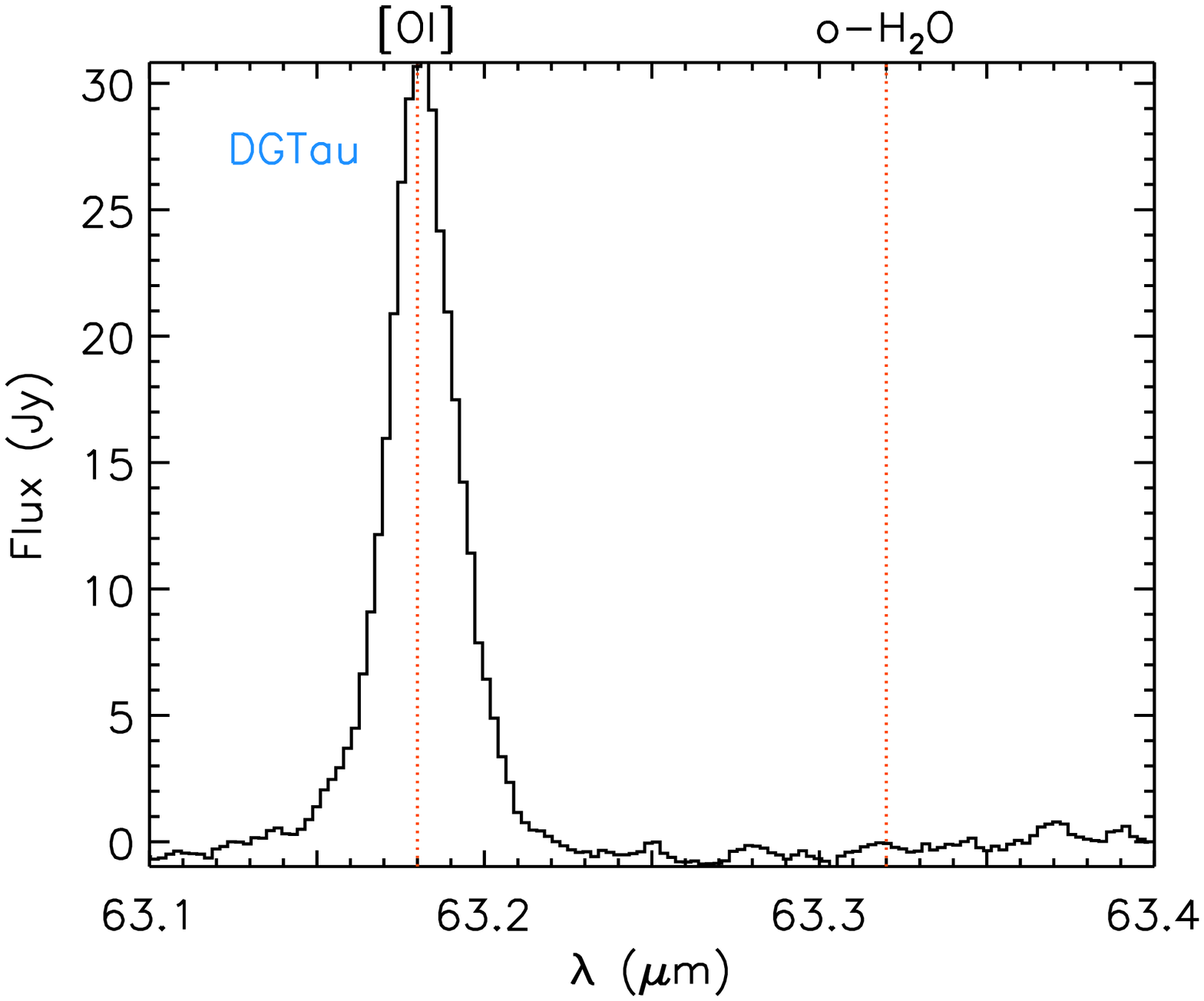}

\includegraphics[width=0.16\textwidth, trim= 0mm 0mm 0mm 0mm, angle=90]{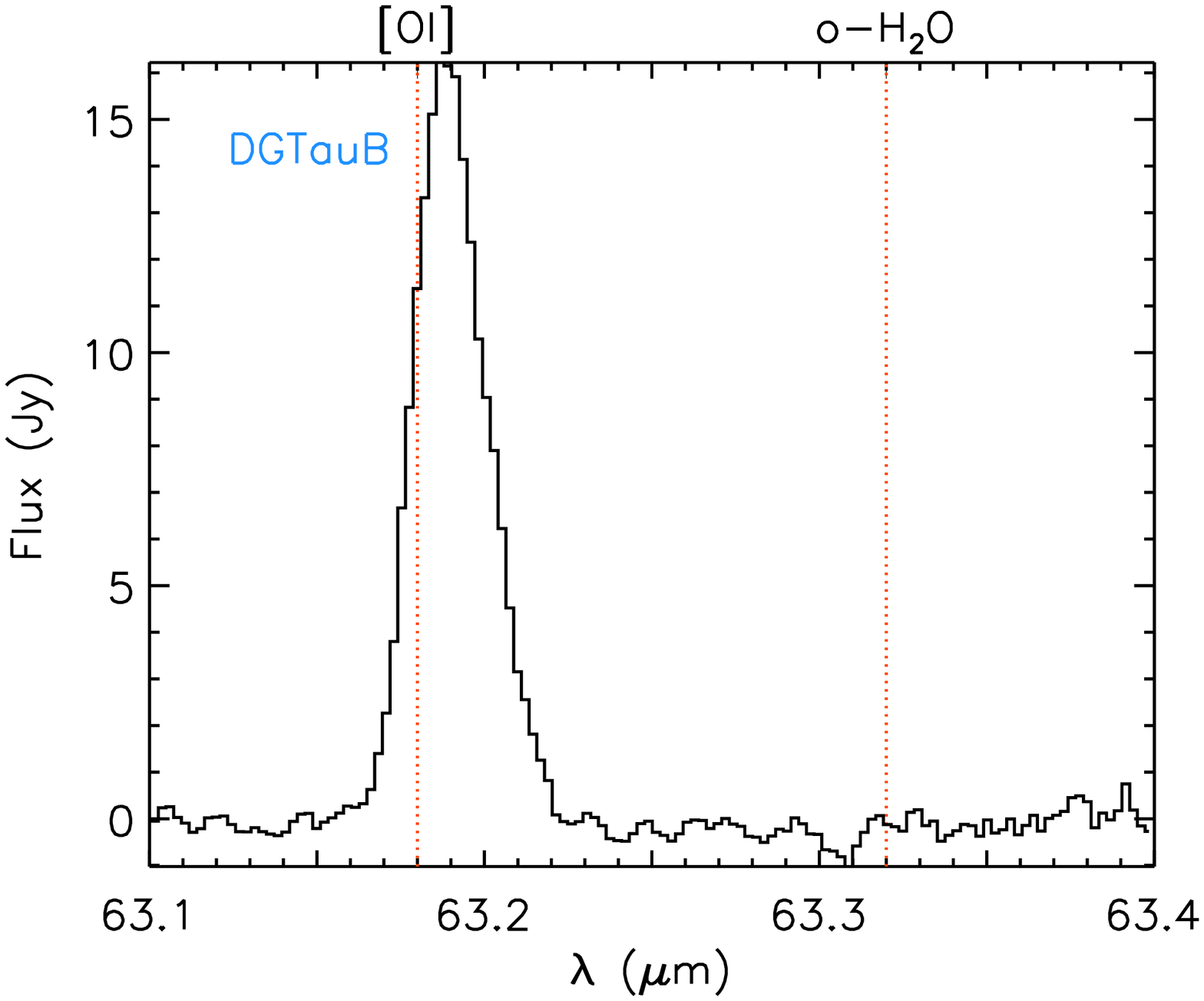}\includegraphics[width=0.16\textwidth, trim= 0mm 0mm 0mm 0mm, angle=90]{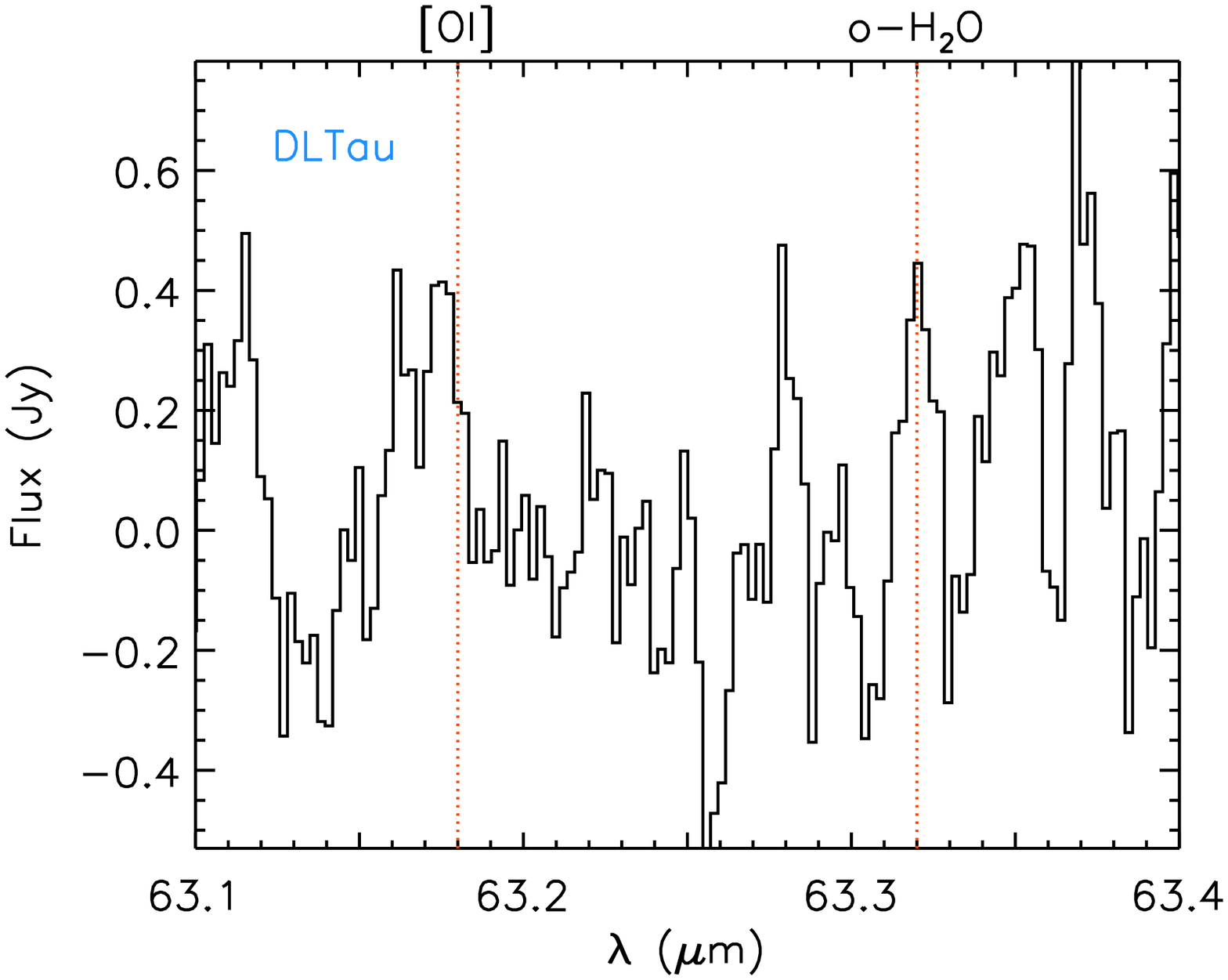}\includegraphics[width=0.16\textwidth, trim= 0mm 0mm 0mm 0mm, angle=90]{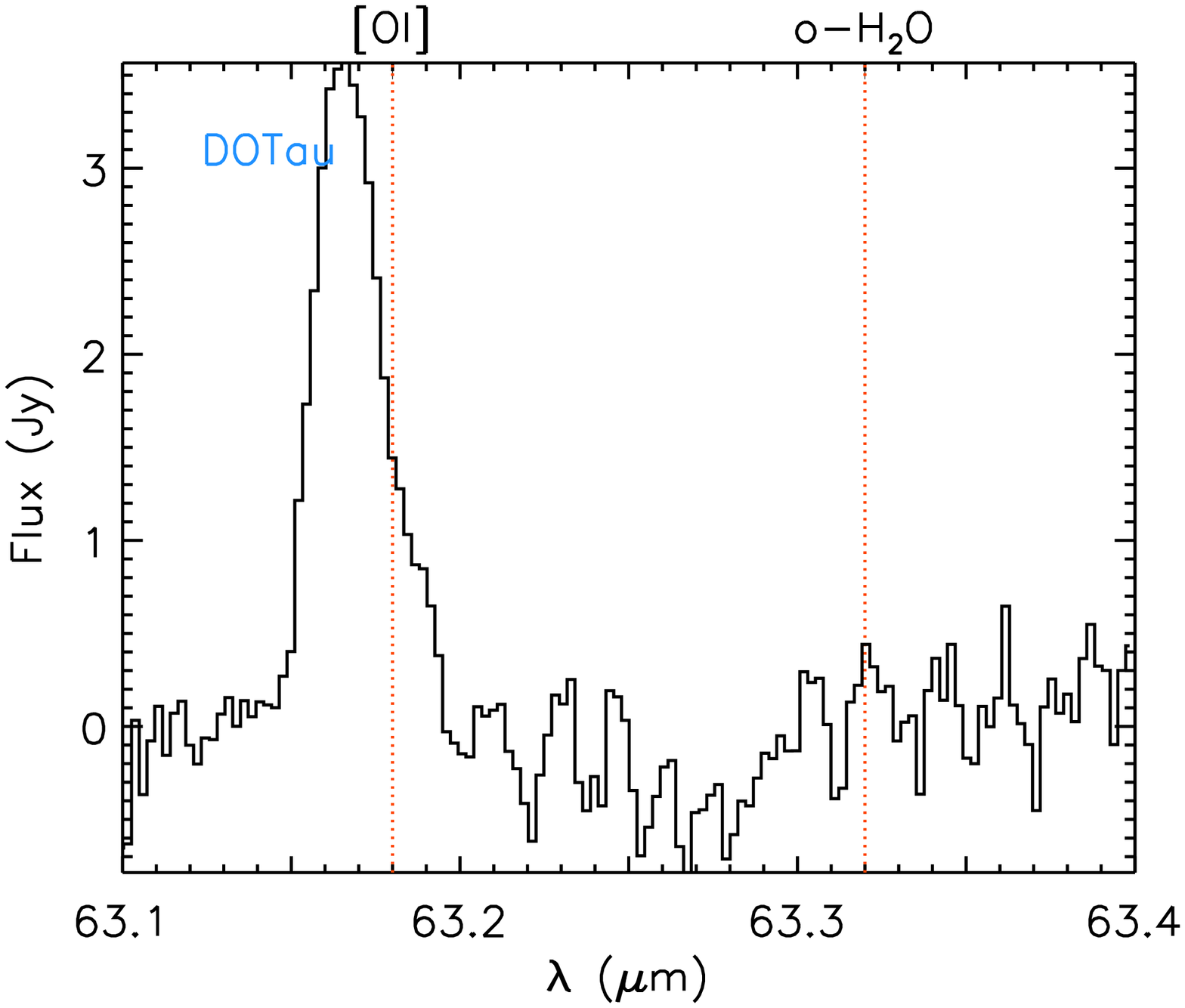}\includegraphics[width=0.16\textwidth, trim= 0mm 0mm 0mm 0mm, angle=90]{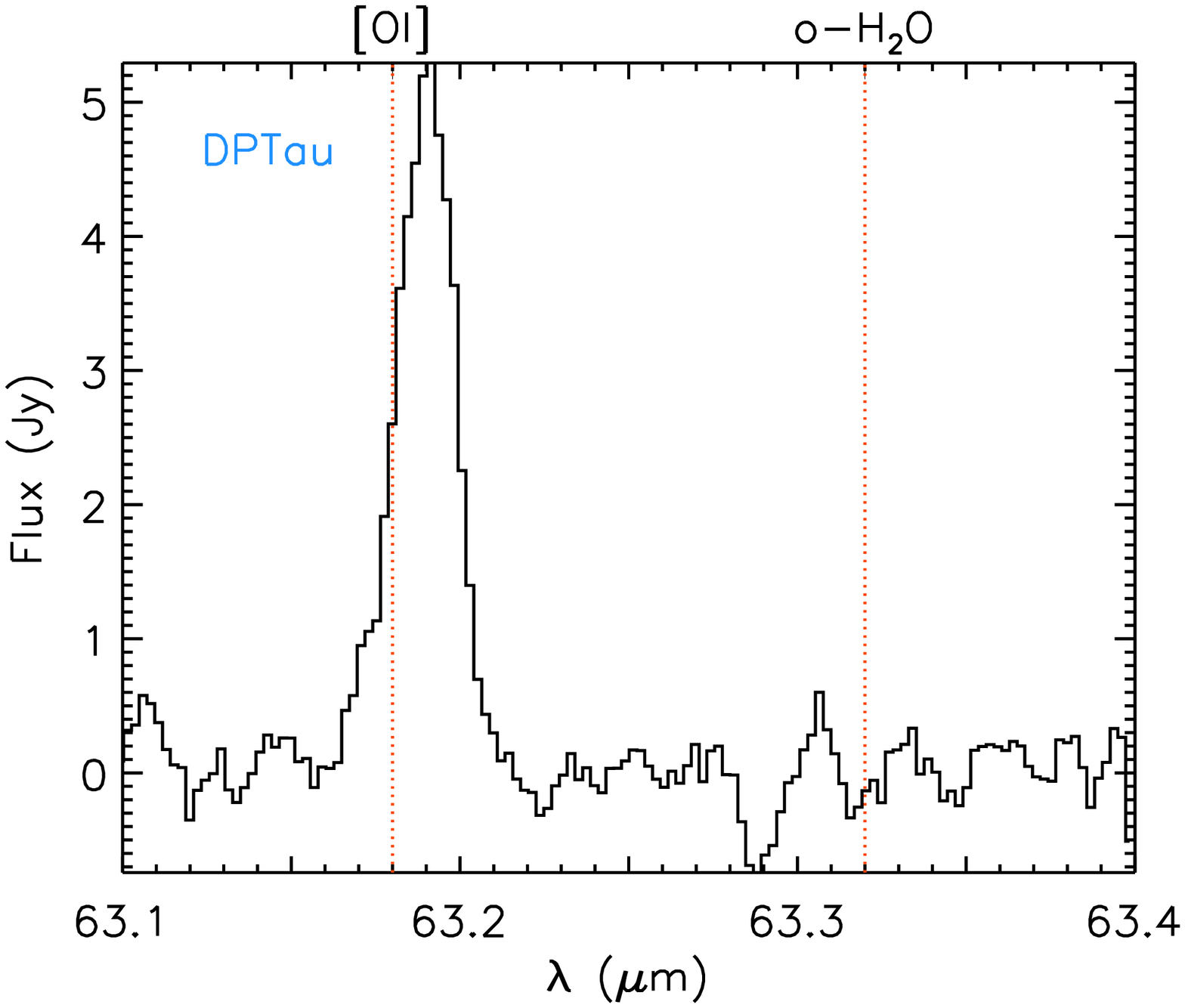}

\includegraphics[width=0.16\textwidth, trim= 0mm 0mm 0mm 0mm, angle=90]{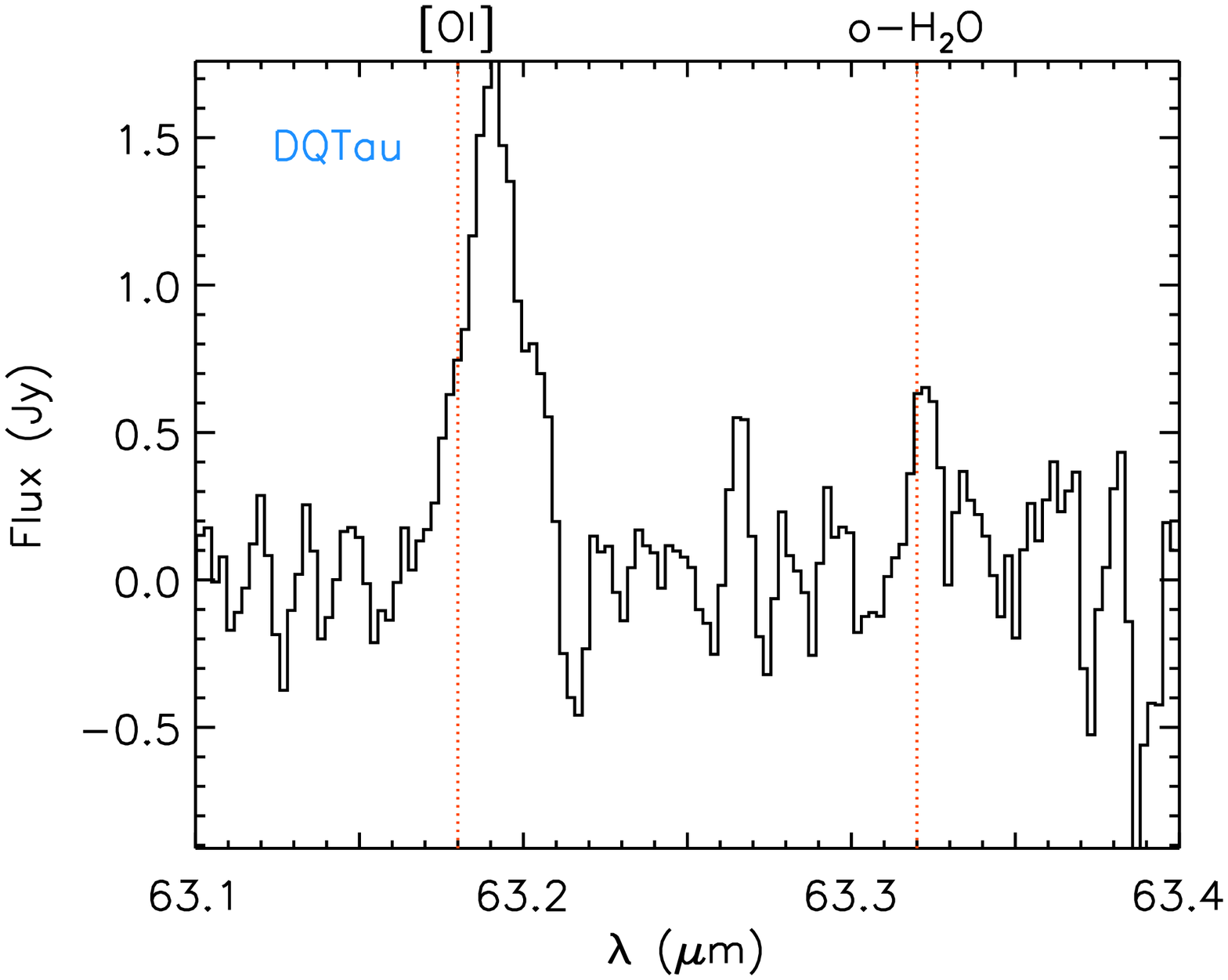}\includegraphics[width=0.16\textwidth, trim= 0mm 0mm 0mm 0mm, angle=90]{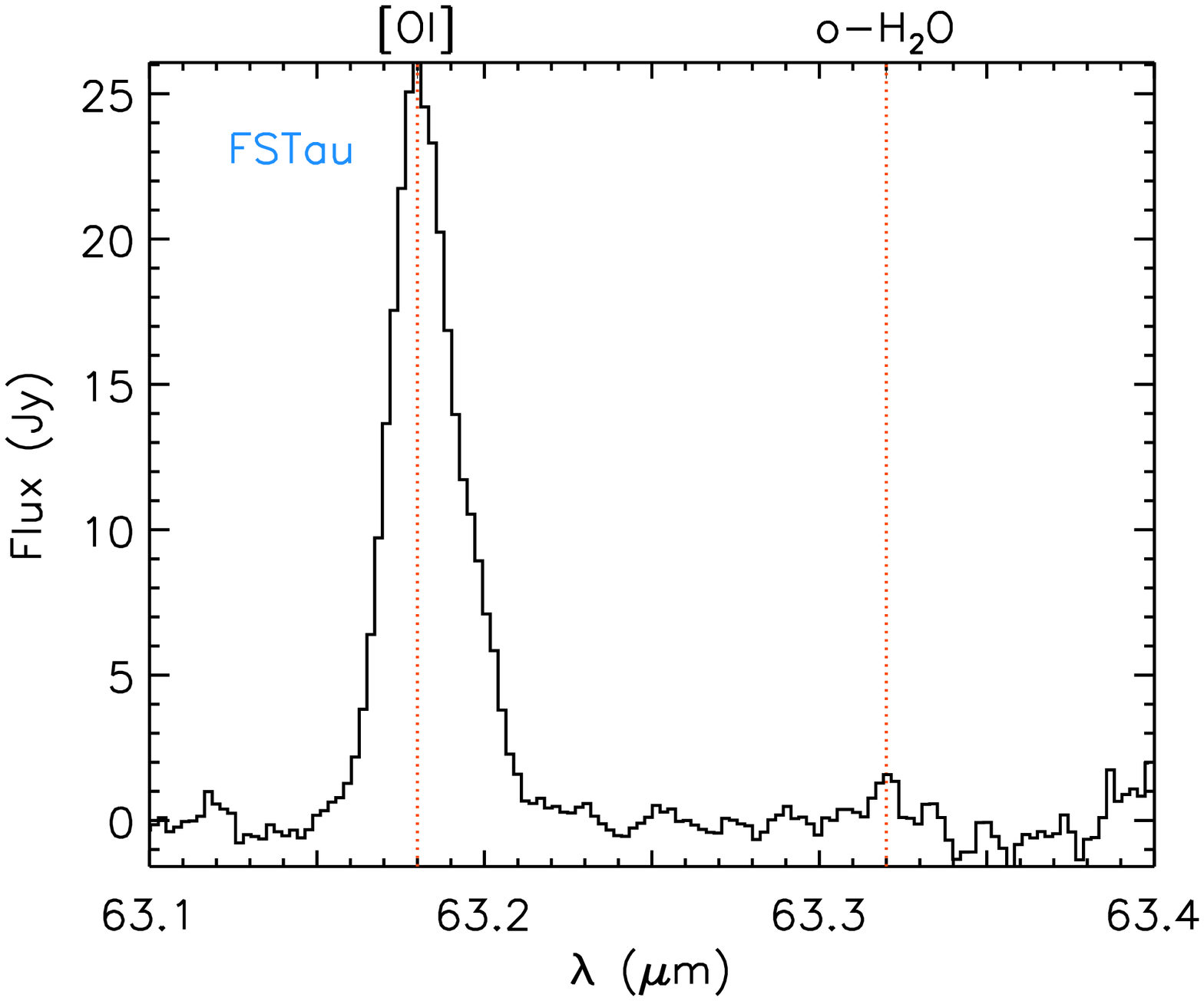}\includegraphics[width=0.16\textwidth, trim= 0mm 0mm 0mm 0mm, angle=90]{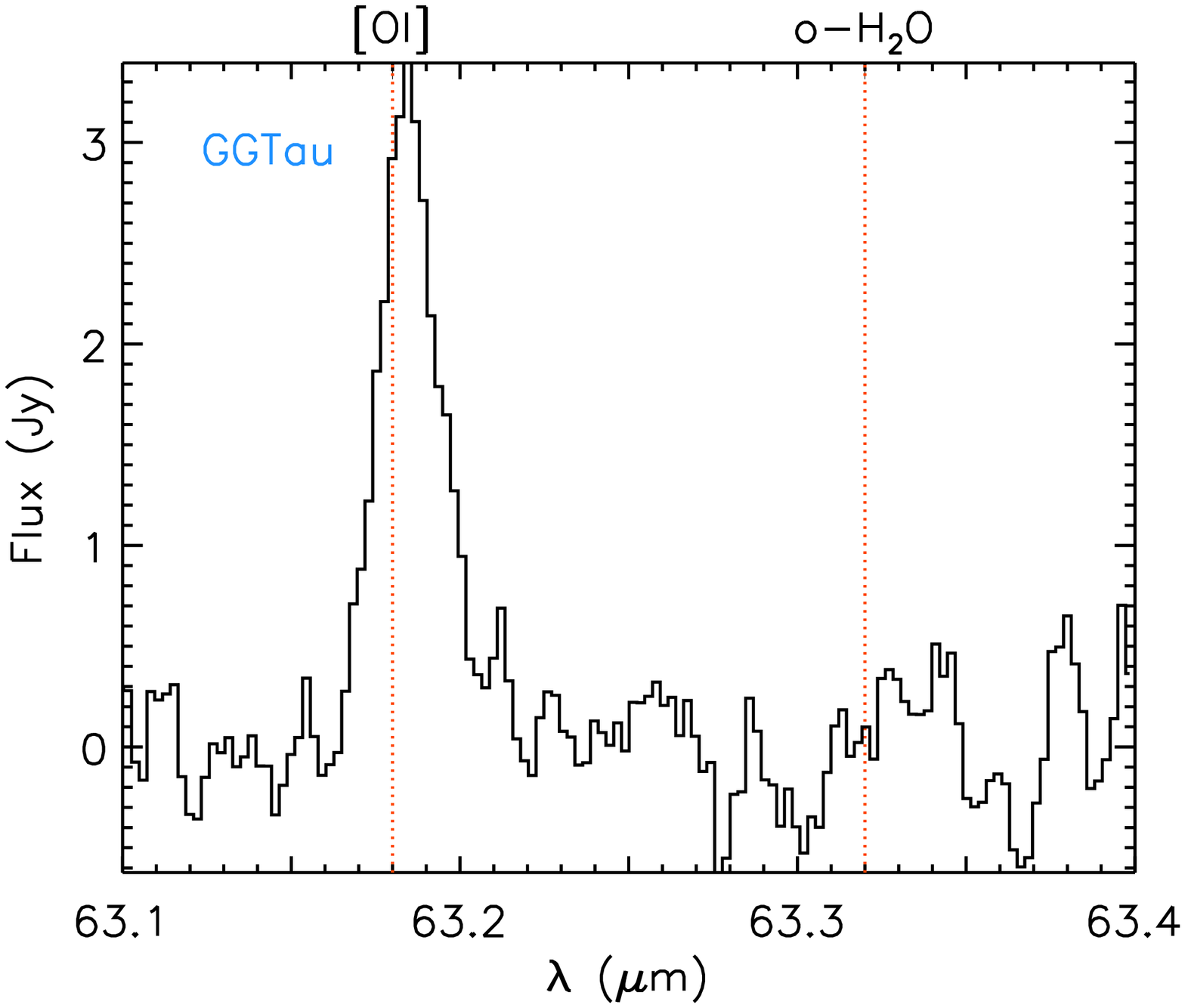}\includegraphics[width=0.16\textwidth, trim= 0mm 0mm 0mm 0mm, angle=90]{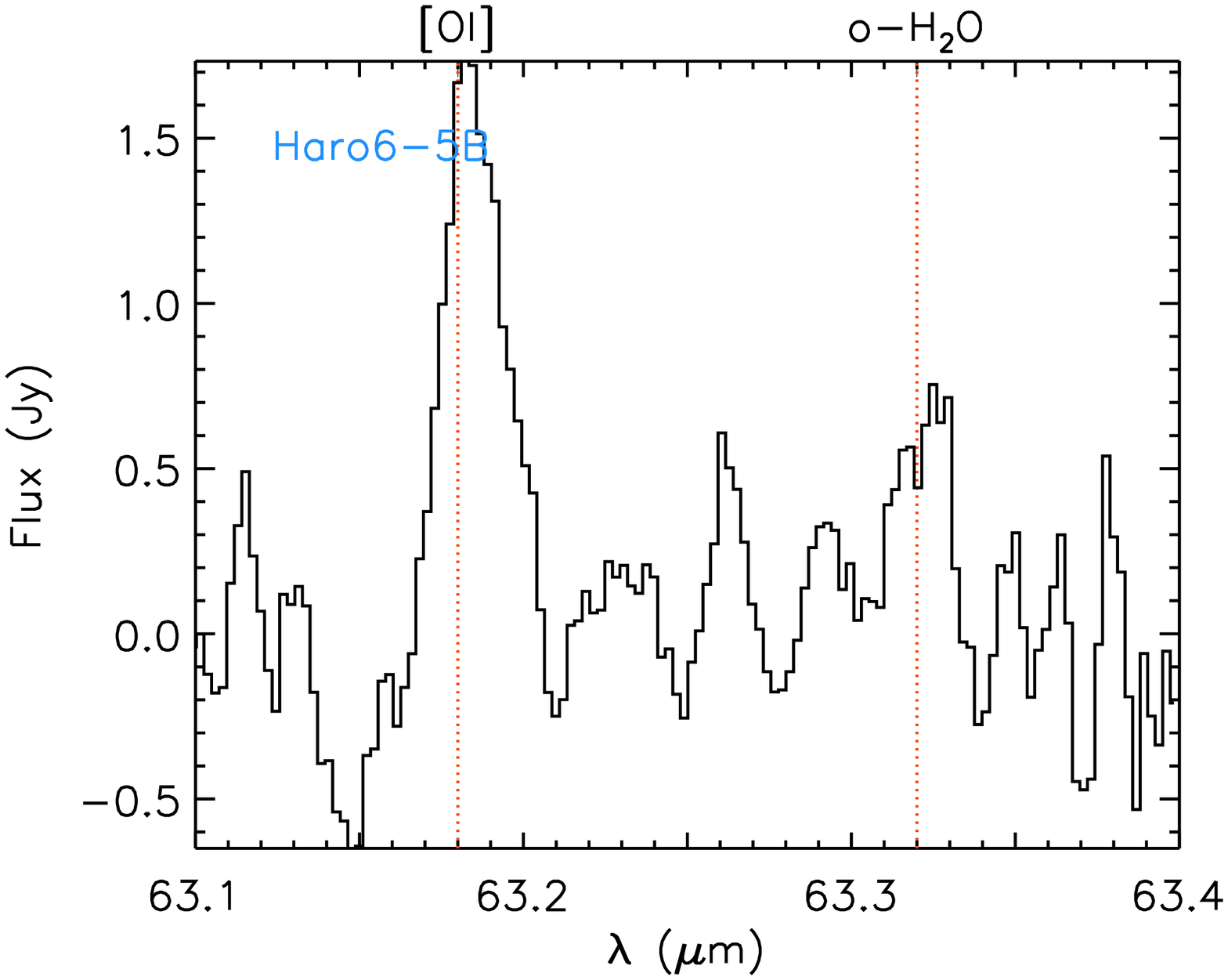}

\includegraphics[width=0.16\textwidth, trim= 0mm 0mm 0mm 0mm, angle=90]{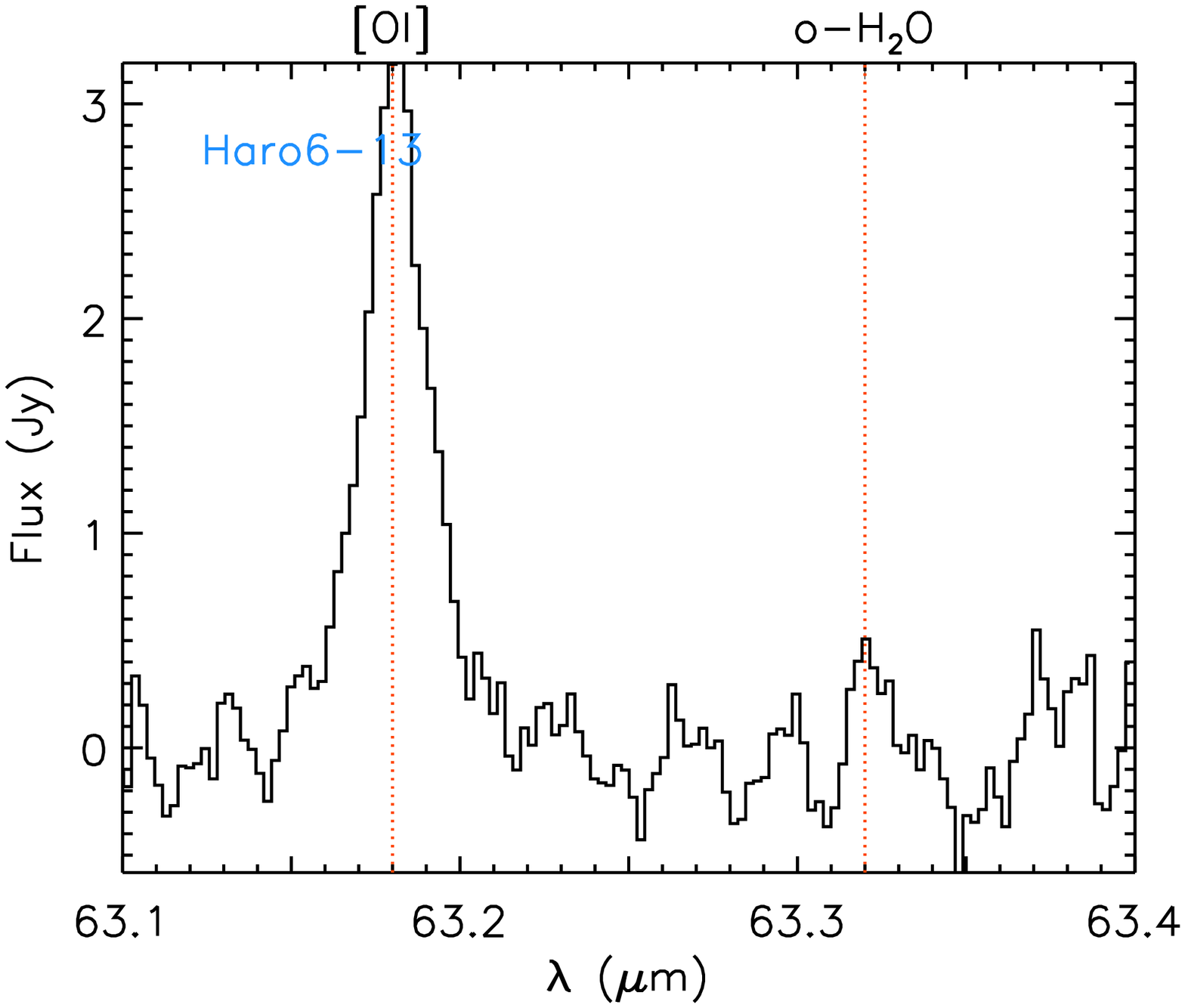}\includegraphics[width=0.16\textwidth, trim= 0mm 0mm 0mm 0mm, angle=90]{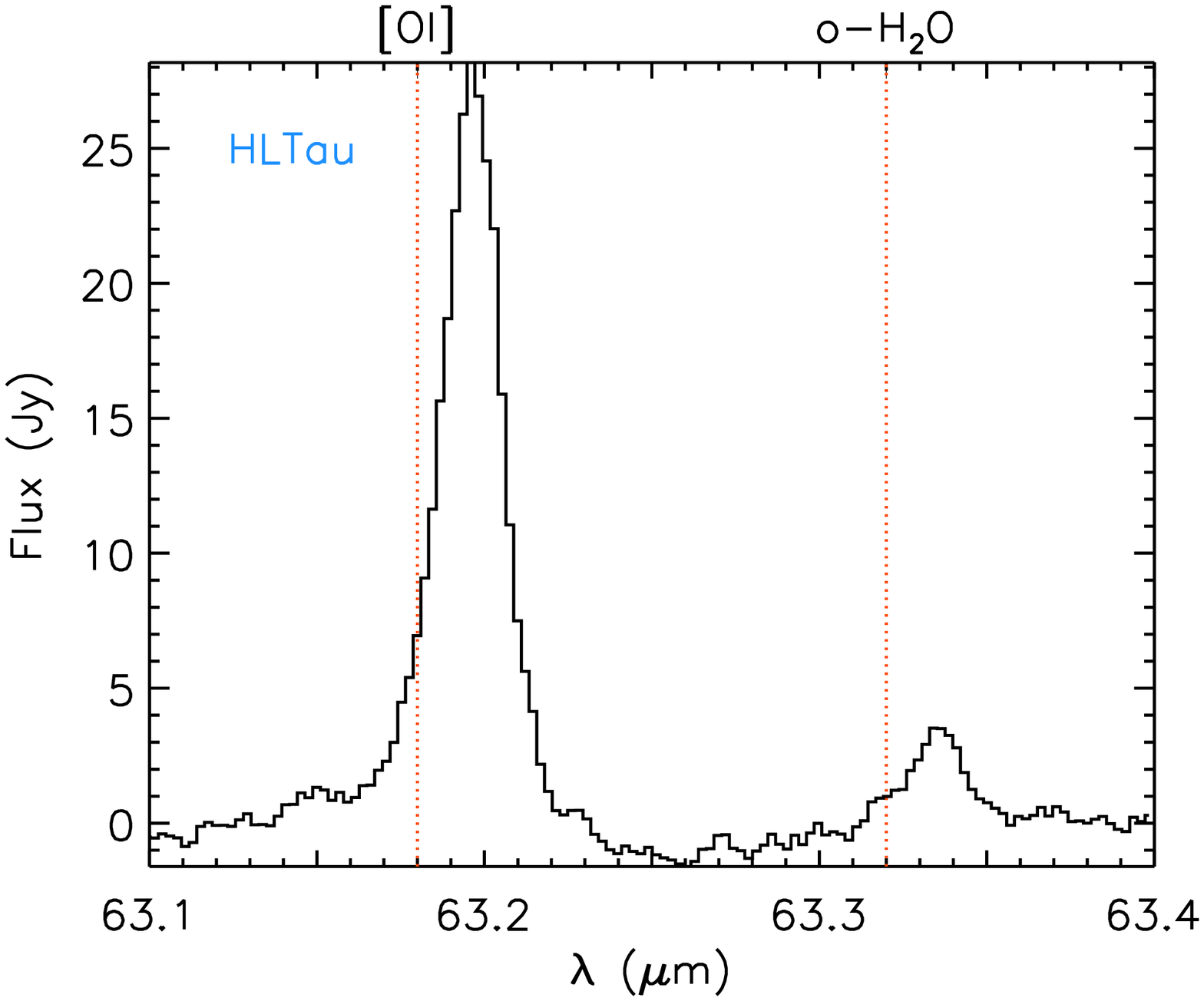}\includegraphics[width=0.16\textwidth, trim= 0mm 0mm 0mm 0mm, angle=90]{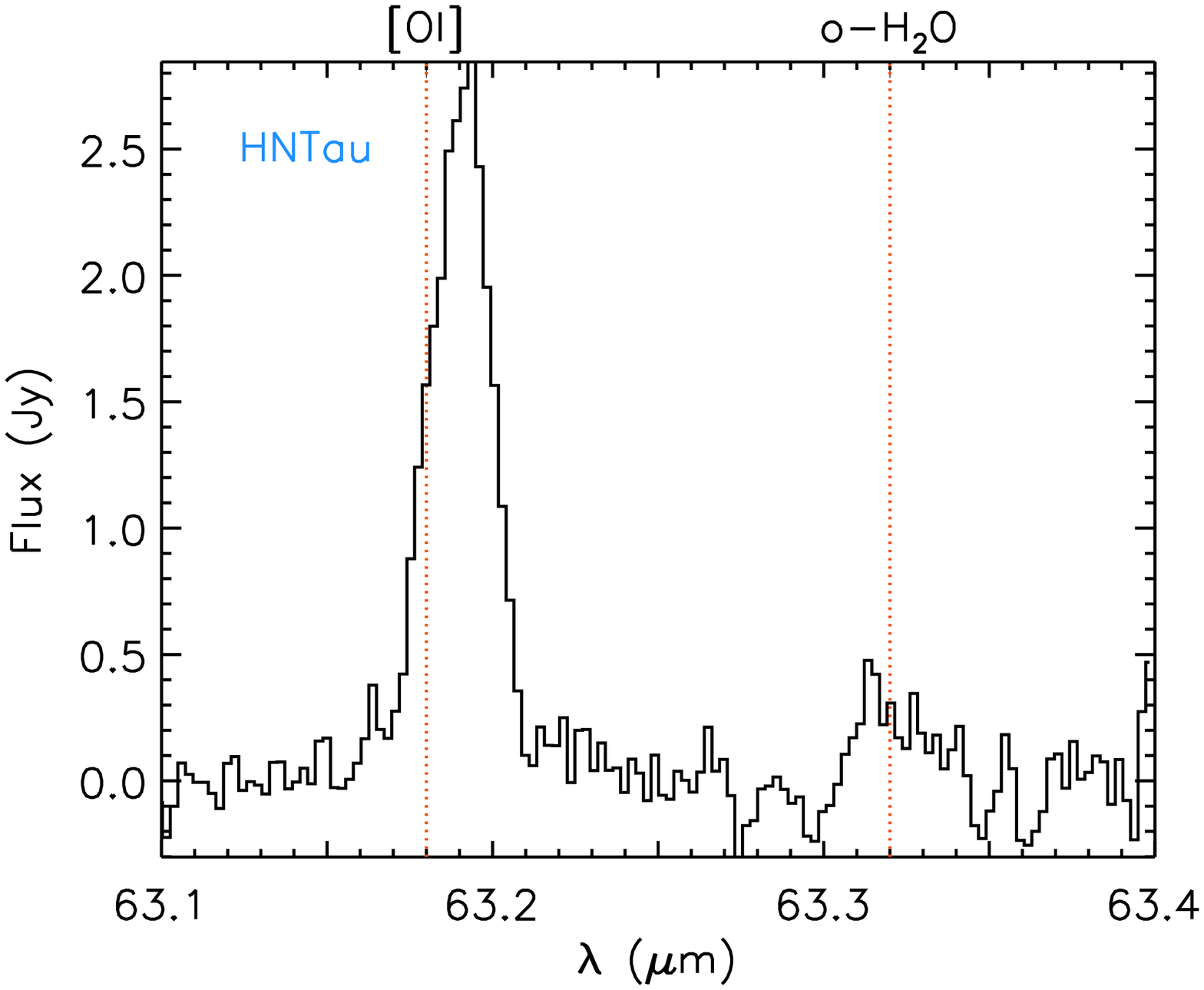}\includegraphics[width=0.16\textwidth, trim= 0mm 0mm 0mm 0mm, angle=90]{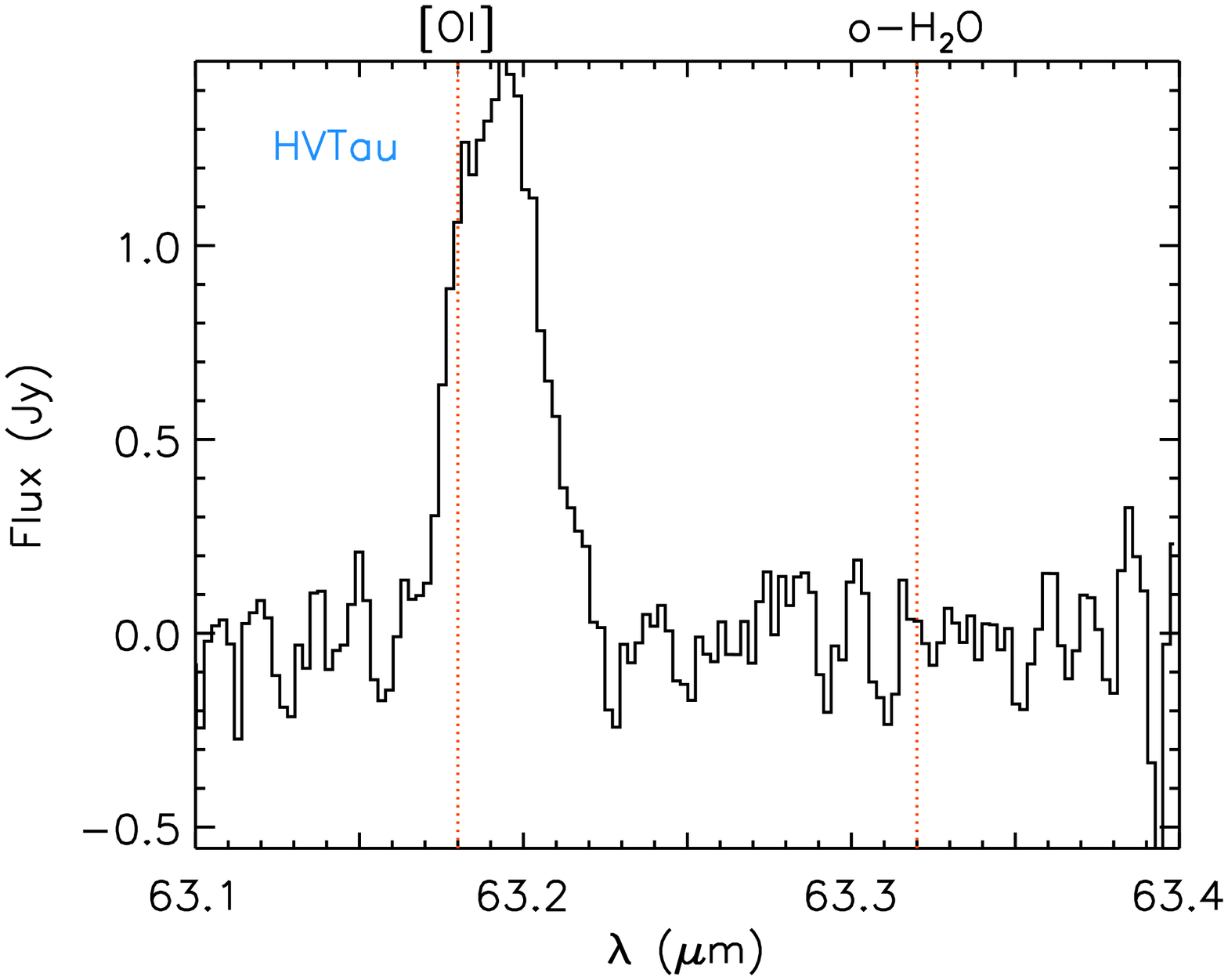}

\includegraphics[width=0.16\textwidth, trim= 0mm 0mm 0mm 0mm, angle=90]{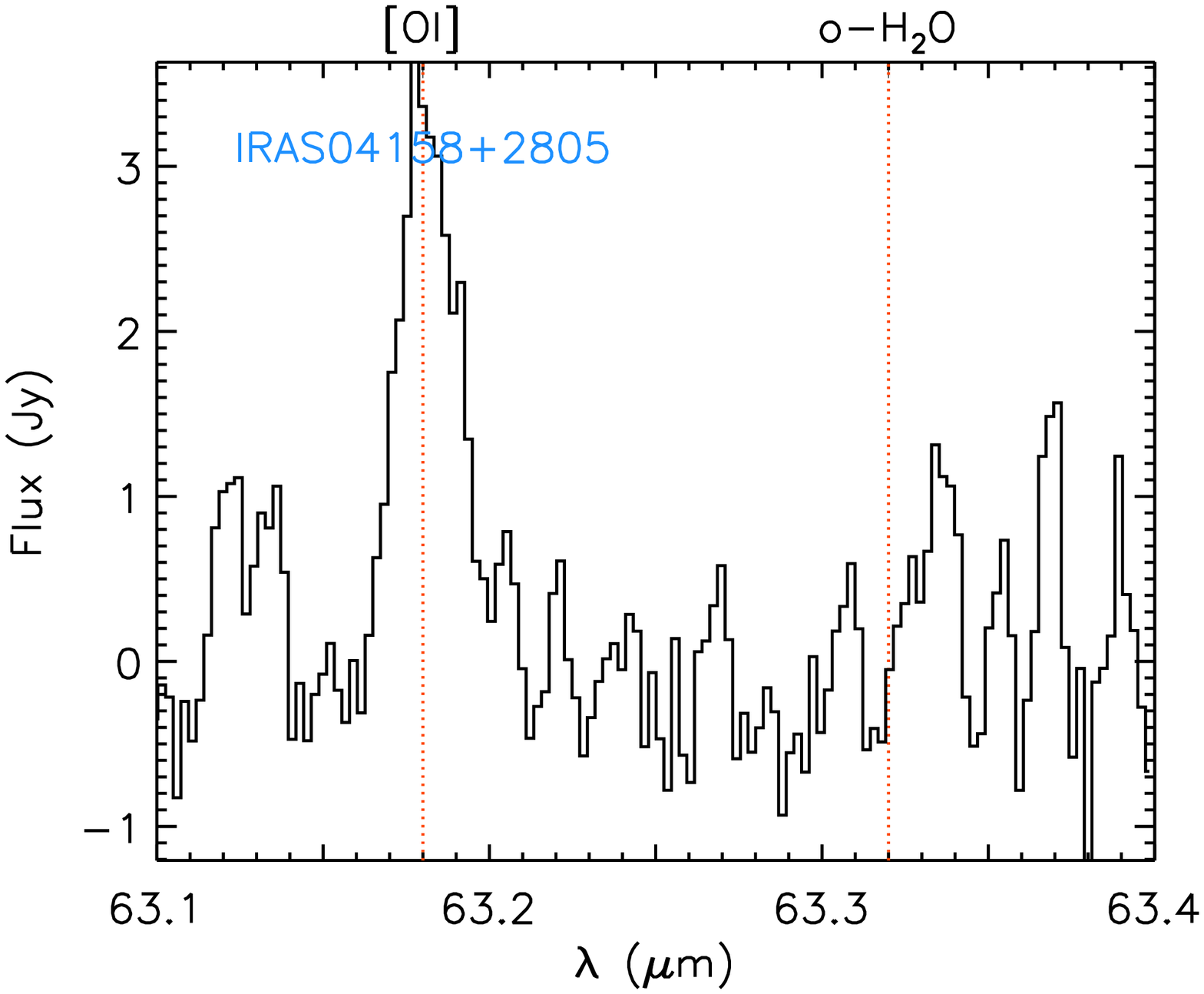}\includegraphics[width=0.16\textwidth, trim= 0mm 0mm 0mm 0mm, angle=90]{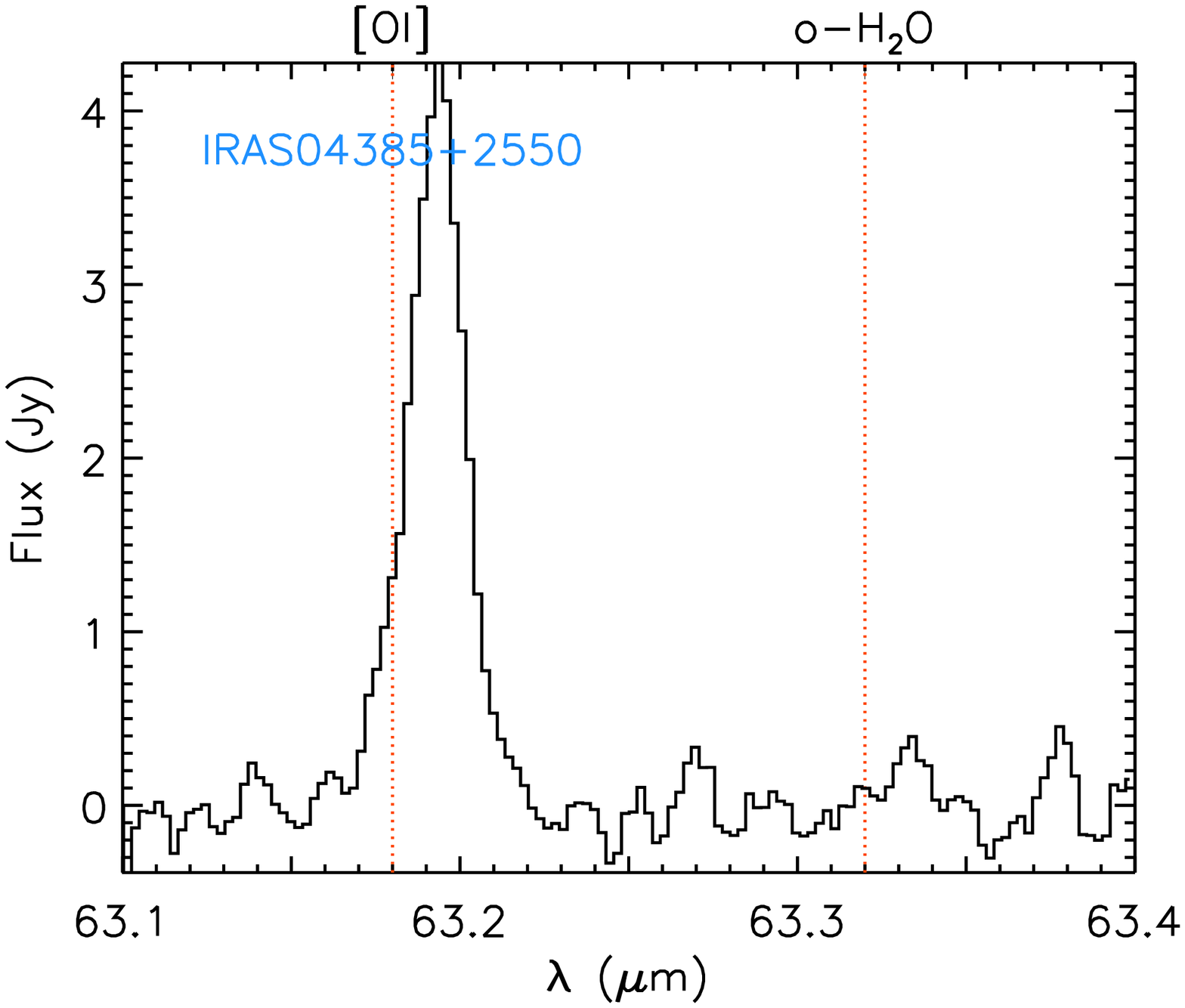}\includegraphics[width=0.16\textwidth, trim= 0mm 0mm 0mm 0mm, angle=90]{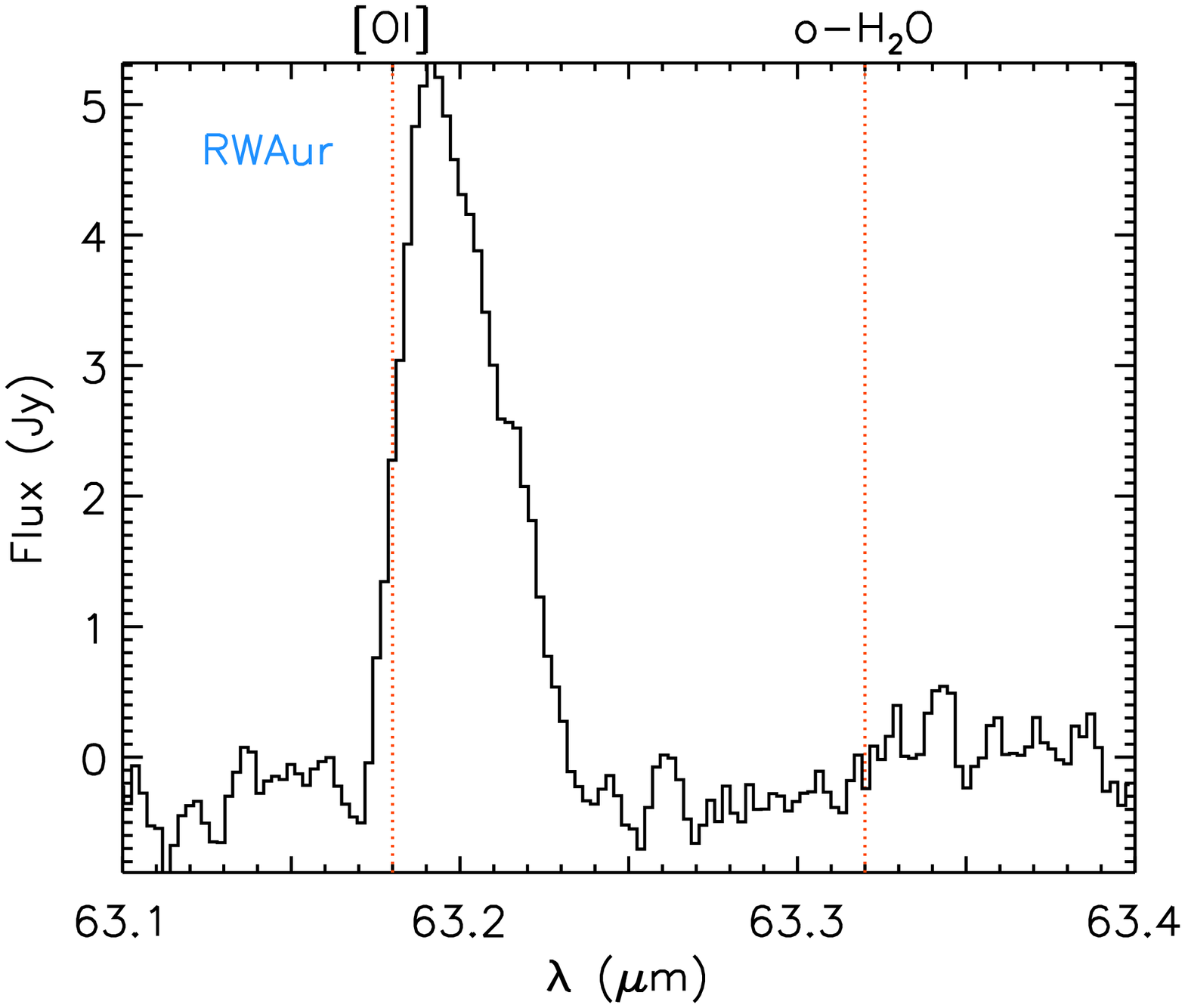}\includegraphics[width=0.16\textwidth, trim= 0mm 0mm 0mm 0mm, angle=90]{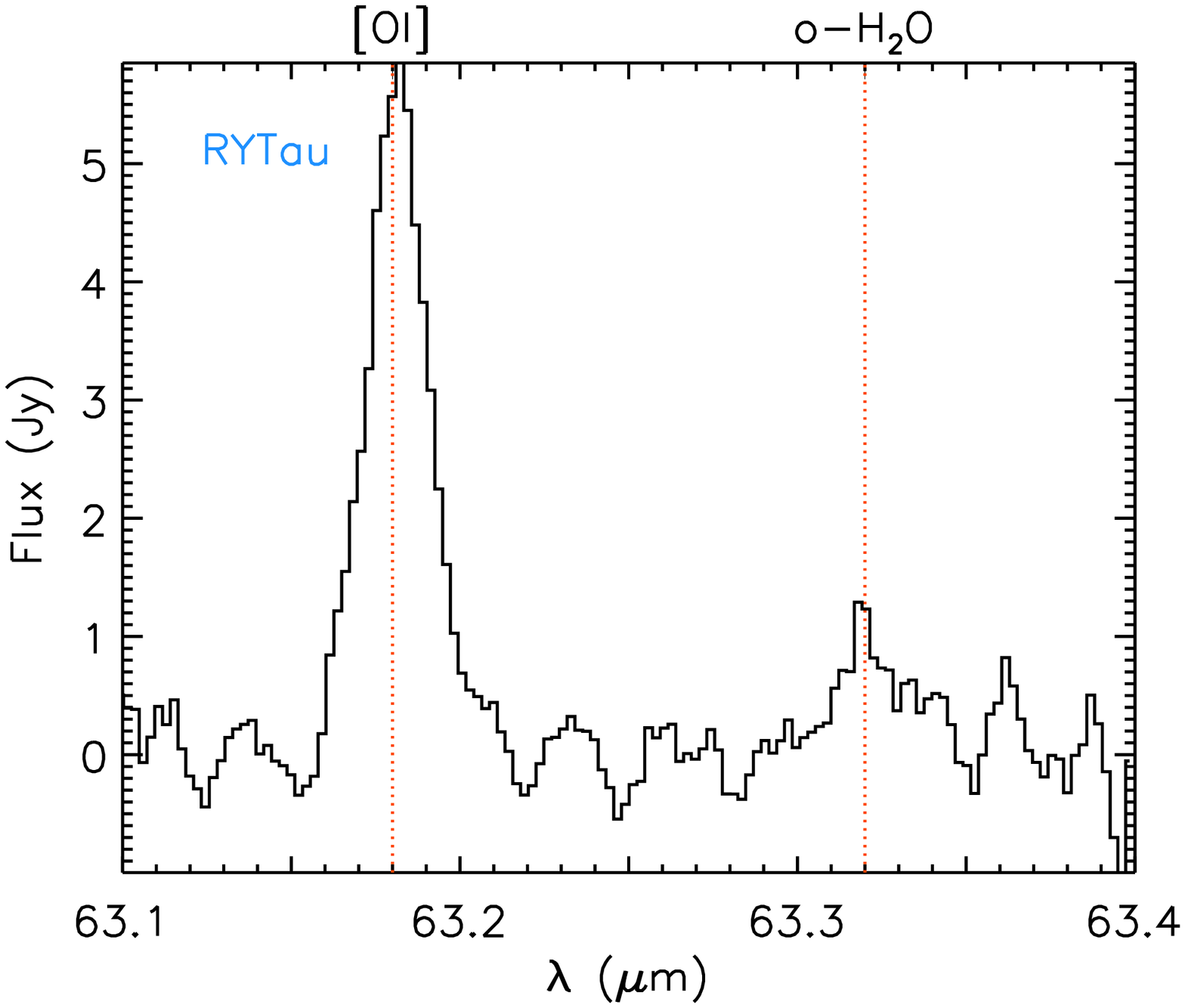}

\includegraphics[width=0.16\textwidth, trim= 0mm 0mm 0mm 0mm, angle=90]{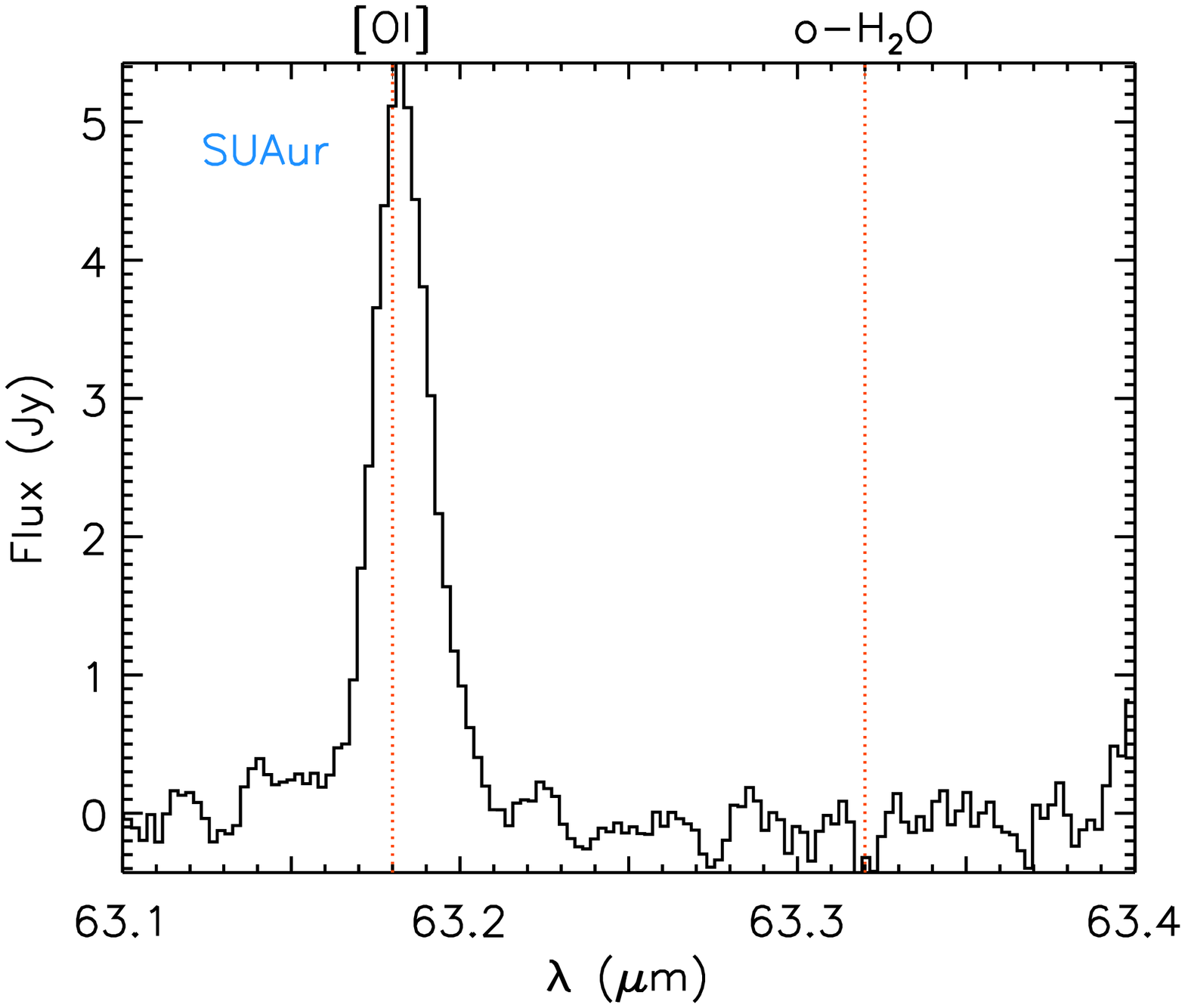}\includegraphics[width=0.16\textwidth, trim= 0mm 0mm 0mm 0mm, angle=90]{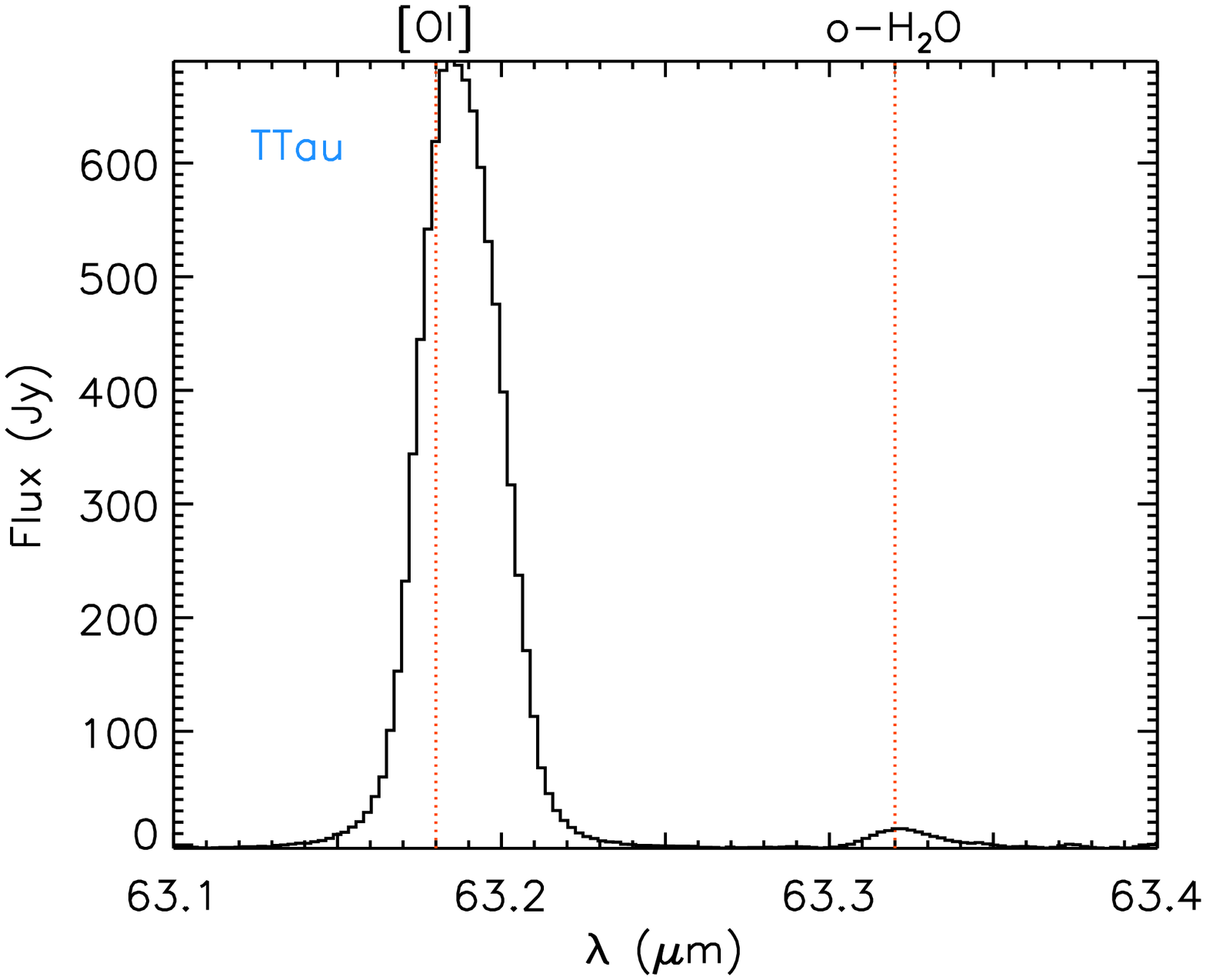}\includegraphics[width=0.16\textwidth, trim= 0mm 0mm 0mm 0mm, angle=90]{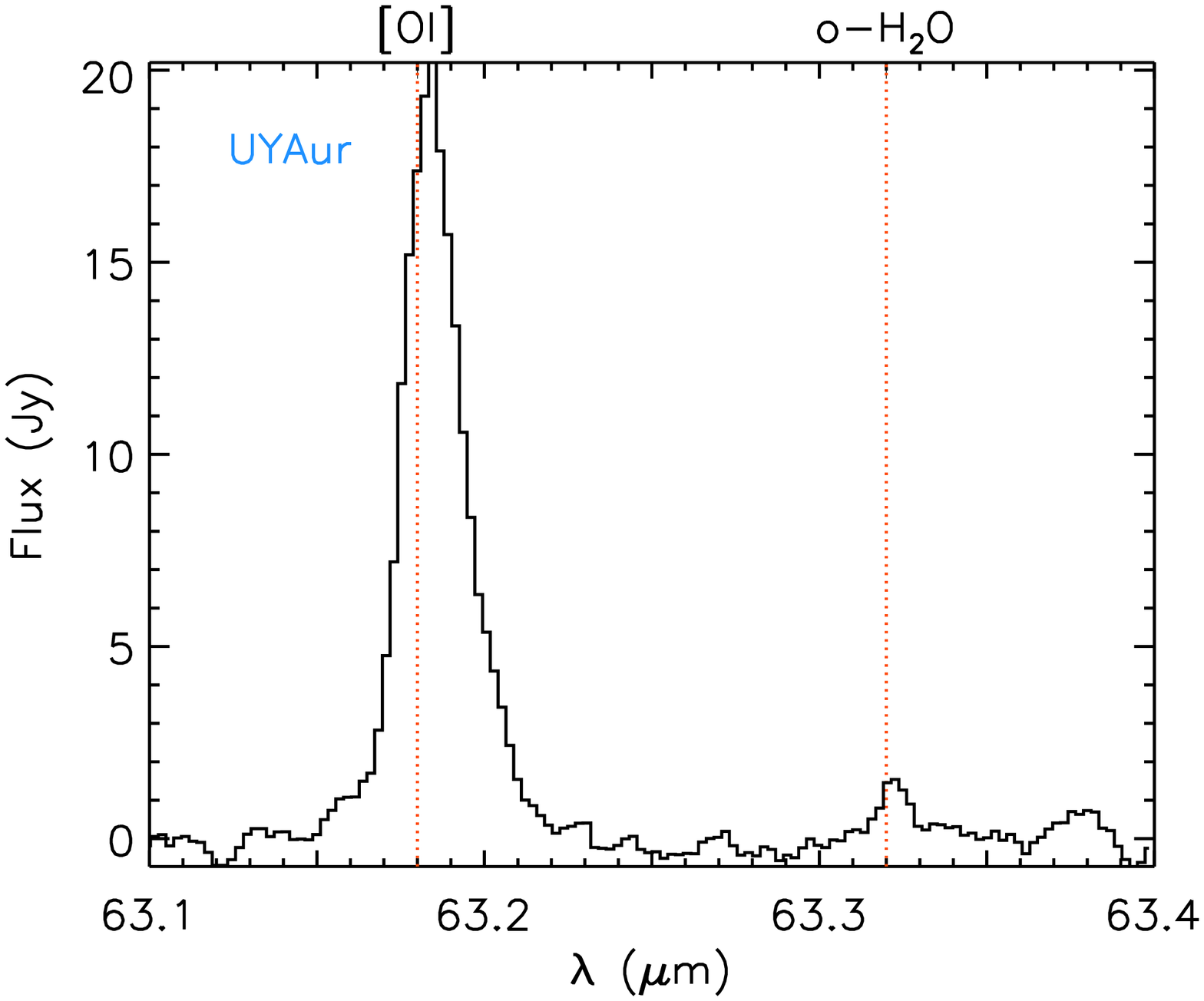}\includegraphics[width=0.16\textwidth, trim= 0mm 0mm 0mm 0mm, angle=90]{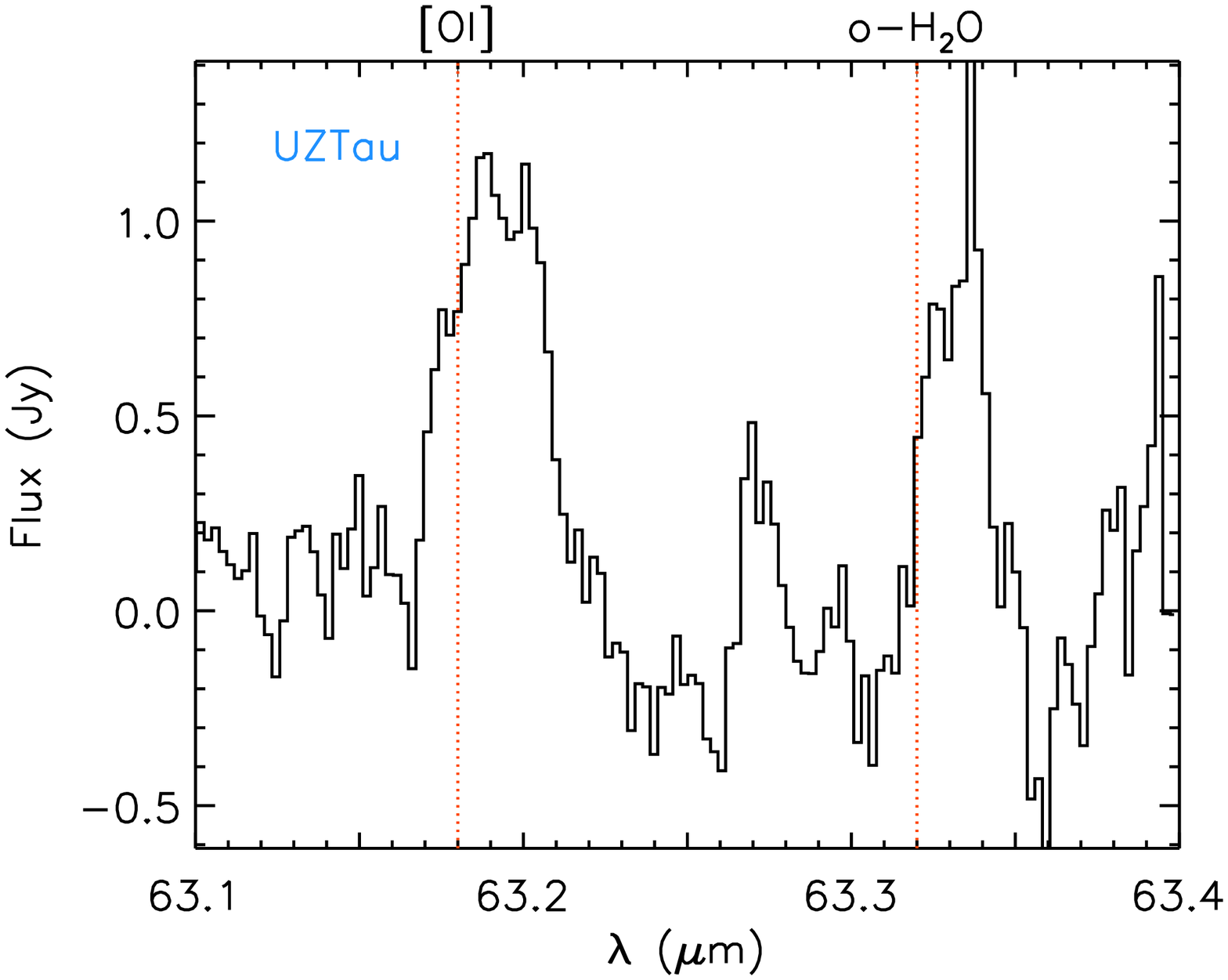}

\includegraphics[width=0.16\textwidth, trim= 0mm 0mm 0mm 0mm, angle=90]{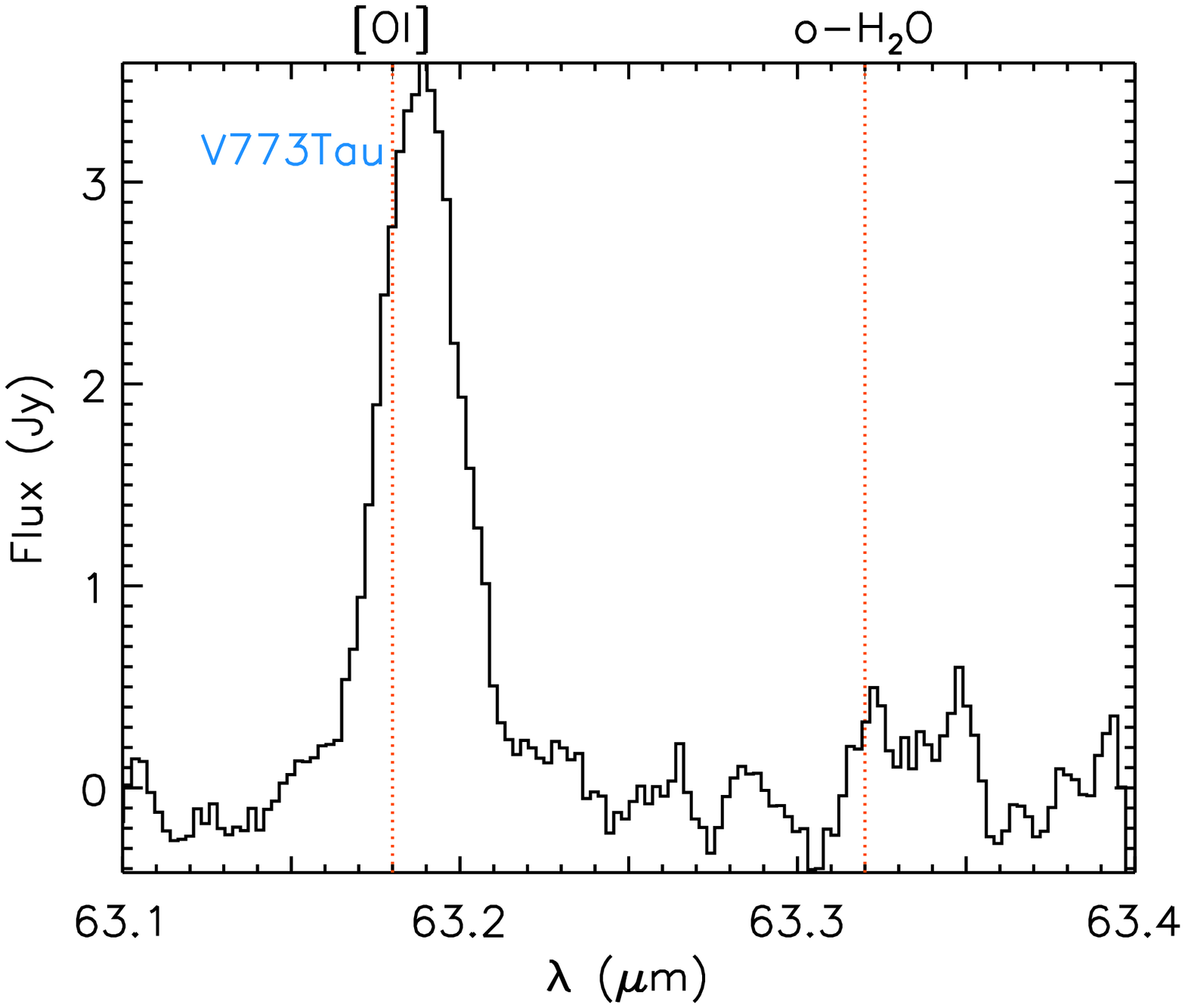}\includegraphics[width=0.16\textwidth, trim= 0mm 0mm 0mm 0mm, angle=90]{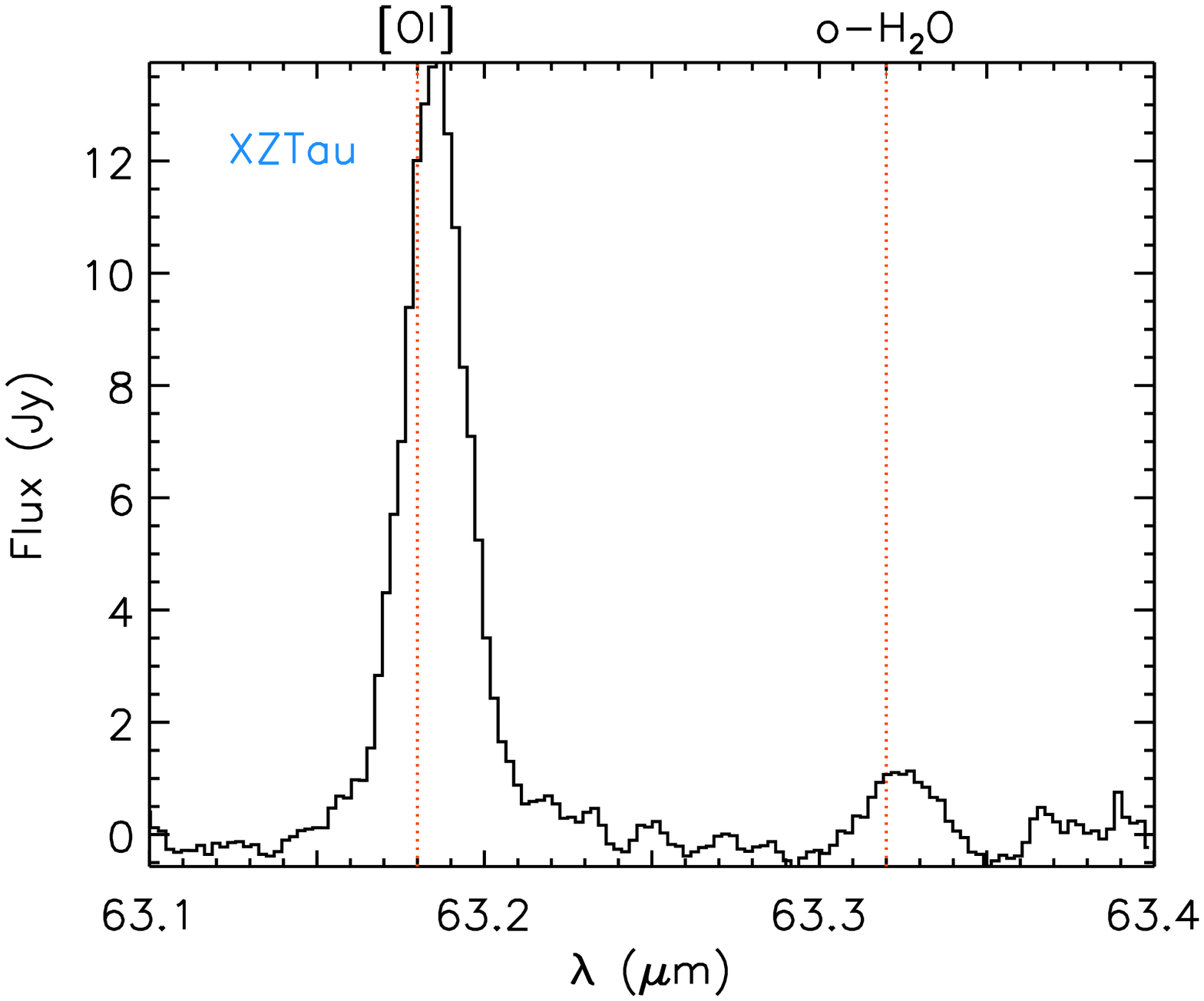}\includegraphics[width=0.16\textwidth, trim= 0mm 0mm 0mm 0mm, angle=90]{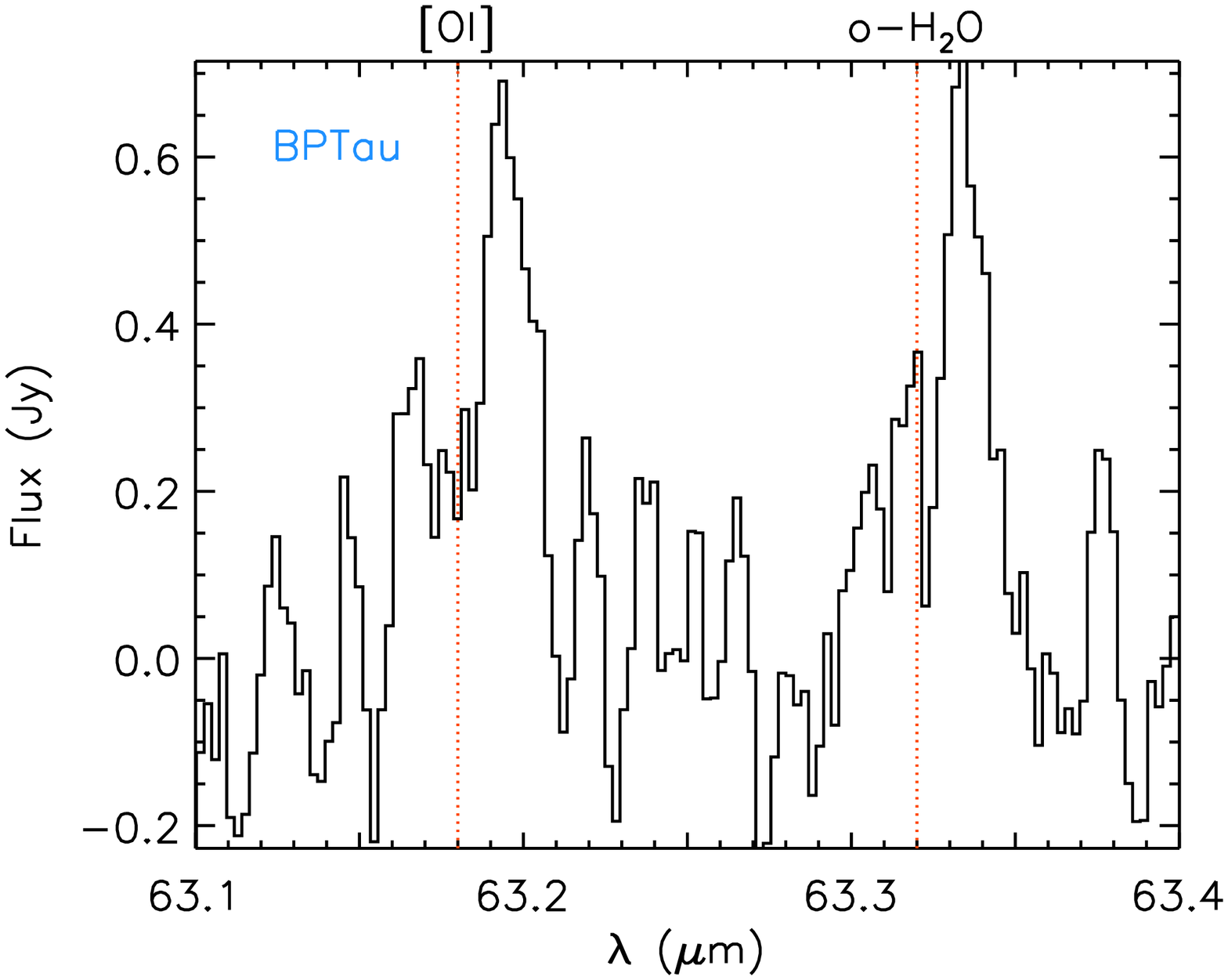}\includegraphics[width=0.16\textwidth, trim= 0mm 0mm 0mm 0mm, angle=90]{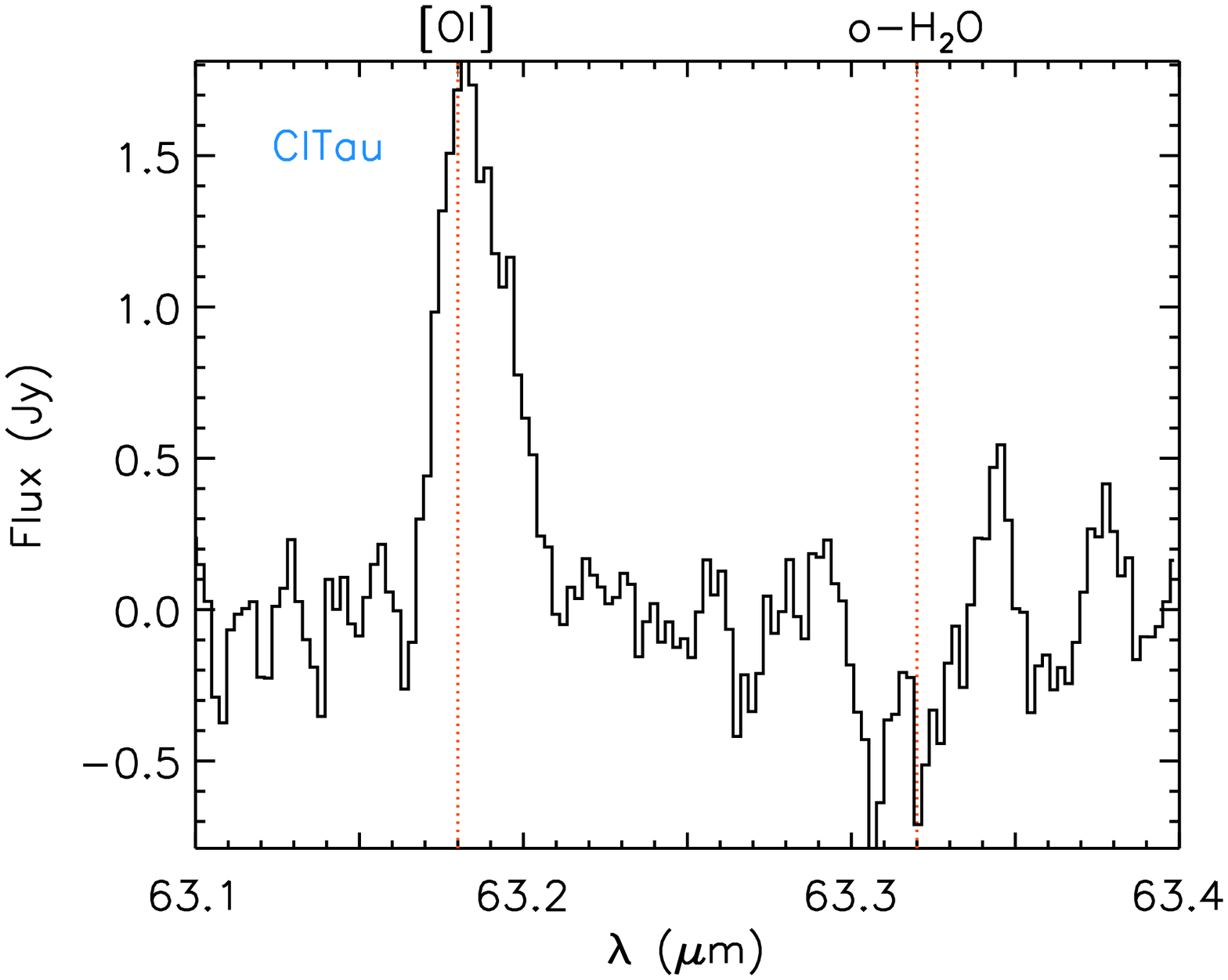}

\includegraphics[width=0.16\textwidth, trim= 0mm 0mm 0mm 0mm, angle=90]{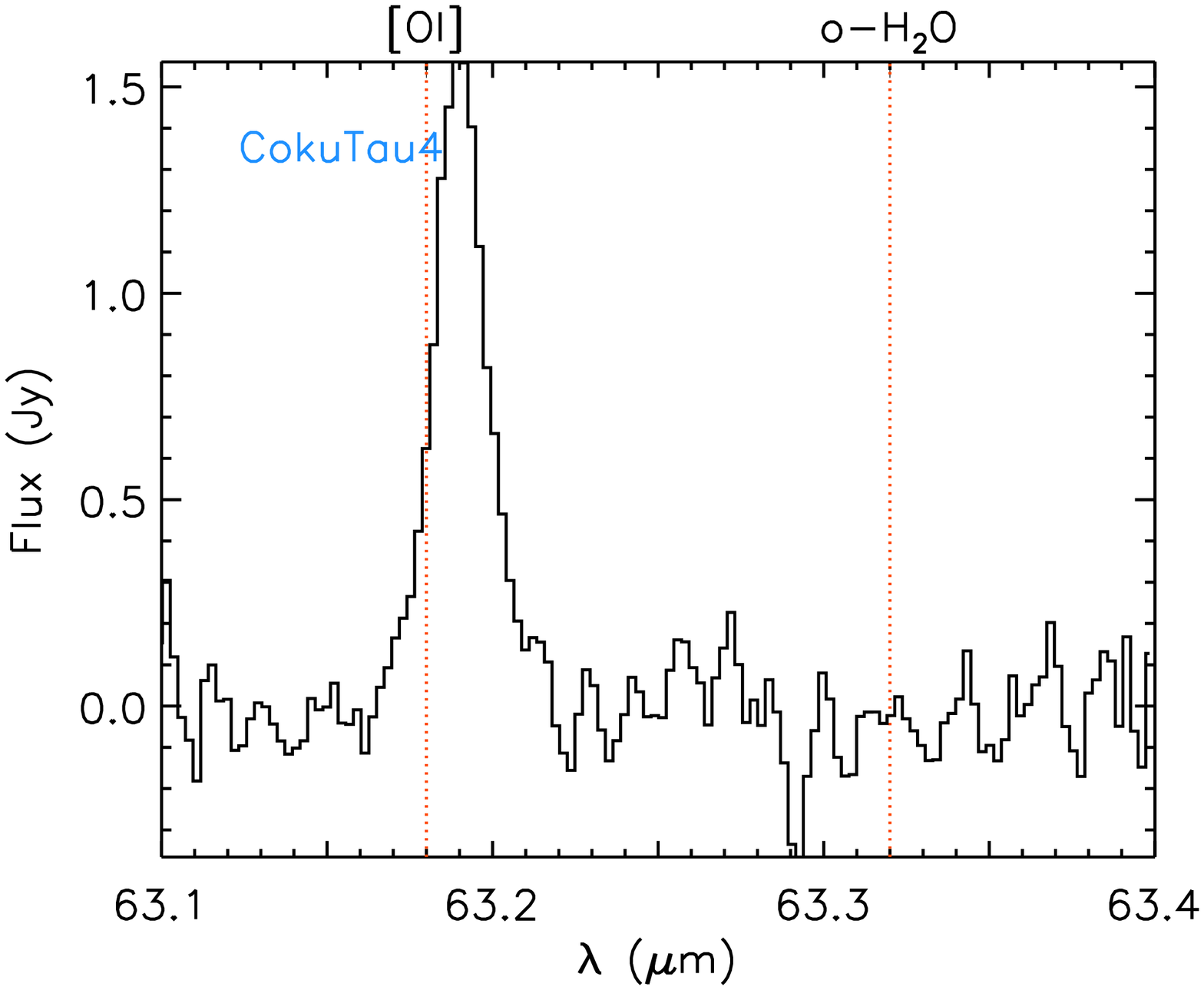}\includegraphics[width=0.16\textwidth, trim= 0mm 0mm 0mm 0mm, angle=90]{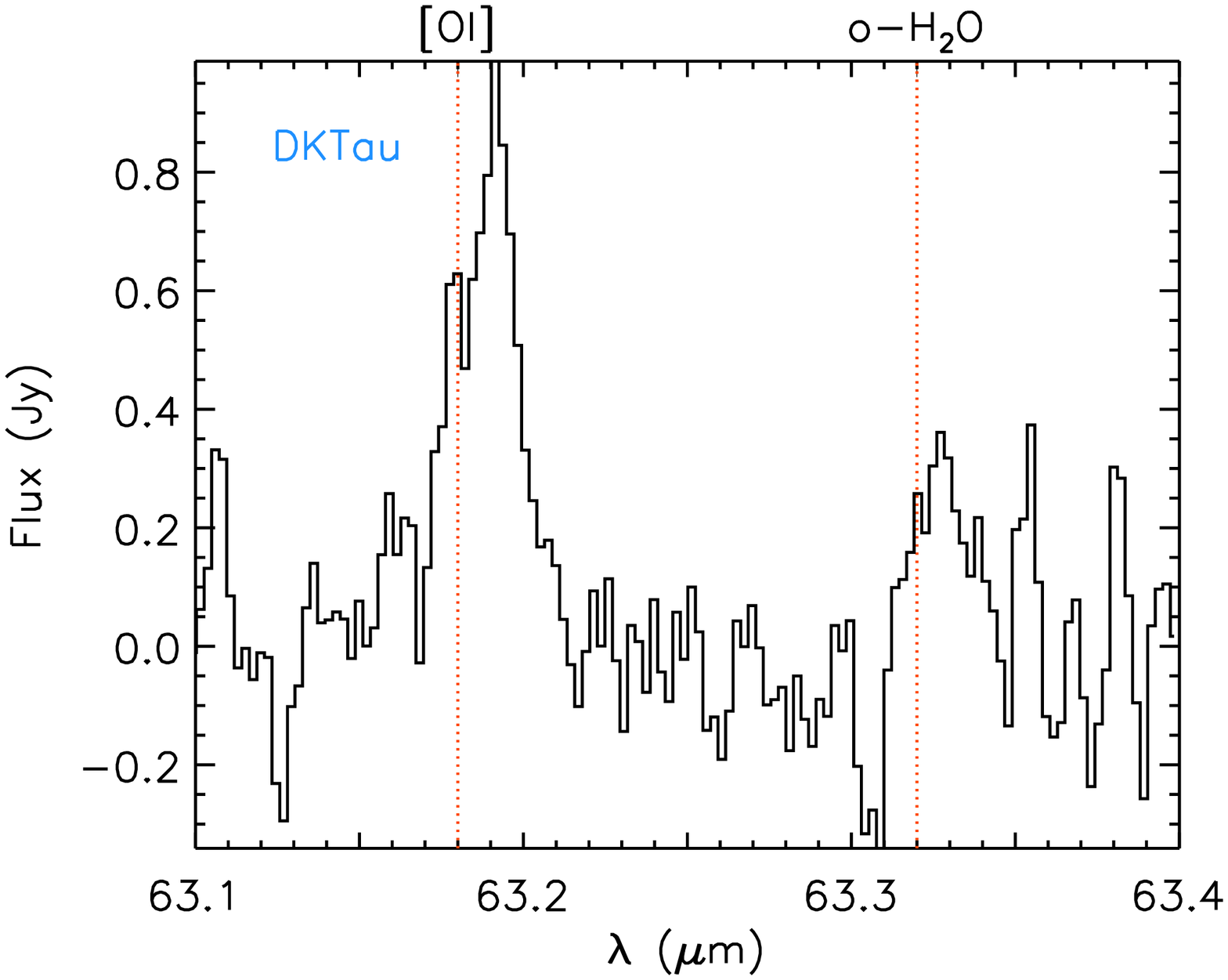}\includegraphics[width=0.16\textwidth, trim= 0mm 0mm 0mm 0mm, angle=90]{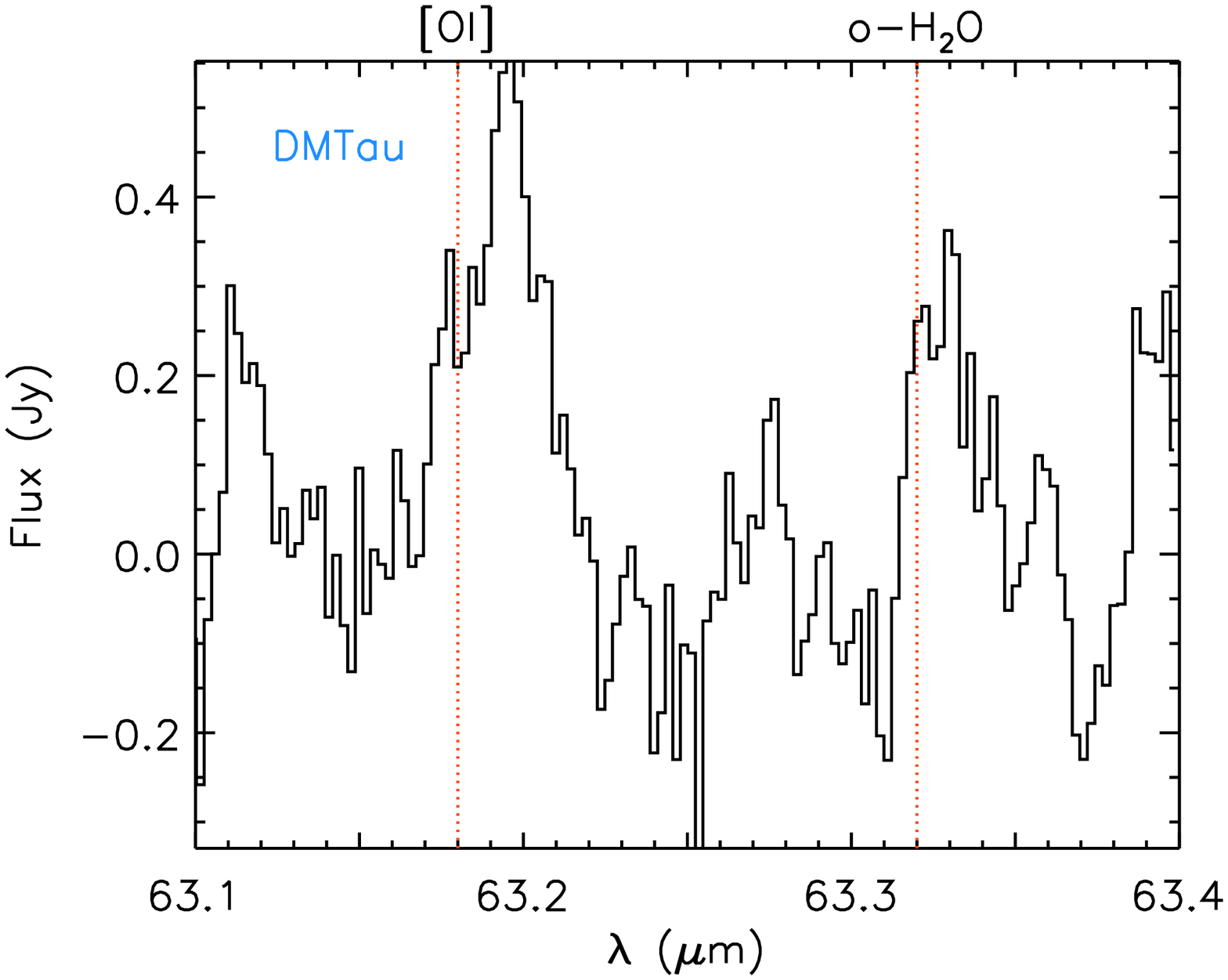}\includegraphics[width=0.16\textwidth, trim= 0mm 0mm 0mm 0mm, angle=90]{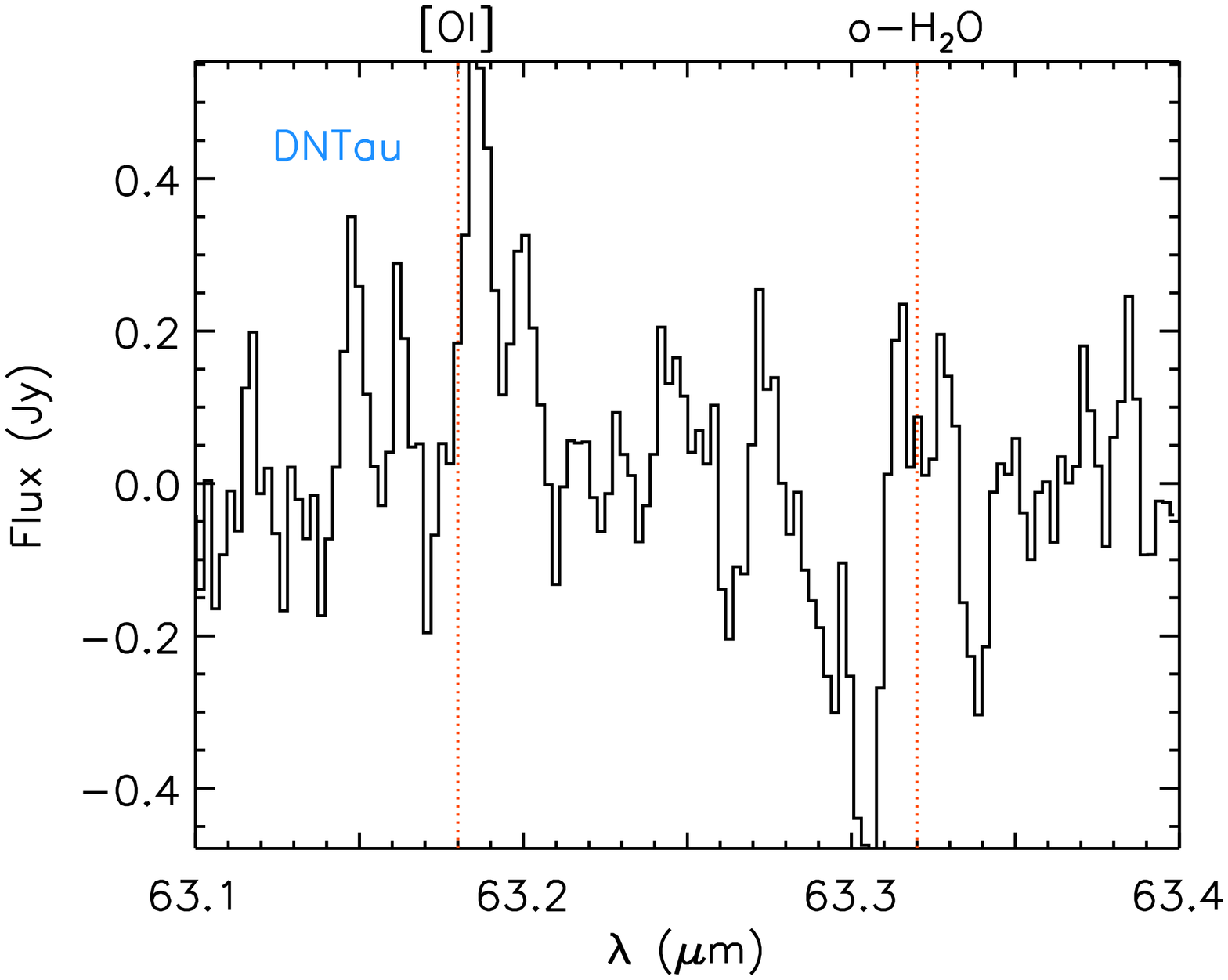}

\caption{Continuum subtracted spectra at 63 $\rm \mu m$ for all objects with [OI] detections. The red vertical lines indicate the positions of [OI] 63.18 $\rm \mu m$ and o-H$_{\rm 2}$O 63.32 $\rm \mu m$. The source GI/GK Tau is counted twice to make a total of 42 [OI] detections.}
\label{figure:allSpec63}
\end{figure*}

\begin{figure*}[htpb]
\centering
\setcounter{figure}{0}
\includegraphics[width=0.16\textwidth, trim= 0mm 0mm 0mm 0mm, angle=90]{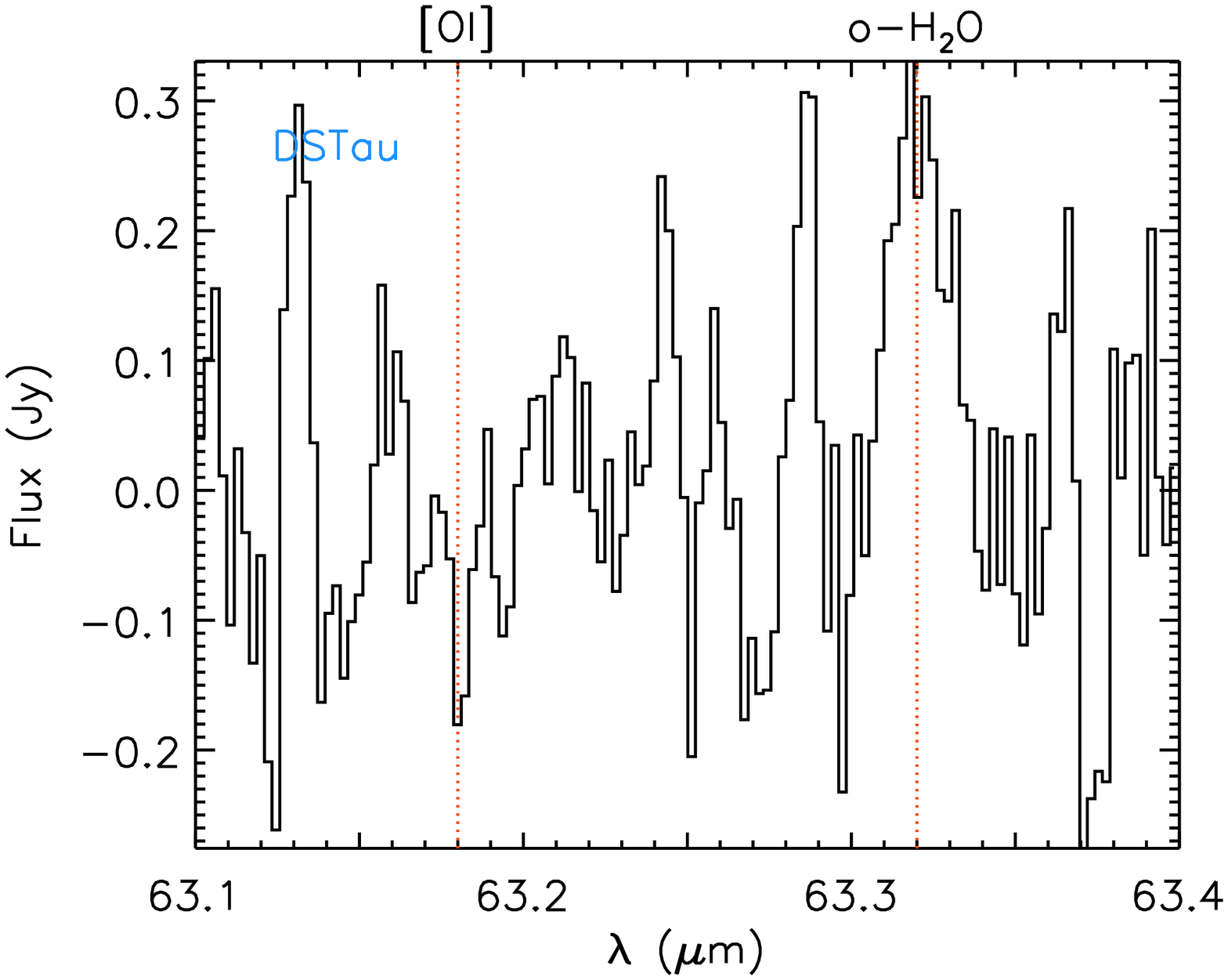}\includegraphics[width=0.16\textwidth, trim= 0mm 0mm 0mm 0mm, angle=90]{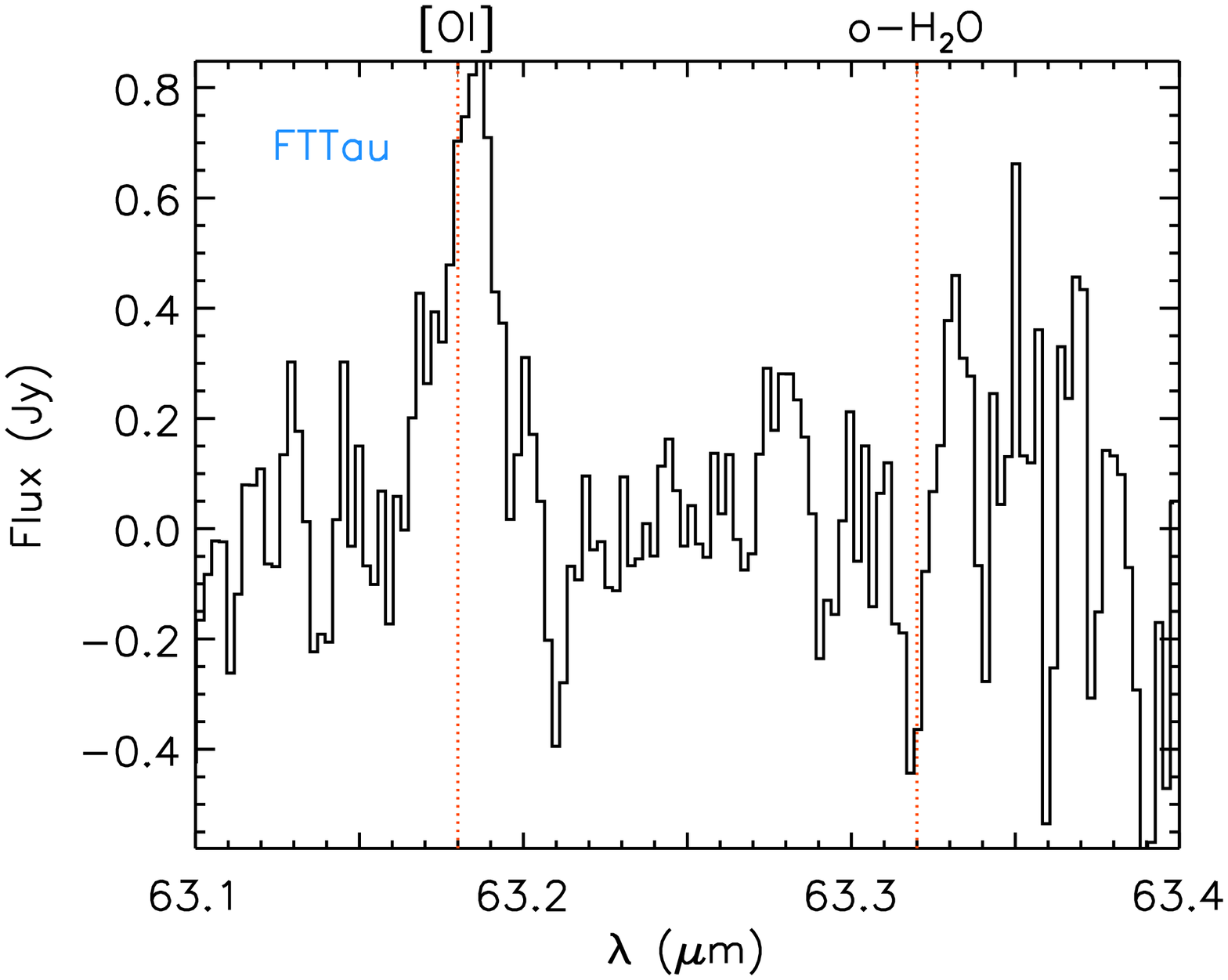}\includegraphics[width=0.16\textwidth, trim= 0mm 0mm 0mm 0mm, angle=90]{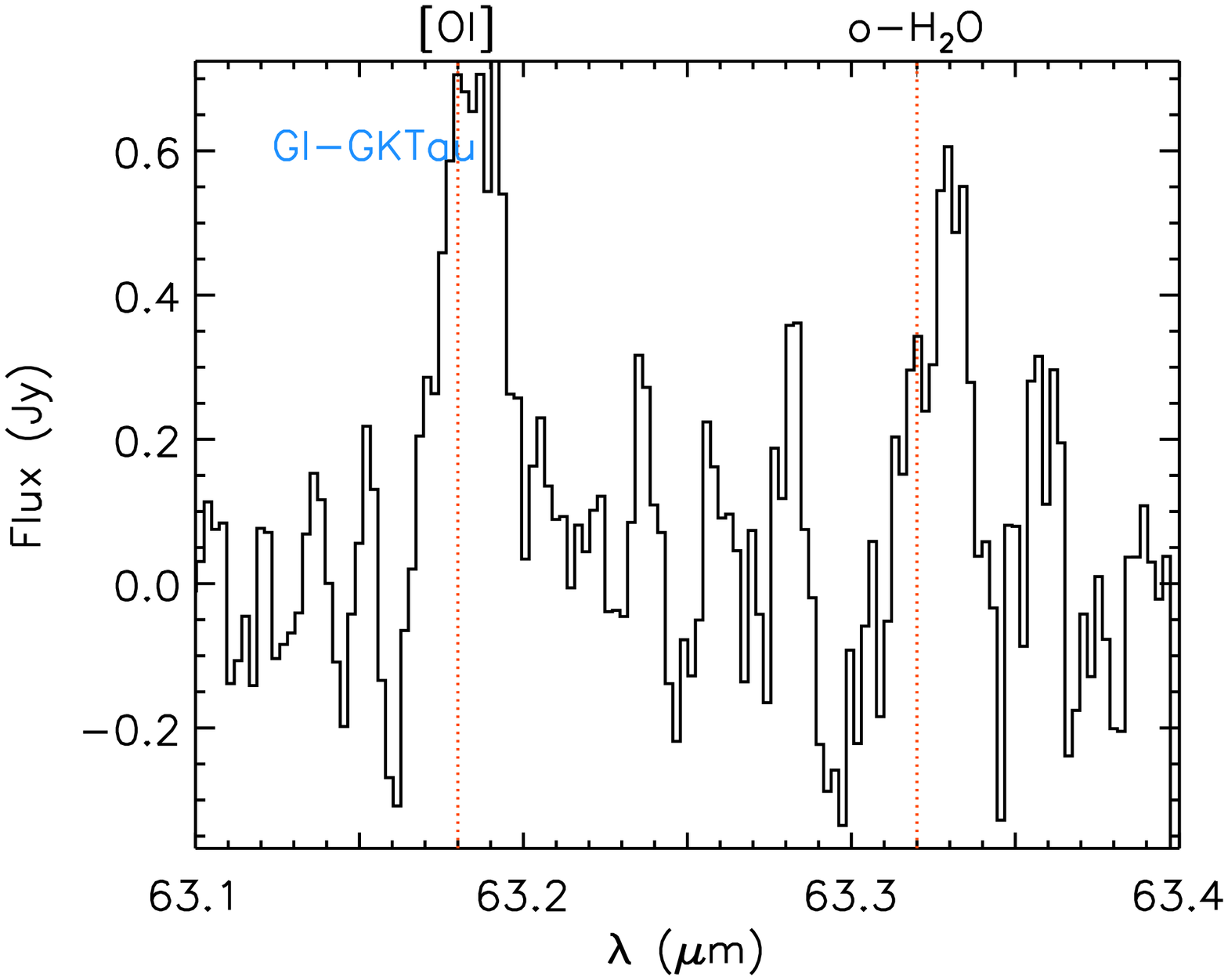}\includegraphics[width=0.16\textwidth, trim= 0mm 0mm 0mm 0mm, angle=90]{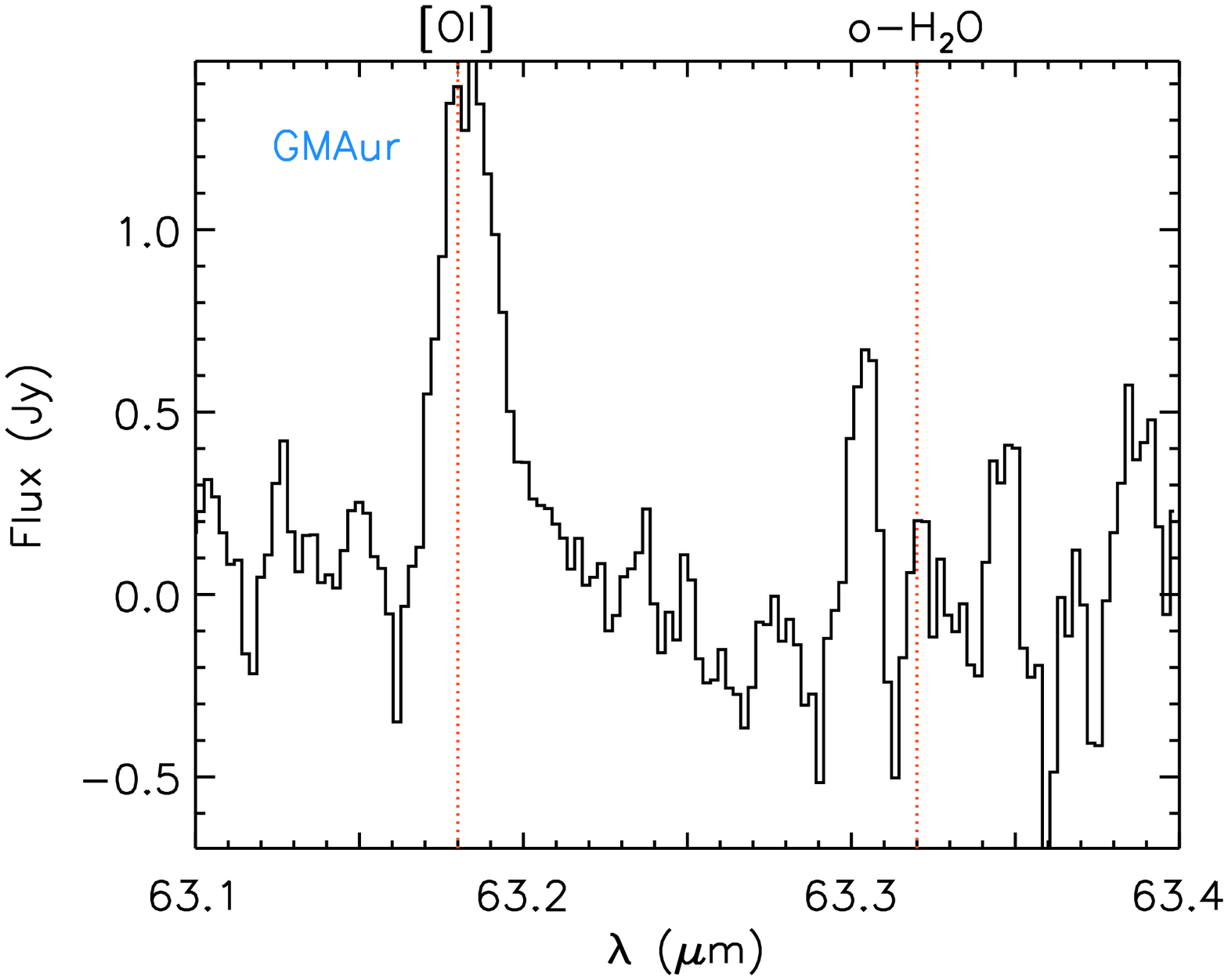}

\includegraphics[width=0.16\textwidth, trim= 0mm 0mm 0mm 0mm, angle=90]{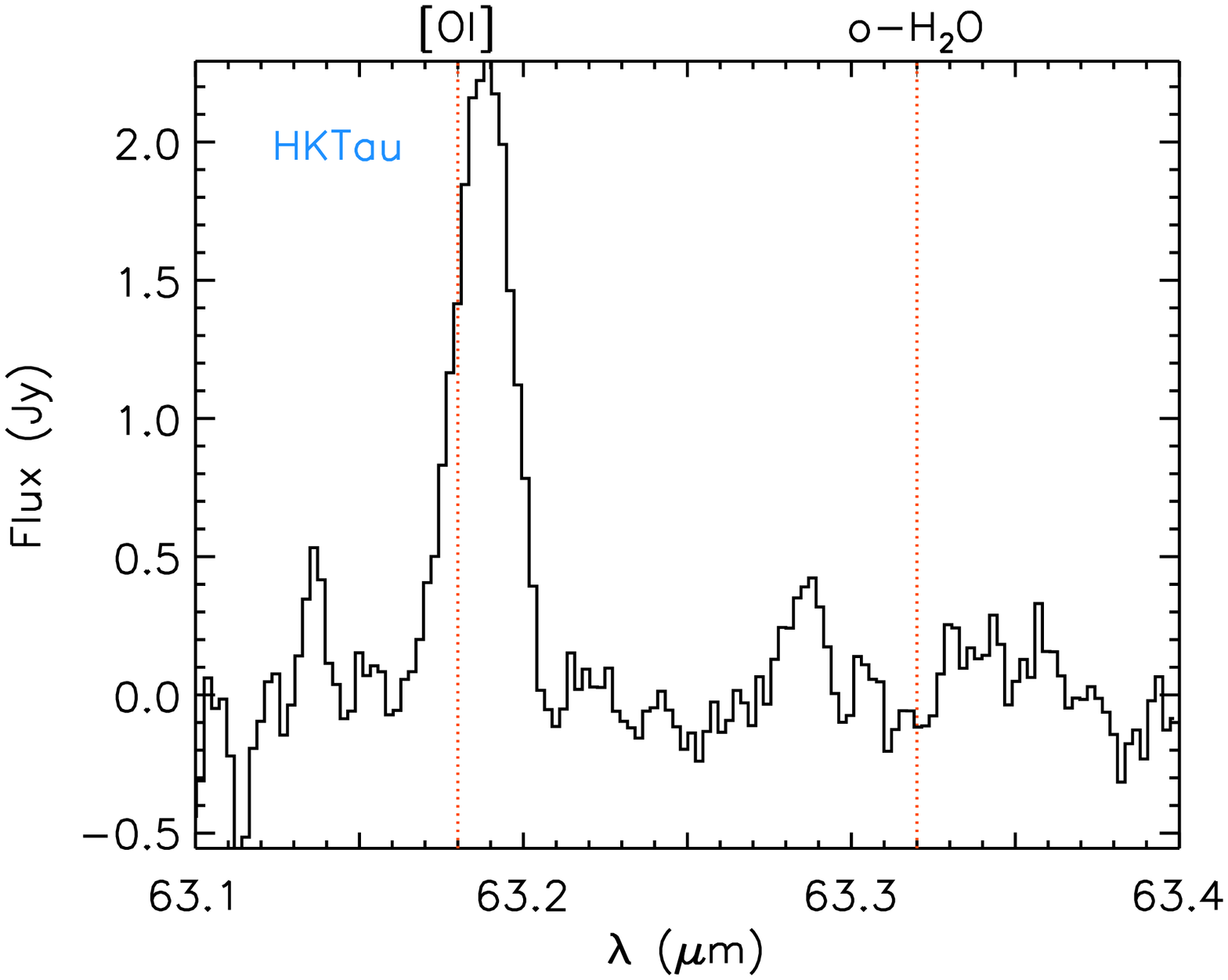}\includegraphics[width=0.16\textwidth, trim= 0mm 0mm 0mm 0mm, angle=90]{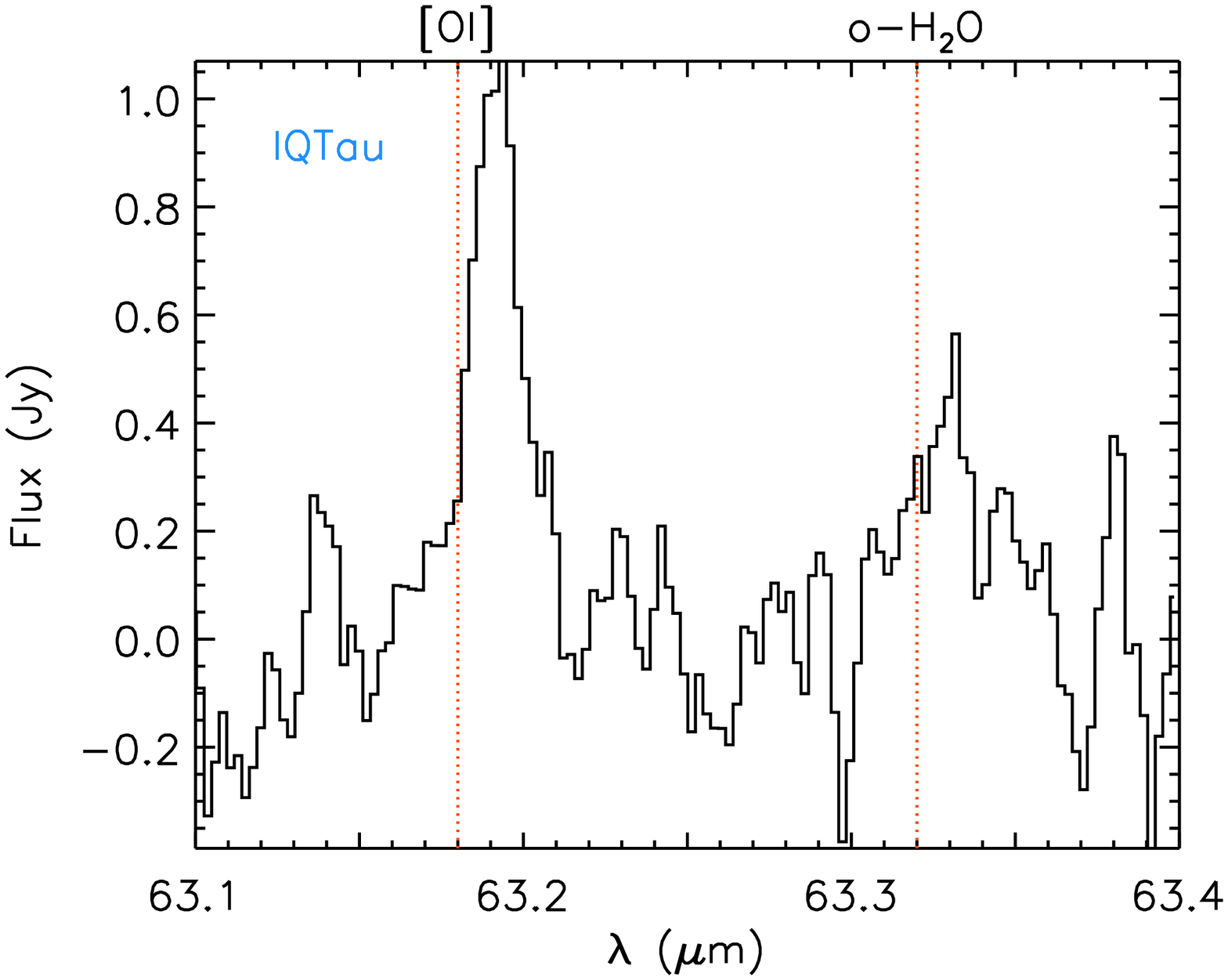}\includegraphics[width=0.16\textwidth, trim= 0mm 0mm 0mm 0mm, angle=90]{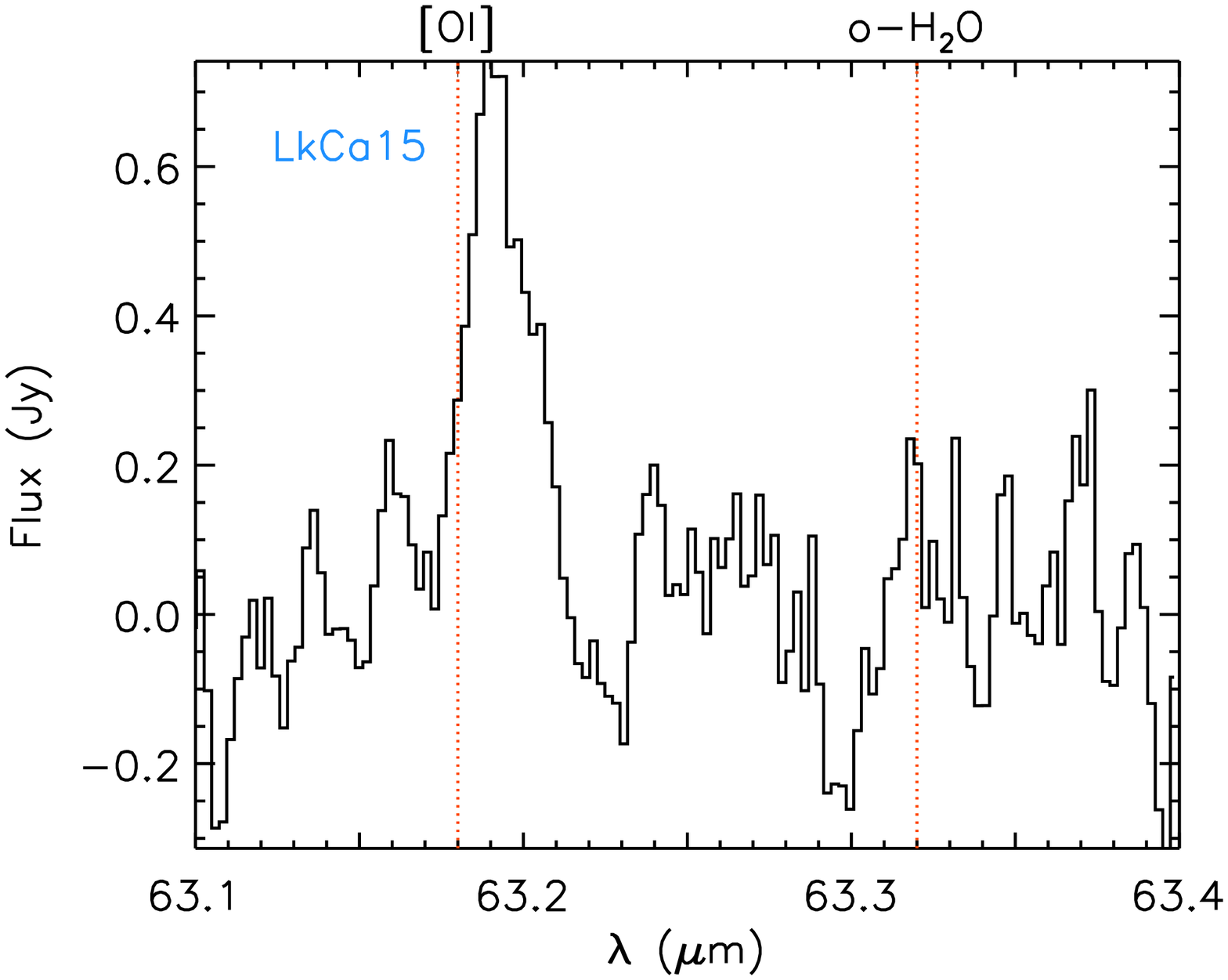}\includegraphics[width=0.16\textwidth, trim= 0mm 0mm 0mm 0mm, angle=90]{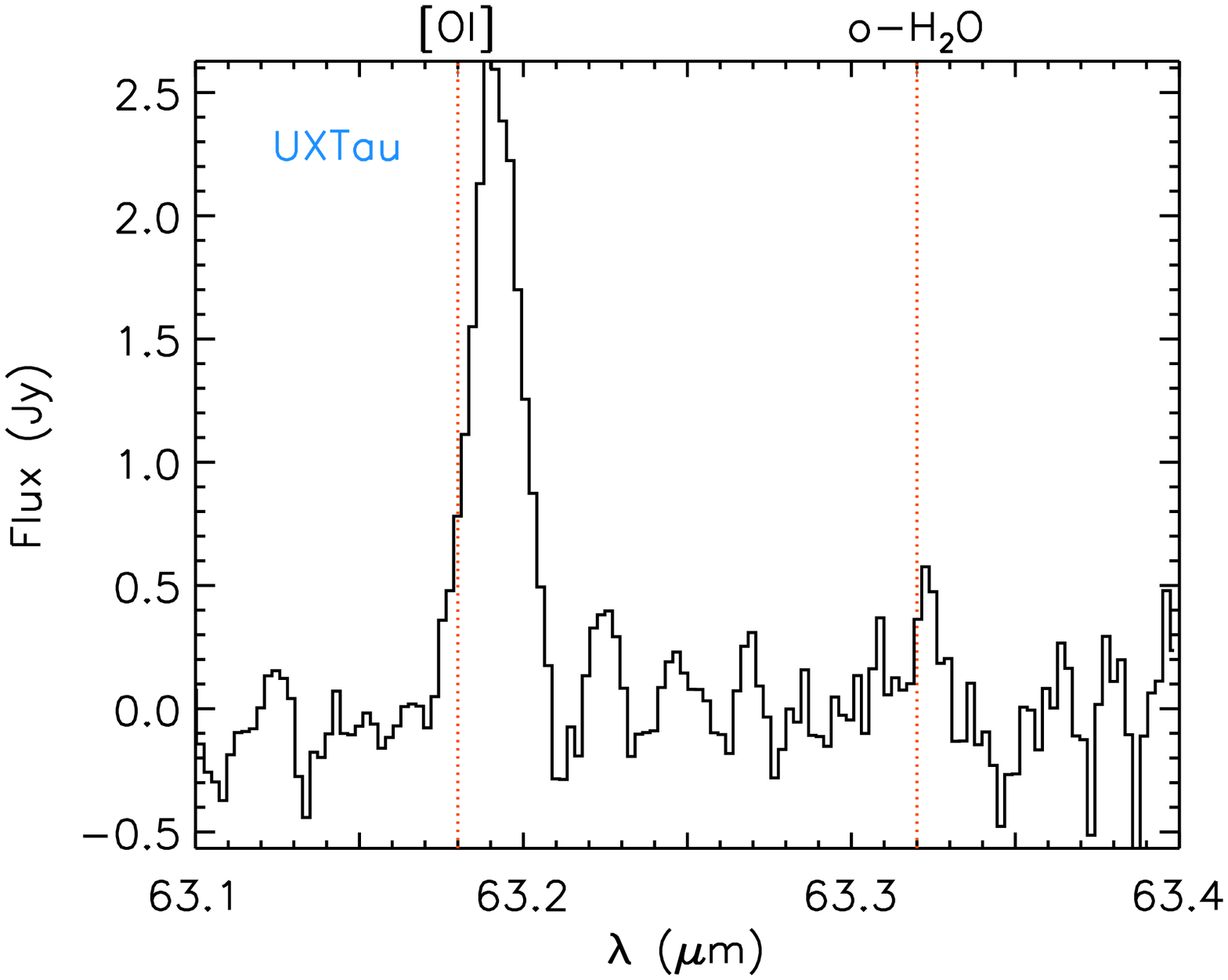}

\includegraphics[width=0.16\textwidth, trim= 0mm 0mm 0mm 0mm, angle=90]{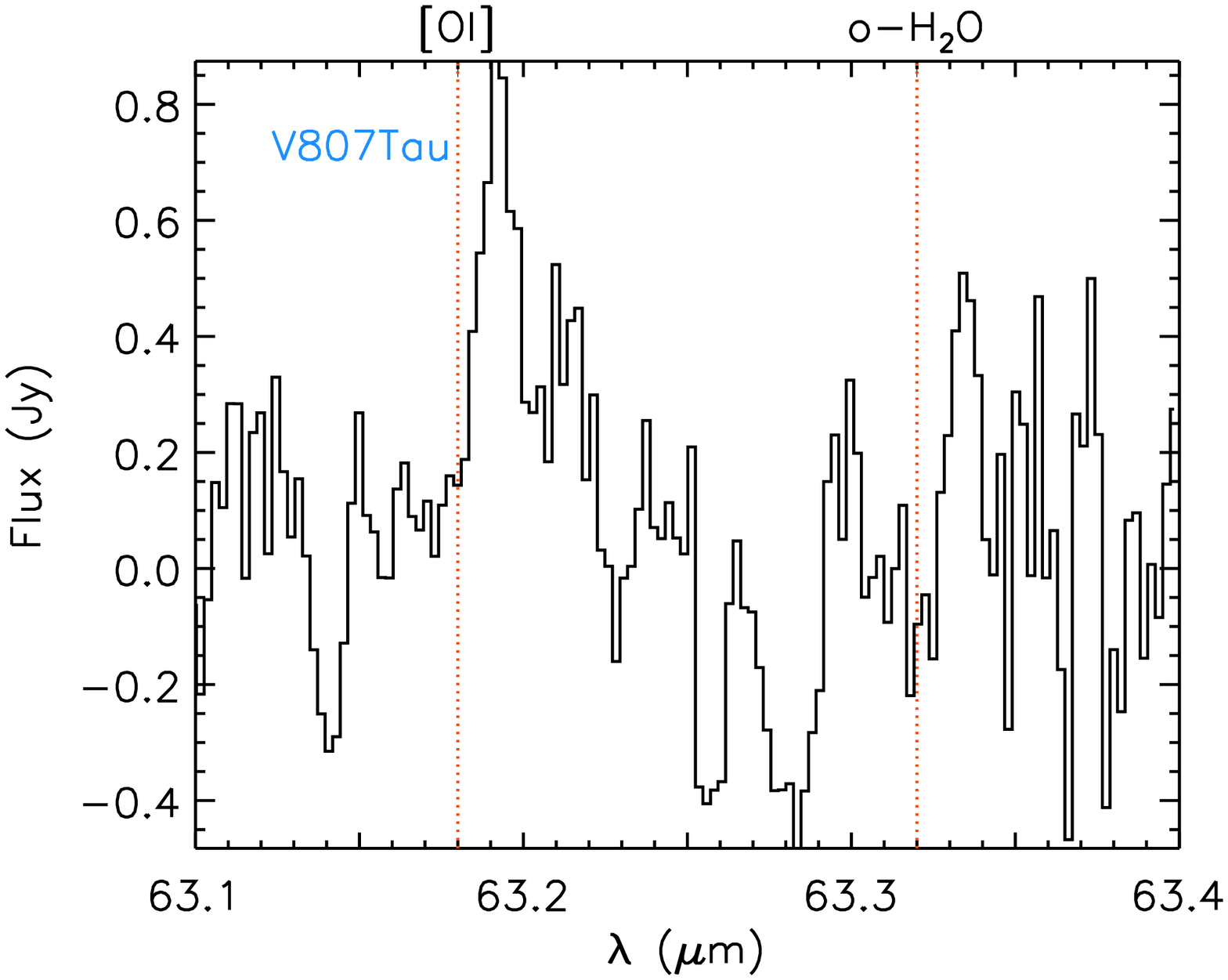}
\caption{\textit{(Continued.)}}
\end{figure*}

%########################################################72um
\begin{figure*}[htpb]
\centering
\setcounter{figure}{1}
\includegraphics[width=0.16\textwidth, trim= 0mm 0mm 0mm 0mm, angle=90]{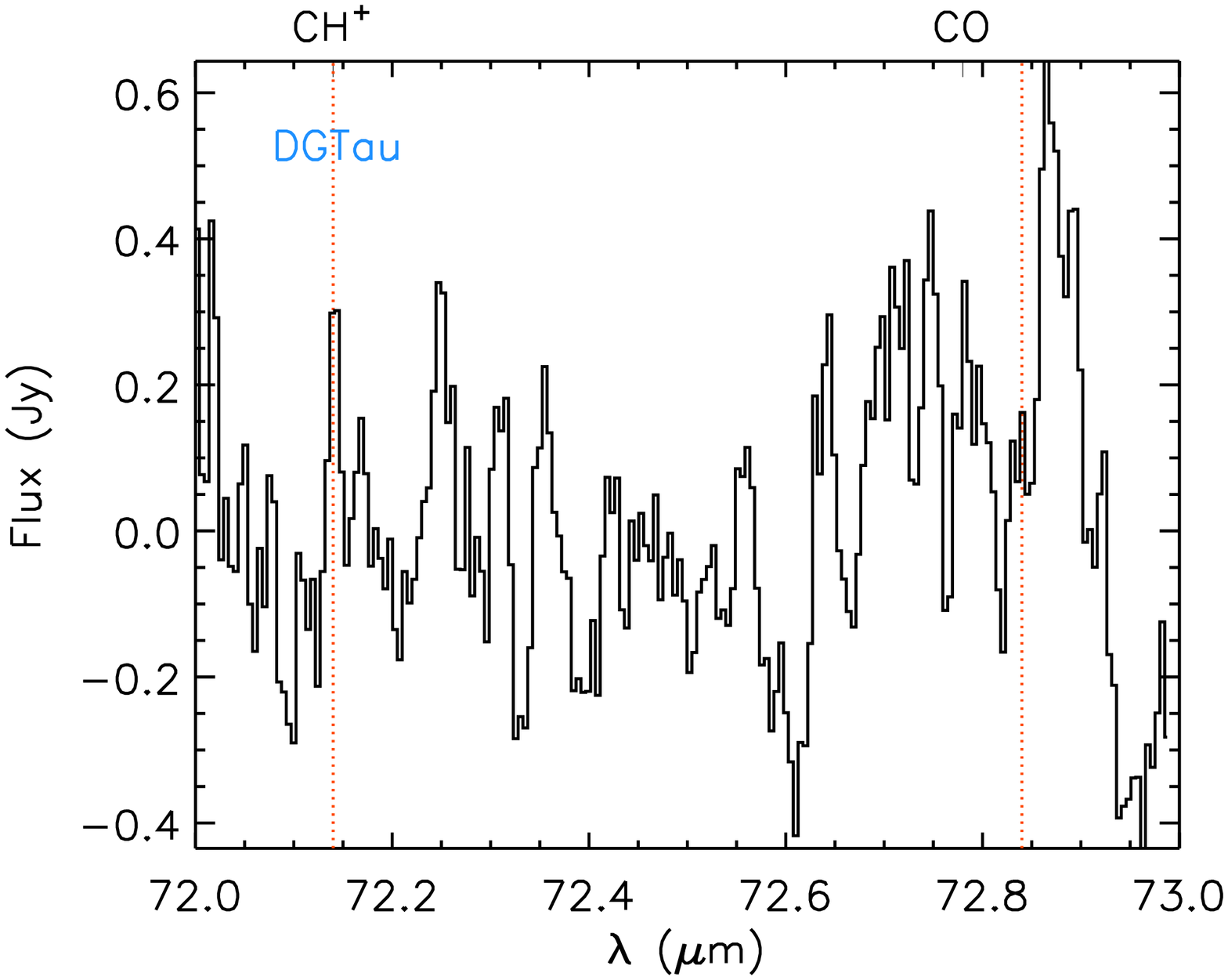}
\includegraphics[width=0.16\textwidth, trim= 0mm 0mm 0mm 0mm, angle=90]{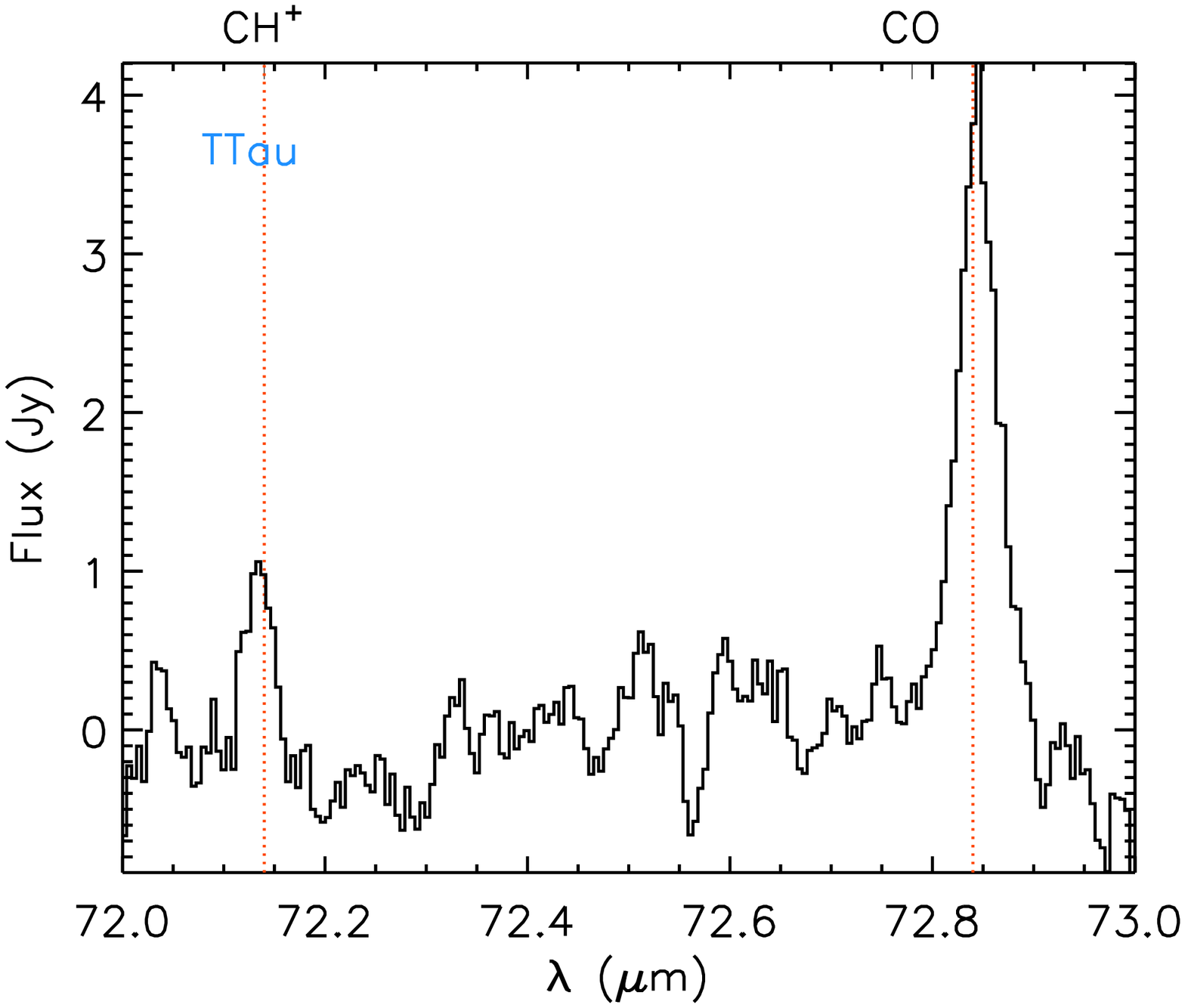}
\caption{Continuum subtracted spectra at 72 $\rm \mu m$ for all objects with CO detections. The red vertical lines indicate the positions of CH$^+$ 72.14 $\rm \mu m$ and CO 72.84 $\rm \mu m$.}
\end{figure*}

%########################################################78um
\begin{figure*}[htpb]
\centering
\setcounter{figure}{2}
\includegraphics[width=0.16\textwidth, trim= 0mm 0mm 0mm 0mm, angle=90]{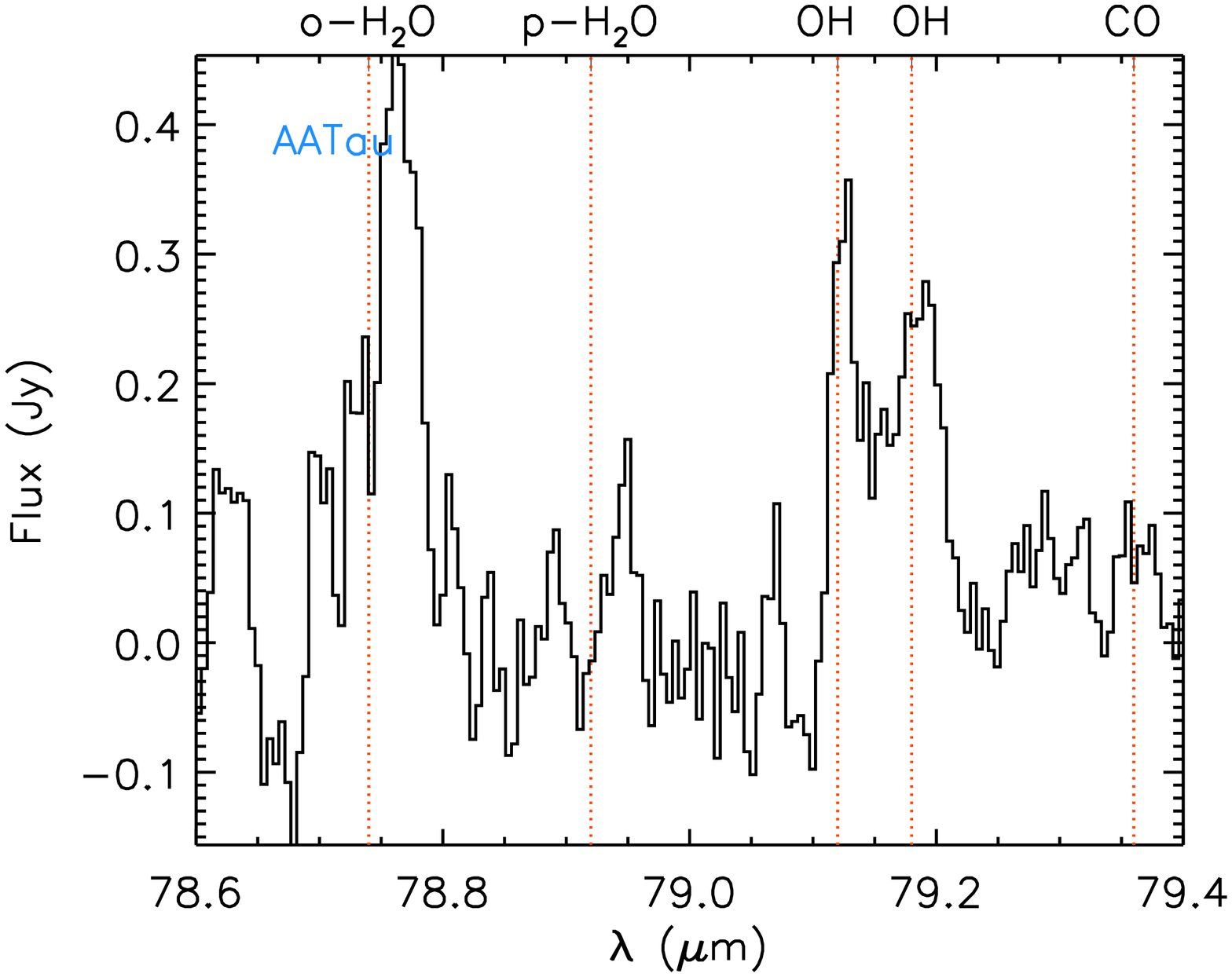}\includegraphics[width=0.16\textwidth, trim= 0mm 0mm 0mm 0mm, angle=90]{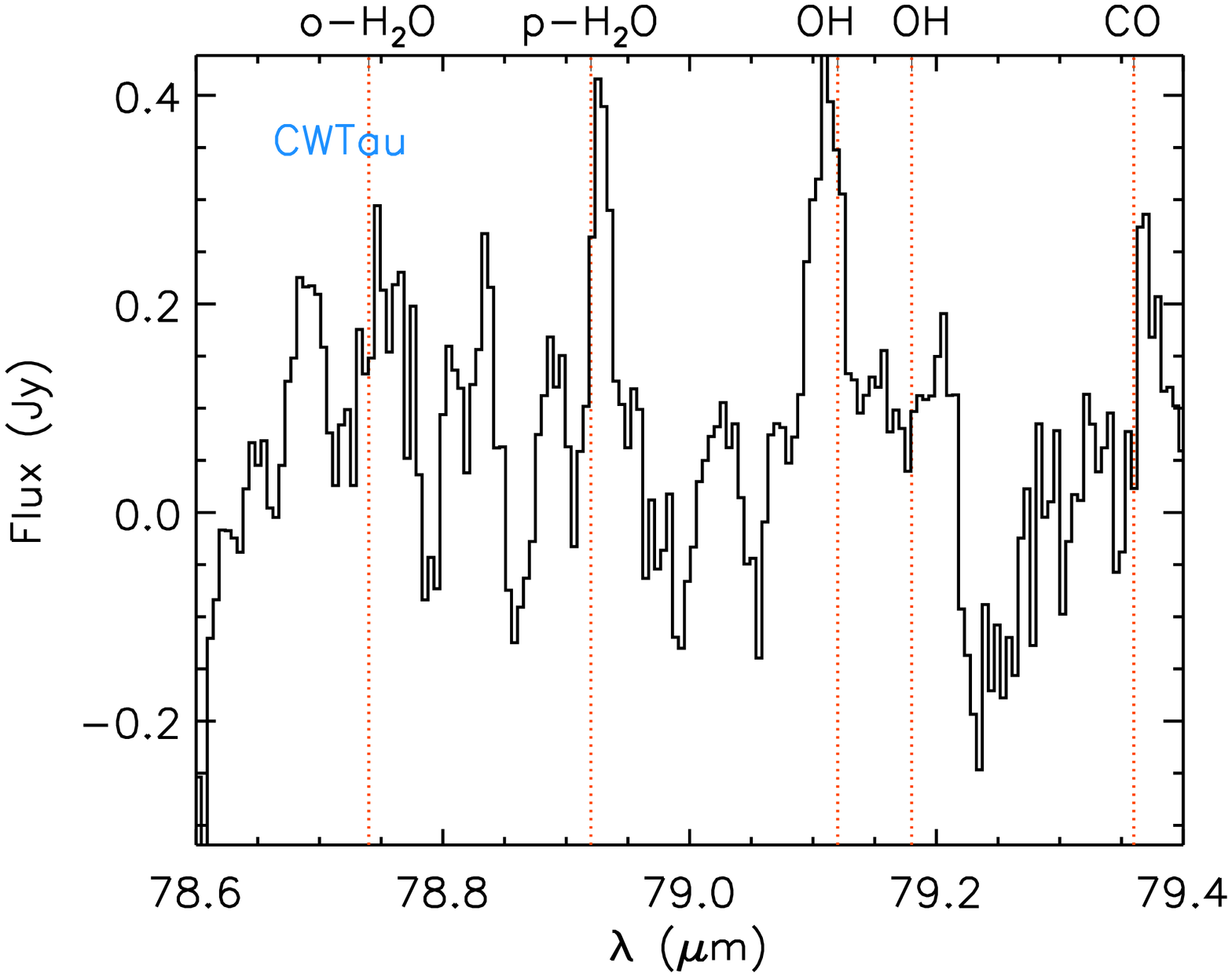}\includegraphics[width=0.16\textwidth, trim= 0mm 0mm 0mm 0mm, angle=90]{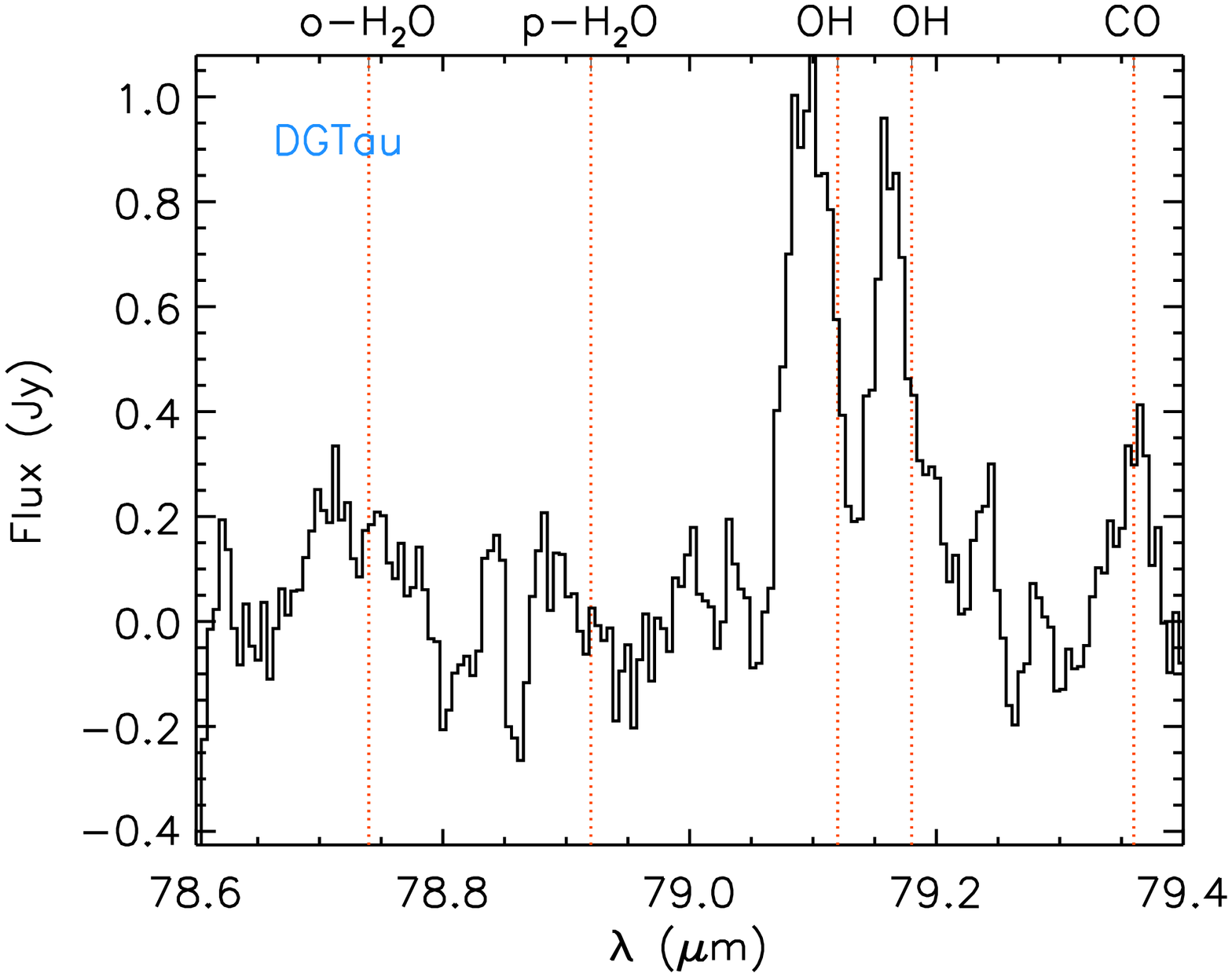}\includegraphics[width=0.16\textwidth, trim= 0mm 0mm 0mm 0mm, angle=90]{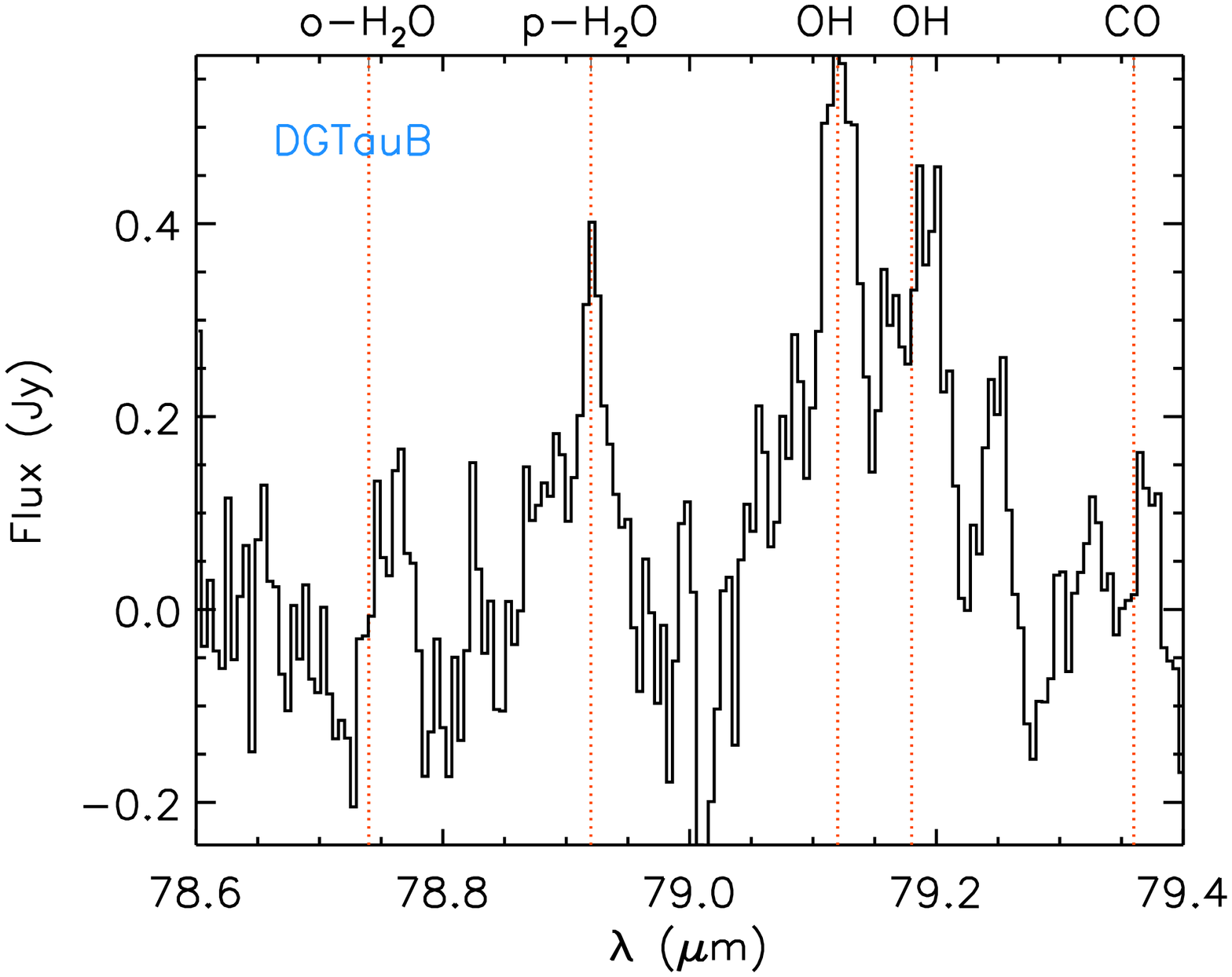}

\includegraphics[width=0.16\textwidth, trim= 0mm 0mm 0mm 0mm, angle=90]{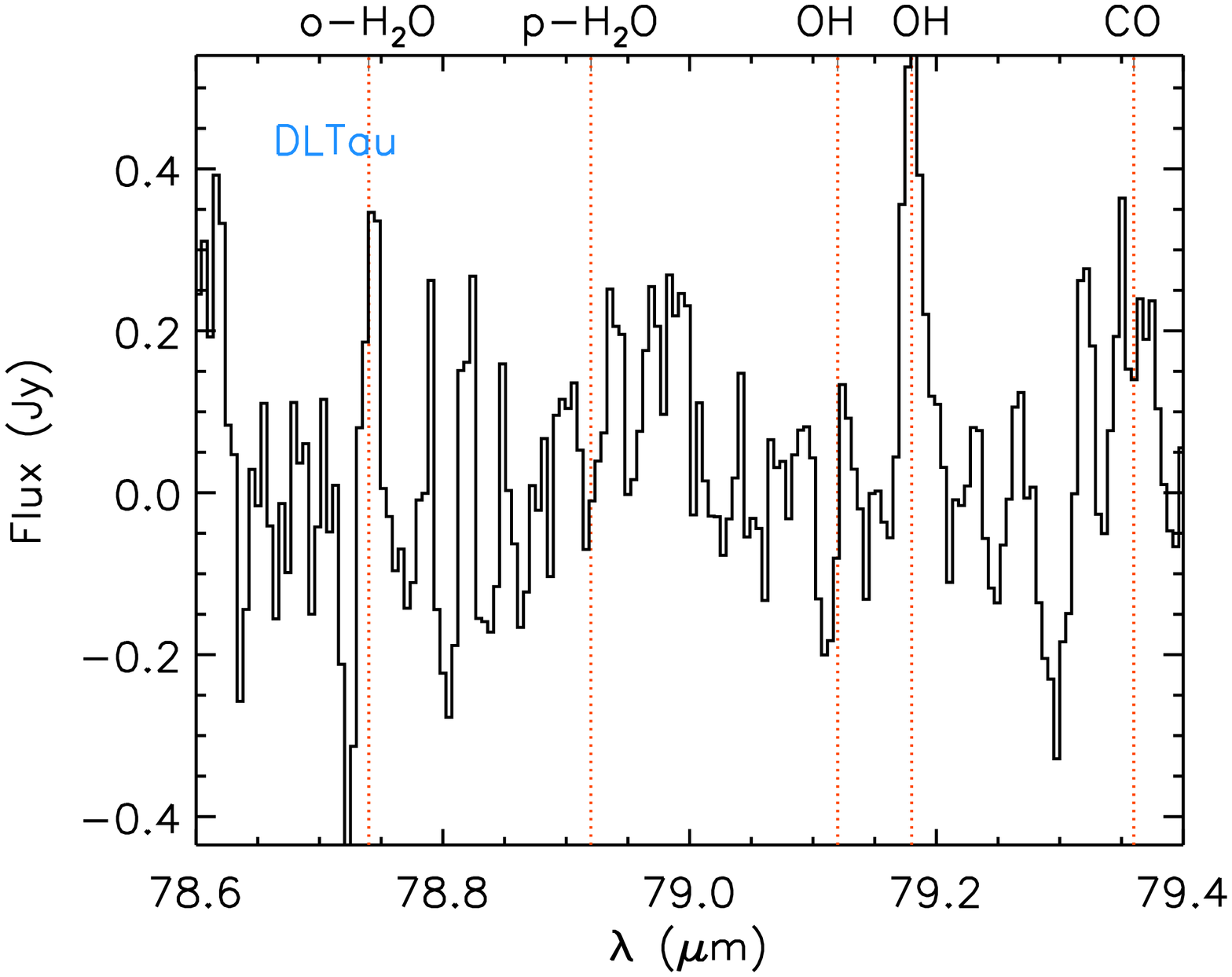}\includegraphics[width=0.16\textwidth, trim= 0mm 0mm 0mm 0mm, angle=90]{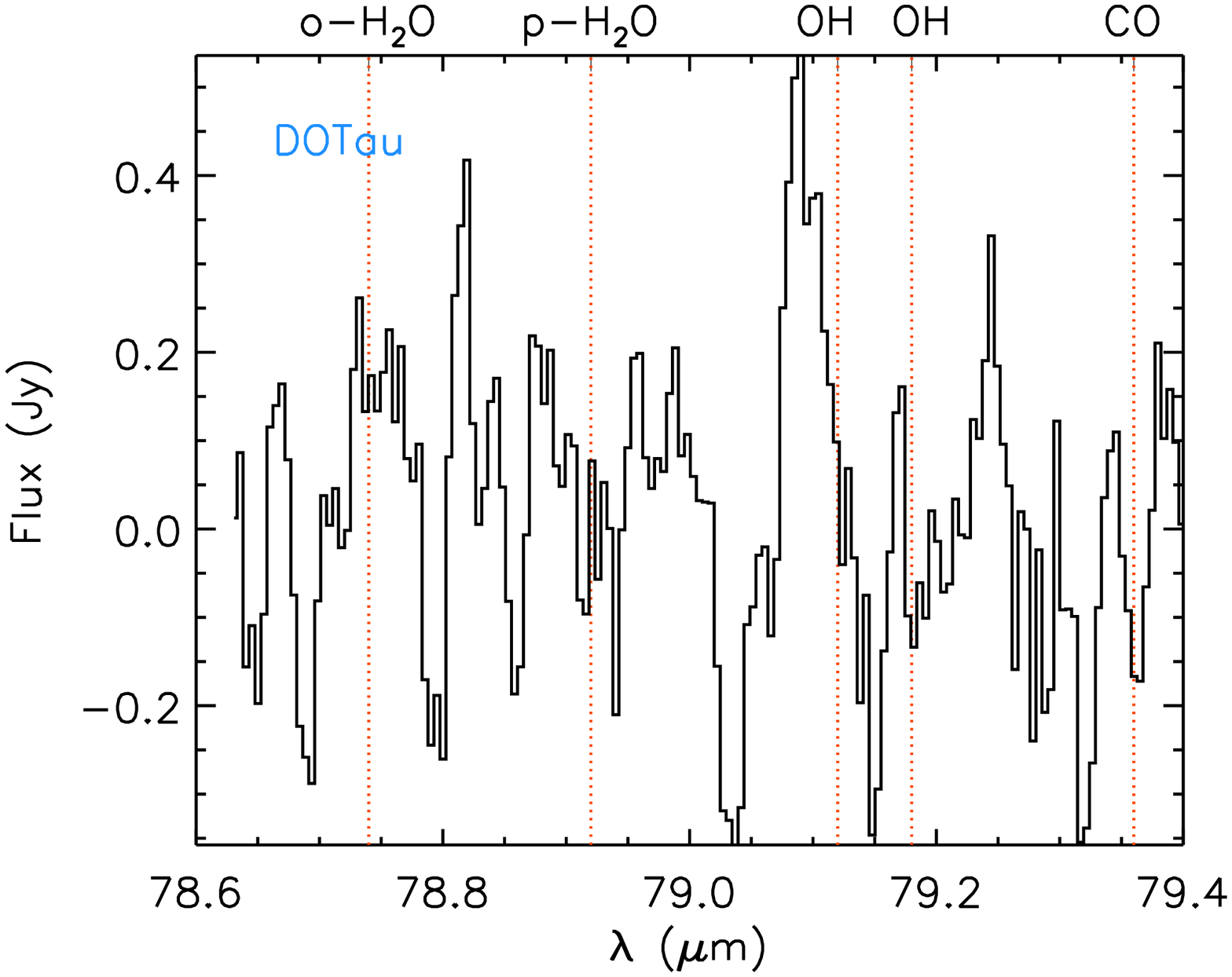}\includegraphics[width=0.16\textwidth, trim= 0mm 0mm 0mm 0mm, angle=90]{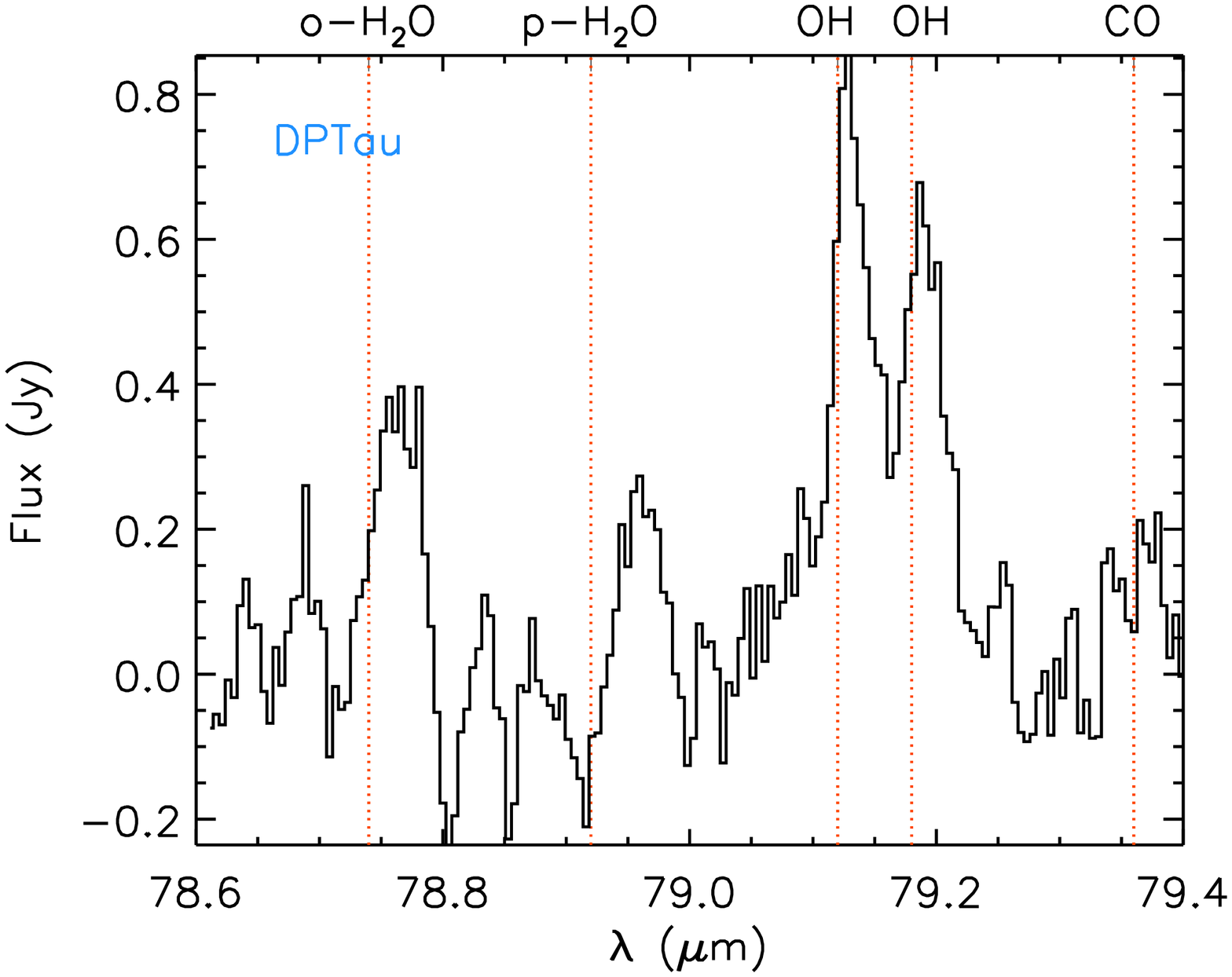}\includegraphics[width=0.16\textwidth, trim= 0mm 0mm 0mm 0mm, angle=90]{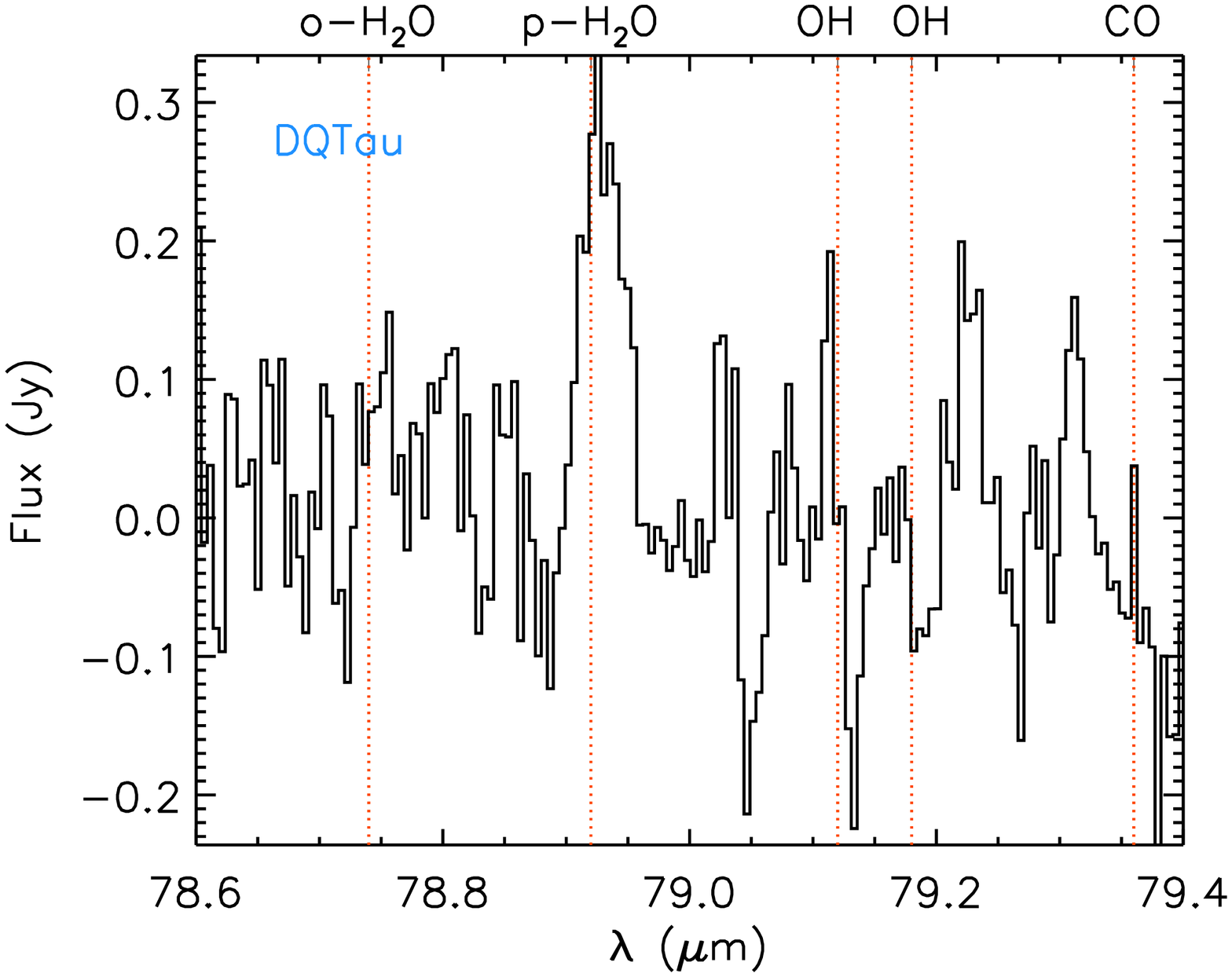}

\includegraphics[width=0.16\textwidth, trim= 0mm 0mm 0mm 0mm, angle=90]{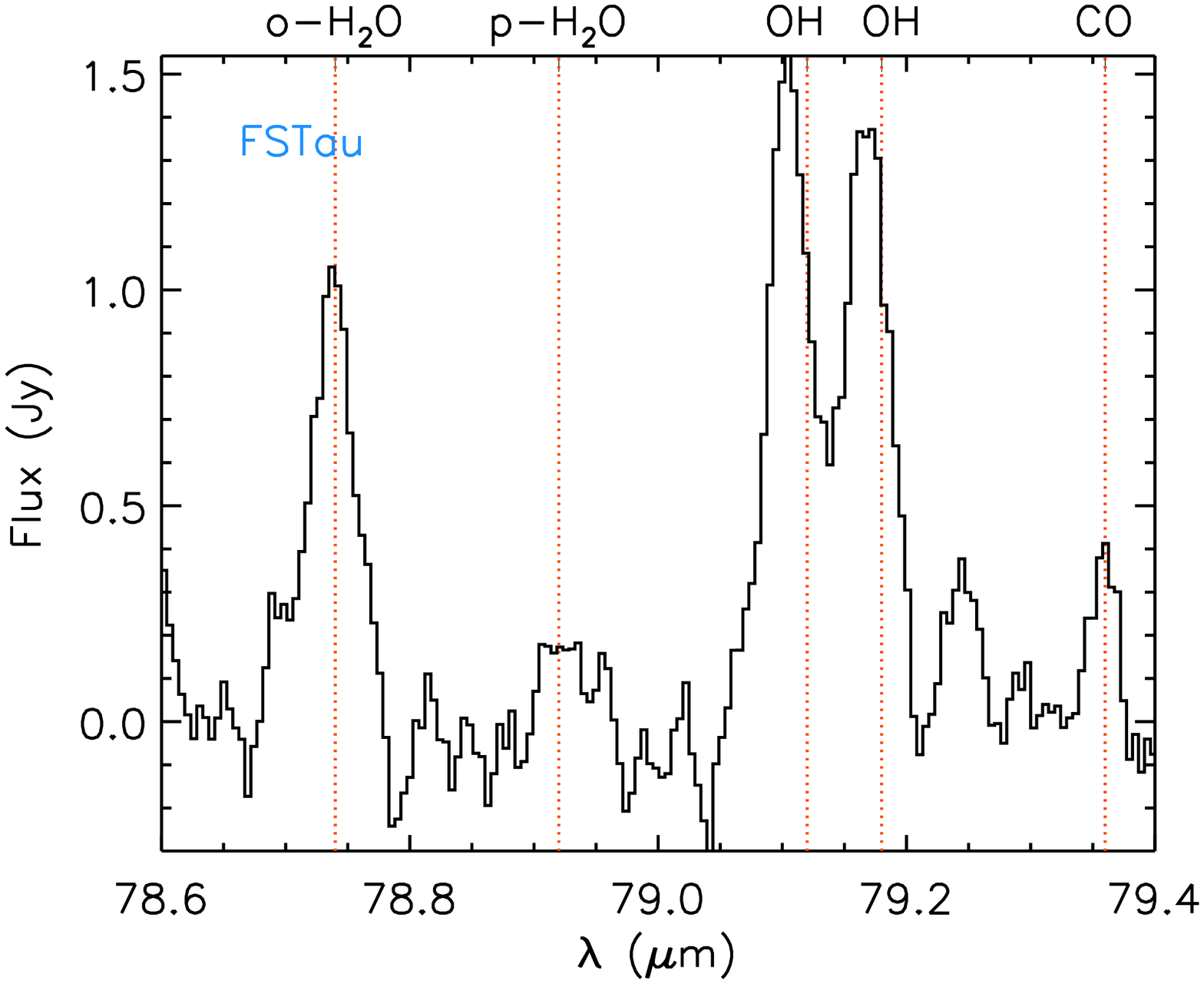}\includegraphics[width=0.16\textwidth, trim= 0mm 0mm 0mm 0mm, angle=90]{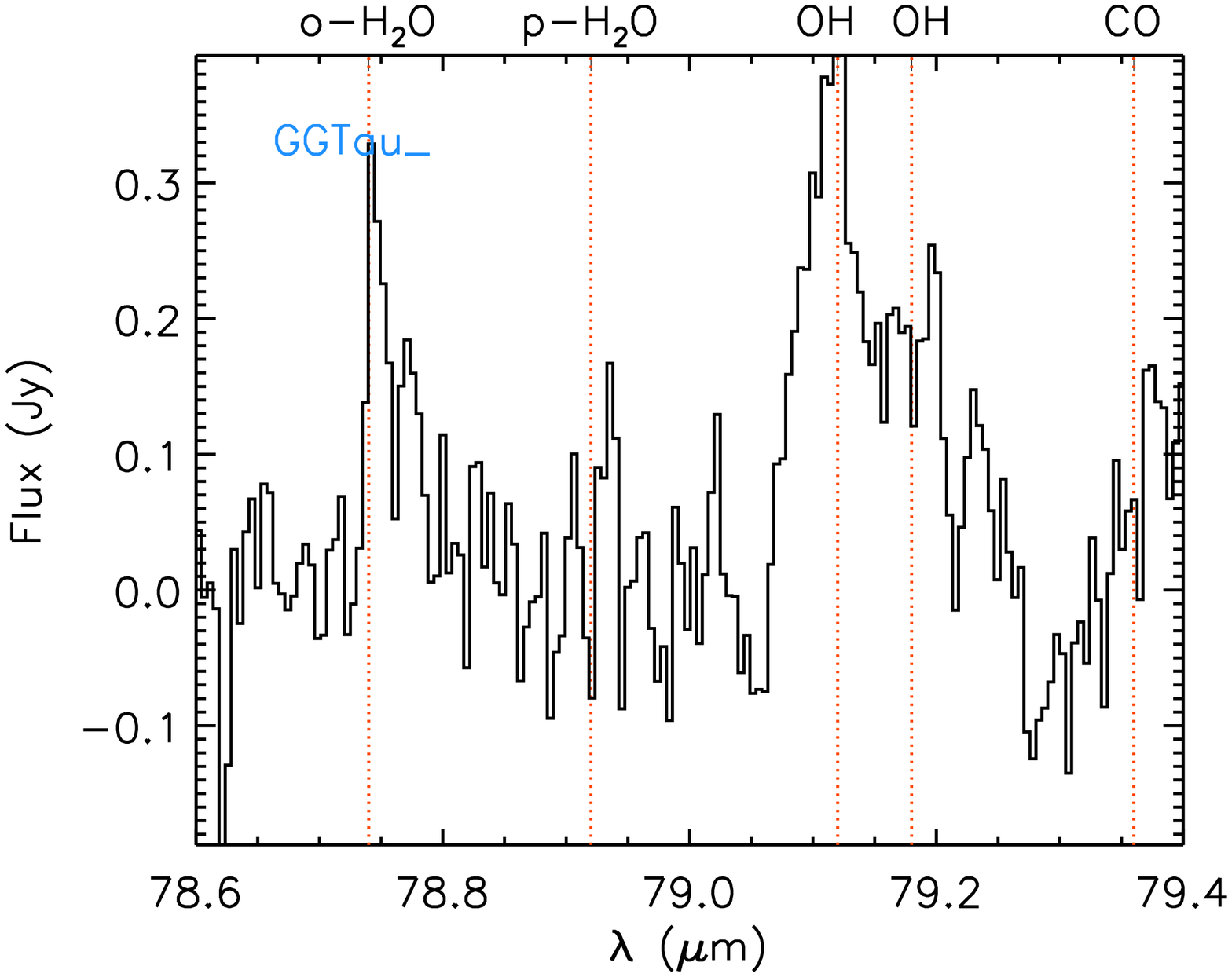}\includegraphics[width=0.16\textwidth, trim= 0mm 0mm 0mm 0mm, angle=90]{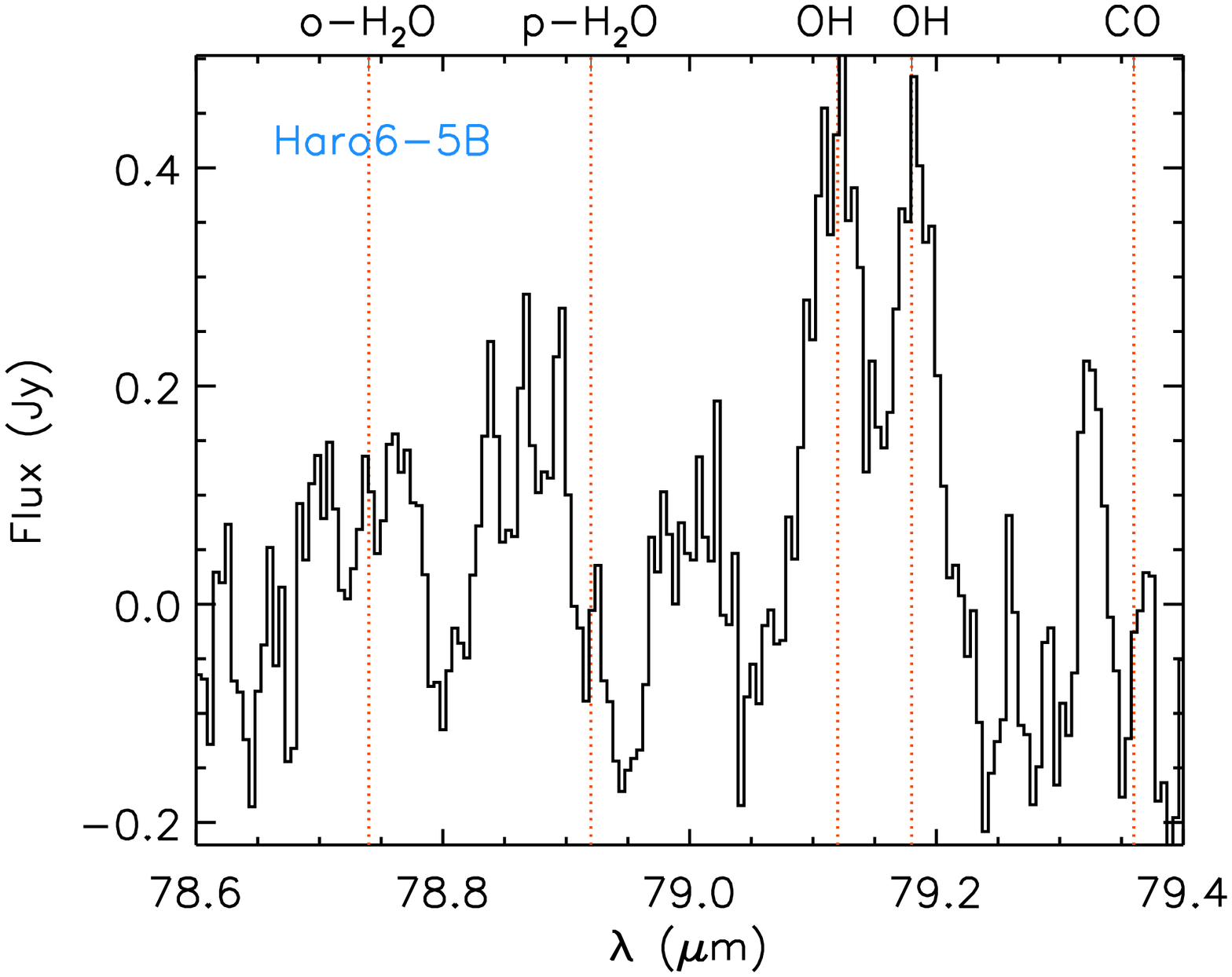}\includegraphics[width=0.16\textwidth, trim= 0mm 0mm 0mm 0mm, angle=90]{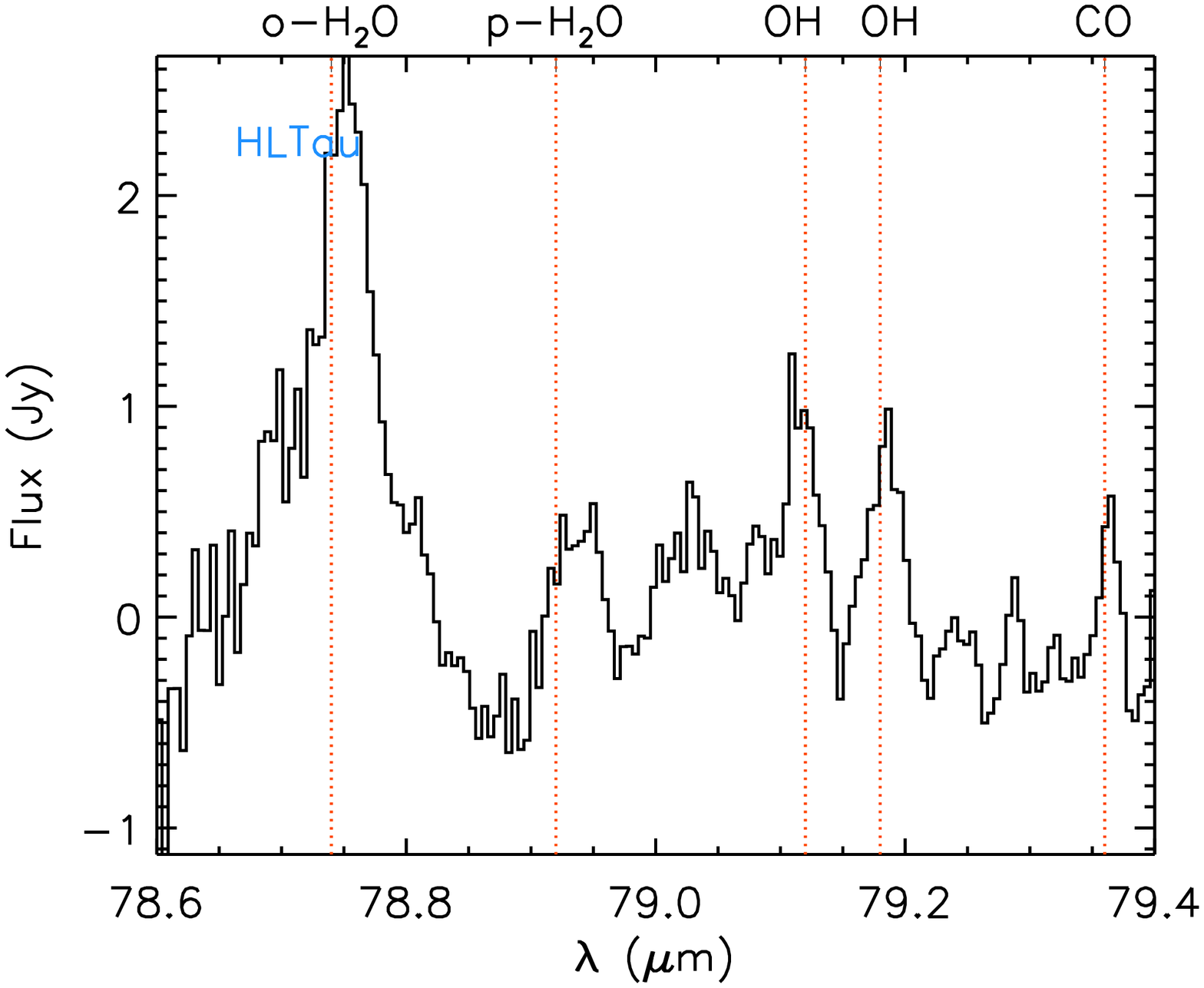}

\includegraphics[width=0.16\textwidth, trim= 0mm 0mm 0mm 0mm, angle=90]{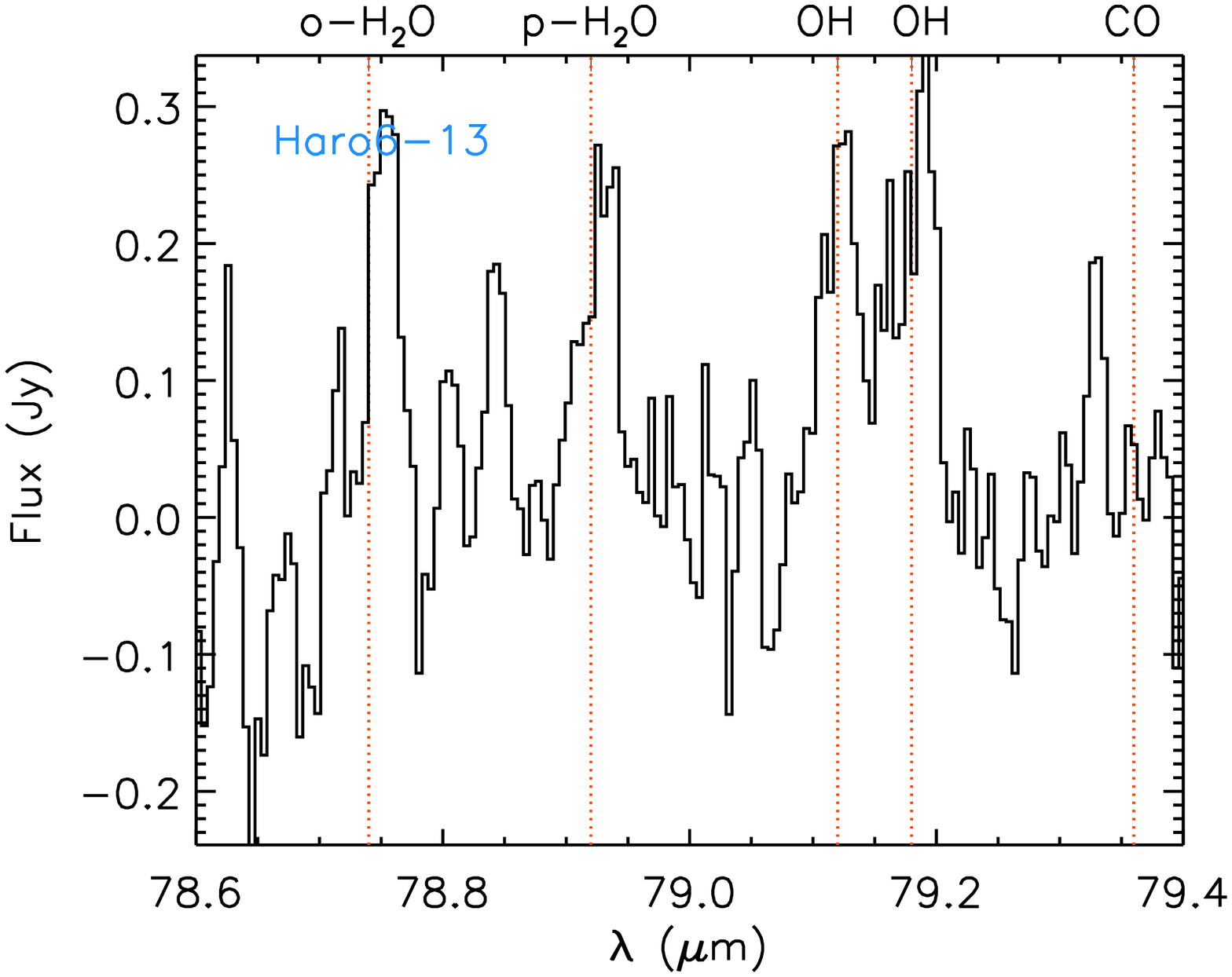}\includegraphics[width=0.16\textwidth, trim= 0mm 0mm 0mm 0mm, angle=90]{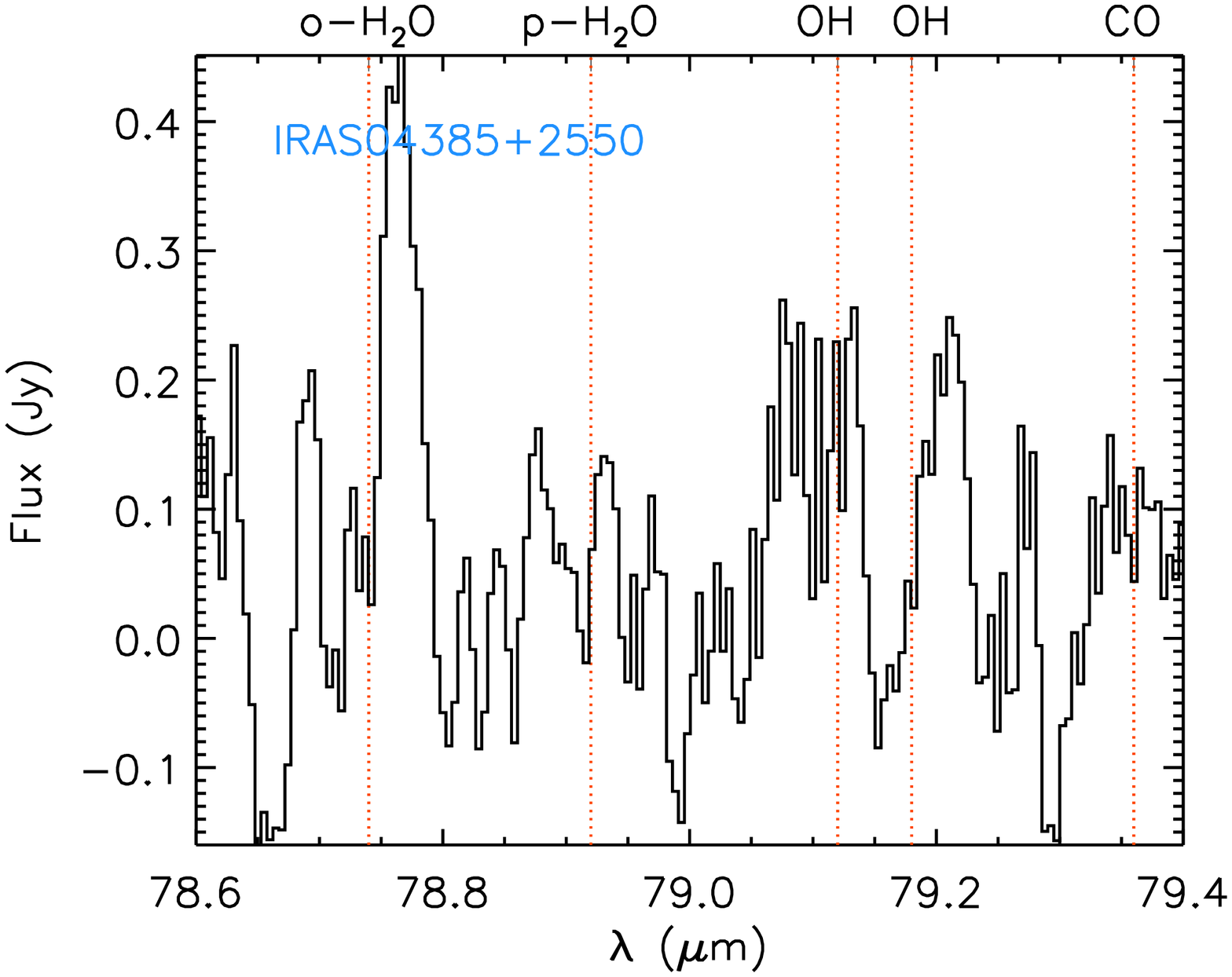}\includegraphics[width=0.16\textwidth, trim= 0mm 0mm 0mm 0mm, angle=90]{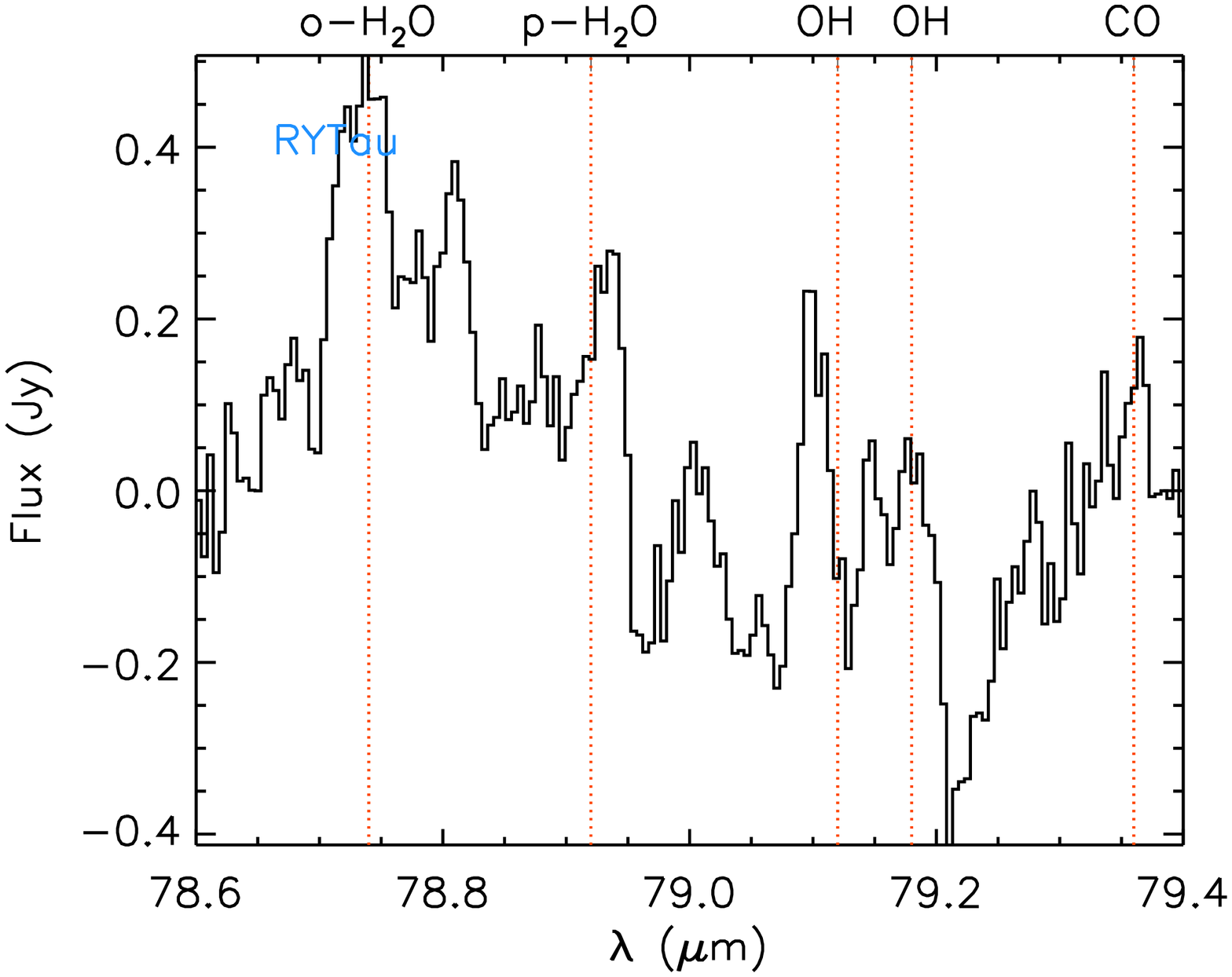}\includegraphics[width=0.16\textwidth, trim= 0mm 0mm 0mm 0mm, angle=90]{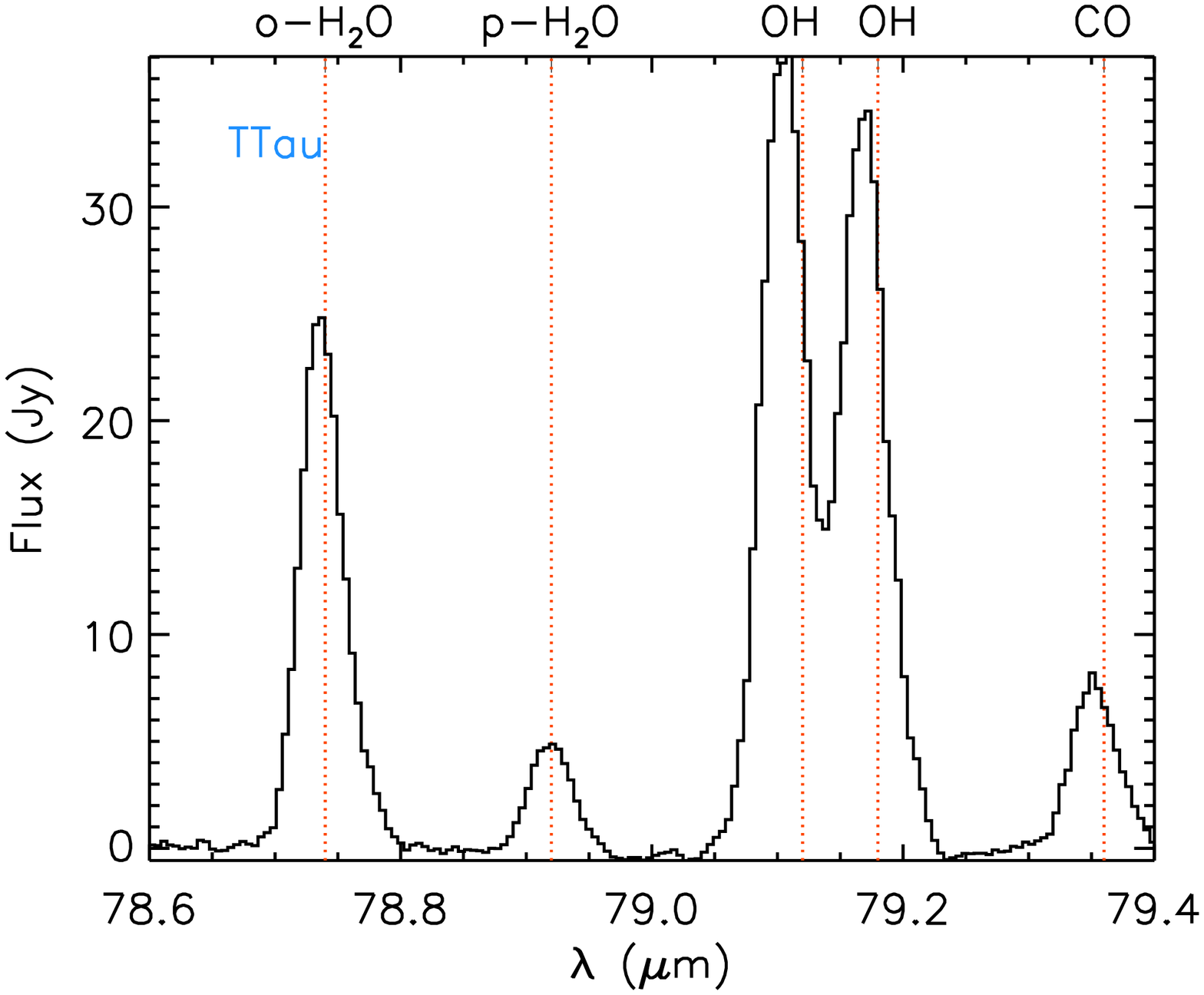}

\includegraphics[width=0.16\textwidth, trim= 0mm 0mm 0mm 0mm, angle=90]{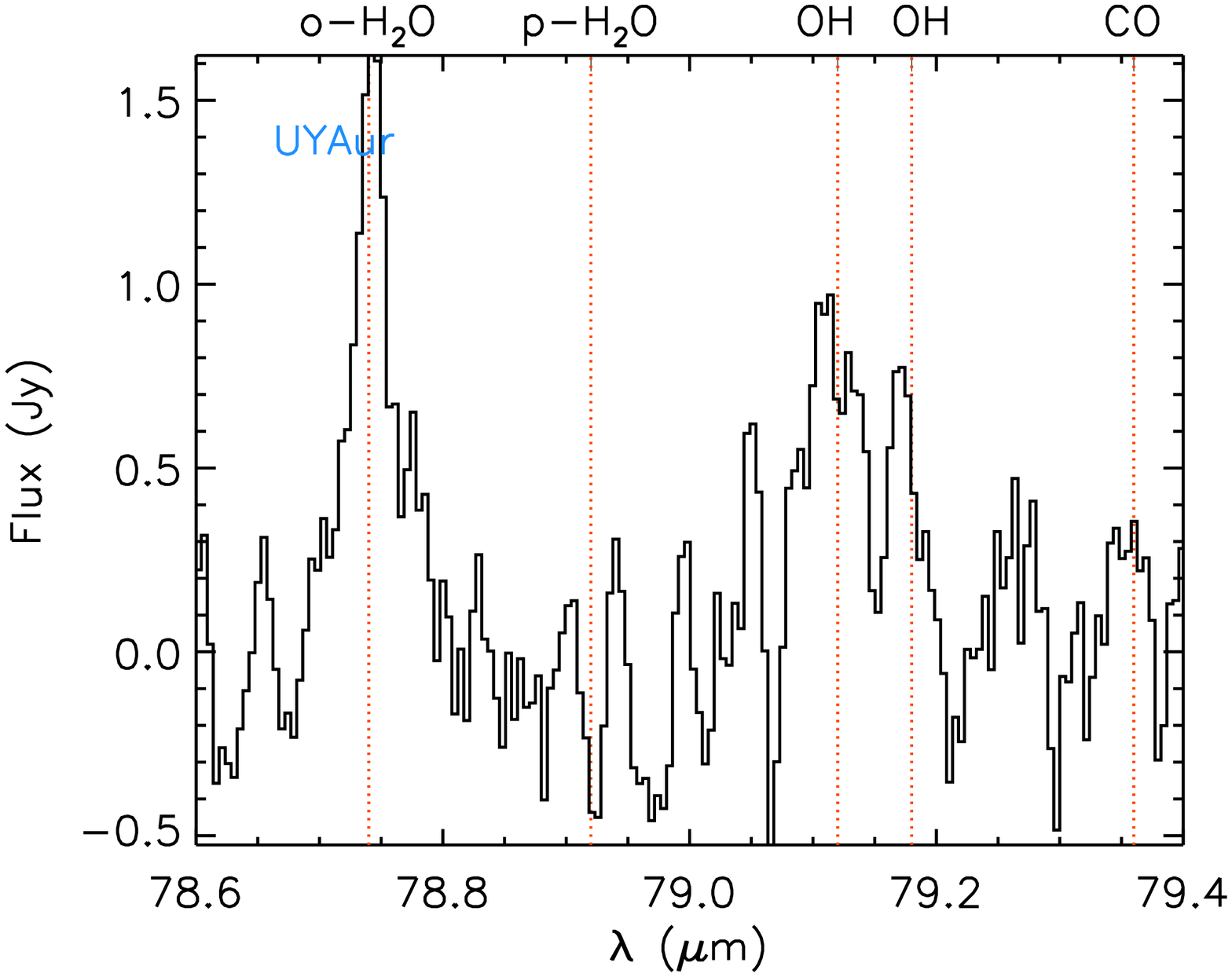}\includegraphics[width=0.16\textwidth, trim= 0mm 0mm 0mm 0mm, angle=90]{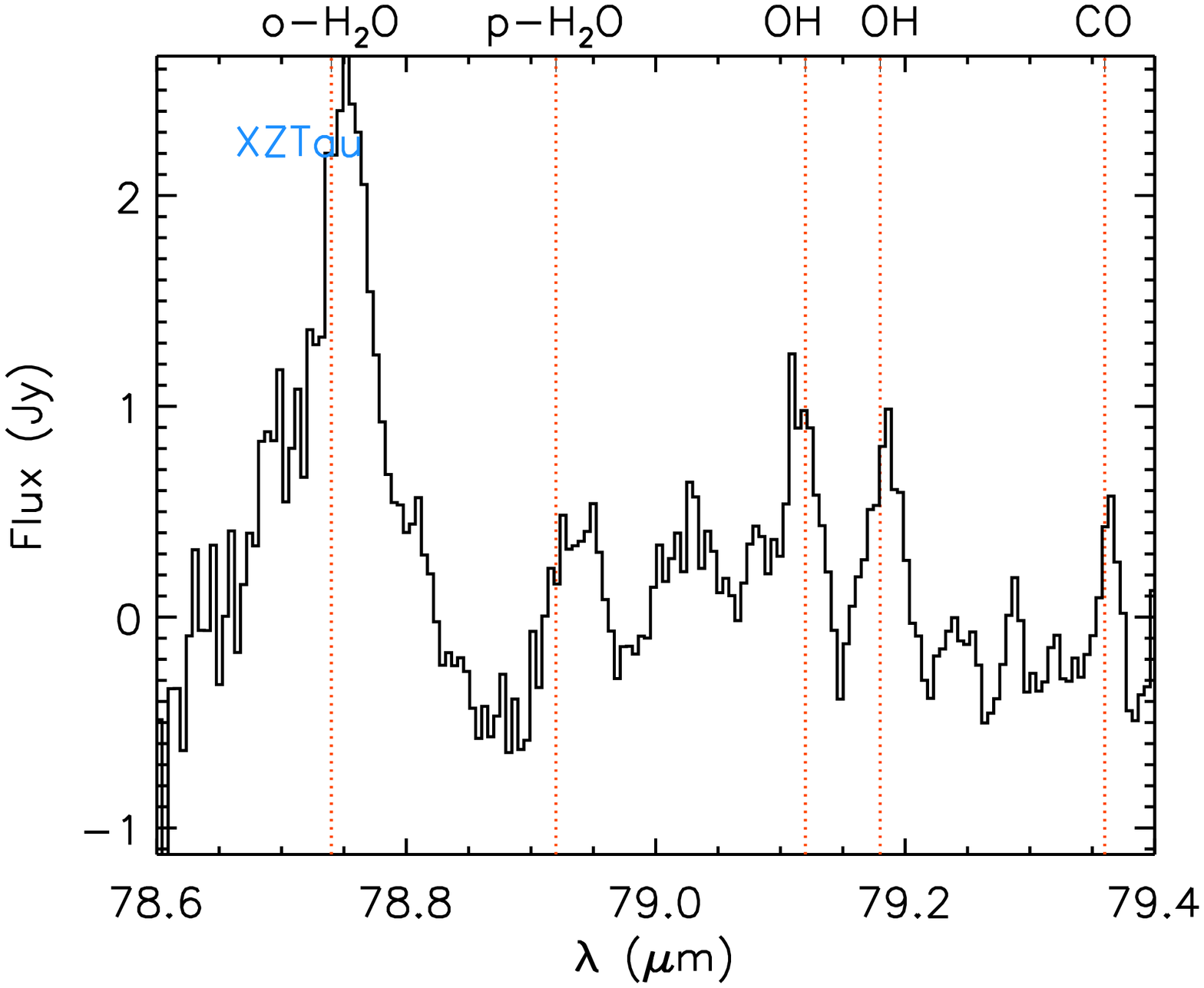}

\caption{Continuum subtracted spectra at 78 $\rm \mu m$ for all objects with detections. The red vertical lines indicate the positions of o-H$_{\rm 2}$O 78.74 $\rm \mu m$, p-H$_{\rm 2}$O 78.92 $\rm \mu m$, OH 78.12+78.18 $\rm \mu m$, and  CO 79.36 $\rm \mu m$.}
\end{figure*}

%########################################################90um
\begin{figure*}
\centering
\setcounter{figure}{3}
\includegraphics[width=0.16\textwidth, trim= 0mm 0mm 0mm 0mm, angle=90]{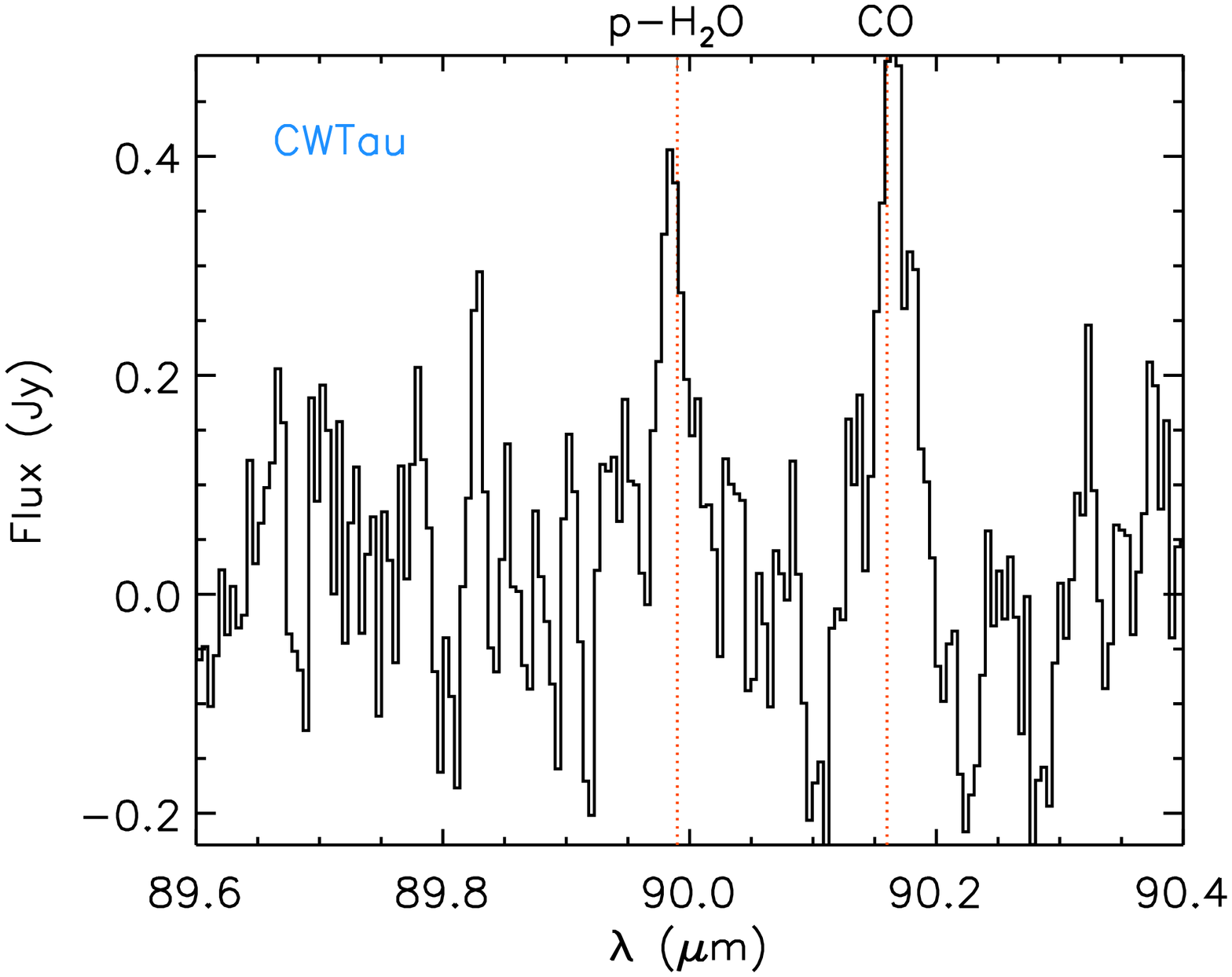}\includegraphics[width=0.16\textwidth, trim= 0mm 0mm 0mm 0mm, angle=90]{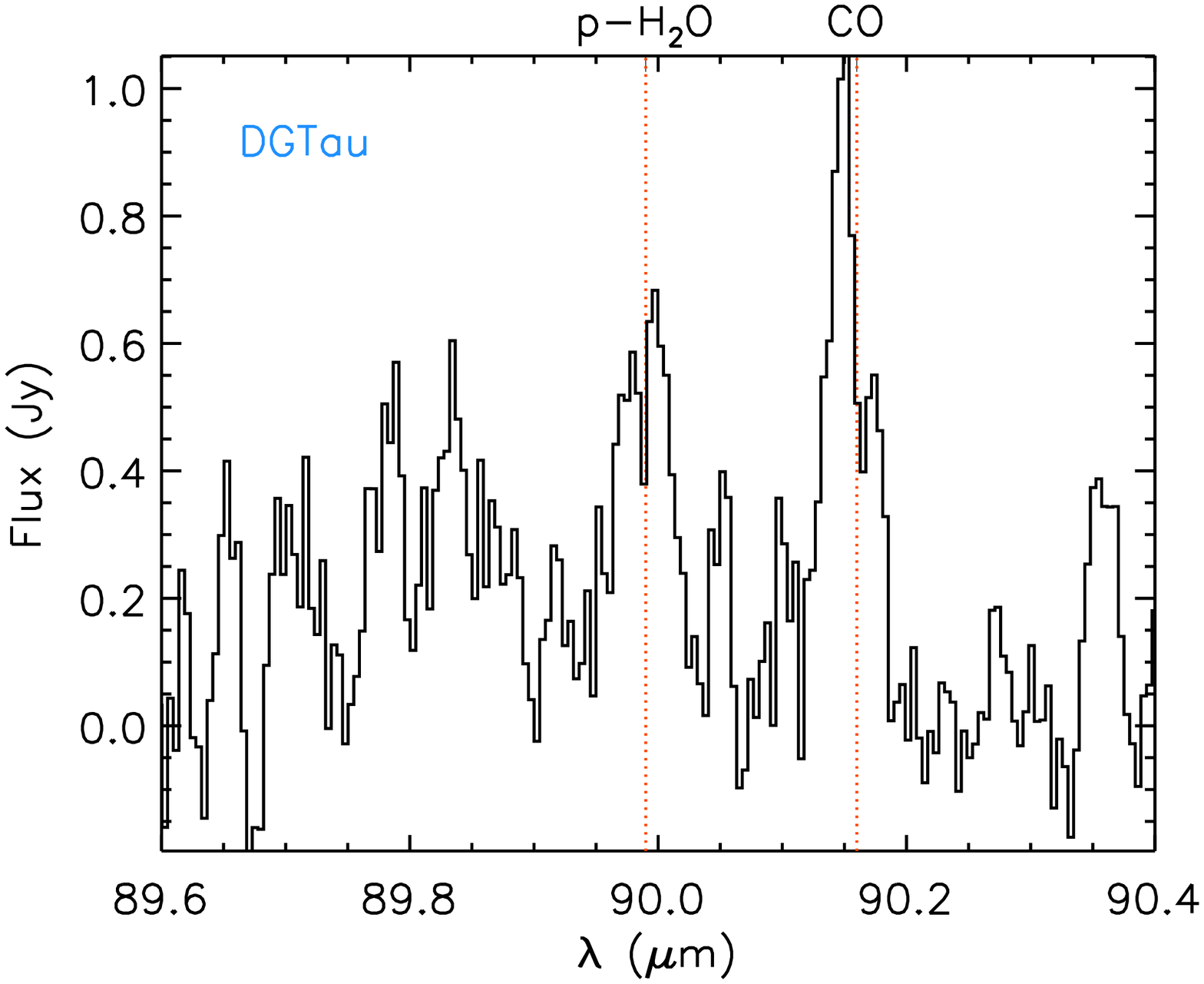}\includegraphics[width=0.16\textwidth, trim= 0mm 0mm 0mm 0mm, angle=90]{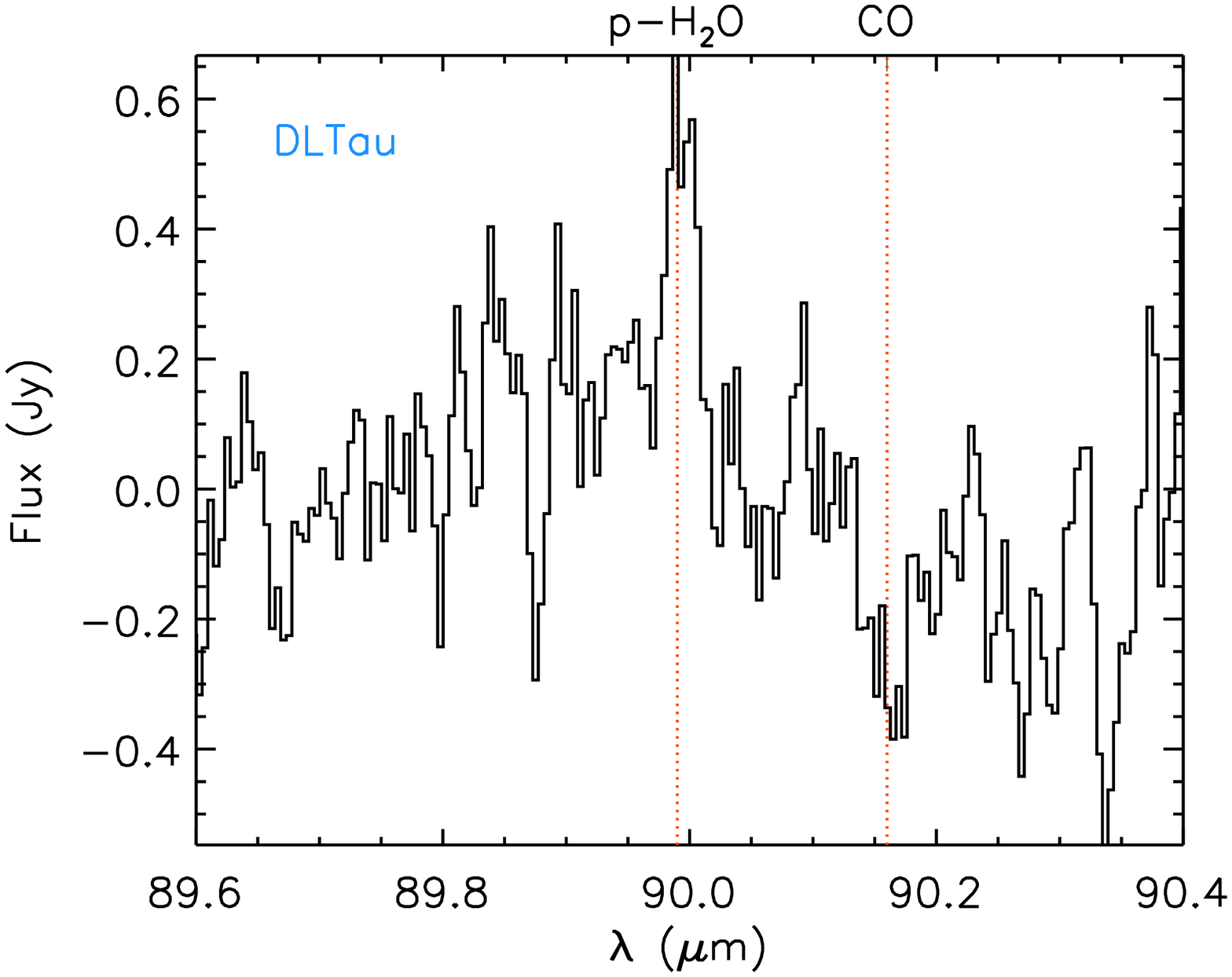}\includegraphics[width=0.16\textwidth, trim= 0mm 0mm 0mm 0mm, angle=90]{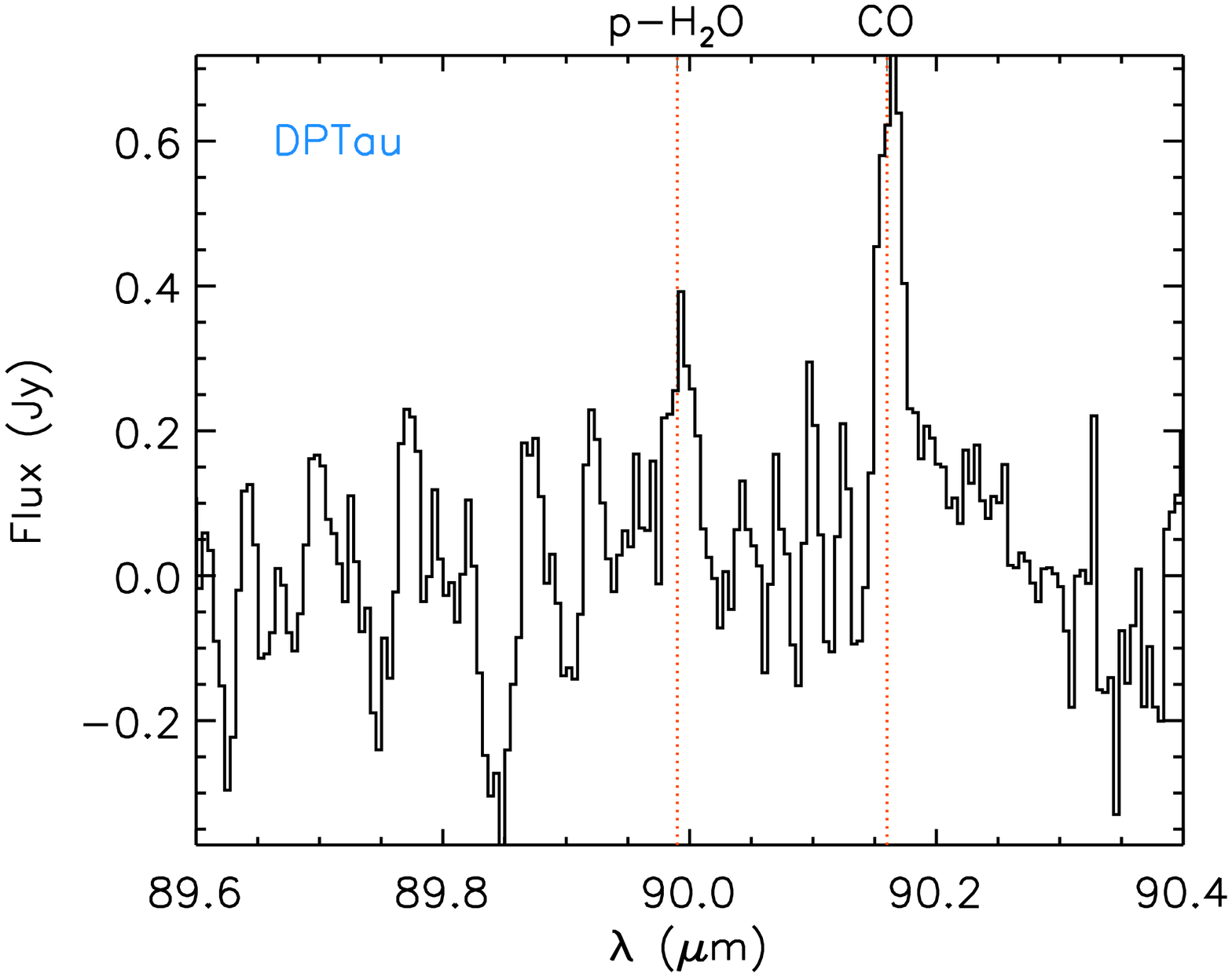}

\includegraphics[width=0.16\textwidth, trim= 0mm 0mm 0mm 0mm, angle=90]{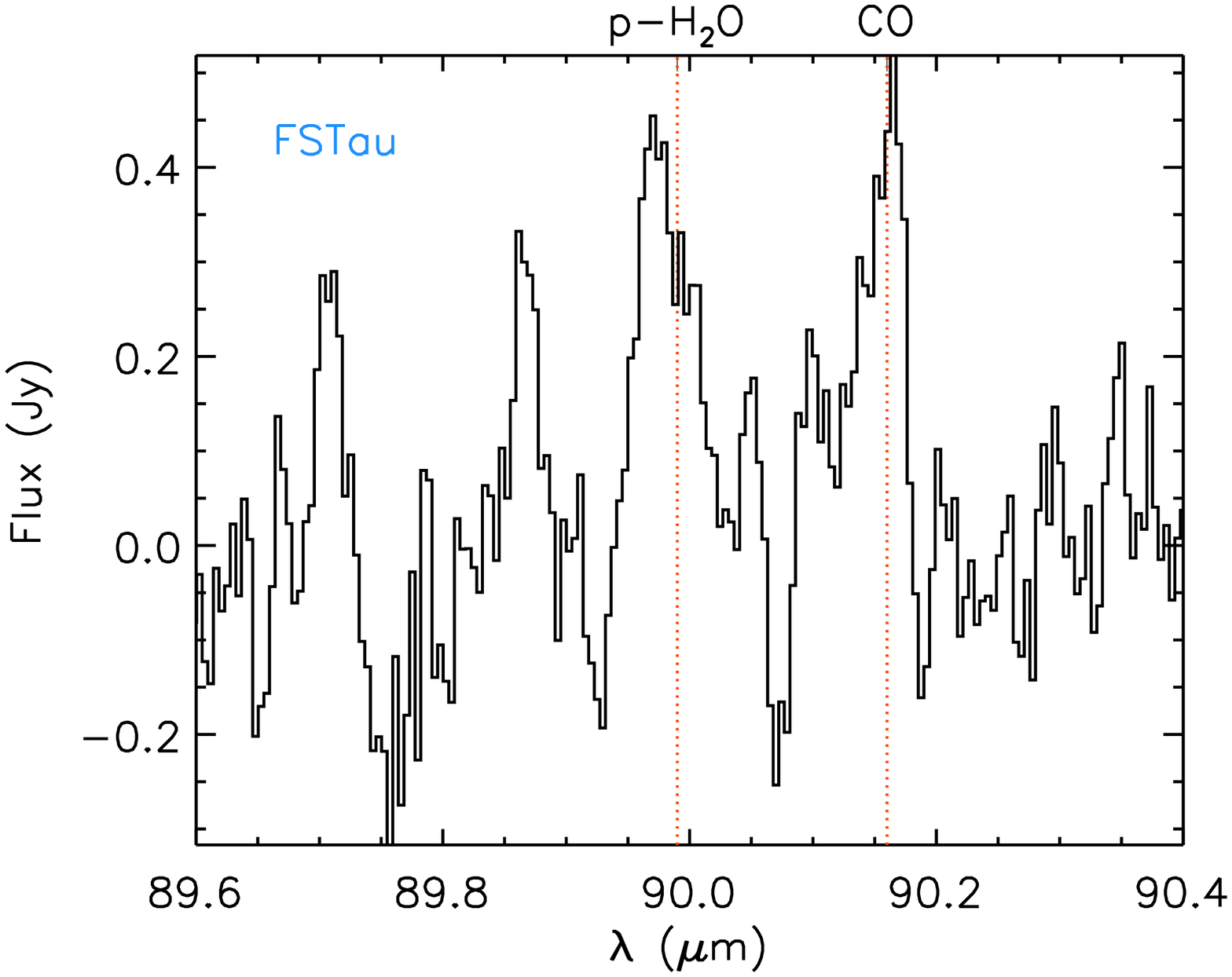}\includegraphics[width=0.16\textwidth, trim= 0mm 0mm 0mm 0mm, angle=90]{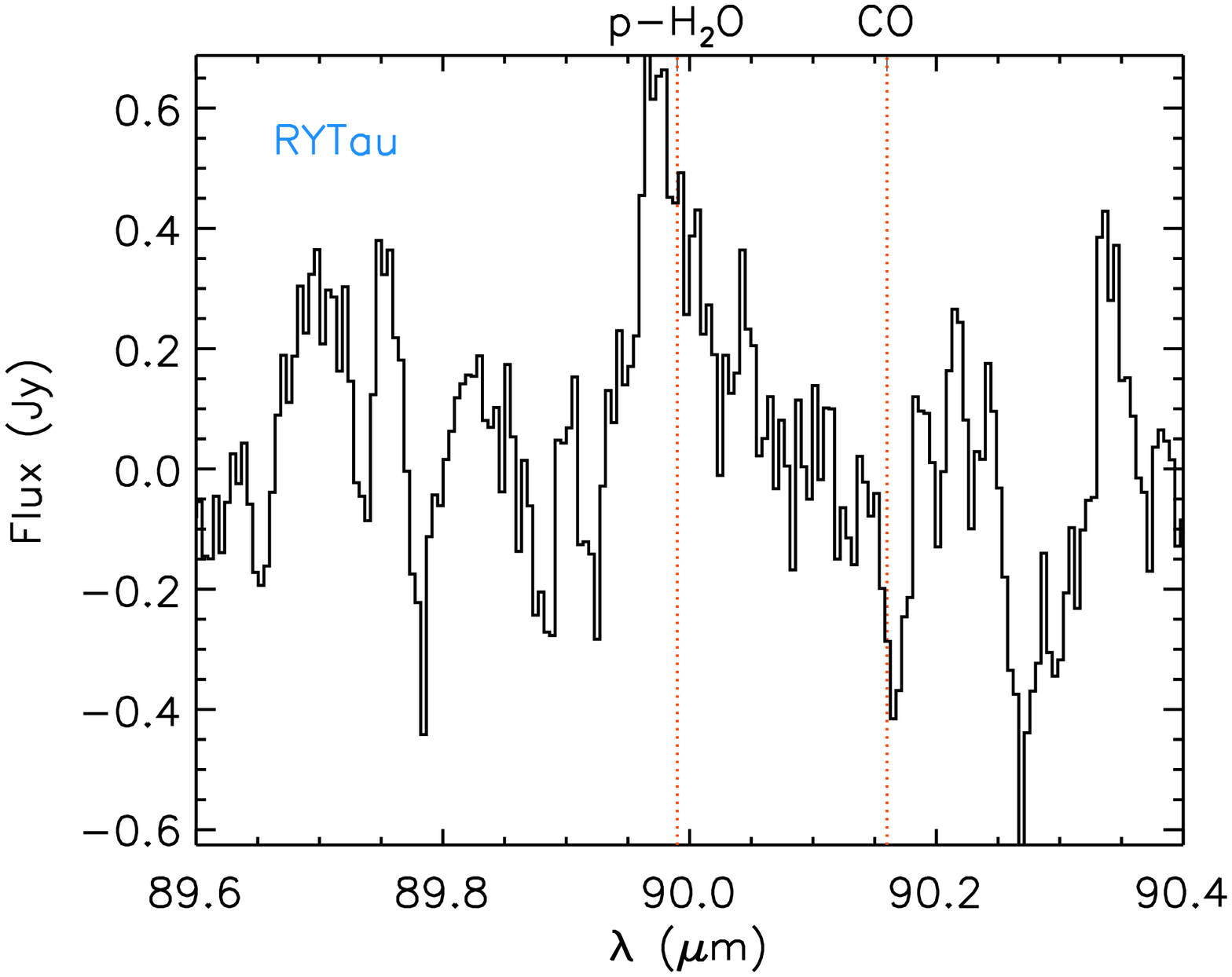}\includegraphics[width=0.16\textwidth, trim= 0mm 0mm 0mm 0mm, angle=90]{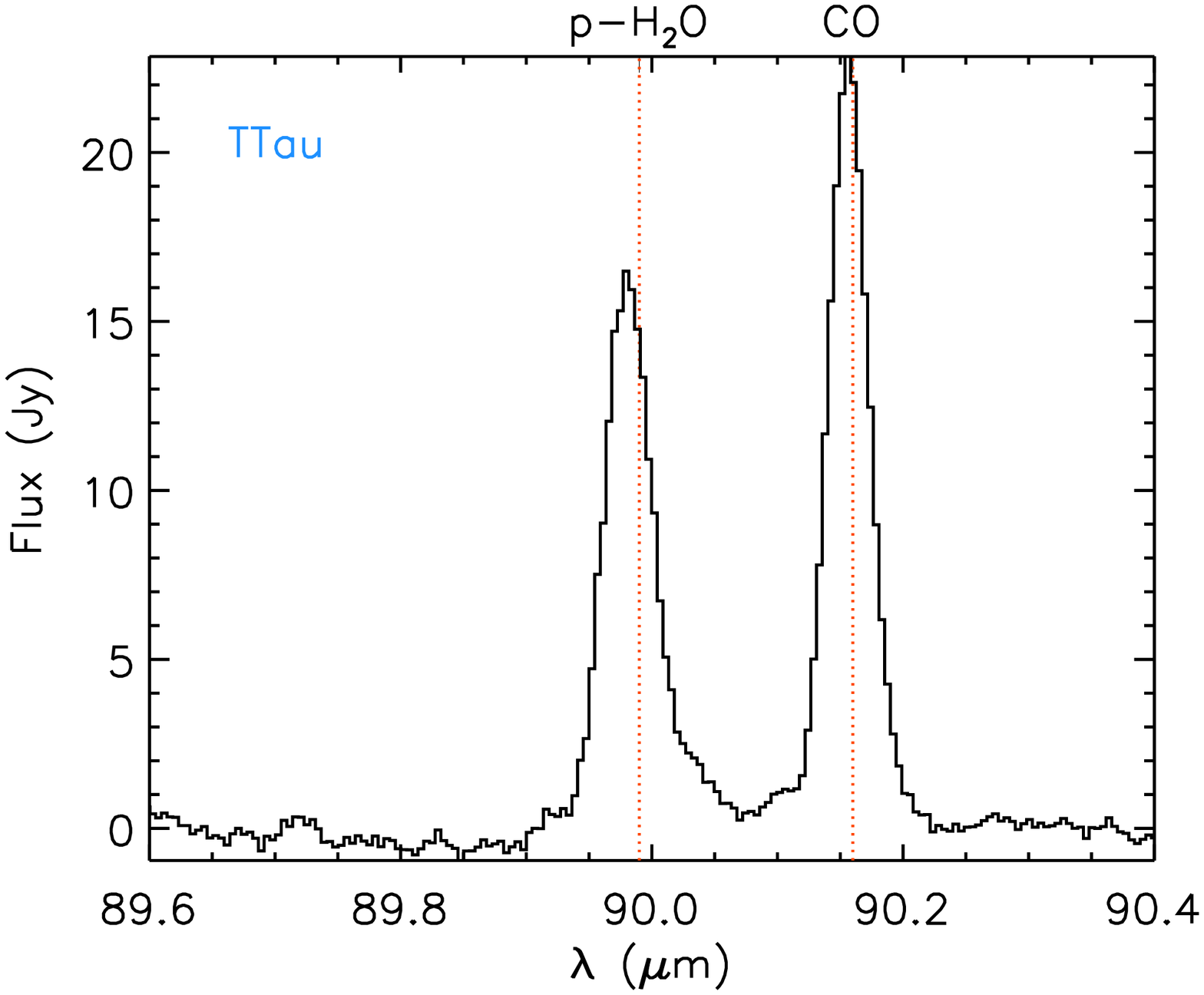}\includegraphics[width=0.16\textwidth, trim= 0mm 0mm 0mm 0mm, angle=90]{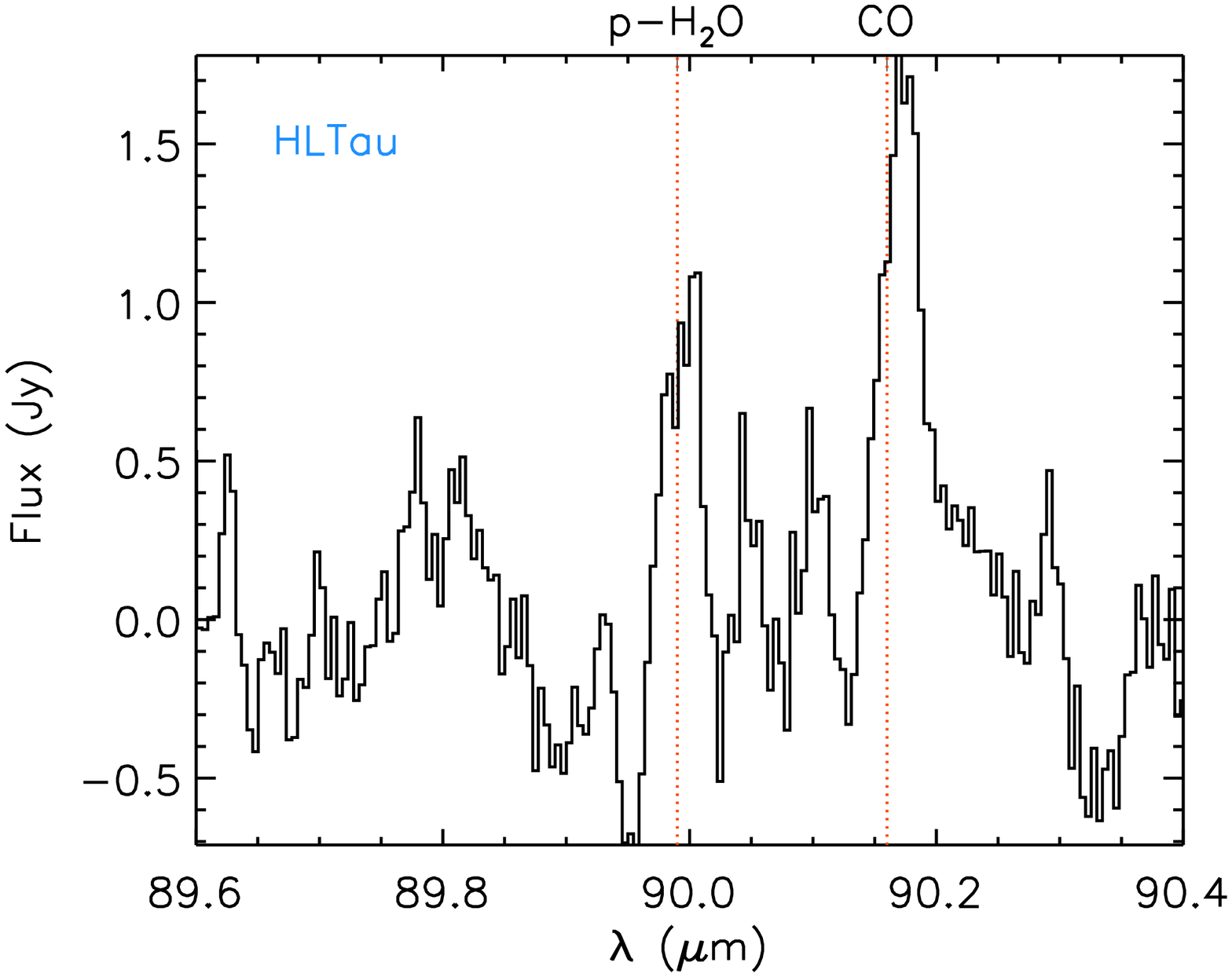}

\includegraphics[width=0.16\textwidth, trim= 0mm 0mm 0mm 0mm, angle=90]{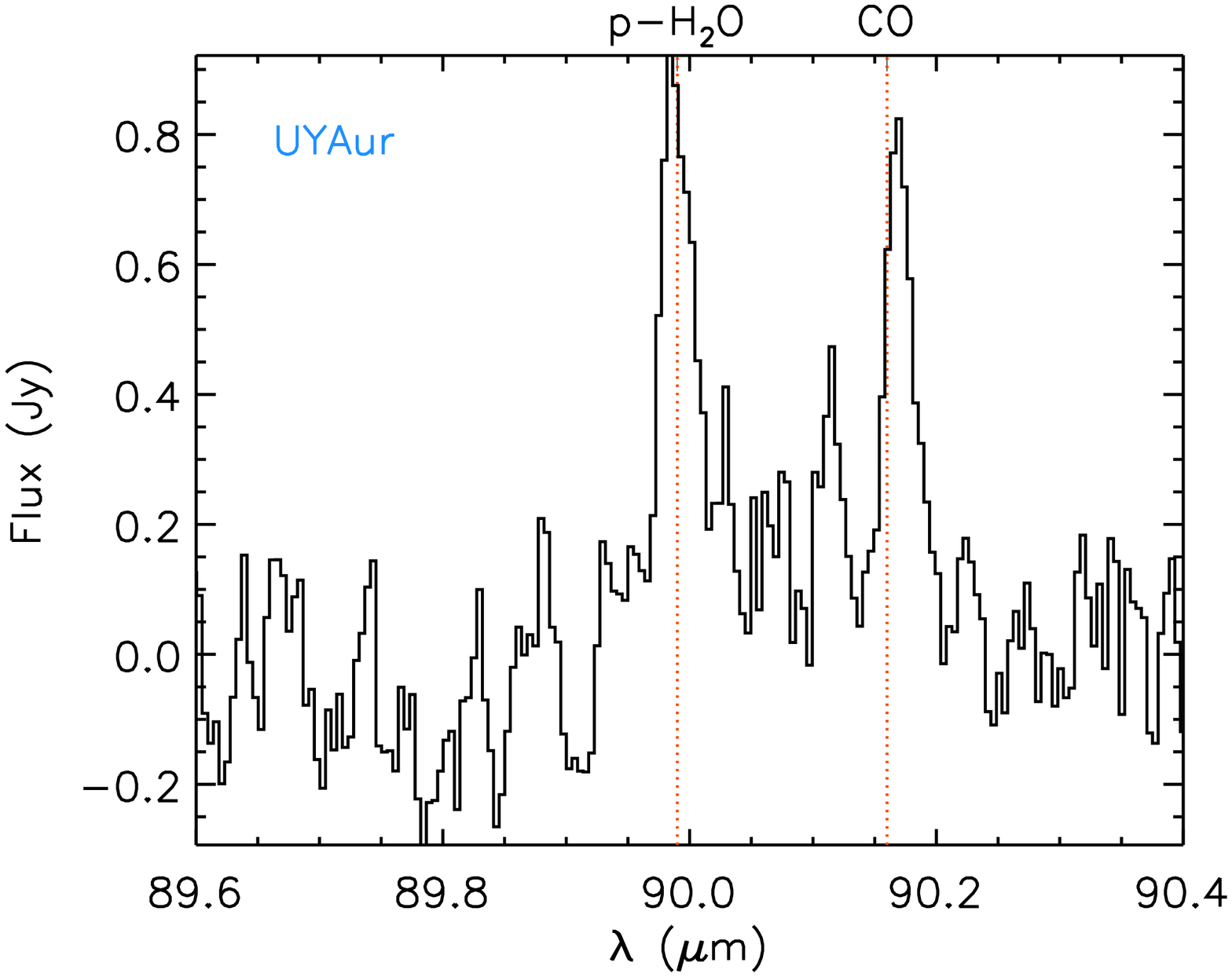}\includegraphics[width=0.16\textwidth, trim= 0mm 0mm 0mm 0mm, angle=90]{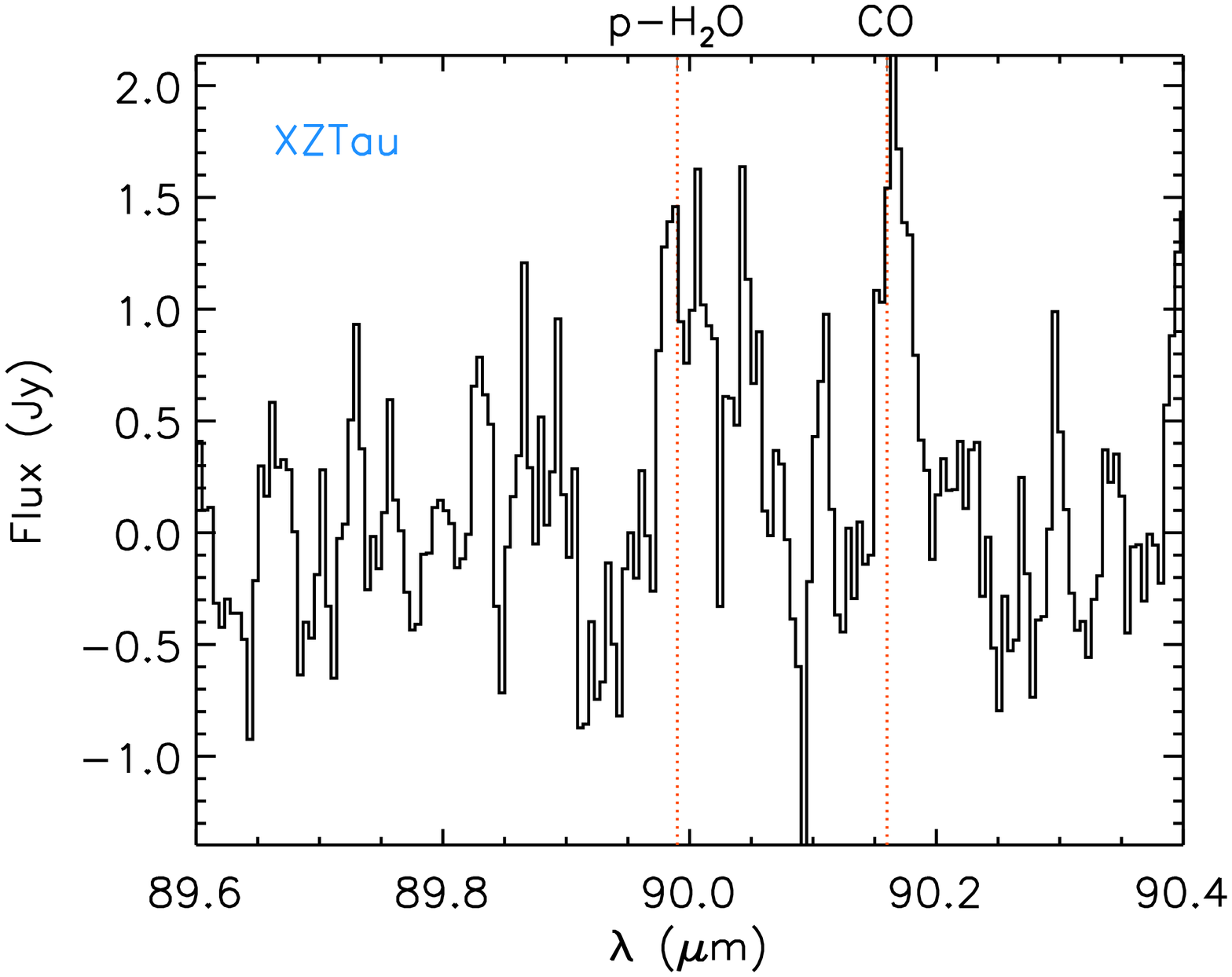}

\caption{Continuum subtracted spectra at 90 $\rm \mu m$ for all objects with detections. The red vertical lines indicate the positions of p-H$_{\rm 2}$O 89.99 $\rm \mu m$ and  CO 90.16 $\rm \mu m$.}
\end{figure*}

\begin{figure*}[htpb]
\centering
\setcounter{figure}{4}
\includegraphics[width=0.16\textwidth, trim= 0mm 0mm 0mm 0mm, angle=90]{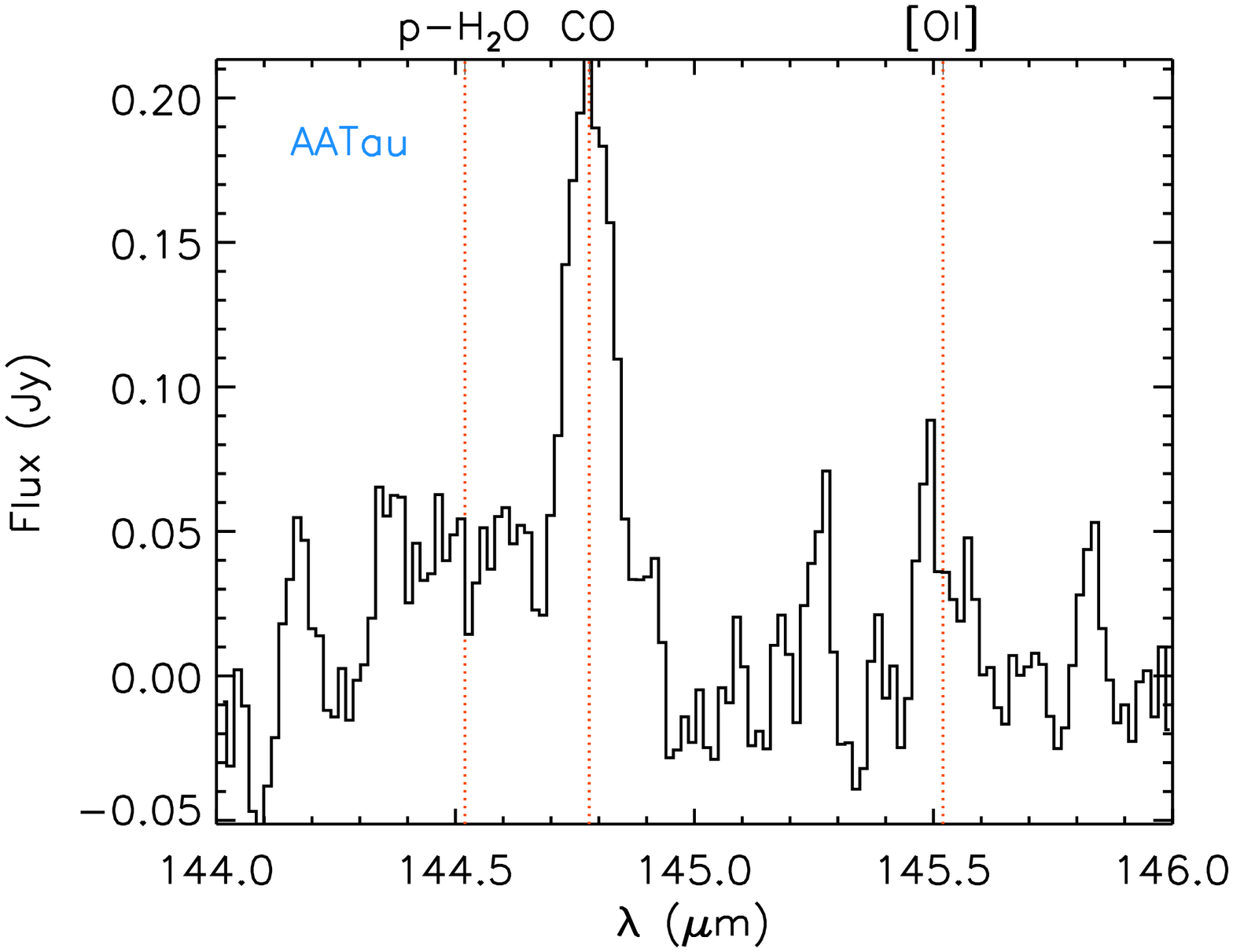}\includegraphics[width=0.16\textwidth, trim= 0mm 0mm 0mm 0mm, angle=90]{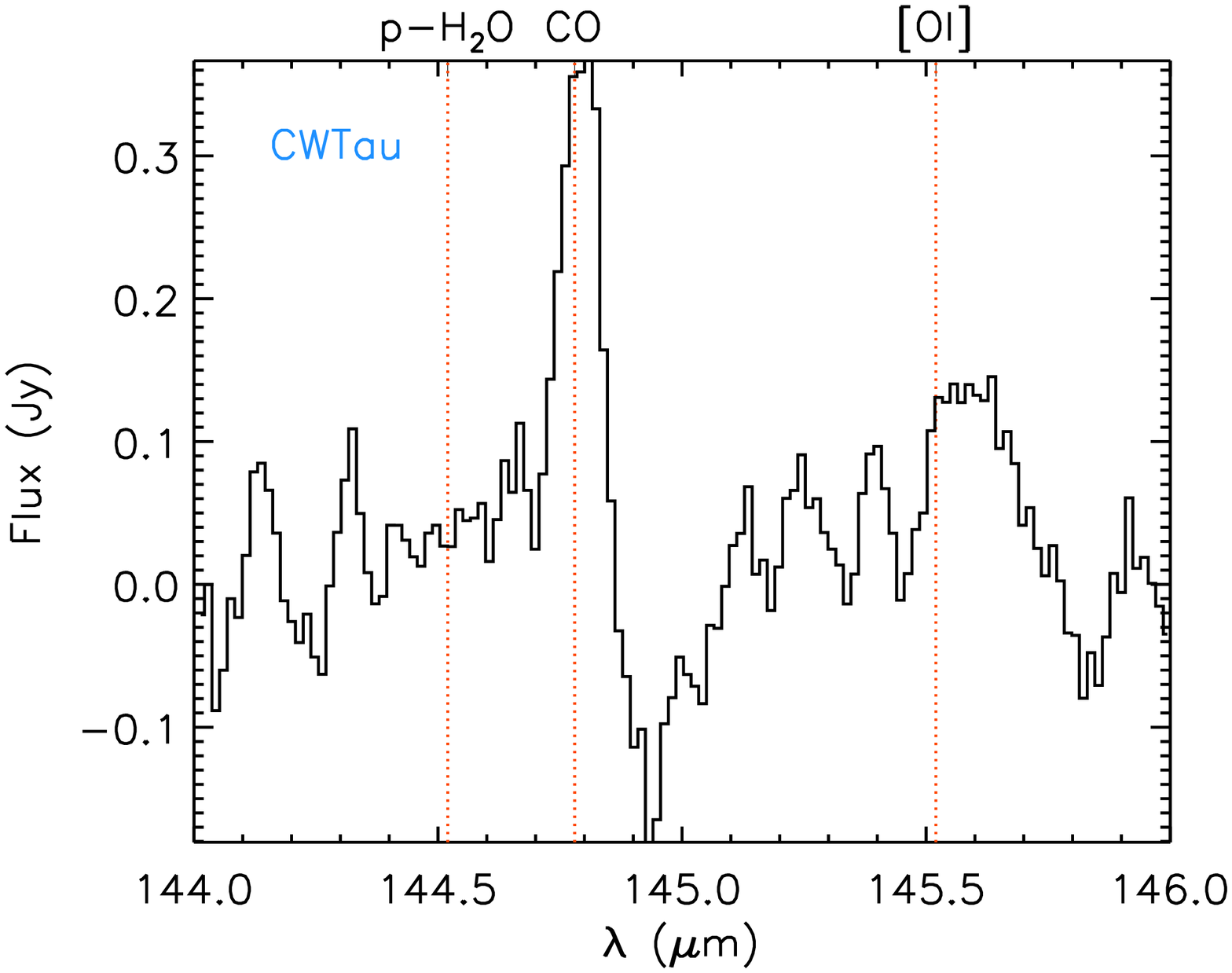}\includegraphics[width=0.16\textwidth, trim= 0mm 0mm 0mm 0mm, angle=90]{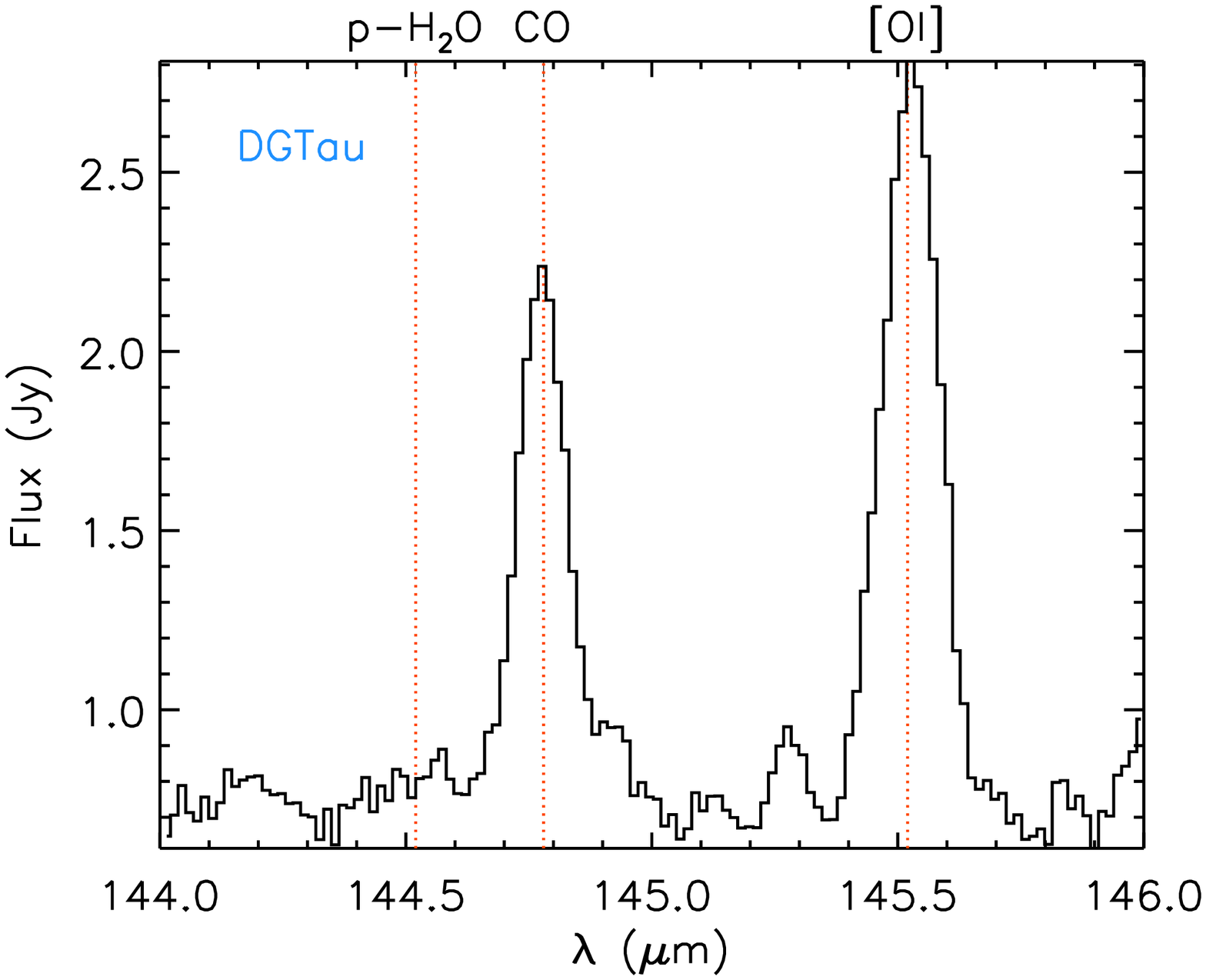}\includegraphics[width=0.16\textwidth, trim= 0mm 0mm 0mm 0mm, angle=90]{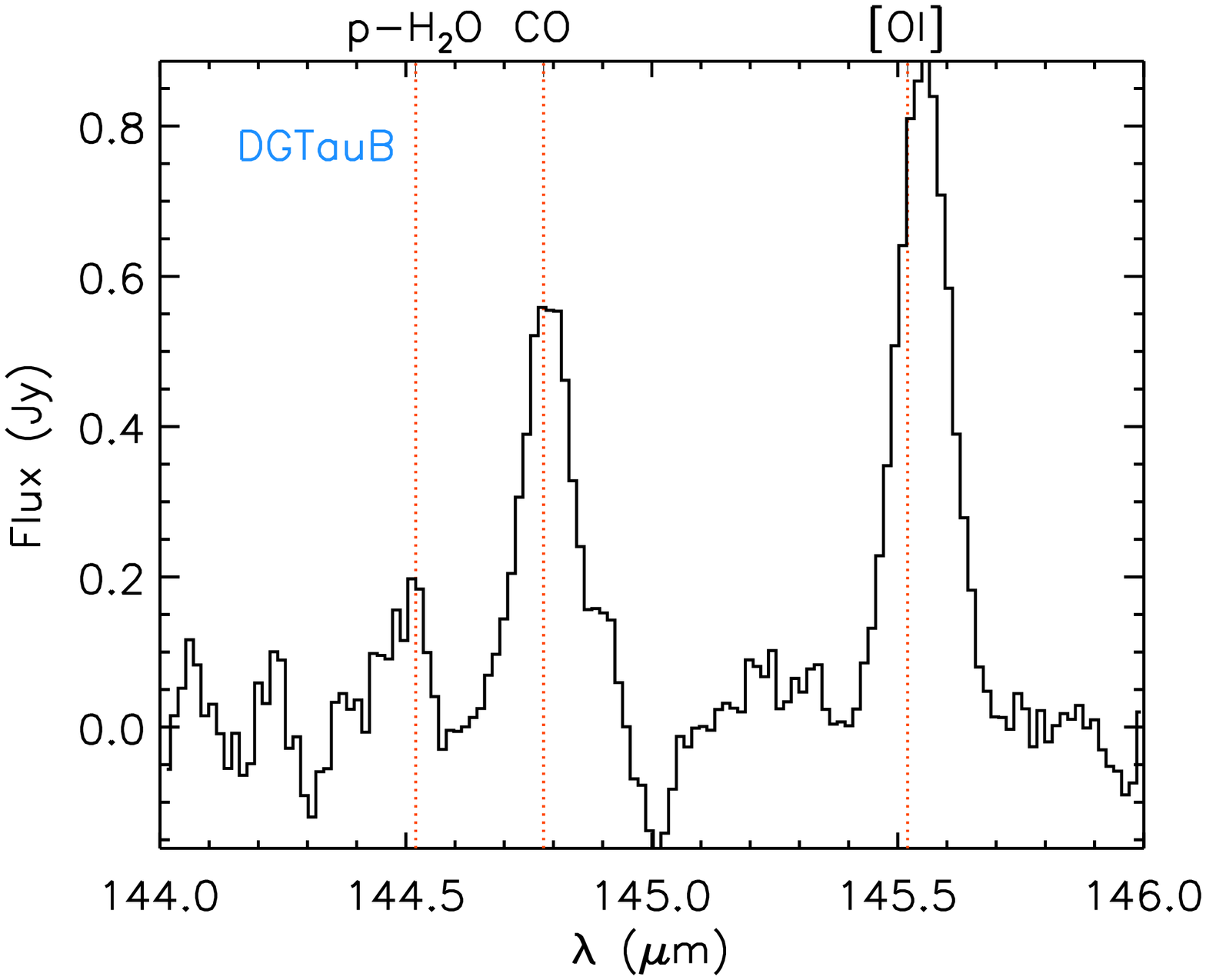}

\includegraphics[width=0.16\textwidth, trim= 0mm 0mm 0mm 0mm, angle=90]{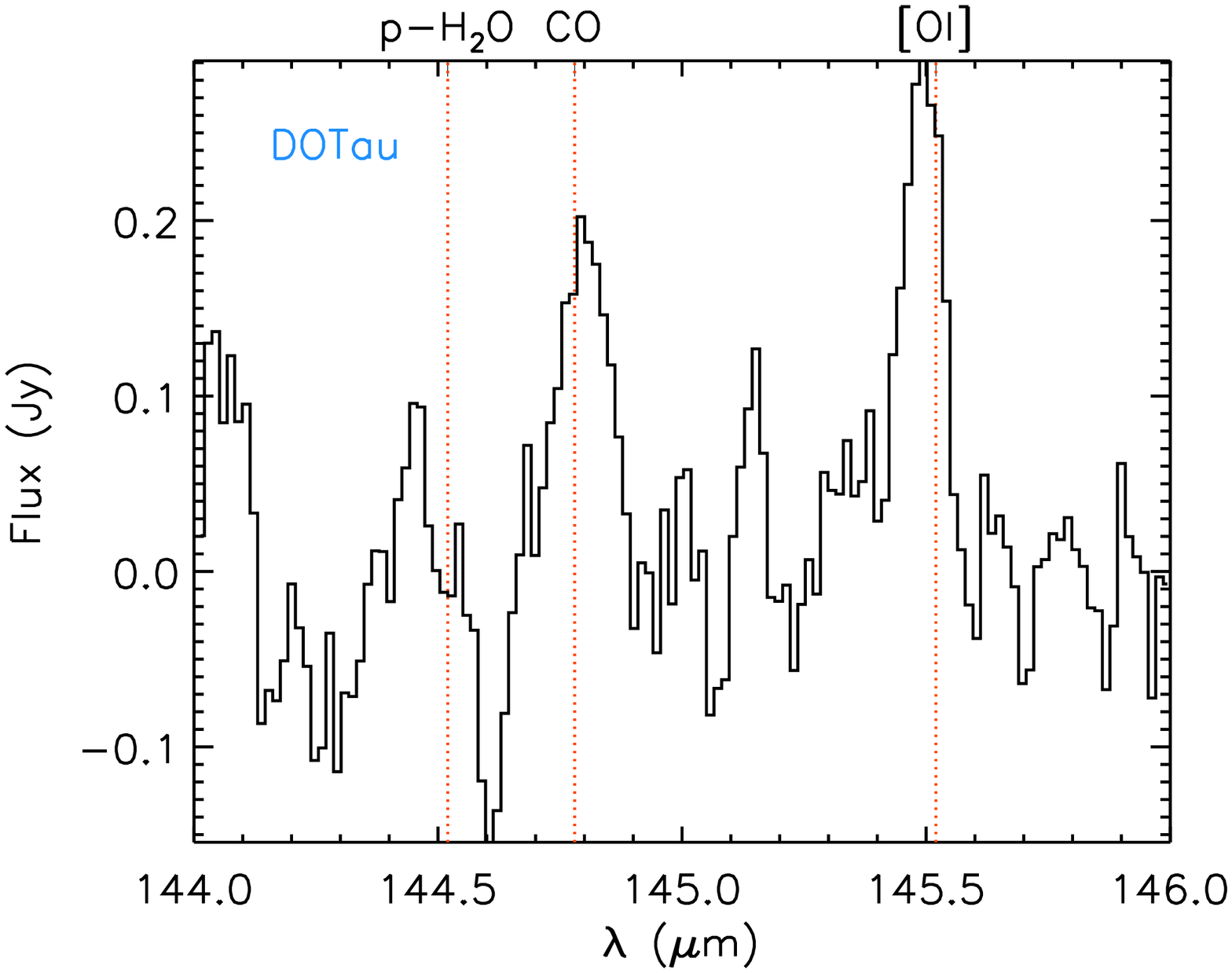}\includegraphics[width=0.16\textwidth, trim= 0mm 0mm 0mm 0mm, angle=90]{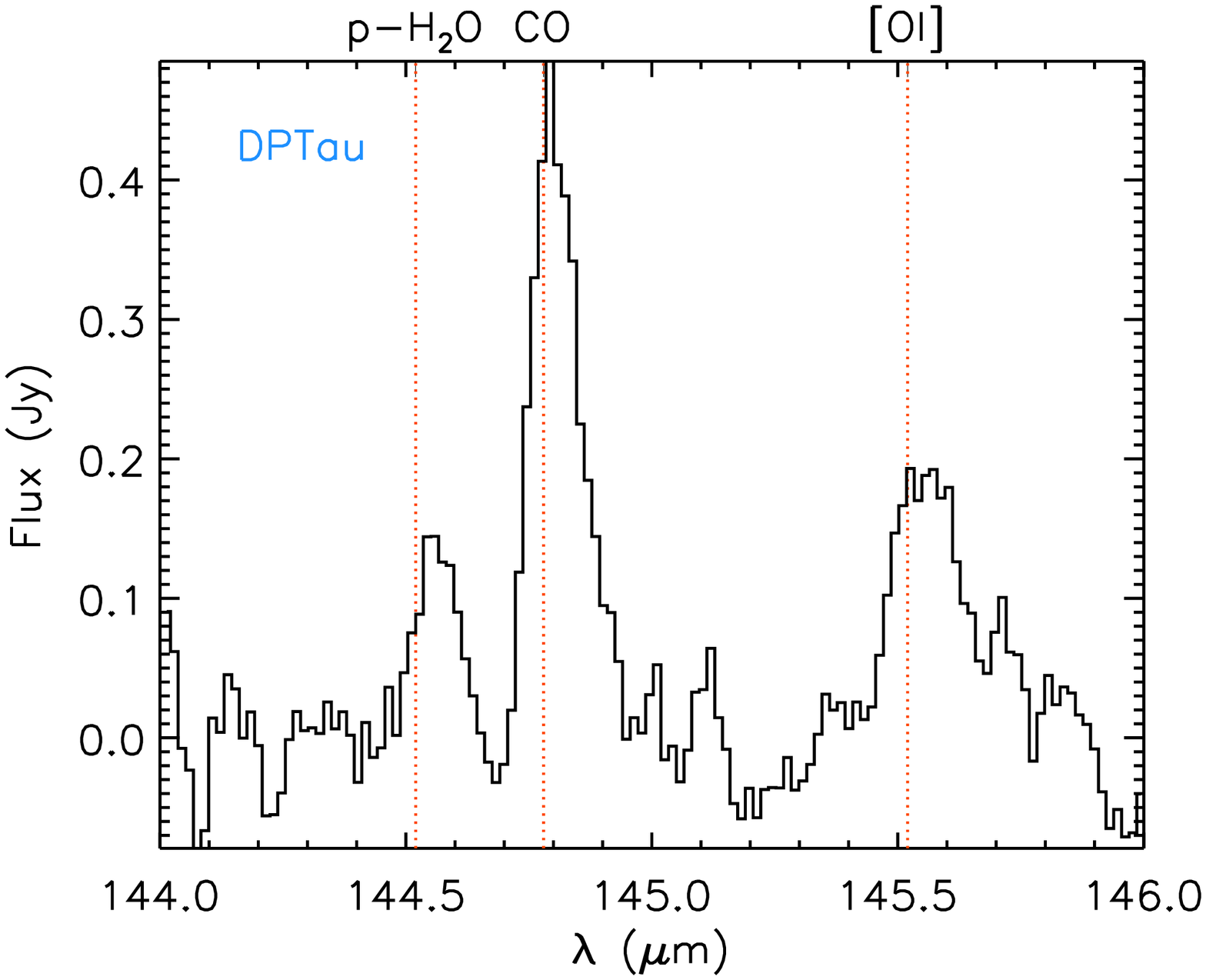}\includegraphics[width=0.16\textwidth, trim= 0mm 0mm 0mm 0mm, angle=90]{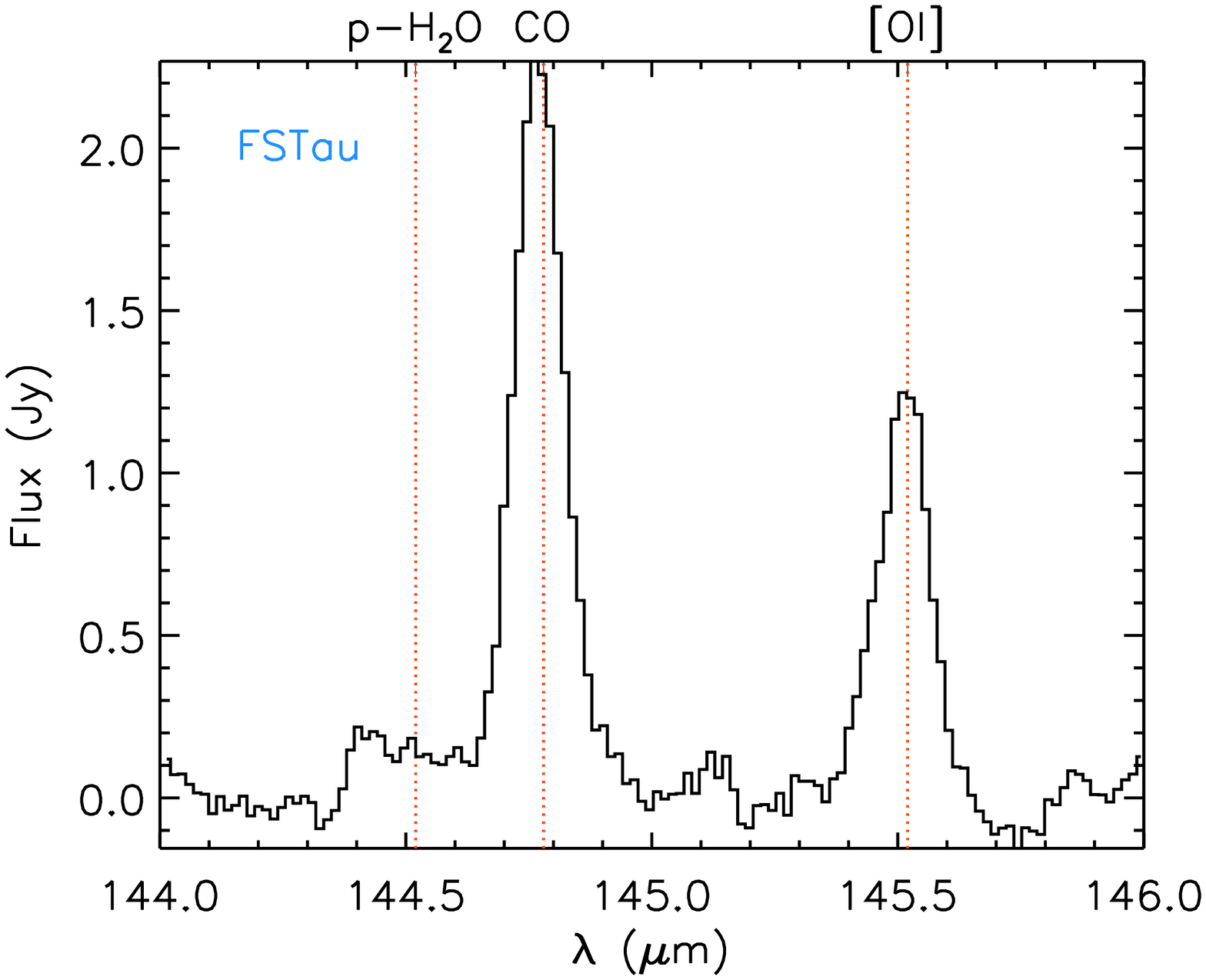}\includegraphics[width=0.16\textwidth, trim= 0mm 0mm 0mm 0mm, angle=90]{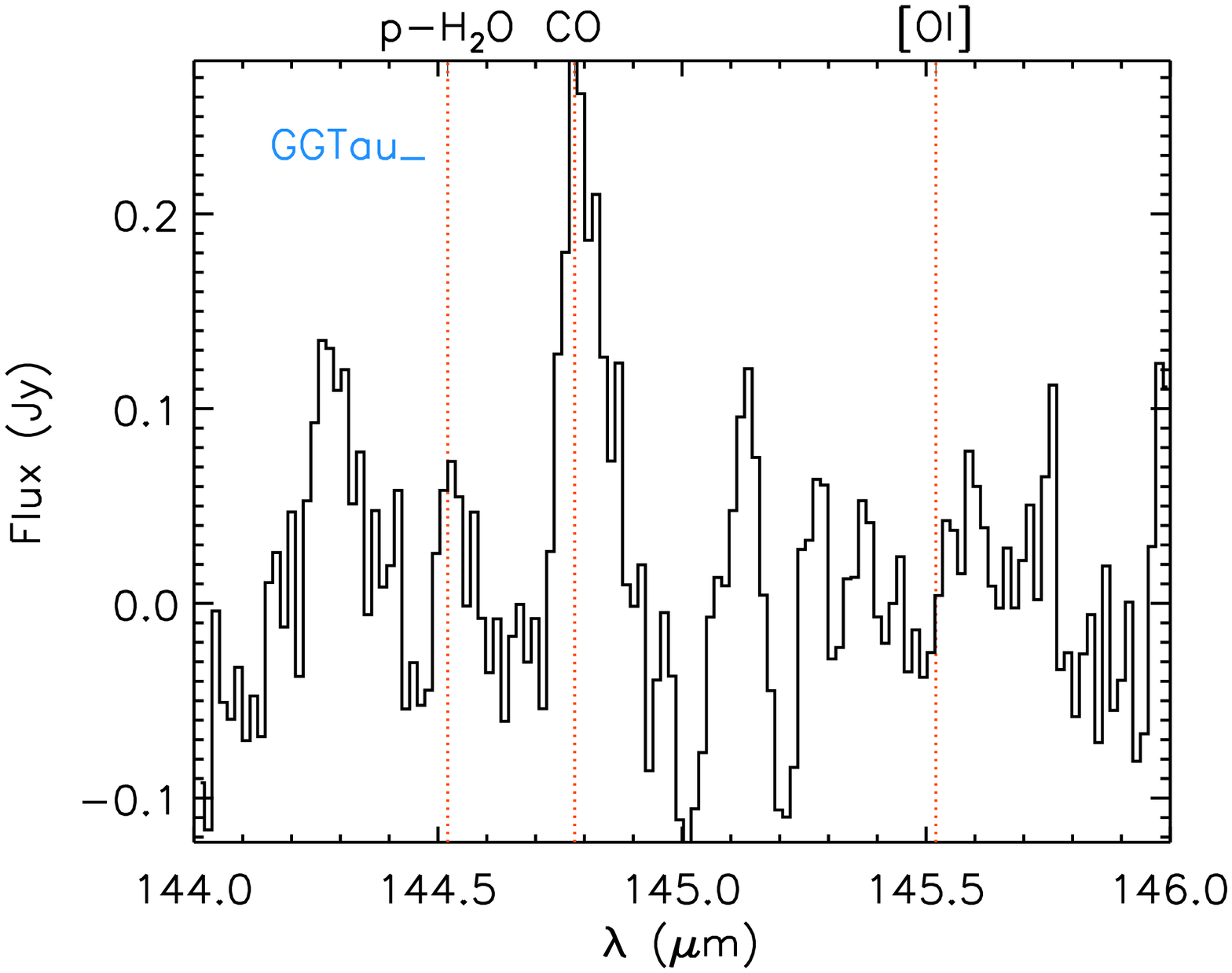}

\includegraphics[width=0.16\textwidth, trim= 0mm 0mm 0mm 0mm, angle=90]{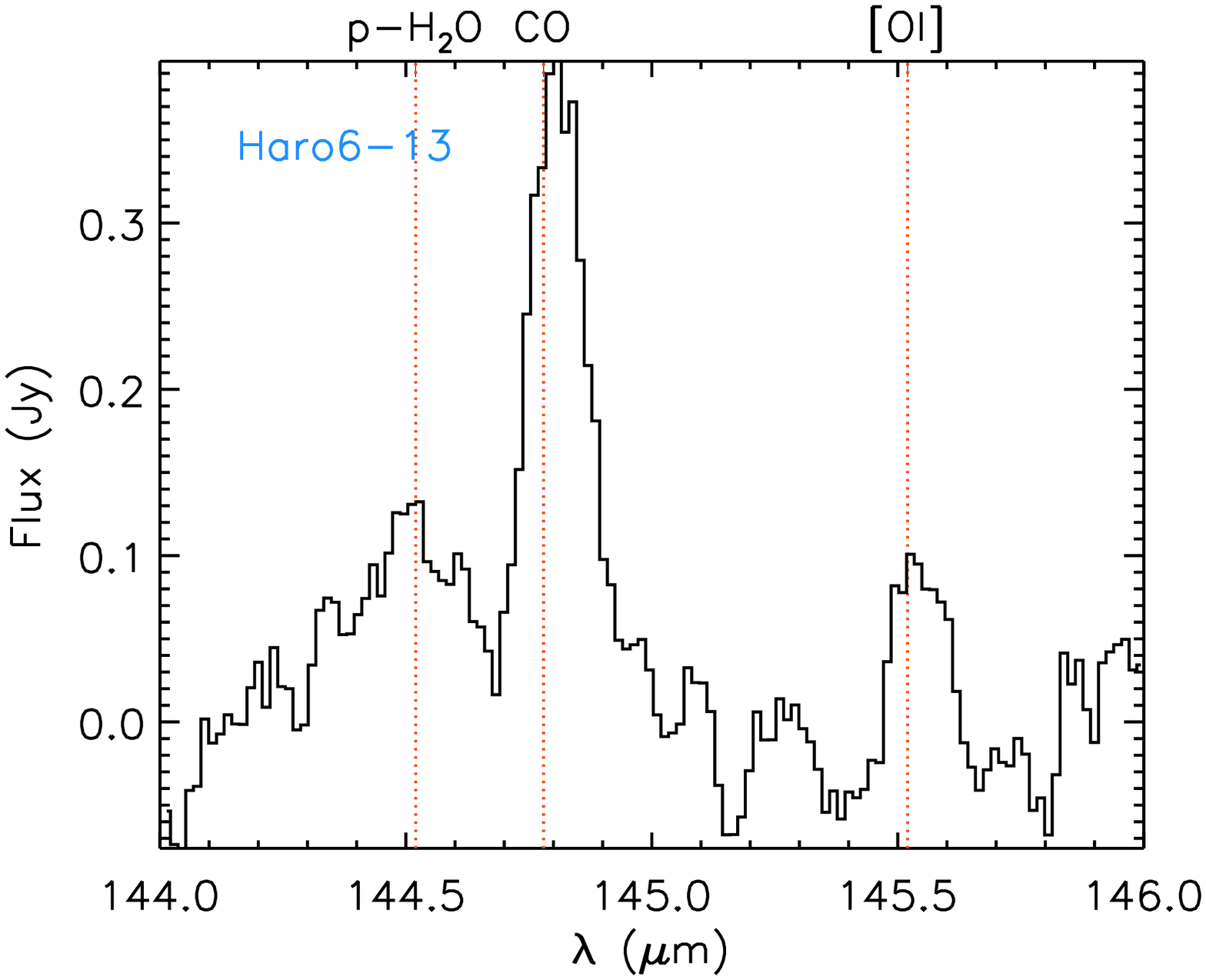}\includegraphics[width=0.16\textwidth, trim= 0mm 0mm 0mm 0mm, angle=90]{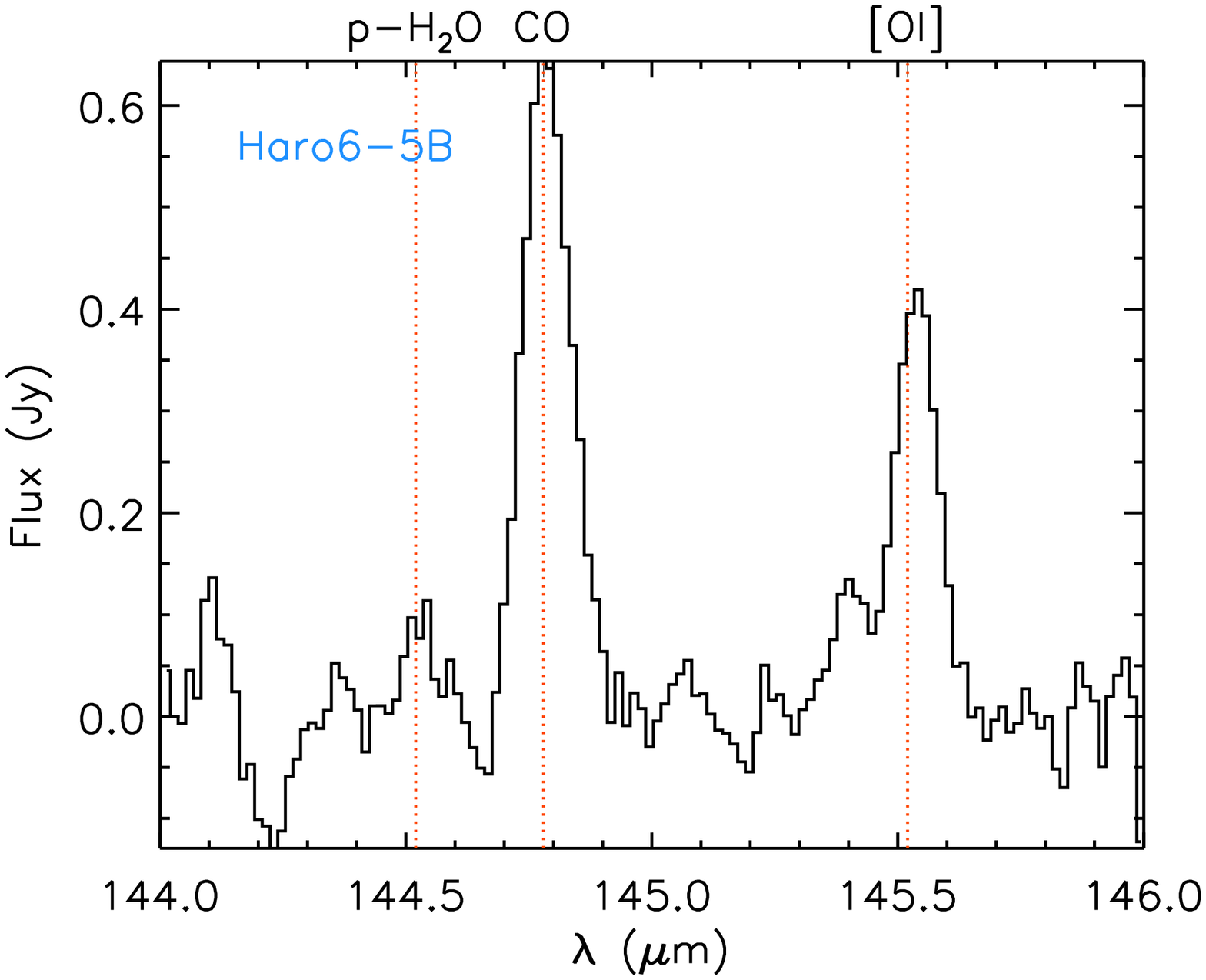}\includegraphics[width=0.16\textwidth, trim= 0mm 0mm 0mm 0mm, angle=90]{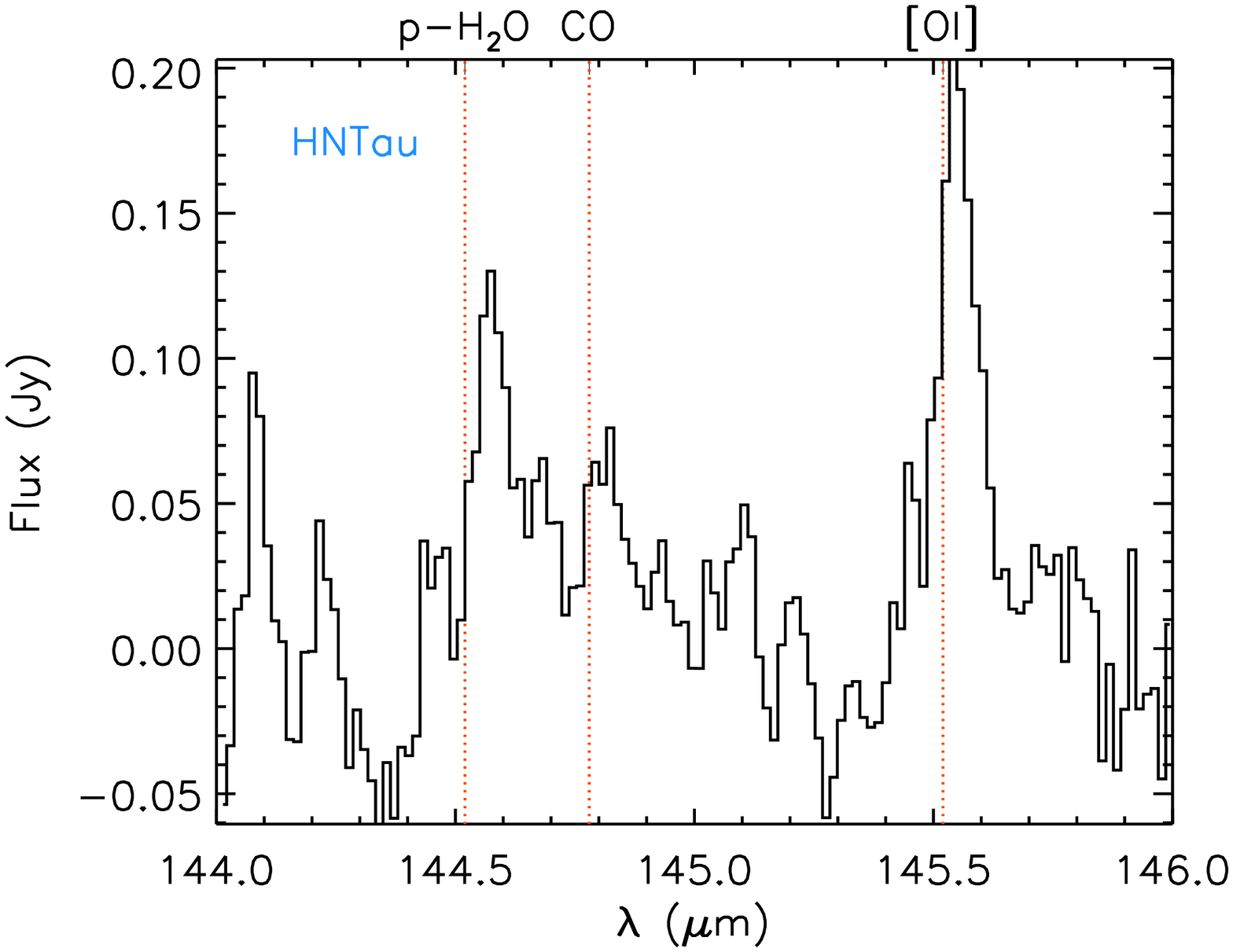}\includegraphics[width=0.16\textwidth, trim= 0mm 0mm 0mm 0mm, angle=90]{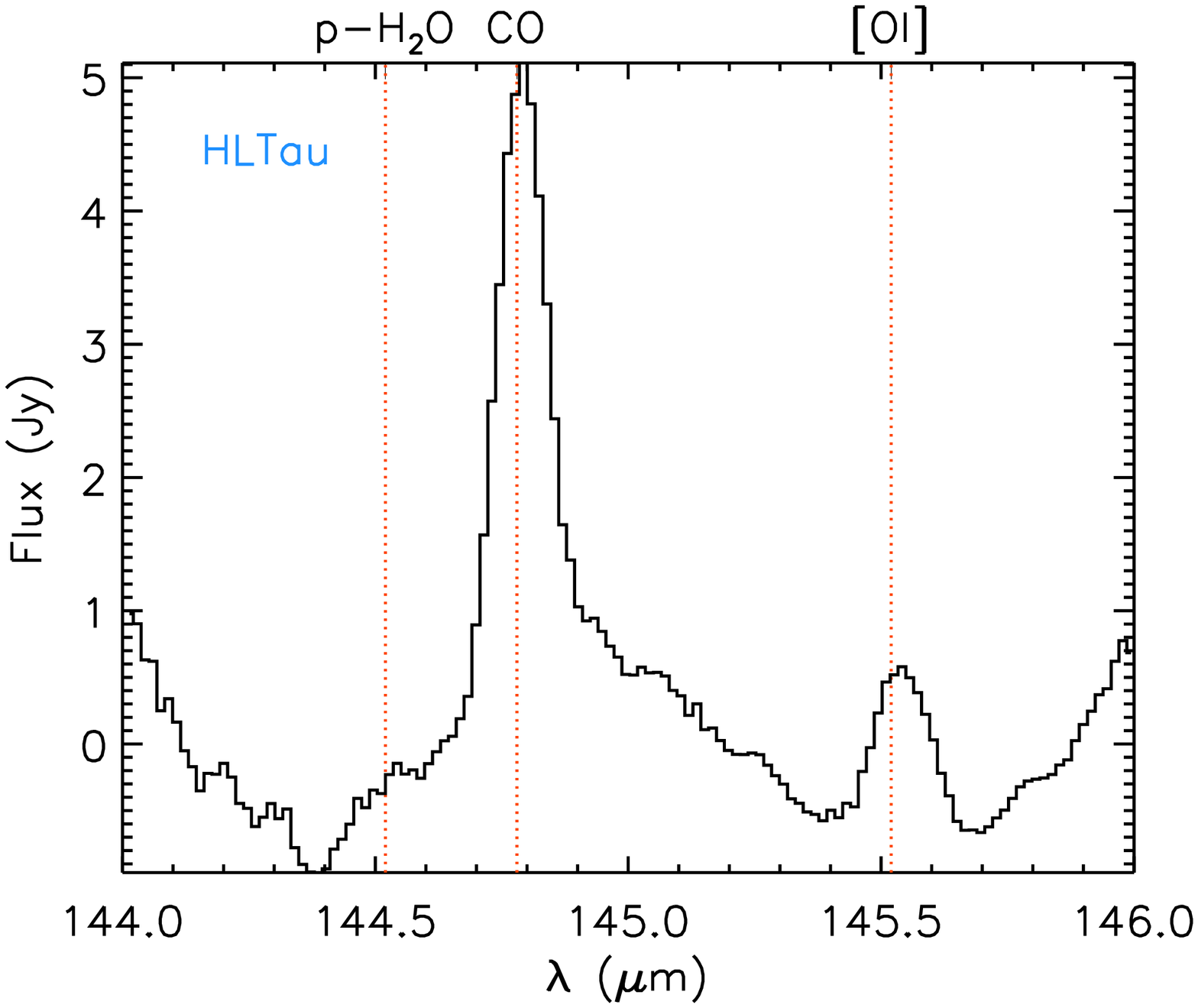}

\includegraphics[width=0.16\textwidth, trim= 0mm 0mm 0mm 0mm, angle=90]{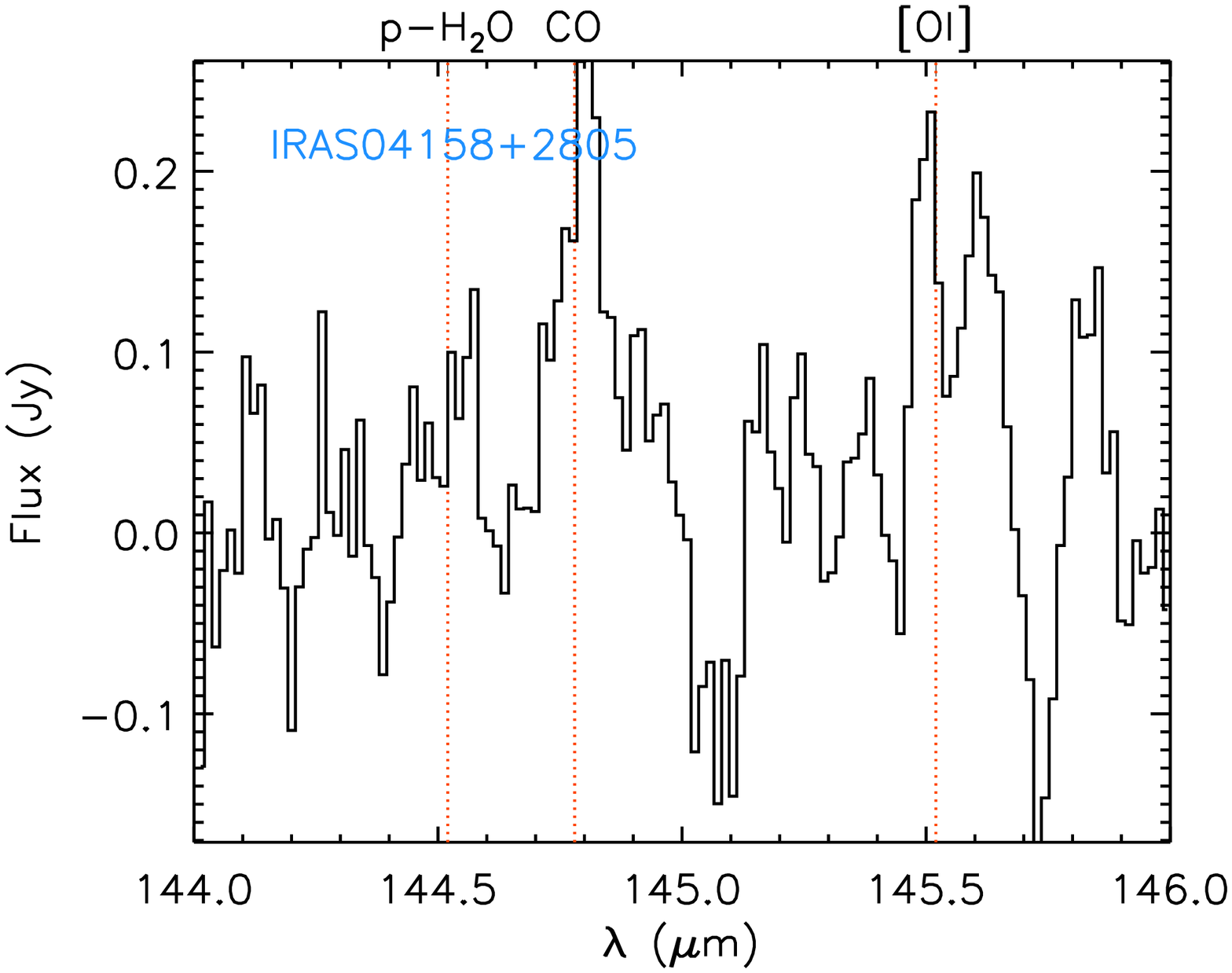}\includegraphics[width=0.16\textwidth, trim= 0mm 0mm 0mm 0mm, angle=90]{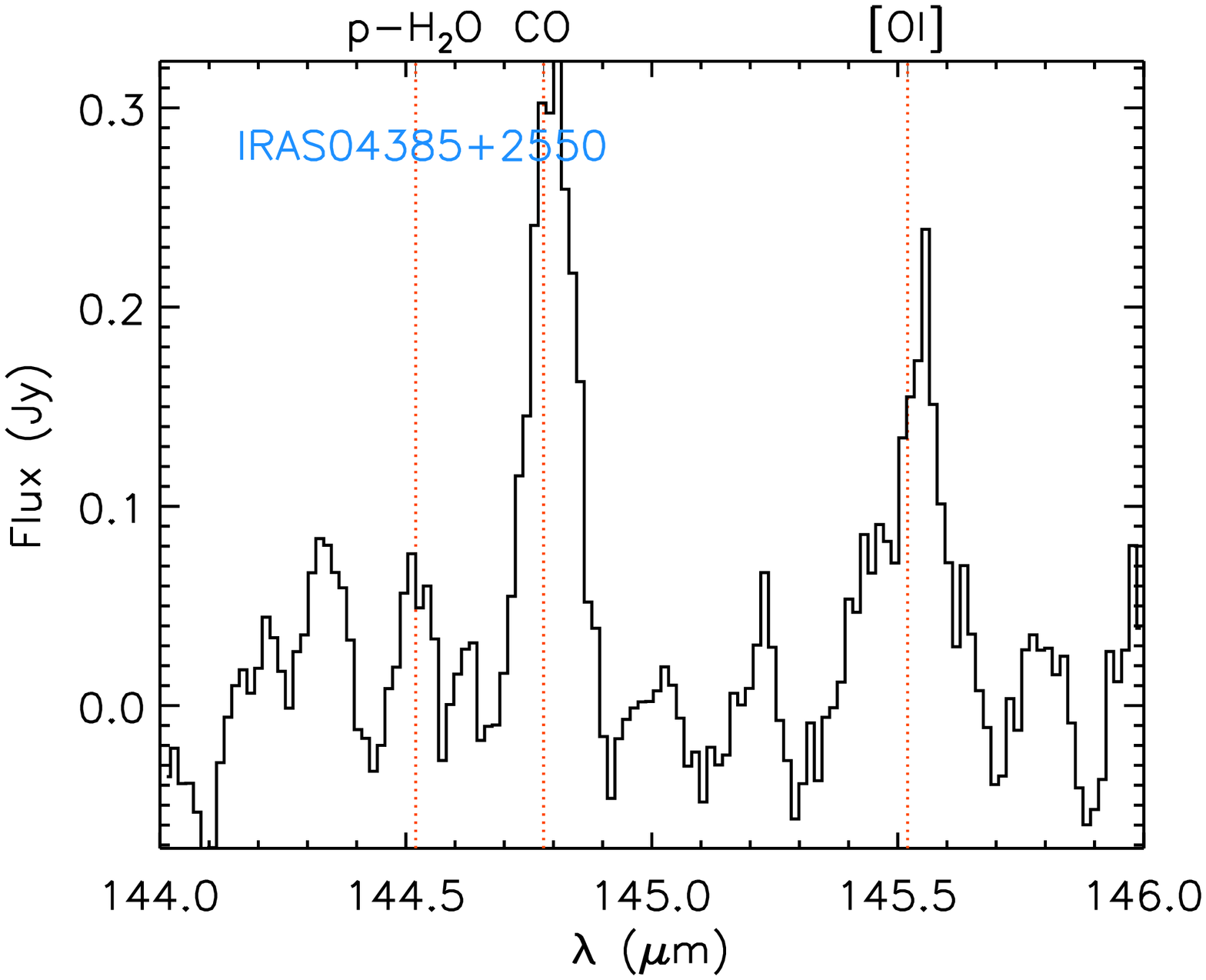}\includegraphics[width=0.16\textwidth, trim= 0mm 0mm 0mm 0mm, angle=90]{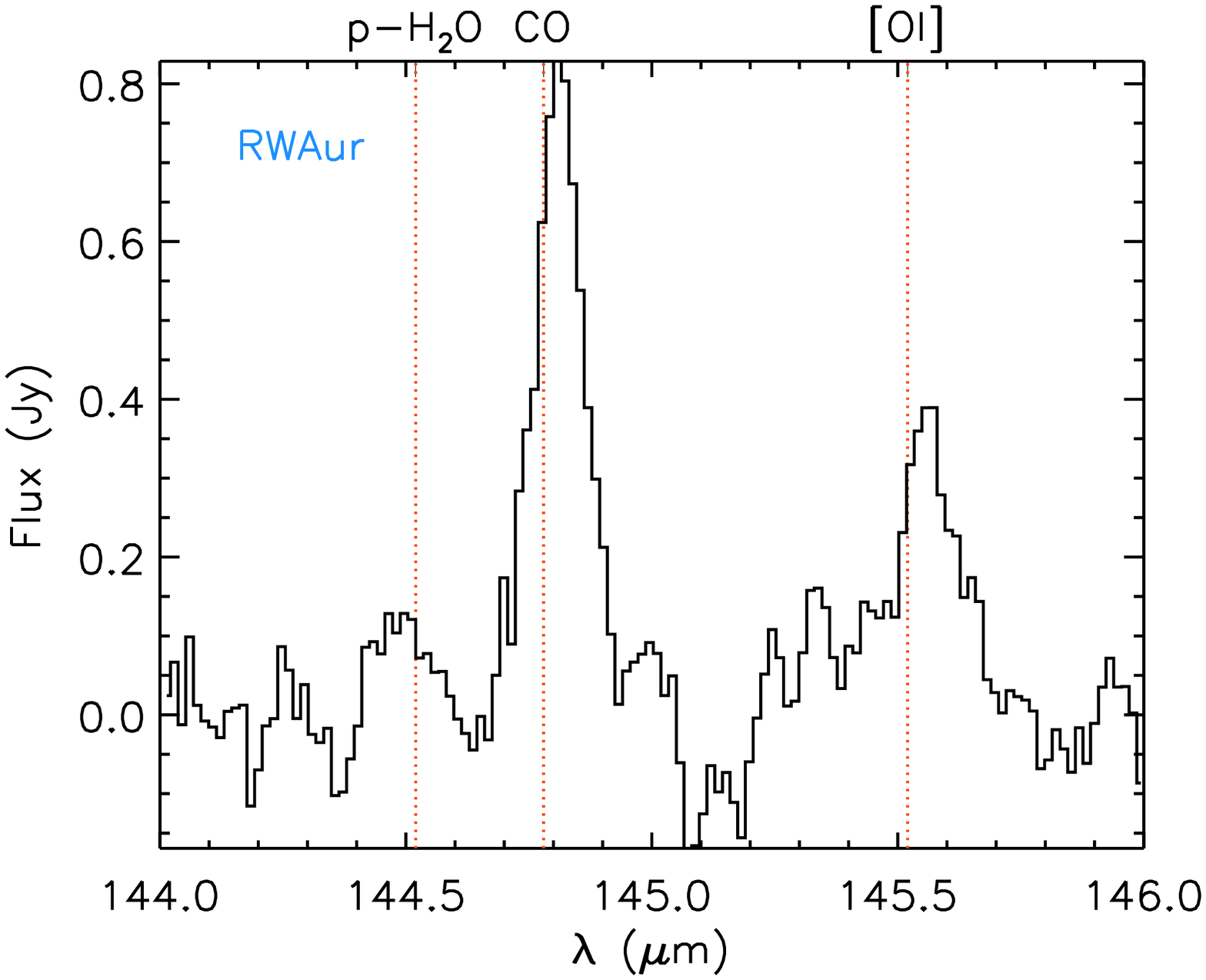}\includegraphics[width=0.16\textwidth, trim= 0mm 0mm 0mm 0mm, angle=90]{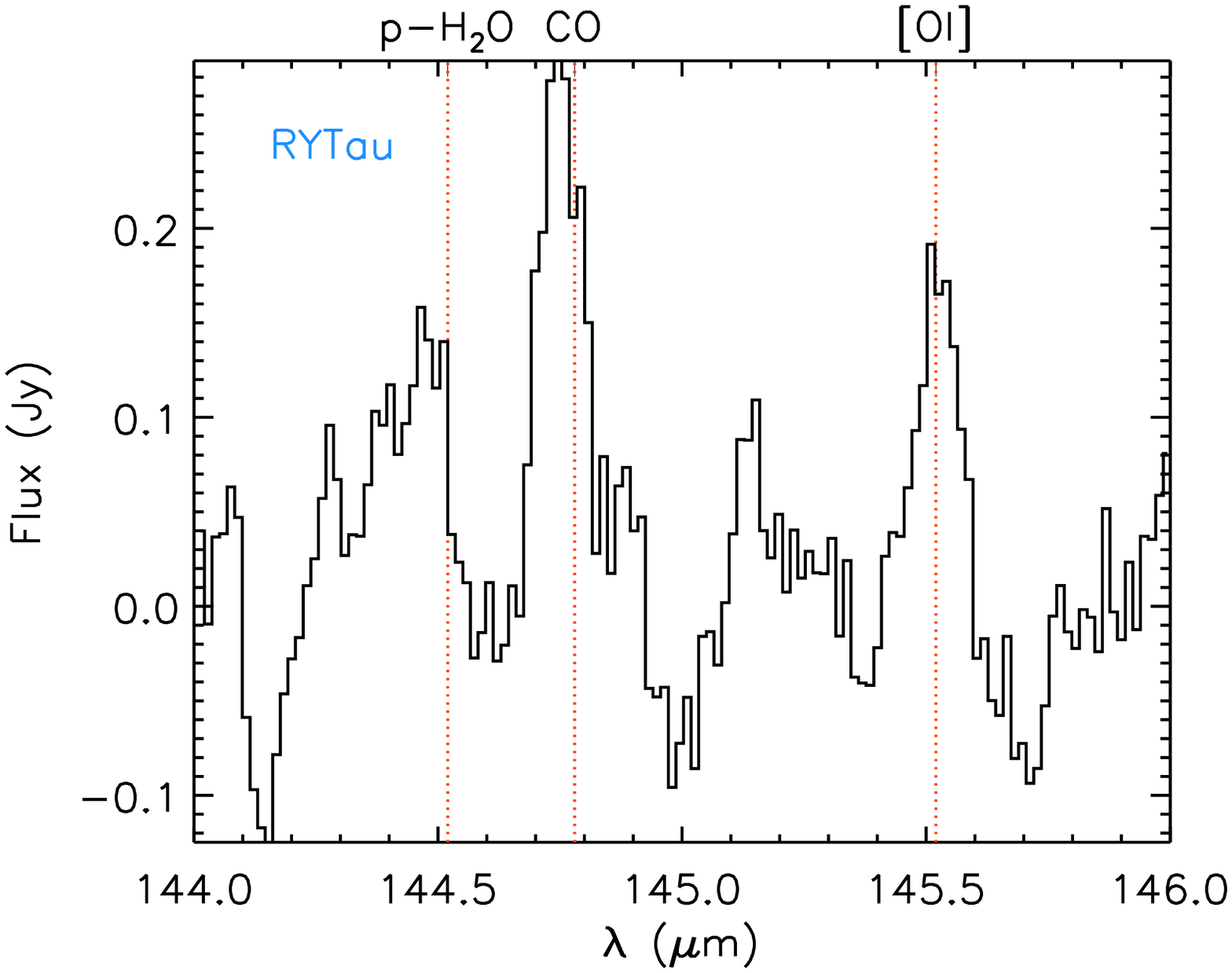}

\includegraphics[width=0.16\textwidth, trim= 0mm 0mm 0mm 0mm, angle=90]{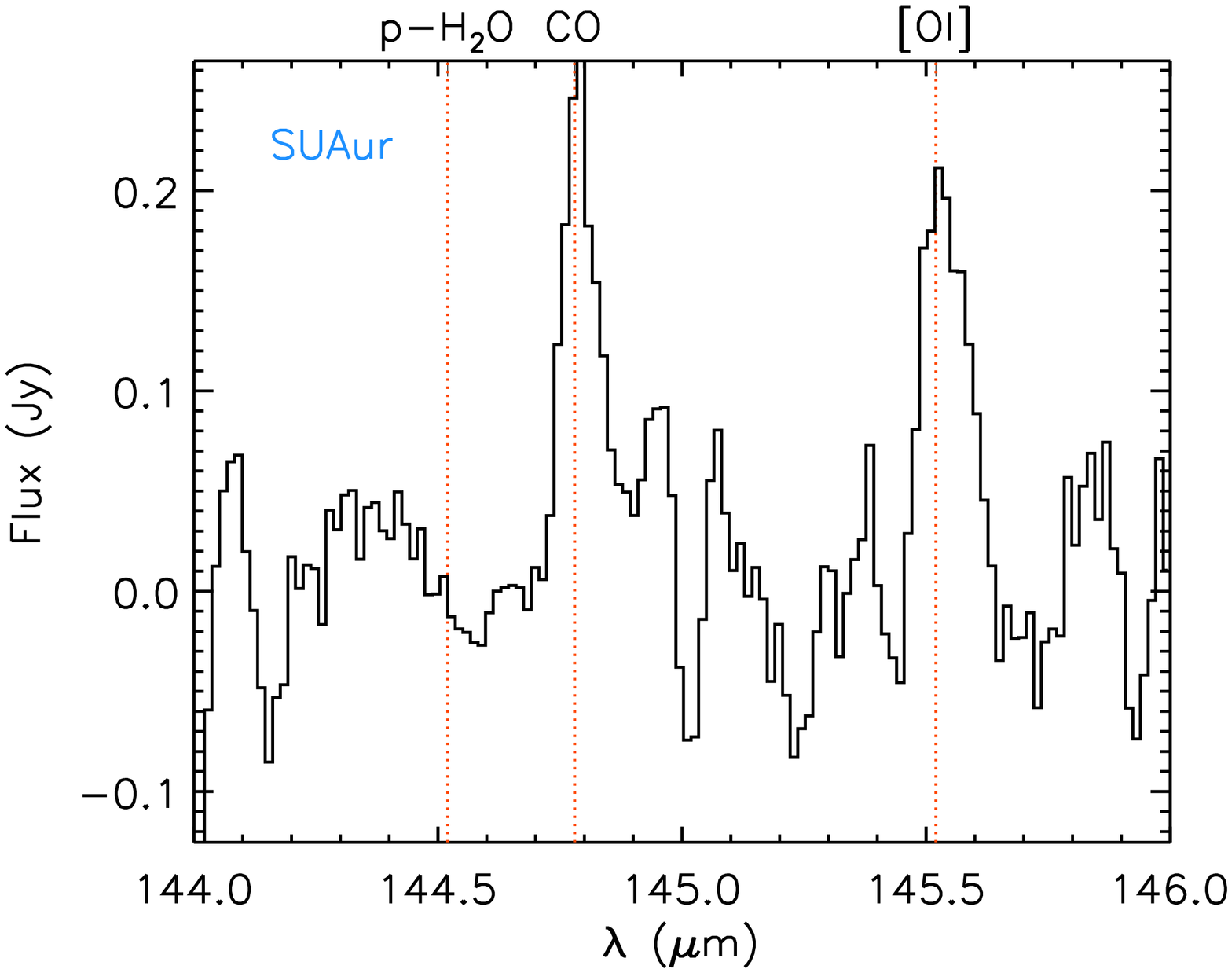}\includegraphics[width=0.16\textwidth, trim= 0mm 0mm 0mm 0mm, angle=90]{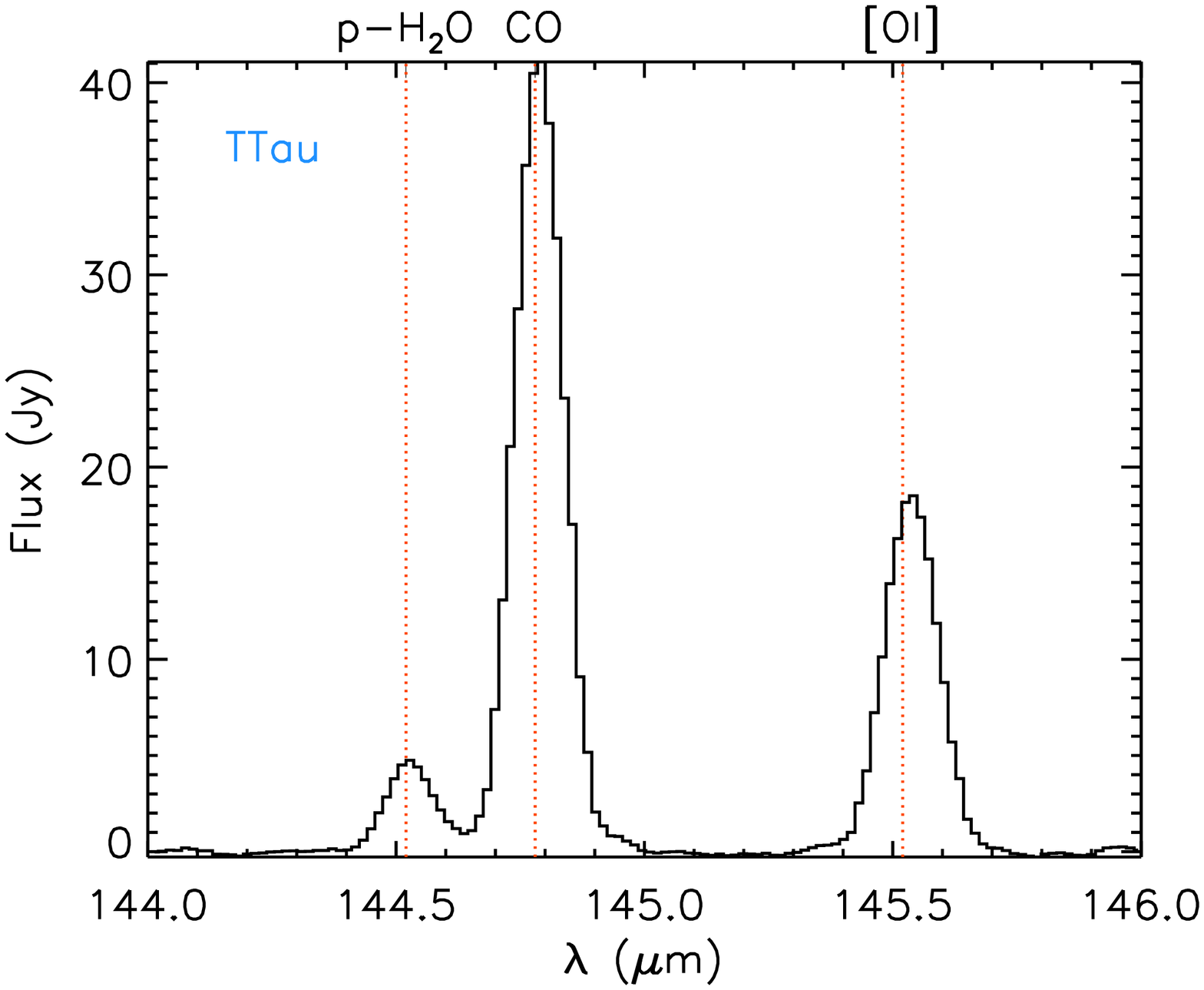}\includegraphics[width=0.16\textwidth, trim= 0mm 0mm 0mm 0mm, angle=90]{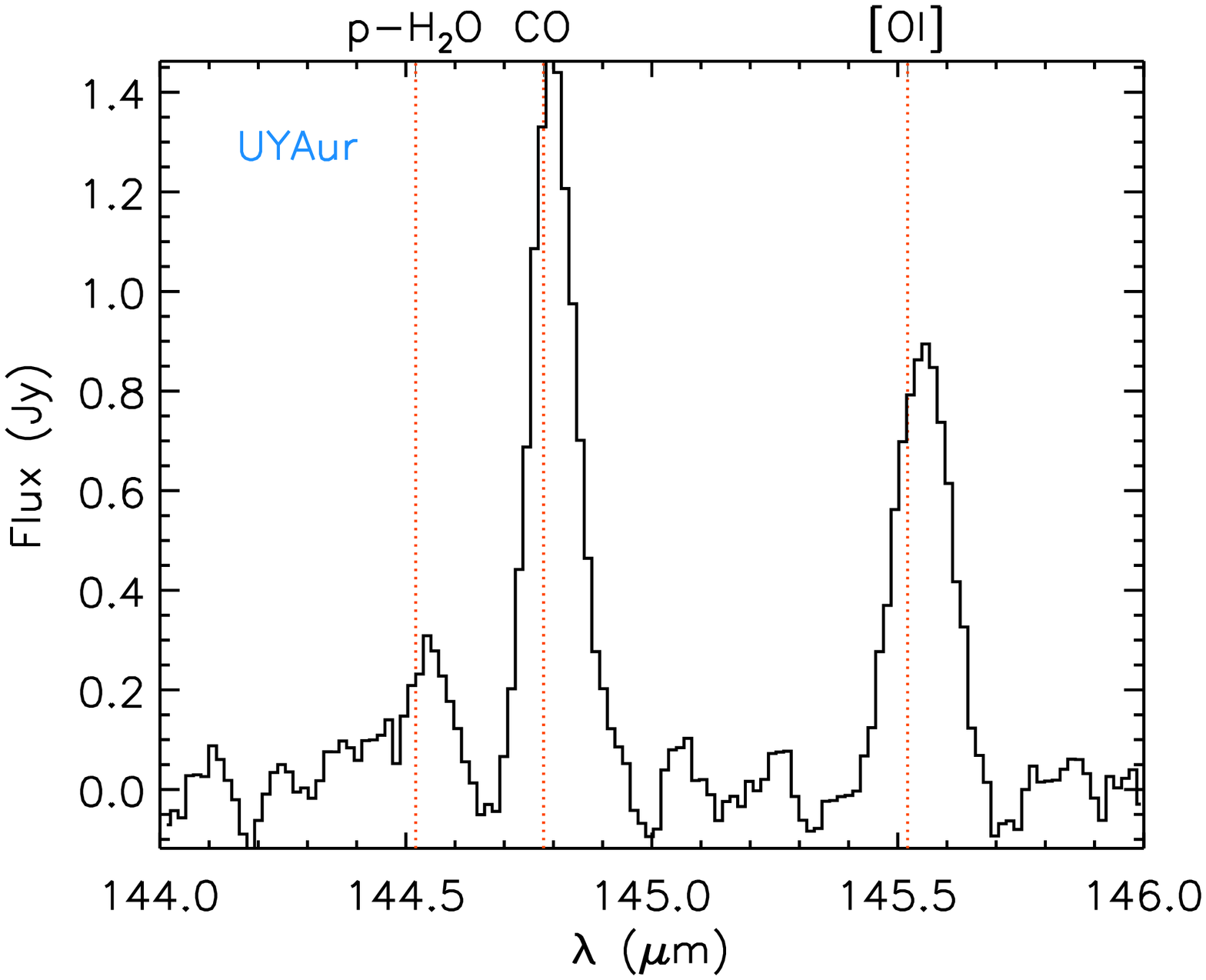}\includegraphics[width=0.16\textwidth, trim= 0mm 0mm 0mm 0mm, angle=90]{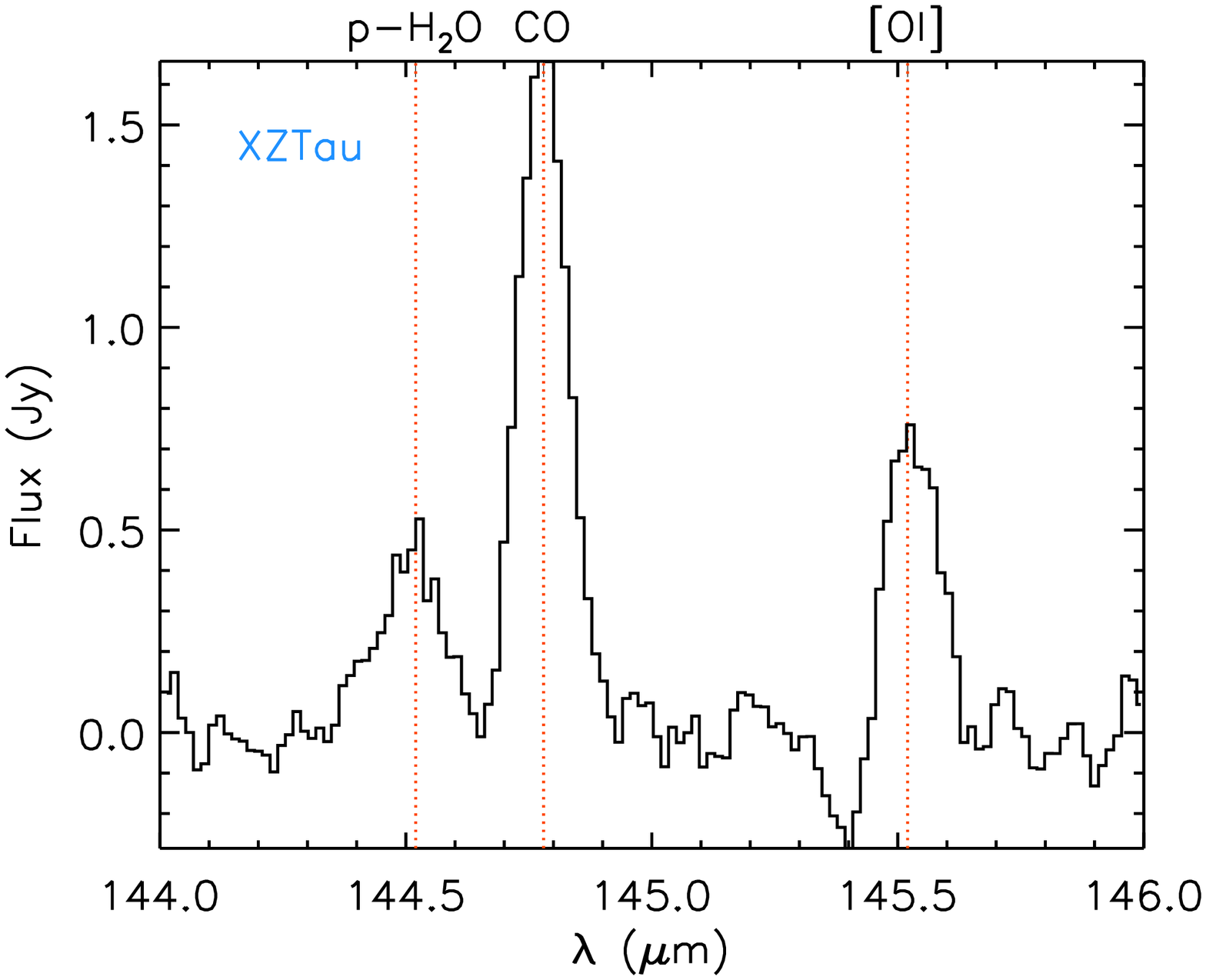}

\includegraphics[width=0.16\textwidth, trim= 0mm 0mm 0mm 0mm, angle=90]{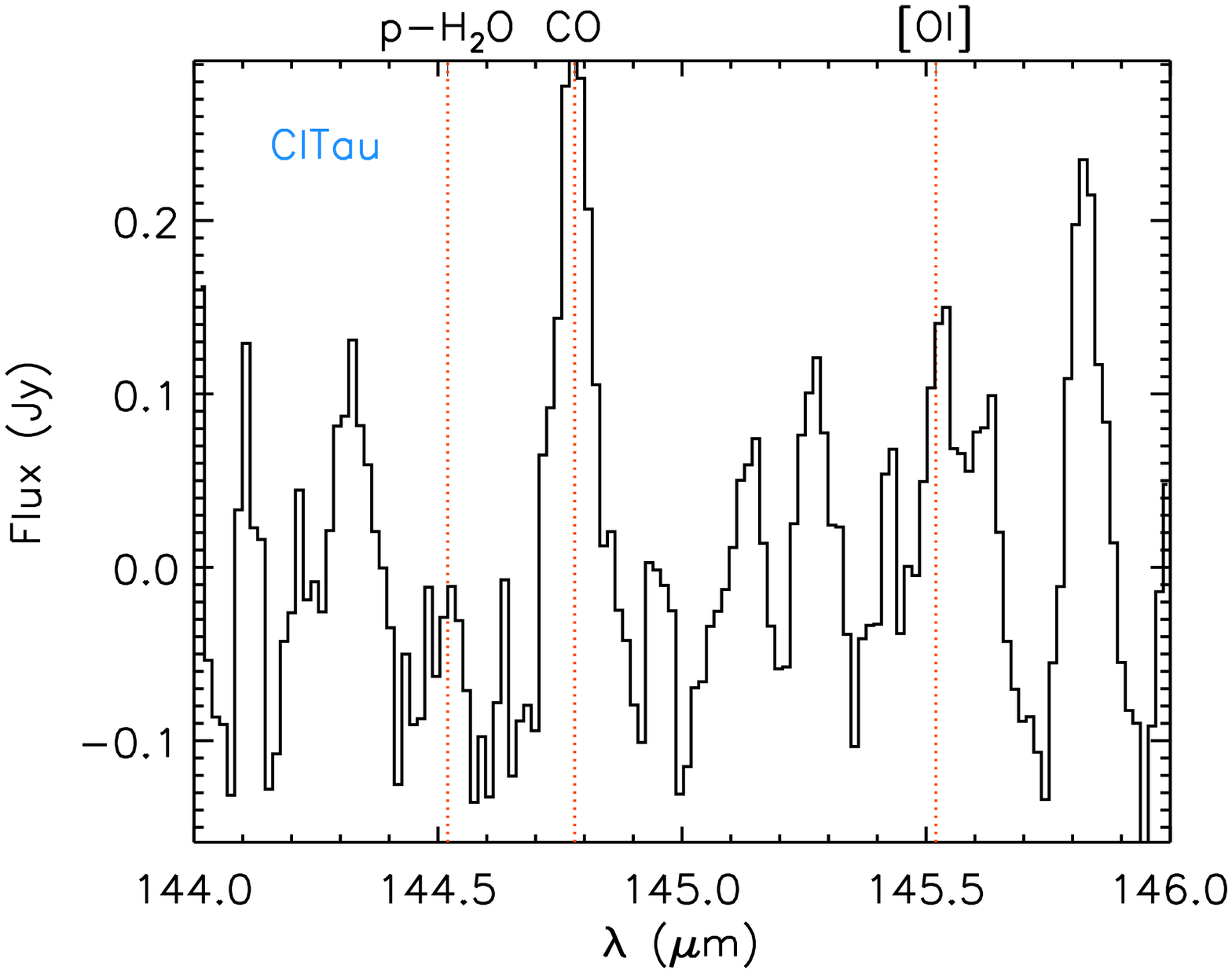}\includegraphics[width=0.16\textwidth, trim= 0mm 0mm 0mm 0mm, angle=90]{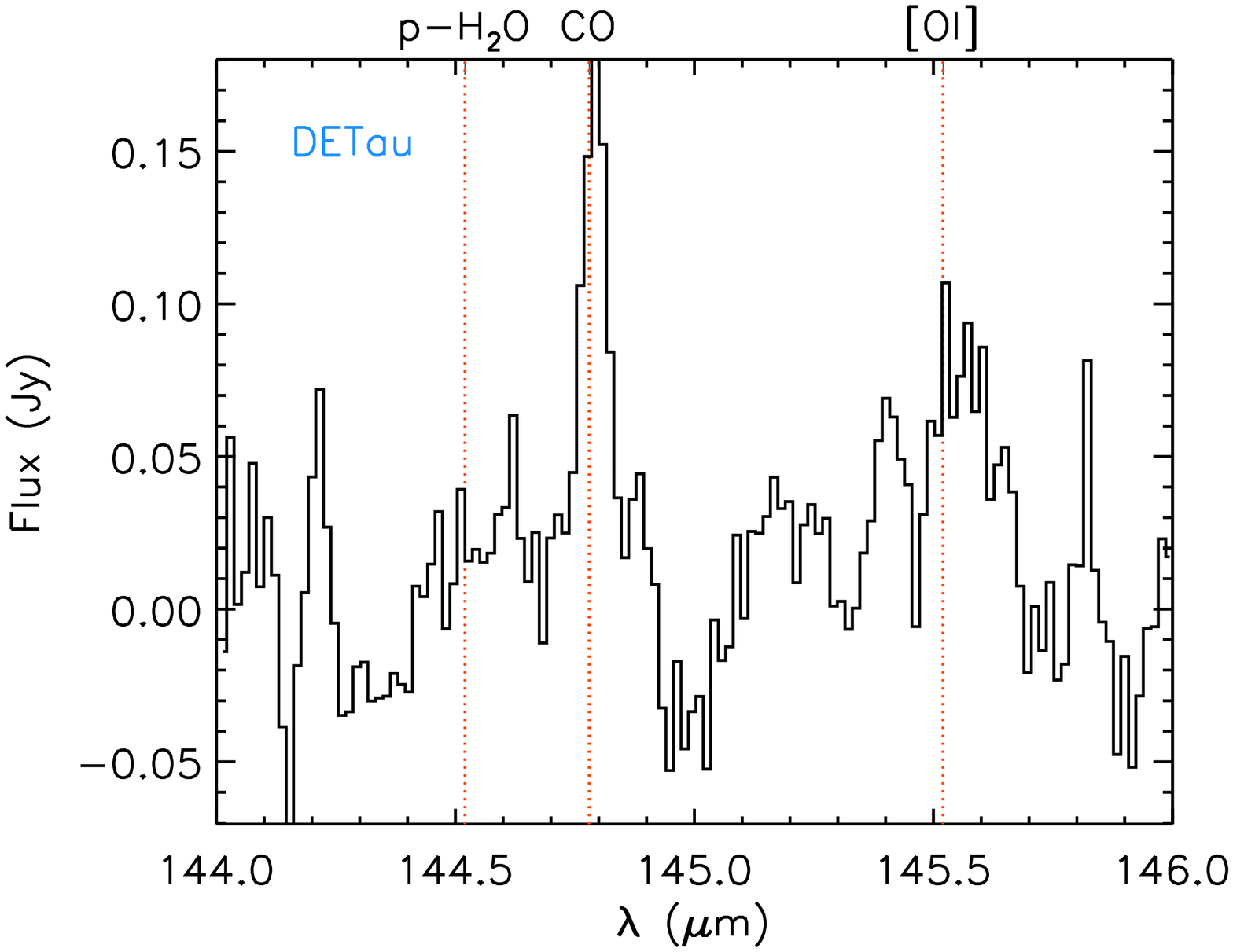}\includegraphics[width=0.16\textwidth, trim= 0mm 0mm 0mm 0mm, angle=90]{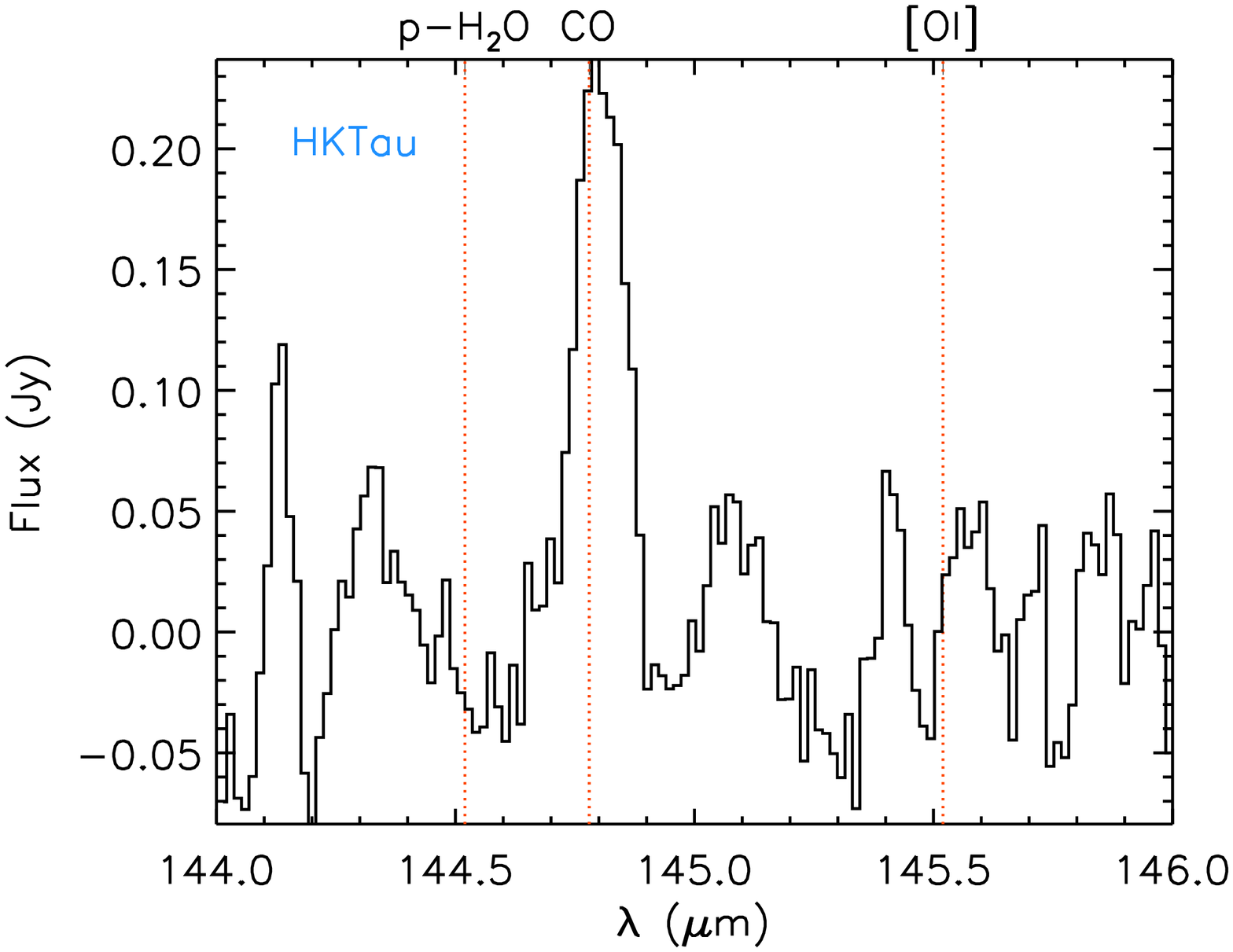}

\caption{Continuum subtracted spectra at 145 $\rm \mu m$ for all objects with detections. The red vertical lines indicate the positions of p-H$_{\rm 2}$O 144.52 $\rm \mu m$, CO 144.78 $\rm \mu m$, and  [OI] 145.52 $\rm \mu m$.}
\end{figure*}

\begin{figure*}[htpb]
\centering
\setcounter{figure}{5}
\includegraphics[width=0.16\textwidth, trim= 0mm 0mm 0mm 0mm, angle=90]{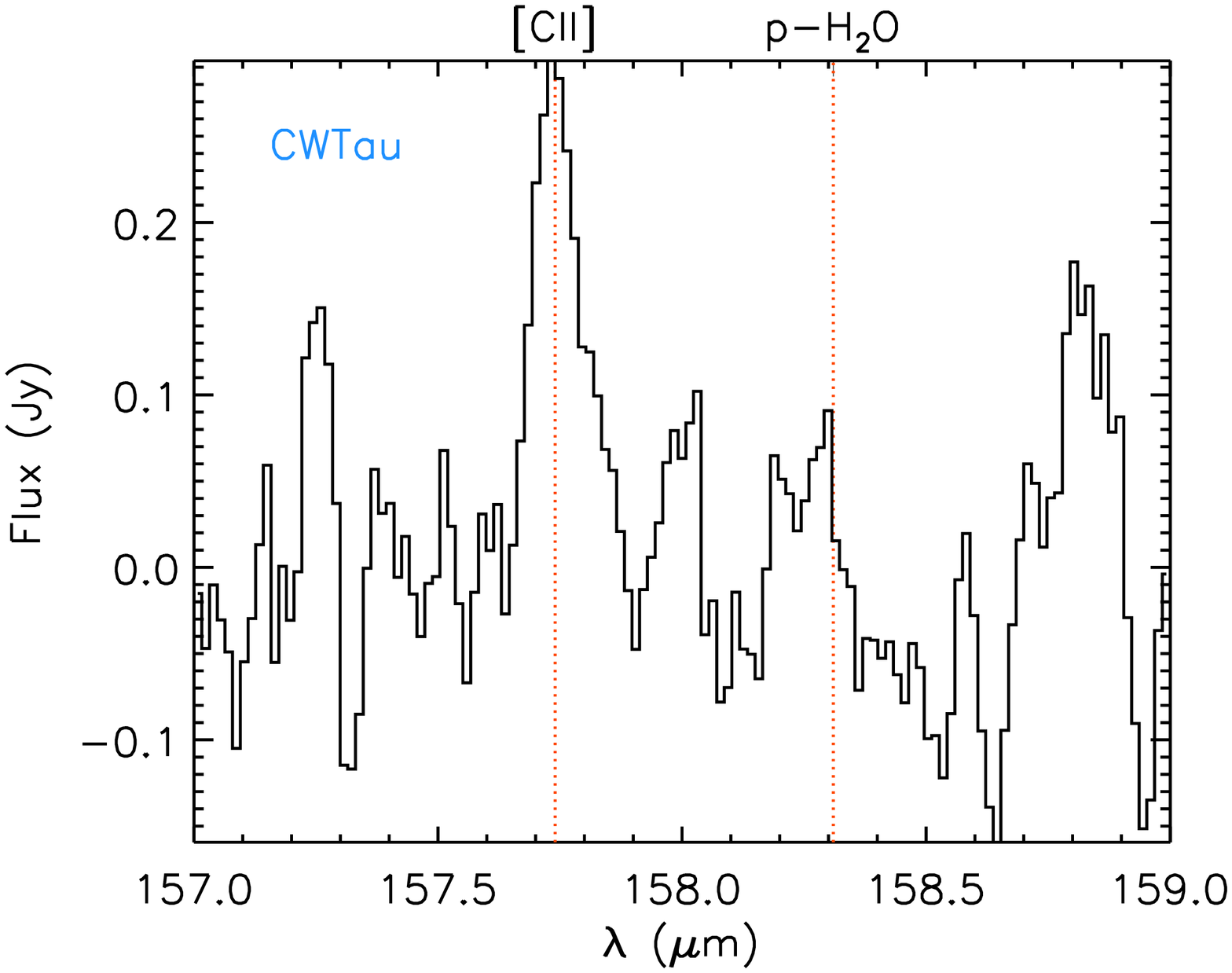}\includegraphics[width=0.16\textwidth, trim= 0mm 0mm 0mm 0mm, angle=90]{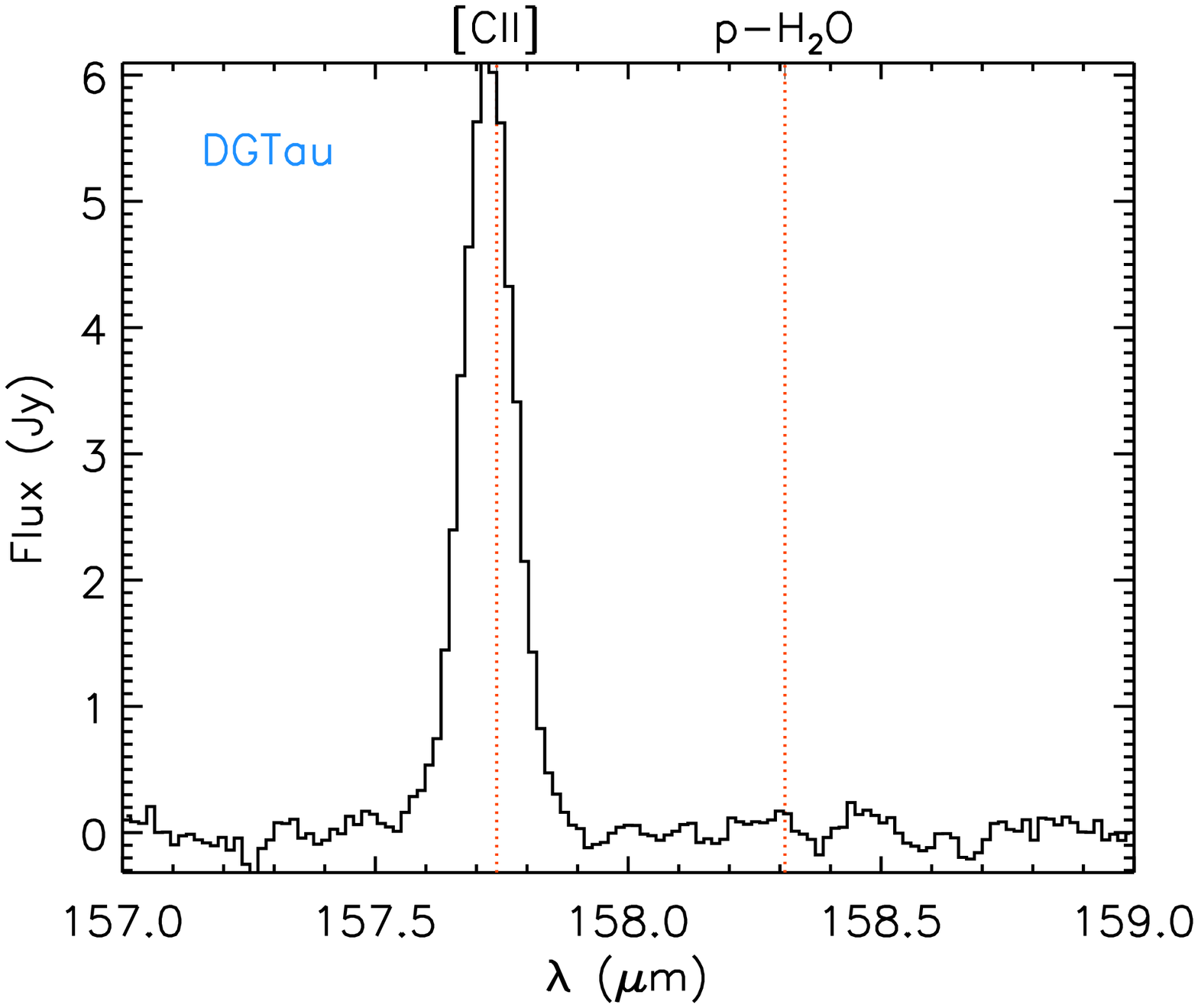}\includegraphics[width=0.16\textwidth, trim= 0mm 0mm 0mm 0mm, angle=90]{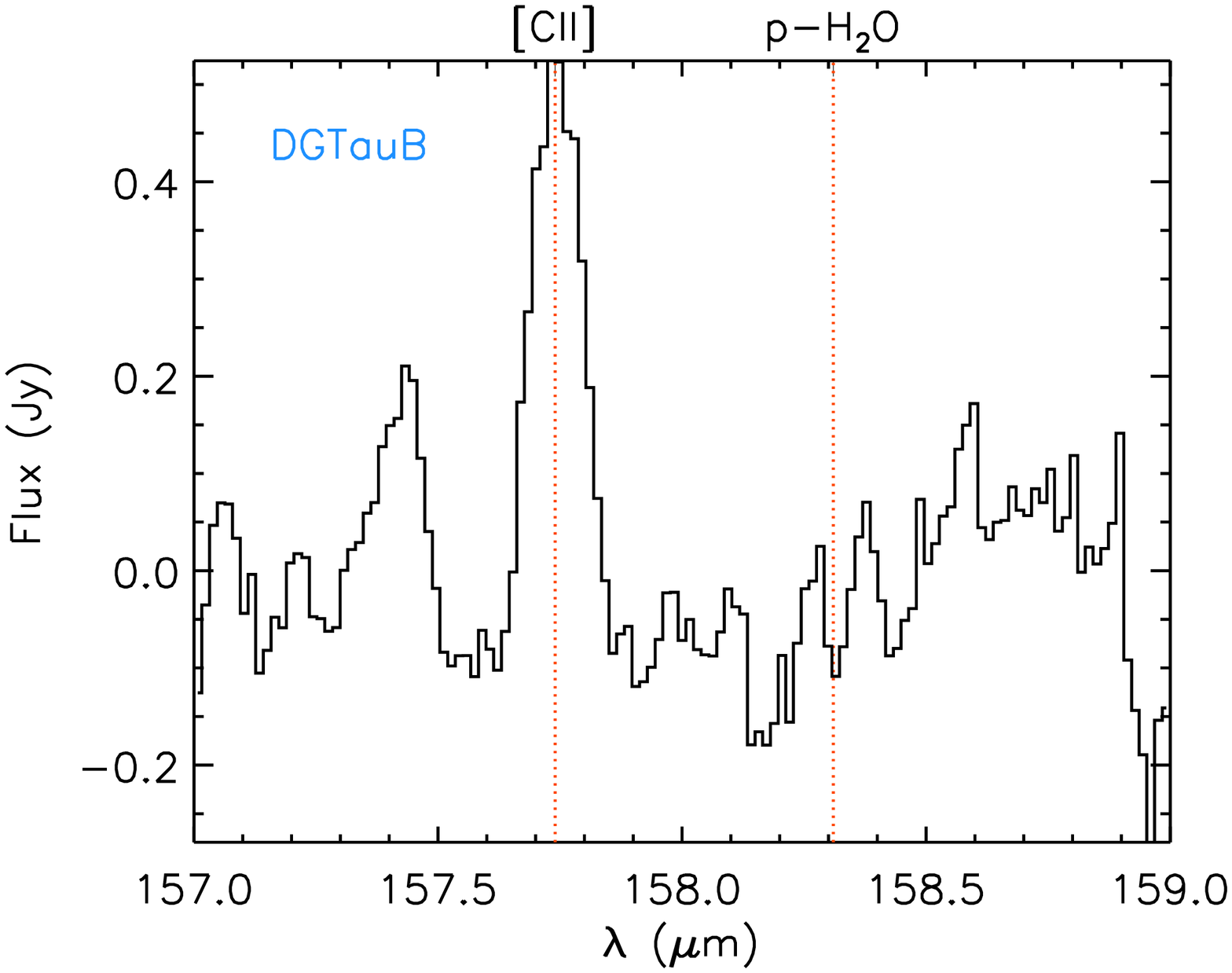}\includegraphics[width=0.16\textwidth, trim= 0mm 0mm 0mm 0mm, angle=90]{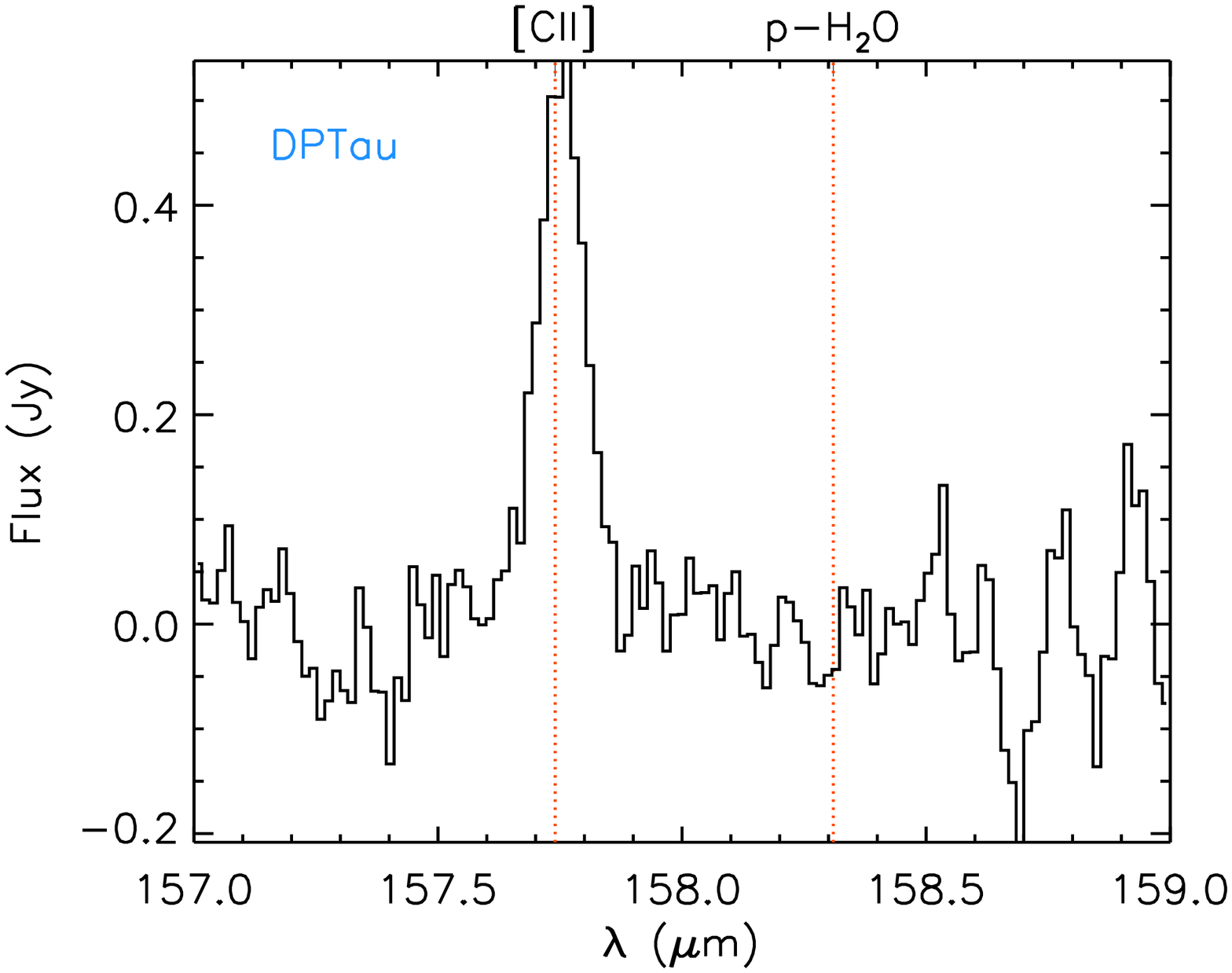}

\includegraphics[width=0.16\textwidth, trim= 0mm 0mm 0mm 0mm, angle=90]{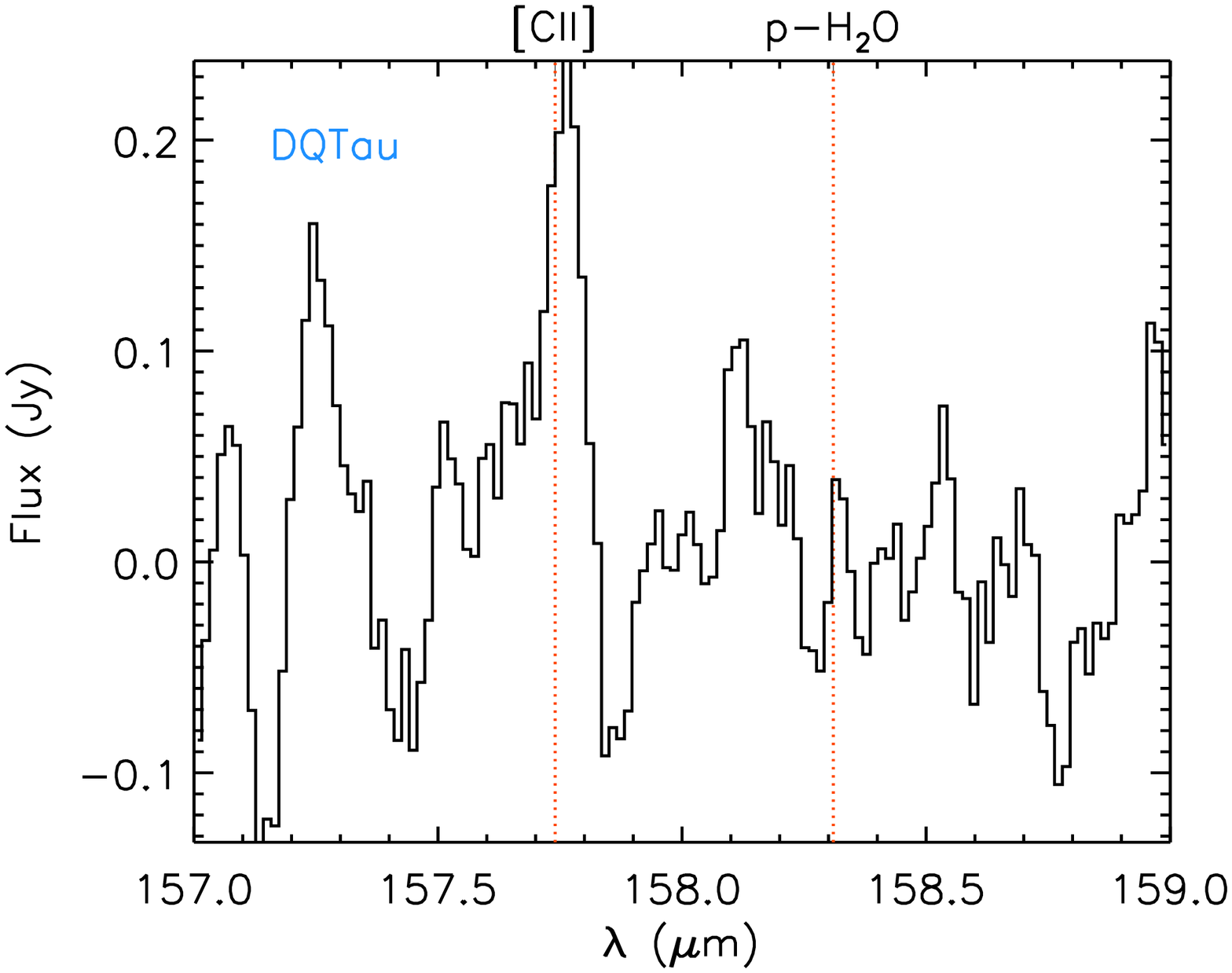}\includegraphics[width=0.16\textwidth, trim= 0mm 0mm 0mm 0mm, angle=90]{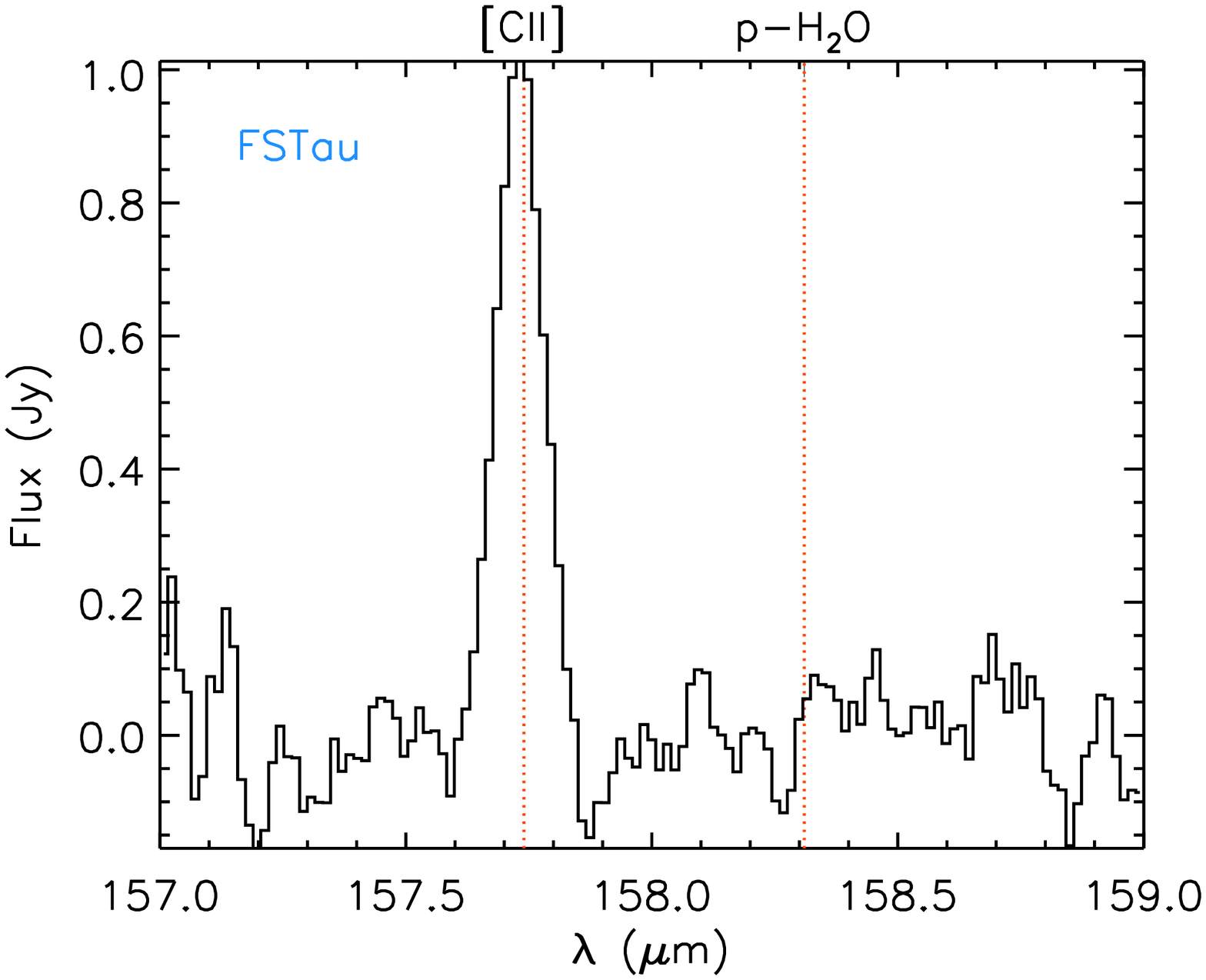}\includegraphics[width=0.16\textwidth, trim= 0mm 0mm 0mm 0mm, angle=90]{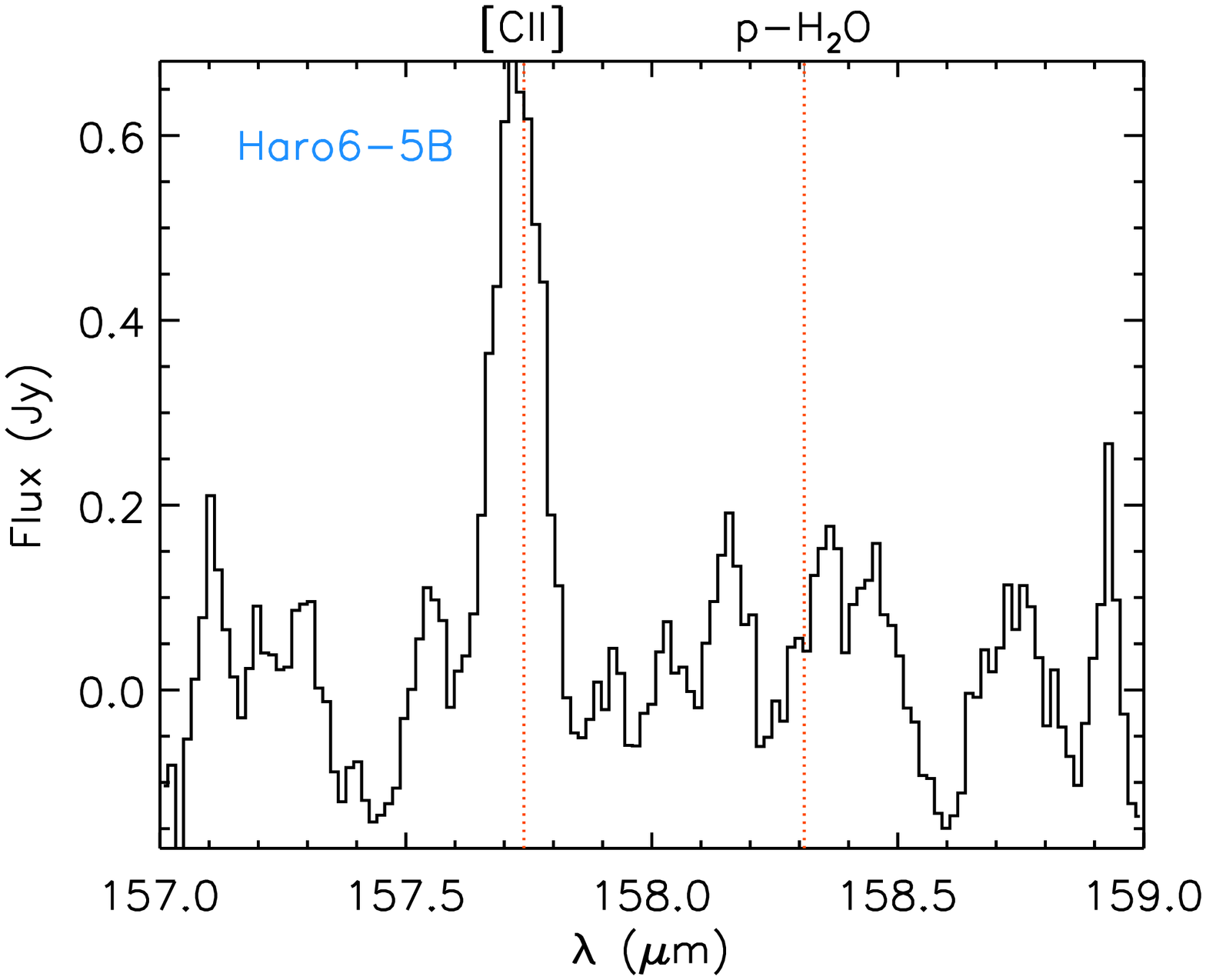}\includegraphics[width=0.16\textwidth, trim= 0mm 0mm 0mm 0mm, angle=90]{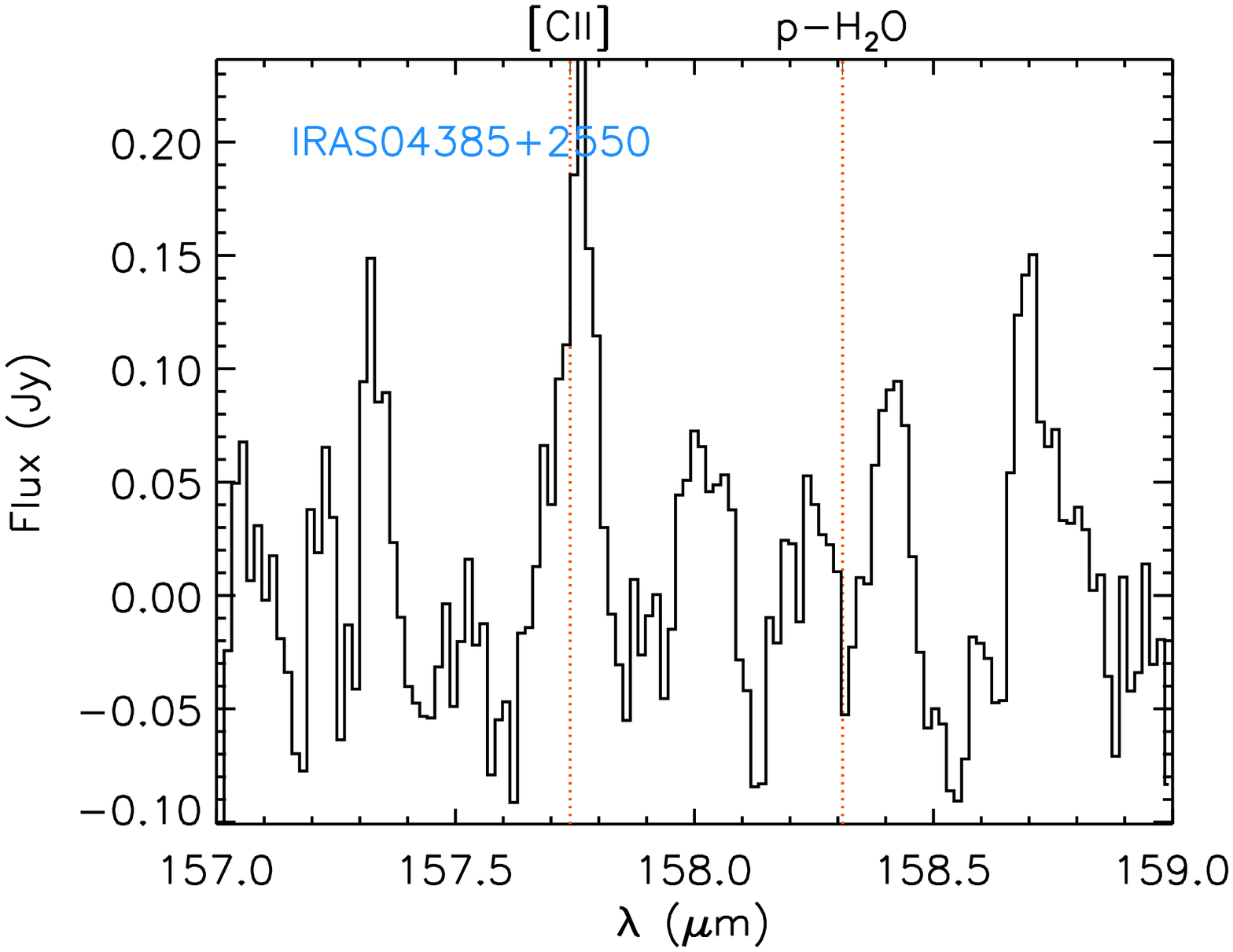}

\includegraphics[width=0.16\textwidth, trim= 0mm 0mm 0mm 0mm, angle=90]{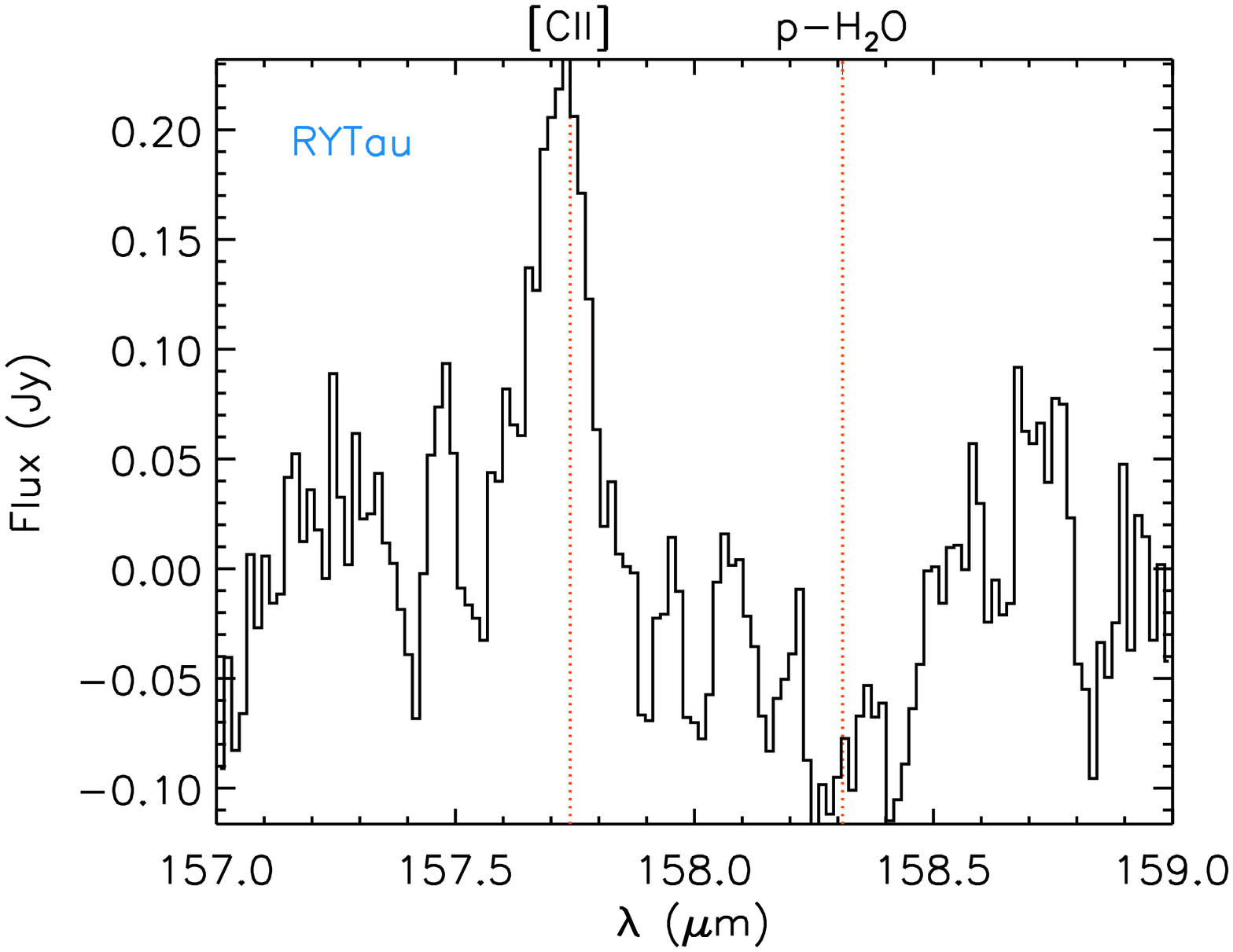}\includegraphics[width=0.16\textwidth, trim= 0mm 0mm 0mm 0mm, angle=90]{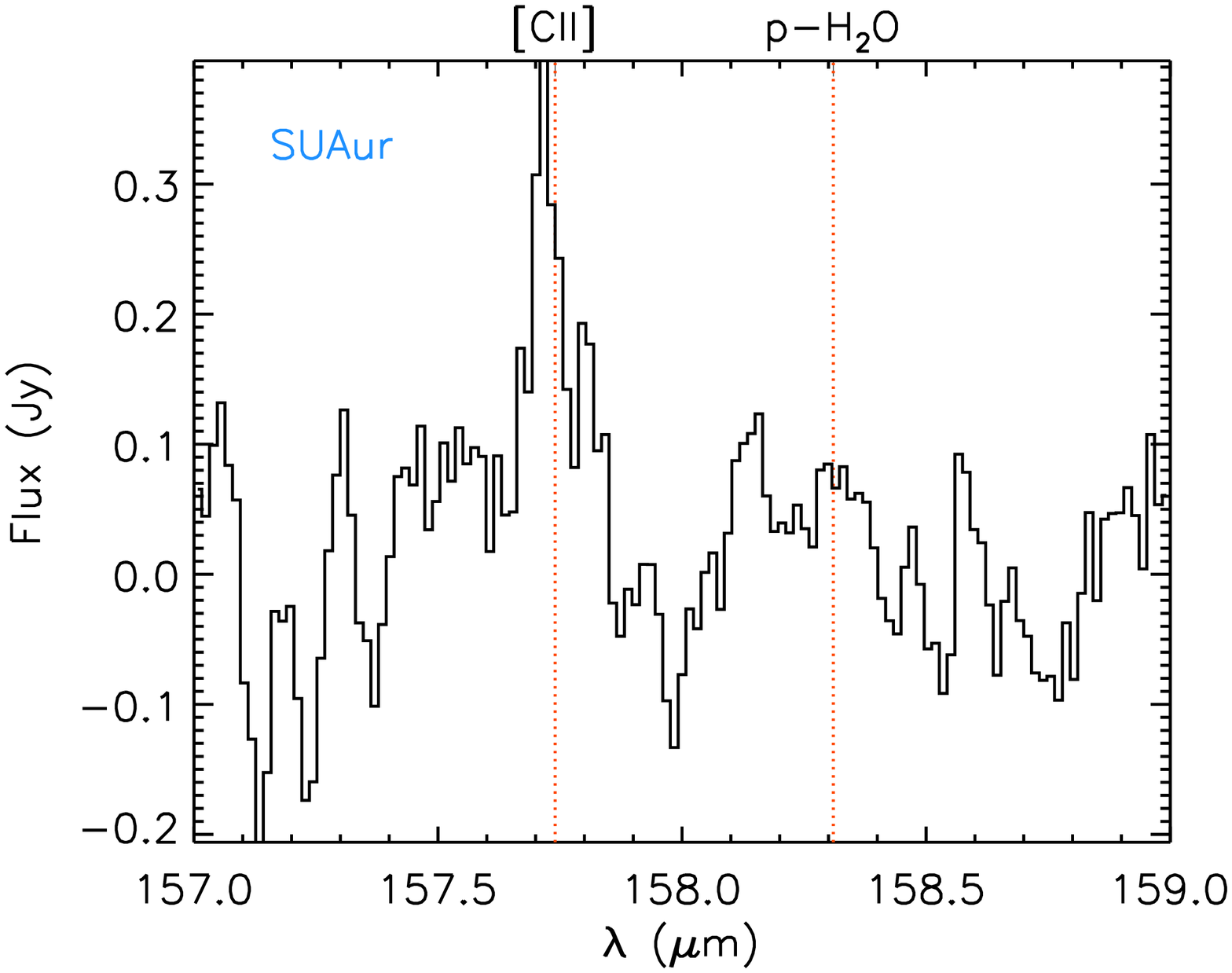}\includegraphics[width=0.16\textwidth, trim= 0mm 0mm 0mm 0mm, angle=90]{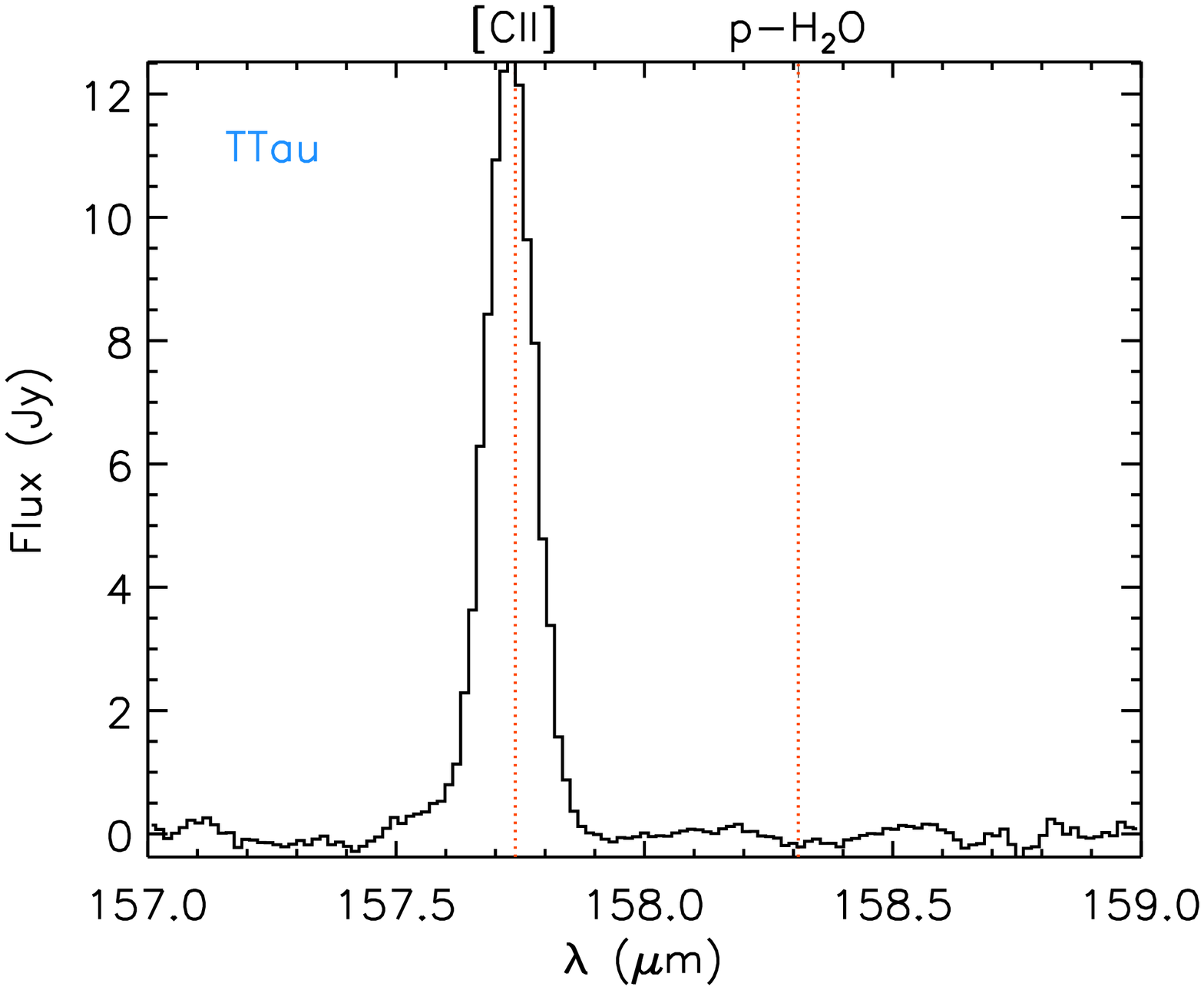}\includegraphics[width=0.16\textwidth, trim= 0mm 0mm 0mm 0mm, angle=90]{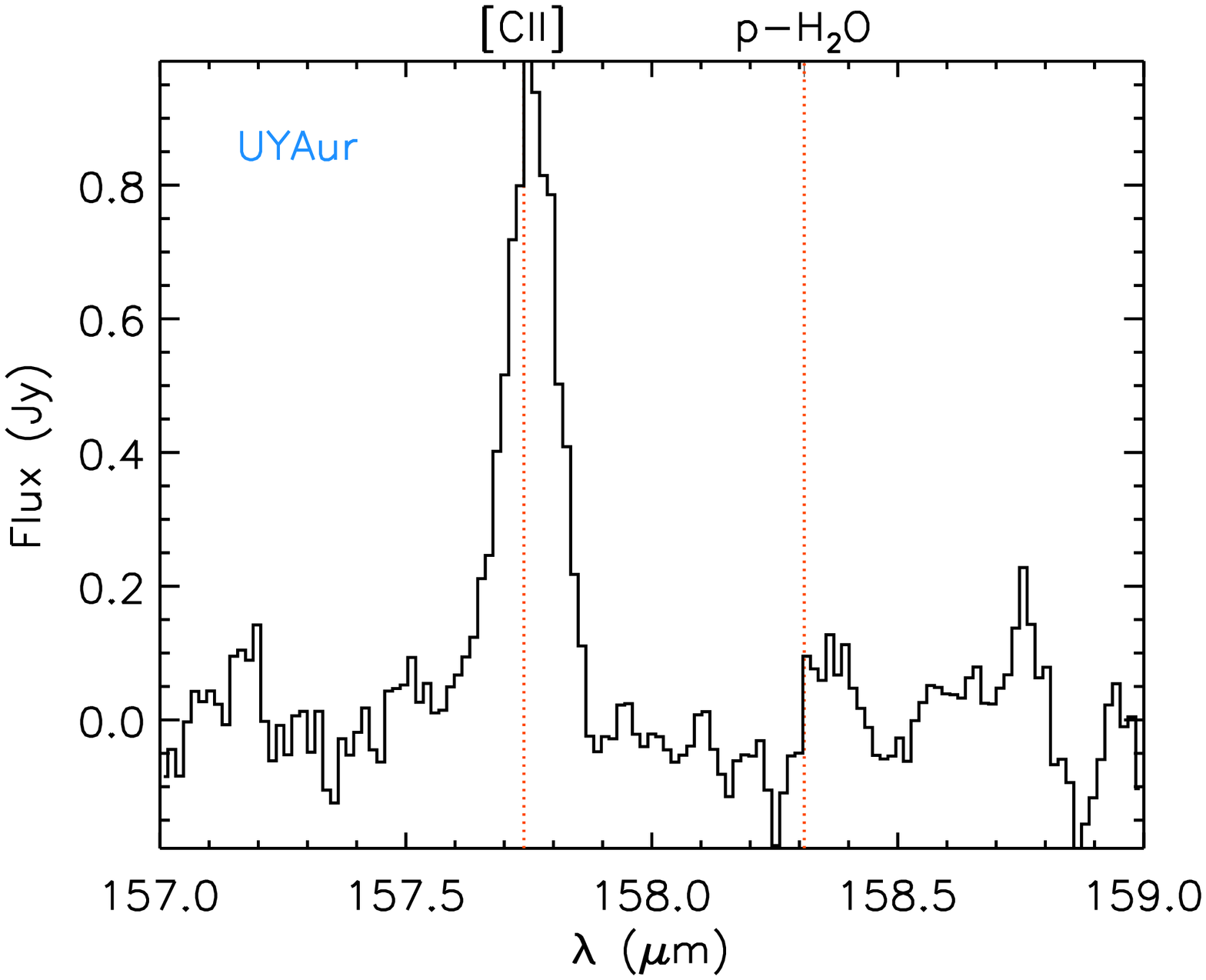}

\includegraphics[width=0.16\textwidth, trim= 0mm 0mm 0mm 0mm, angle=90]{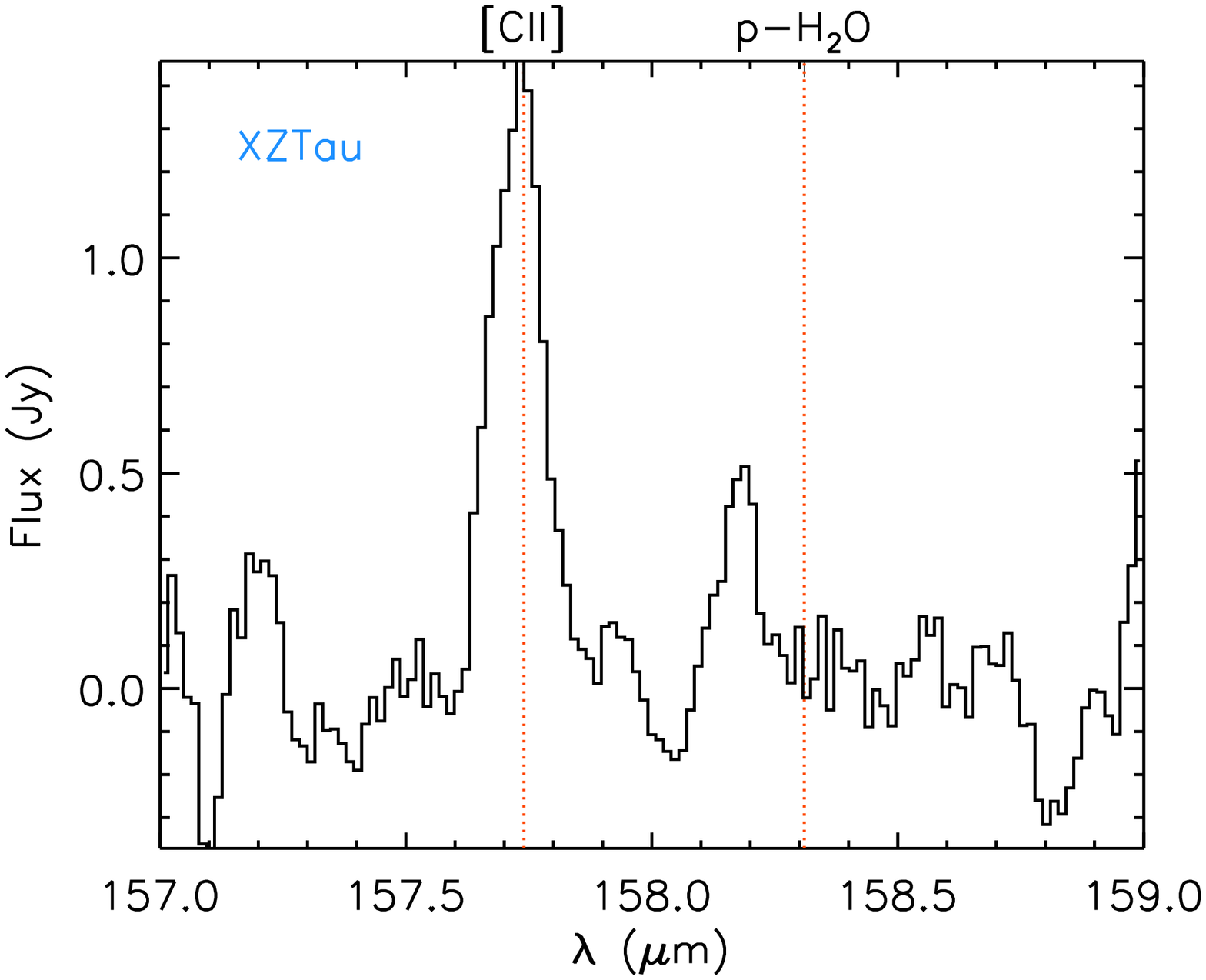}

\caption{Continuum subtracted spectra at 145 $\rm \mu m$ for all objects with detections. The red vertical lines indicate the positions of [CII] 157.74 $\rm \mu m$ and p-H$_{\rm 2}$O 158.31 $\rm \mu m$.}
\end{figure*}

\begin{figure*}[htpb]
\centering
\setcounter{figure}{6}
\includegraphics[width=0.16\textwidth, trim= 0mm 0mm 0mm 0mm, angle=90]{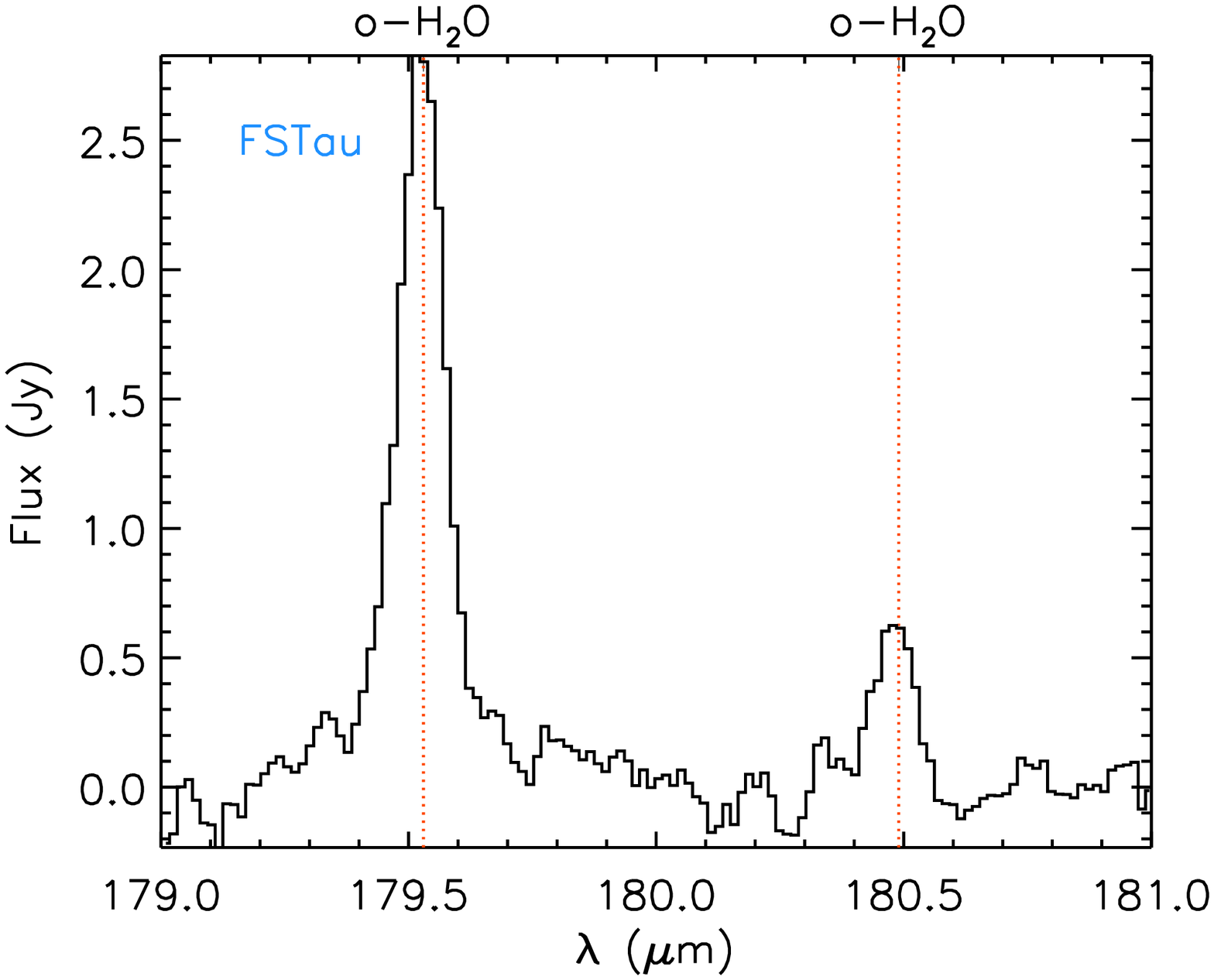}\includegraphics[width=0.16\textwidth, trim= 0mm 0mm 0mm 0mm, angle=90]{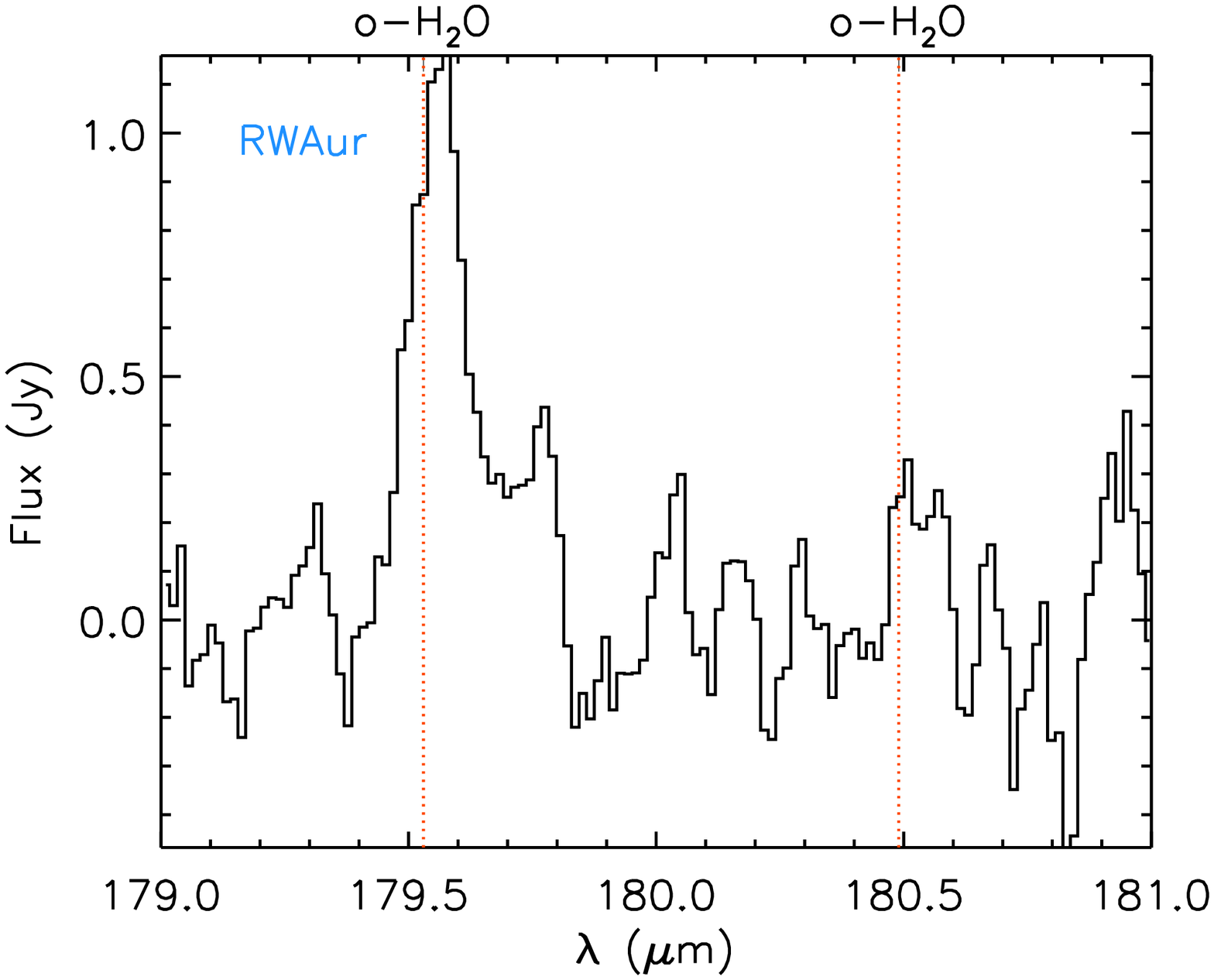}\includegraphics[width=0.16\textwidth, trim= 0mm 0mm 0mm 0mm, angle=90]{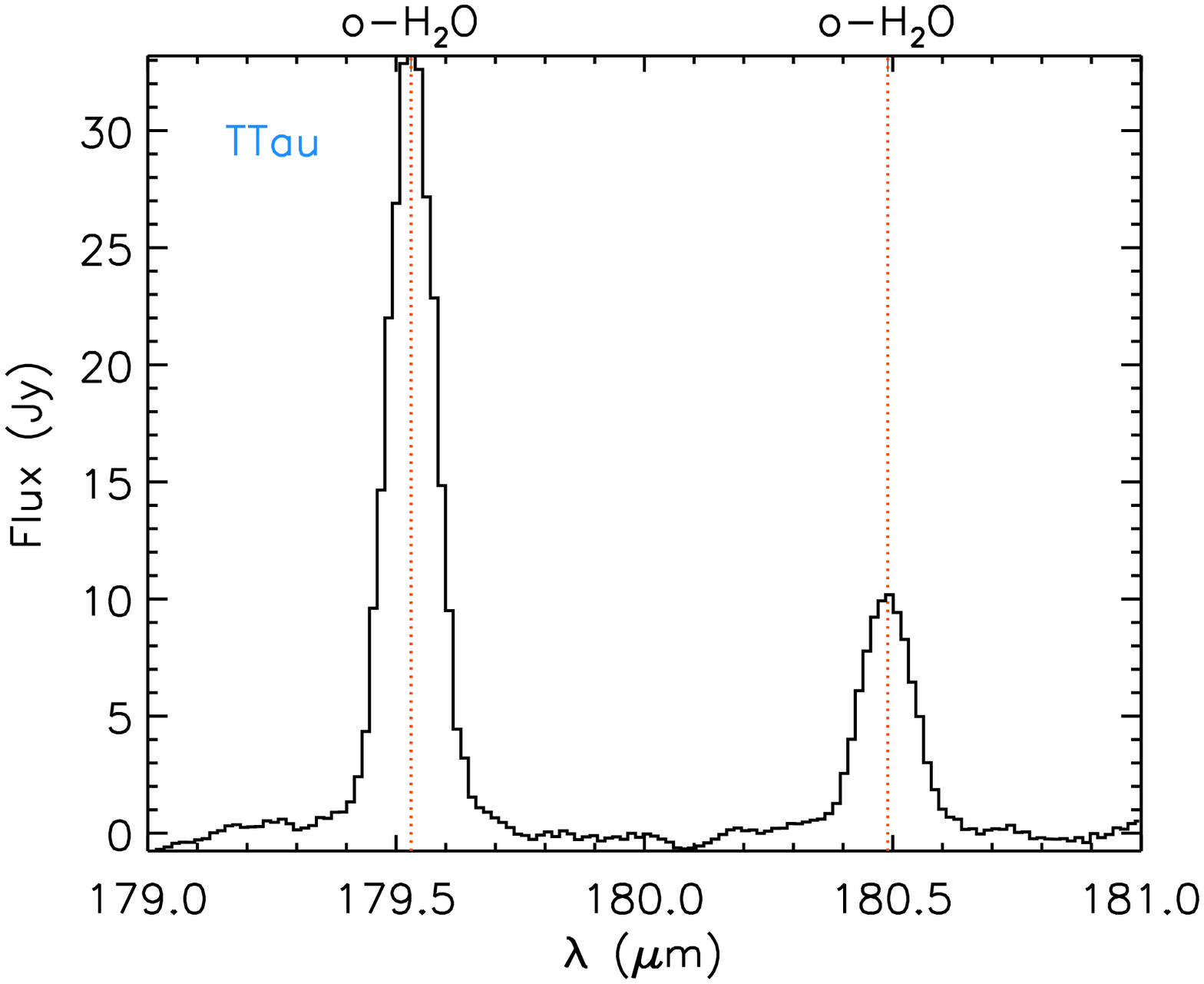}\includegraphics[width=0.16\textwidth, trim= 0mm 0mm 0mm 0mm, angle=90]{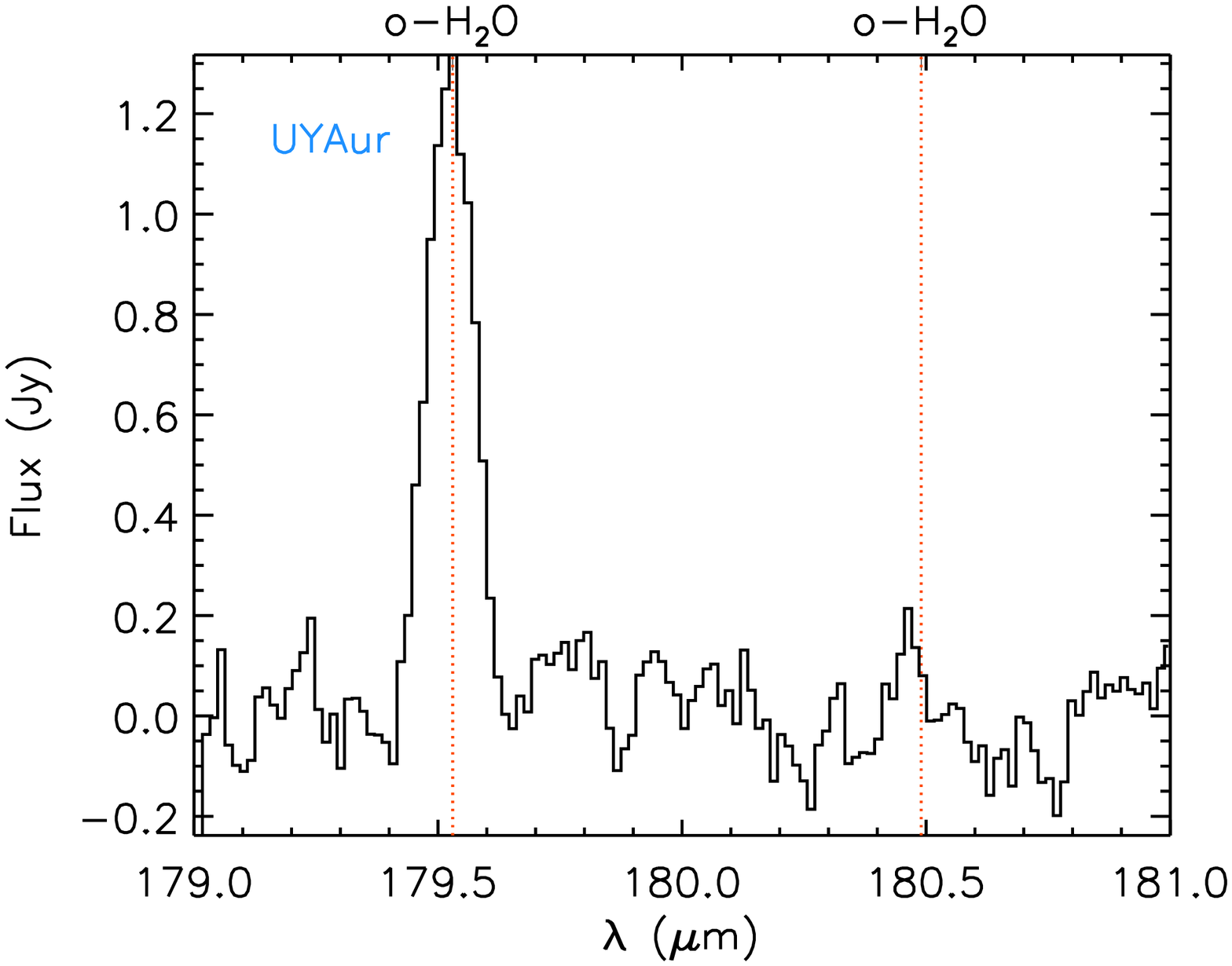}
\includegraphics[width=0.16\textwidth, trim= 0mm 0mm 0mm 0mm, angle=90]{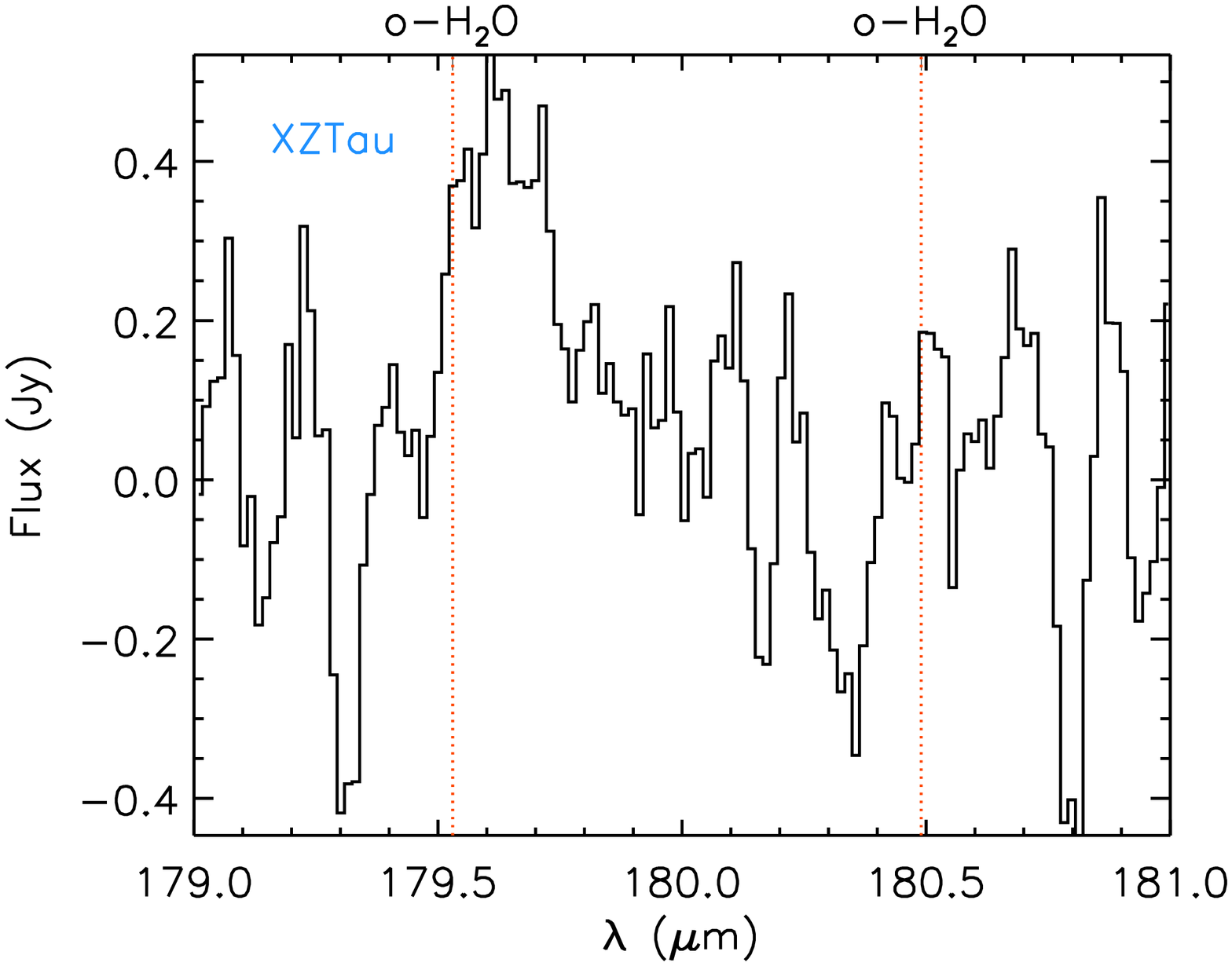}
\caption{Continuum subtracted spectra at 180 $\rm \mu m$ for all objects with detections. The red vertical lines indicate the positions of  o-H$_{\rm 2}$O 179.53  $\rm \mu m$ and o-H$_{\rm 2}$O 180.49 $\rm \mu m$.}
\label{figure:allSpec180}
\end{figure*}

\end{appendix}
\end{document}